\documentclass[12pt]{article}
\usepackage{rotate}
\usepackage{graphics}
\usepackage{array}
\usepackage{longtable}
\usepackage{amsmath}
\usepackage{amssymb}
\newlength{\pubnumber} 

\catcode`\@=11
\@addtoreset{equation}{section}
\def\eqr#1{{Eq.\ (\ref{#1})}}
\def\wh{\widehat}
\def\wt{\widetilde}
\def\pri{^{\prime}}

\def\section{\@startsection{section}{1}{\z@}{3.5ex plus 1ex minus .2ex}
 {2.3ex plus .2ex}{\large\bf}}
\def\subsection{\@startsection{subsection}{2}{\z@}{2.3ex plus .2ex}
 {2.3ex plus .2ex}{\bf}}
\newcommand{\notE}{E\kern-0.6em\hbox{/}\kern0.05em_T}
\newcommand{\lsoft}{\mbox{${\mathcal{L}}_{soft}\, \,$}}

\catcode`@=11
\long\def\@caption#1[#2]#3{\par\addcontentsline{\csname
  ext@#1\endcsname}{#1}{\protect\numberline{\csname
  the#1\endcsname}{\ignorespaces #2}}\begingroup
    \small
    \@parboxrestore
    \@makecaption{\csname fnum@#1\endcsname}{\ignorespaces #3}\par
  \endgroup}
\catcode`@=12

\providecommand{\LyX}{L\kern-.1667em\lower.25em\hbox{Y}\kern-.125emX\@}

\input epsf.tex

\begin{document}
\begin{titlepage}
\samepage{
\setcounter{page}{1}
\rightline{MCTP-03-39}
\rightline{SHEP-03/25}
\rightline{FERMILAB-PUB-03/228-T}
\rightline{CERN-TH/2003-182}
\rightline{\tt hep-ph/0312374}
\rightline{(submitted to {\it Physics Reports}\/)}
\vfill
\begin{center}
 {\Large \bf The Soft Supersymmetry-Breaking Lagrangian: \\
       Theory and Applications\\}
\vfill
 {\large D.~J.~H.~Chung${}^{1,2}$, L.~L.~Everett${}^{2,3}$, 
G.~L.~Kane${}^{4}$, S.~F.~King${}^{5}$, J.~Lykken${}^{6}$, and 
Lian-Tao Wang${}^{1}$\\} 
\vspace{.25in}
 {\it 1. Department of Physics, University of Wisconsin, Madison, WI,
53706, USA\\}
{\it 2. CERN, TH Division, CH-1211 Geneva 23, Switzerland\\}
{\it 3. Department of Physics, University of Florida, Gainesville, FL, 
USA\\}
{\it 4. Michigan Center for Theoretical Physics, Ann Arbor, MI, 48109, 
USA\\}
{\it 5. Department of Physics and
Astronomy, University of Southampton, Southhampton, S017 1BJ, UK\\}
{\it 6. Fermi National Accelerator Laboratory, Batavia, IL, USA\\}
\end{center}
\vfill
\begin{abstract}
{\rm
After an introduction recalling the theoretical motivation for low energy
(100 GeV to TeV scale) supersymmetry, this review describes the theory and
experimental implications of the soft supersymmetry-breaking Lagrangian of
the general minimal supersymmetric standard model (MSSM).  Extensions to
include neutrino masses and nonminimal theories are also discussed.  
Topics covered include models of supersymmetry breaking, phenomenological
constraints from electroweak symmetry breaking, flavor/CP violation, 
collider searches, and cosmological constraints including dark matter and
implications for baryogenesis and inflation.}

\end{abstract}
\smallskip}
\vfill
\end{titlepage}

\setcounter{footnote}{0}


\def\beq{\begin{equation}}
\def\eeq{\end{equation}}
\def\beqn{\begin{eqnarray}}
\def\eeqn{\end{eqnarray}}

\def\bone{{\bf 1}}

\def\Str{{{\rm Str}\,}}
\def\rep#1{{\bf{#1}}}
\def\KM{{Ka\v{c}-Moody}}
\def\bQ{{\bf Q}}
\def\bbox#1{\parbox[t]{0.85 in}{{#1}}}
\def\bbreak{\vfill\break}


\def\inbar{\,\vrule height1.5ex width.4pt depth0pt}

\def\IC{\relax\hbox{$\inbar\kern-.3em{\rm C}$}}
\def\IQ{\relax\hbox{$\inbar\kern-.3em{\rm Q}$}}
\def\IR{\relax{\rm I\kern-.18em R}}
 \font\cmss=cmss10 \font\cmsss=cmss10 at 7pt
\def\IZ{\relax\ifmmode\mathchoice
 {\hbox{\cmss Z\kern-.4em Z}}{\hbox{\cmss Z\kern-.4em Z}}
 {\lower.9pt\hbox{\cmsss Z\kern-.4em Z}}
 {\lower1.2pt\hbox{\cmsss Z\kern-.4em Z}}\else{\cmss Z\kern-.4em Z}\fi}

\def\NPB#1#2#3{{\it Nucl.\ Phys.}\/ {\bf B#1} (19#2) #3}
\def\PLB#1#2#3{{\it Phys.\ Lett.}\/ {\bf B#1} (19#2) #3}
\def\PRD#1#2#3{{\it Phys.\ Rev.}\/ {\bf D#1} (19#2) #3}
\def\PRL#1#2#3{{\it Phys.\ Rev.\ Lett.}\/ {\bf #1} (19#2) #3}
\def\PRT#1#2#3{{\it Phys.\ Rep.}\/ {\bf#1} (19#2) #3}
\def\CMP#1#2#3{{\it Commun.\ Math.\ Phys.}\/ {\bf#1} (19#2) #3}
\def\MODA#1#2#3{{\it Mod.\ Phys.\ Lett.}\/ {\bf A#1} (19#2) #3}
\def\IJMP#1#2#3{{\it Int.\ J.\ Mod.\ Phys.}\/ {\bf A#1} (19#2) #3}
\def\NUVC#1#2#3{{\it Nuovo Cimento}\/ {\bf #1A} (#2) #3}
\def\etal{{\it et al.\/}}

\hyphenation{su-per-sym-met-ric non-su-per-sym-met-ric}
\hyphenation{space-time-super-sym-met-ric}


\vfill\eject
\setcounter{page}{2}
\tableofcontents

\vfill\eject
\setcounter{footnote}{0}
\section{Introduction}
\label{sect1}
The Standard Model of elementary particle physics (SM)  
\cite{Weinberg:tq,Glashow:tr,Salam} is a spectacularly successful theory
of the known particles and their electroweak and strong forces. The SM
is a gauge theory, in which the gauge group $SU(3)_c\times
SU(2)_L\times U(1)_Y$ is spontaneously broken to $SU(3)_c\times
U(1)_{EM}$ by the nonvanishing vacuum expectation value (VEV) of a
fundamental scalar field, the Higgs field, at energies of order 100 
GeV. Although the SM provides a correct description of virtually all known
microphysical nongravitational phenomena, there are a number of
theoretical and phenomenological issues that the SM fails to address
adequately:

\begin{itemize}

\item {\it Hierarchy problem.} Phenomenologically the mass of
the Higgs boson associated with electroweak symmetry breaking must be in 
the electroweak range.  However, radiative corrections to the Higgs
mass are quadratically dependent on the UV cutoff $\Lambda$,
since the masses of fundamental scalar fields are not protected by 
chiral or gauge symmetries.  The ``natural"  value of the Higgs mass is 
therefore of $O(\Lambda)$ rather than $O(100$ GeV), leading to a 
destabilization of the hierarchy of
the mass scales in the SM.\footnote{In other words, to achieve $m\sim
O(100$ GeV) it is necessary to fine-tune the scalar mass-squared parameter
$m^2_0\sim\Lambda^2$ of the fundamental ultraviolet theory to a precision
of $m^2/\Lambda^2$.  If, for example, $\Lambda = 10^{16}$ GeV and $m=100$
GeV, the precision of tuning must be $10^{-28}$.}

\item {\it Electroweak symmetry breaking (EWSB).} In the SM, electroweak
symmetry breaking is parameterized by the Higgs boson $h$ and its
potential $V(h)$. However, the Higgs sector is not constrained by any 
symmetry principles, and it must be put into the theory by
hand.

\item {\it Gauge coupling unification.} The idea that the gauge couplings
undergo renormalization group evolution in such a way that they meet at a
point at a high scale lends credence to the picture of grand unified
theories (GUTs) and certain string theories.  However, precise
measurements of the low energy values of the gauge couplings demonstrated
that the SM cannot describe gauge coupling unification (see {\it e.g.}
\cite{Marciano}) accurately enough to imply it is more than an accident.

\item {\it Family structure and fermion masses.} 
The SM does not explain the existence of three families and can only 
parameterize the strongly hierarchical values of the fermion masses. 
Massive neutrinos imply that the theory has to be extended, as in the 
SM the neutrinos are strictly left-handed and massless.  Right-handed
neutrinos can be added, but achieving ultralight neutrino masses from the
seesaw mechanism \cite{seesaw1,seesaw2} requires the introduction of a new
scale much larger than $O(100$ GeV).

\item {\it Cosmological challenges.} 
 Several difficulties are encountered when trying to build cosmological
models based solely on the SM particle content.  The SM cannot explain the
baryon asymmetry of the universe; although the Sakharov criteria
\cite{Sakharov:dj} for baryogenesis 
can be met, the baryon asymmetry generated at the electroweak phase 
transition is too small. 
The SM also does not have a viable candidate for the cold
dark matter of the universe, nor a viable inflaton.  The most difficult
problem the SM has when trying to connect with the gravitational sector
is the absence of the expected scale of the cosmological constant.

\end{itemize}
Therefore, the Standard Model must be extended and its
foundations strengthened.  Theories with {\bf low energy supersymmetry} 
have emerged as the strongest candidates for physics beyond the SM. There 
are strong reasons to expect that low energy supersymmetry is the probable 
outcome of experimental and theoretical progress and that it will soon be 
directly confirmed by experiment.  In the simplest supersymmetric world, 
each particle has a {\it superpartner} which differs in spin by $1/2$ and 
is related to the original particle by a supersymmetry transformation.  
Since supersymmetry relates the scalar and fermionic sectors, the chiral 
symmetries which protect the masses of the fermions also protect the 
masses of the scalars from quadratic divergences, leading to an elegant 
resolution of the hierarchy problem.

Supersymmetry must be a broken symmetry, because exact supersymmetry
dictates that every superpartner is degenerate in mass with its
corresponding SM particle, a possibility which is decisively ruled out by
experiment. Possible ways to achieve a spontaneous breaking of 
supersymmetry breaking depend on the form of the high energy theory. In 
many ways, it is not surprising that supersymmetry breaking is not yet 
understood --- the symmetry breaking was the last thing understood for the 
Standard Model too (assuming it is indeed understood). Supersymmetry may 
even be explicitly broken without losing some of its attractive features 
if the breaking is of a certain type known as {\it soft breaking}. If 
supersymmetry is broken in this way, the superpartner masses can be lifted 
to a phenomenologically
acceptable range.  Furthermore, the scale of the mass splitting should be
of order the $Z$ mass to TeV range because it can be tied to the scale of
electroweak symmetry breaking.

Whether supersymmetry is explicitly or spontaneously broken, the effective
Lagrangian at the electroweak scale is expected to be parameterized by a
general set of soft supersymmetry-breaking terms if the attractive
features of supersymmetry are to be a part of the physics beyond the SM.  
The subject of this review is the phenomenological implications of this
assumption and the resulting constraints on the parameters of the soft
supersymmetry-breaking Lagrangian \lsoft from both particle physics and
cosmology.

For our purposes, the phrase low energy supersymmetry will always mean
softly broken $N=1$ supersymmetry with an effective soft supersymmetry-breaking
Lagrangian containing mass parameters that are typically of order the
electroweak to TeV scale but otherwise not {\it a priori} special nor
constrained.  The minimal extension of the SM with low energy
supersymmetry, known as the minimal supersymmetric standard model (MSSM),
is the primary concern of this review.  Generic predictions of the MSSM
include a plethora of new particles, the superpartners of the SM fields,
which have masses in the electroweak to TeV range, set by the scale of 
the \lsoft parameters. If low energy supersymmetry is indeed the 
resolution of the hierarchy problem chosen by nature, direct 
evidence of the existence of the superpartners should be discovered within 
the next decade, either at current experiments at the upgraded 
$p\overline{p}$ Fermilab Tevatron collider or at the forthcoming Large 
Hadron Collider (LHC) at CERN.

Sometimes people suggest that supersymmetry advocates have been overly
optimistic in arguing for the observability of superpartners.  In that
connection it's perhaps amusing to quote from a review published in 1985
\cite{Haber:1984rc}: ``We only want to conclude that (1) the physics of
supersymmetry is nice enough so that experimenters should take it very
seriously and really search for evidence of supersymmetry, (2) theorists
should take supersymmetry seriously enough to help think of better ways to
search, and (3) fortunately, if nature is not supersymmetric on the weak
scale, it will be possible to know this definitively with the accelerators
and detectors that should be available within about the next decade and
the kinds of analysis we have discussed."  At that time, of course, the
SSC development was underway.

Low energy supersymmetry has long been considered the best-motivated
possibility for new physics at the TeV scale.  The main reasons that low
energy supersymmetry is taken very seriously are not its elegance or its
likely theoretical motivations, but its successful explanations and
predictions.  Of course, these successes may just be remarkable
coincidences because there is as yet no direct experimental evidence for
supersymmetry. Either superpartners and a light Higgs boson must be
discovered or demonstrated not to exist at collider energies, in which
case low energy supersymmetry does not describe nature.  The main
successes are as follows:
\begin{itemize}

\item {\it Hierarchy problem.} 
The SM Higgs sector has two ``naturalness'' problems.  One is the 
technical naturalness problem associated with the absence of a symmetry 
protecting the Higgs mass at the electroweak scale when the natural 
cutoff scale is at or above the GUT scale.\footnote{In 
other words, the radiative corrections naturally give the Higgs a mass of 
order the GUT scale or a similarly large cutoff scale; unlike the 
fermions, there is no chiral symmetry protecting the scalar sector.} The 
second problem is associated with explaining the origin of the electroweak 
scale, when a more ``fundamental'' embedding theory such as a GUT or 
string theory typically is defined at a scale which is at least 
$10^{13}$ times larger than the electroweak scale.  This is typically 
referred to as the gauge hierarchy problem. The unavoidable
nature of the hierarchy problem is explained in detail in Martin's
pedagogical review \cite{Martin:1997ns}.

Supersymmetry provides a solution to the technical hierarchy problem
\cite{Maiani:cx}, as the Higgs mass parameter is not renormalized as long
as supersymmetry is unbroken.  Supersymmetry also mitigates the gauge
hierarchy problem by breaking the electroweak symmetry radiatively through
logarithmic running, which explains the large number $\sim 10^{13}$.
 
\item {\it Radiative electroweak symmetry breaking.} 
With plausible boundary conditions at a high 
scale (certain couplings such as the top quark Yukawa of 
$O(1)$ and no bare Higgs mass parameter $\mu$ in the superpotential), low 
energy supersymmetry can provide 
the explanation of the origin of electroweak symmetry breaking  
\cite{Ibanez:fr,Alvarez-Gaume:1981wy,Inoue:1982ej,Inoue:1982pi,Inoue:1983pp}.  
To oversimplify a little (this will be expanded in 
Section~\ref{EWSBsect}), the SM effective Higgs potential has 
the form $V=m^2h^2+\lambda h^4$.
First, supersymmetry requires that the quartic coupling 
$\lambda$ is a function of the $U(1)_Y$ and 
$SU(2)$ gauge couplings 
$\lambda=({g'}^2+g^2)/2$.  Second, the $m^2$ parameter runs to negative 
values at the electroweak scale,
driven by the large top quark Yukawa coupling. Thus the ``Mexican
hat'' potential with a minimum away from $h =0$ is derived rather than
assumed.  As is typical for progress in physics, this explanation is not
from first principles, but it is an explanation in terms of the next
level of the effective theory which depends on the crucial 
assumption that the \lsoft mass parameters 
have values of order the electroweak scale.  Once superpartners are 
discovered, 
the question of supersymmetry breaking must be answered in any event and 
it is a genuine success of the theory that whatever explains 
supersymmetry breaking is also capable of resolving 
the crucial issue of $SU(2)\times U(1)$ breaking.

\item {\it Gauge coupling unification.} 
In contrast to the SM, the MSSM allows for the unification of
the gauge couplings, as first pointed out in the context of GUT models by 
\cite{Dimopoulos:1981yj,Dimopoulos:1981zb,Sakai:1981gr}. The extrapolation
of the low energy values of the gauge couplings using renormalization
group equations and the MSSM particle content shows that the gauge
couplings unify at the scale $M_G \simeq 3 \times 10^{16}$ GeV 
\cite{Giunti:ta,Amaldi:1991cn,Langacker:1991an,Ellis:1990wk}.  Gauge 
coupling unification and electroweak symmetry breaking depend on
essentially the same physics since each needs the soft masses and $\mu$ to 
be of order the electroweak scale.

\item {\it Cold dark matter.} In supersymmetic theories, the lightest
superpartner (LSP) can be stable.  This stable superpartner provides a 
nice cold dark matter candidate \cite{Pagels:ke,Goldberg:1983nd}.  Simple 
estimates of its relic density are of the right order of magnitude to 
provide the observed amount. LSPs were noticed as good candidates before 
the need for nonbaryonic cold dark matter was established.


\end{itemize}
\noindent Supersymmetry has also made several correct predictions:

\begin{enumerate}
\item  Supersymmetry predicted in the early 1980s that the 
top quark would be heavy \cite{Ibanez:wd,Pendleton:as}, because this was a   
necessary condition for the validity of the electroweak 
symmetry breaking explanation.

\item Supersymmetric grand unified theories with a high fundamental
scale accurately predicted the present experimental value of $\sin
^{2}\theta _{W}$ before it was measured
\cite{Dimopoulos:1981zb,Dimopoulos:1981yj,Ibanez:yh,Einhorn:1981sx}.

\item Supersymmetry requires a light Higgs boson to exist
\cite{Kane:1992kq,Espinosa:1992hp}, consistent with current precision
measurements, which suggest $M_{h} < 200$ GeV \cite{lepewwg}.

\item When LEP began to run in 1989 it was recognized that either LEP
would discover superpartners if they were very light or, because all
supersymmetry effects at LEP are loop effects and supersymmetry effects
decouple as superpartners get heavier, there would be no significant
deviations from the SM discovered at LEP.  That is, it is only possible to
have loop effects large enough to measure at LEP + SLC if superpartners
are light enough to observe directly.  In nonsupersymmetric approaches
with strong interactions near the electroweak scale it was natural to
expect significant deviations from the Standard Model at LEP.  

\end{enumerate}
Together these successes provide powerful indirect evidence that low
energy supersymmetry is indeed part of the correct description of nature.

Remarkably, supersymmetry was not invented to explain any of the above
physics.  Supersymmetry was discovered as a beautiful property of string 
theories and was studied for its own sake in the early 1970s
\cite{Ramond:gb,Neveu:iv,Volkov:jx,Golfand:iw,Wess:tw}.  Only after
several years of studying the theory did it become clear that
supersymmetry solved the above problems, one by one. Furthermore, all of
the above successes can be achieved simultaneously, with one consistent
form of the theory and its parameters. Low energy supersymmetry also
has no known incorrect predictions; it is not easy to construct a theory
that explains and predicts certain phenomena and has no conflict with
other experimental observations.

People unfamiliar with supersymmetry may think supersymmetric theories 
have too many degrees of freedom because of large parameter spaces.
Here we just remark that the parameter structure is the same as that of 
the SM. Particle masses, flavor rotation angles and phases, and
Higgs VEVs have to be measured. Everything else is determined by the
symmetries and the assumption of soft supersymmetry breaking.

The physics is analogous to that in the SM with the quark masses and the
Cabibbo-Kobayashi-Maskawa (CKM) matrix which contains three flavor mixing
angles and one phase.  In supersymmetric models there are parameters
that are masses, flavor rotation angles, and phases.  Just as for the CKM
matrix, all of these parameters have to be measured, unless a compelling
theory determines them eventually. Before the top quark mass was known, in
order to study top physics a value for the top quark mass was assumed.
Then its production cross section, decay branching ratios and signatures,
and all aspects of its behavior could be calculated.  Since the other
needed SM parameters were measured, only the top mass was unknown; if 
other SM parameters had not yet been measured various values for them
would also have to be assumed.  The situation for superpartners is similar
--- for any given set of superpartner masses and flavor mixing angles and
phases the observable aspects of superpartner behavior can be calculated.  
Any tentative supersymmetry signal can be then studied to decide if it is
consistent with the theory. Furthermore, predictions can be made which can
help to plan future facilities. 

We will see that in the MSSM, ${\mathcal{L}}_{soft}$ will contain at least 
105 new parameters, depending on what is included.  While that might
seem like a lot, most arise from flavor physics and all of the parameters
have clear physical interpretations.  
Once there is data most will be measured, and their patterns may 
provide hints about the form of the high energy theory.
In the historical development of the SM, once it was known that 
the effective Lagrangian was $V-A$ many parameters disappeared and the
structure led to recognizing it was a gauge theory which reduced the
number more. Probably the situation will be similar for 
supersymmetry.\footnote{Counting parameters depends on assumptions.  One
reasonable way to count the SM parameters for comparison with
supersymmetry is to assume that all of the particles are known, but not
their masses or interactions. Then the $W$ and $Z$ vertices can each have
a spacetime tensor character of scalar, vector, etc (S, V, T, A, P) and
each can be complex (so multiply by 2).  Conserving electric charge, the
$Z$ can have 12 different flavor-conserving vertices for the 12 quarks and
leptons (e,$\mu, \tau, \nu_e, \nu_{\mu}, \nu_{\tau}, u, c, t, d, s, b$),
plus 12 additional flavor-changing vertices ($e\mu, e\tau, \mu \tau,$
{\it etc.}). This gives 240 parameters ($12\times 5\times 2\times 2$).  Similar
counting for the $W$ gives 180.  There are 12 masses.  Self-couplings of
$W$ and $Z$ allowing CP violation give 10.  The total here is 442
parameters.}

It is often argued that gauge coupling unification is the most important
success of supersymmetry and it is indeed a major result. But the issue of
how to break the electroweak symmetry is the more fundamental problem.
Explaining the mechanism of electroweak symmetry breaking is the deepest
reason why low energy supersymmetry should be expected in nature.  No
other approach should be taken to be of comparable interest for
understanding physics beyond the SM unless it can provide an appropriate
explanation of electroweak symmetry breaking. Actually, the gauge coupling
unification and the explanation of electroweak symmetry breaking basically
are equivalent. Both require the same input --- soft 
supersymmetry-breaking
parameters and a $\mu$ parameter of order the electroweak scale --- except
that the electroweak symmetry breaking mechanism also needs a Yukawa
coupling of order unity (in practice, the top quark coupling).

The success of gauge coupling unification and the explanation of 
electroweak symmetry breaking have two implications that should be kept in 
mind. First, they suggest the theory is perturbative up to scales of order 
the unification scale.  They do not imply a desert, but only that whatever 
is in the desert does not make the theory nonperturbative or change the
logarithmic slope.  Second, they suggest that physics has a larger
symmetry at the unification scale than at the electroweak scale.  

One way to view the logic of the successes of supersymmetry is as
follows.  There are really two hierarchy problems, the sensitivity of the
Higgs mass to all higher scales, and the need for $\mu$ to have a
weak-scale value instead of a unification scale value.  If supersymmetry 
is an effective theory of the zero modes of an underlying theory, then 
$\mu=0$ at the high scale since it enters as a mass term.  The 
nonrenormalization theorem guarantees no high scale value is 
generated by quantum corrections.  Once supersymmetry is broken, an 
effective $\mu$ of the order of the soft masses can be generated. 
Next assume the Higgs mass hierarchy problem is understood because
all the superpartner masses, which depend on the effective $\mu$ term as
well as the soft supersymmetry-breaking parameters, are below the TeV
scale.  Once this information is put into the theory, then radiative
electroweak symmetry breaking and gauge coupling unification both occur
automatically without further input and the other successes of
supersymmetry follow as well.

The framework for this review is the
traditional one with the Planck scale $M_{Pl}=1.2\times 10^{19}$ GeV and 
gauge coupling unification somewhat above $10^{16}$ GeV.  Specifically, in 
this review attention is mostly confined to the standard picture in which 
all extra dimensions are assumed to be small. This traditional picture 
based on having a primary theory at the Planck scale, with the hierarchy 
of scales protected by supersymmetry, has the advantage of providing
beautiful, understandable explanations for electroweak symmetry breaking
and the other results already mentioned.  While a consistent quantum
theory of gravity and the SM forces appears to require extra
dimensions in some sense, they are certainly not required to be larger
than the inverse of the unification scale.  Within the superstring 
framework, our discussion thus applies to scenarios with a high string 
scale. 
At present, alternative approaches ({\it e.g.} involving low 
fundamental scales and large extra dimensions) have not been able to 
reproduce all of the successes of supersymmetric theories, in particular 
at the level of detailed model building.  While alternative approaches are 
certainly worthy of further exploration, low energy supersymmetry is on 
stronger theoretical ground.

The main result that will emerge from any fundamental theory
which predicts low energy supersymmetry is the soft supersymmetry-breaking
Lagrangian, ${\mathcal{L}}_{soft}$ 
\cite{Dimopoulos:1981zb,Girardello:1981wz}.  As an example, consider
string theory, which provides a consistent quantum theory of
gravitational and gauge interactions.  However, string theory is
formulated with extra dimensions.  It must be compactified to 4D
and supersymmetry must be broken to give an effective theory at the
unification scale or other appropriate high scale. 4D string models have
been built which can incorporate the known forces and fundamental
particles, although fully realistic models are still lacking.  The origin
and dynamical mechanism of supersymmetry breaking in string theory is 
still not known, and despite extensive investigations no compelling 
scenario has emerged from the top-down approach.  Therefore, it is our 
belief that until ${\mathcal{L}}_{soft}$ is at least partly measured, it 
will not be possible to recognize the structure of the underlying theory.

After ${\mathcal{L}}_{soft}$ is measured, it must be translated to the
unification scale. This is a significant challenge because it necessarily 
will involve assumptions as to the nature of physics at higher energy 
scales.  This is in part because the region between
the weak or collider scale and the unification scale need not be 
empty; other obstacles exist, as will be discussed.  Indeed, a variety of 
states in that region are expected, including right-handed neutrinos 
involved in generating neutrino masses, possible axion scales, possible 
vector or SU(5) multiplets, {\it etc.}  One generally assumes that the 
theory remains perturbative in the region from about a TeV to the 
unification scale. There is strong evidence for this assumption --- both 
the unification of the gauge couplings and the explanation of electroweak 
symmetry breaking independently imply that the theory is indeed 
perturbative in this region. The hope is that the measured patterns of 
the  ${\mathcal{L}}_{soft}$ parameters will lead to further 
advances in understanding Planck scale physics, {\it e.g.} for string theorists 
to recognize how to find the correct string vacuum (assuming string theory 
is the correct approach to the underlying theory).

Most of what is not yet known about supersymmetry is parameterized by the
soft supersymmetry-breaking Lagrangian ${\mathcal{L}}_{soft}$.  In the 
following,  
several possible patterns of the ${\mathcal{L}}_{soft}$
parameters will be investigated, with the goal of describing how the
parameters can be measured in model-independent ways and their subsequent
implications for ultraviolet physics.  Our goal in writing this
review is to gather in one place a summary of much that is known about
${\mathcal{L}}_{soft}$. Our intended readers are not experts, but
theorists or experimenters who want to learn more about what will become
the central area of activity once superpartners are discovered, and those
entering the field from other areas or as students.

We have chosen to put the review in the form where the main text is
smoothly readable, and to put a number of technical details and
complicated pedagogy in appendices.  In particular, the appendices contain
a full listing, in a uniform notation, of the soft supersymmetry-breaking
Lagrangian, the associated mass matrices and mass eigenstate observable
particles, the renormalization group equations, and the Feynman rules, in
a general form without approximations and with full inclusion of phases.  
We hope that this uniform treatment will help both in saving time in the
future for many workers, and in reducing translation errors. 

Finally we repeat that this is a review focused on the soft 
supersymmetry-breaking Lagrangian.  Since the soft supersymmetry-breaking 
Lagrangian is central to all physics
beyond the Standard Model, we must cover many topics, from flavor to
colliders to cosmology.  Each of these topics could and often does have
its own review.  We have tried to balance the treatments and emphasize
mainly the connections of each topic to \lsoft, and we hope the reader
understands that we are not reviewing each of the subfields more fully.  
We have always given references that point to other reviews and recent
literature.

\setcounter{footnote}{0}
\section{The soft supersymmetry-breaking Lagrangian}
\label{lsoftgeneralsect}
This section of the review is organized as follows.  We begin with a brief 
overview of $N=1$, $D=4$ supersymmetry, for those unfamiliar with its 
basic features and terminology.  We then introduce the minimal 
supersymmetric standard model (MSSM) in Section~\ref{mssmintrosect},  
before presenting the soft supersymmetry-breaking parameters in 
Section~\ref{paramsect}.  A careful count of the parameters is given in 
Section~\ref{paramcountsect}.  Finally, a general overview of the 
parameter space of the MSSM is provided in Section~\ref{validmssmsect}; 
this section also includes an outline of the remaining sections of the 
review.

\subsection{Brief introduction to $N=1$, $D=4$ supersymmetry}
\label{briefsusyintro}
The purpose of this subsection is to introduce basic notions of $N=1$,
$D=4$ supersymmetry, enough for readers new to the topic to be able
to understand the presentation of the MSSM and many of its
phenomenological implications. While certain details of the construction
of supersymmetric theories are discussed in Appendix~\ref{backgroundsect},
no attempt is made here to provide a detailed pedagogical introduction to
supersymmetry.  For more detailed theoretical approaches and the reasons
for supersymmetry's technical appeal, we direct the interested reader to
the many existing and forthcoming textbooks
\cite{Bagger:1990qh,West:tg,Weinberg:cr,Gates:nr,Ramond:vh} and reviews
\cite{Nilles:1983ge,Haber:1984rc,Sohnius:qm,Argyreslect,Ramond:1994fm,Lykken:1996xt,Drees:1996ca,Martin:1997ns}.

We start with global supersymmetry, beginning once again with the 
definition of supersymmetry presented in the introduction.  Supersymmetry 
is defined to be a symmetry which relates bosonic and fermionic 
degrees of freedom:
\begin{equation}   
\label{susytmn}
Q|B>\simeq |F>;\;\;\;\;
Q|F>\simeq |B>,
\end{equation}
in which $Q$ denotes the spin $1/2$ generator of the
supersymmetry algebra.  We focus here exclusively on $N=1$ 
supersymmetry in four dimensional spacetime, for which the supersymmetry 
algebra is given by the anticommutator
\begin{equation}
\label{susygen2}
\{Q_{\alpha}, \overline{Q}_{\beta}\}=2\sigma^{\mu}_{\alpha \beta} P_{\mu},
\end{equation}
where $\sigma^{\mu}$ are Pauli matrices, $\alpha,\beta$ are spinor 
indices, and $P_{\mu}$ denotes the momentum.  
Eq.~(\ref{susygen2}) demonstrates that the supersymmetry algebra also 
includes the usual Poincare algebra of spacetime. Both the momentum and 
angular momentum generators have vanishing commutators with the 
supersymmetry generators.

Given the supersymmetry algebra, its irreducible representations, or {\it
supermultiplets}, can be constructed systematically; this procedure is
described {\it e.g.} in \cite{Bagger:1990qh,Sohnius:qm}.  Supermultiplets
by definition contain an equal number of bosonic and fermionic degrees of
freedom. Supersymmetry representations are either {\it on-shell}
multiplets, in which the equations of motion of the fields are used, or
{\it off-shell} representations.  The off-shell multiplets contain
additional nonpropagating degrees of freedom required for the closure of
the supersymmetry algebra.  These nondynamical auxiliary fields can be
eliminated through their equations of motion.  However, we keep them here
because they are useful in certain mnemonic devices in the construction of
the Lagrangian, and also because they are the order parameters of
supersymmetry breaking (see Section~\ref{lsoftmodelsect}).
 
Within $N=1$, $D=4$ supersymmetry, two types of representations, the {\it
chiral} and {\it vector} supermultiplets, are most useful for
phenomenological purposes:
\begin{itemize}

\item {\it Chiral supermultiplets}.  Each chiral supermultiplet contains
one complex scalar $\phi$, one two-component chiral fermion $\psi$, and an
auxiliary scalar field $F$.  
 
\item {\it Vector supermultiplets}. Each massless vector multiplet
contains a spin 1 vector gauge boson $V^{a}_{\mu}$ a Majorana spinor
$\lambda^a$ called the {\it gaugino}, and a scalar auxiliary field $D^a$, 
($a$ labels the gauge group generators). 

\end{itemize} 
In the construction of supersymmetric theories, it is often more 
convenient to work with entities known as {\it superfields} 
\cite{Salam:1974yz}.  For our purposes the terms 
superfield and supermultiplet can be used interchangeably. 
A chiral superfield will be denoted by $\hat{\Phi}=\{ \phi$, $ \psi $, 
$F\}$, and a vector superfield by $\hat{V}=\{ V^a_{\mu}, \lambda^a, D^a 
\}$. 

Let us now turn to the interactions of supersymmetric theories.  The 
main feature is that many of the terms present in a general 
nonsupersymmetric Lagrangian are related by supersymmetry
transformations, and hence the number of independent coupling constants is
greatly reduced.  Many of the interactions can be encoded within
certain functions of the superfields which contain all the independent
couplings and act as generating functions for the Lagrangian. 
Given these functions, it is straightforward to write down the 
complete (usually quite lengthy) Lagrangian following a given set of 
rules.  These rules are presented in many of the standard reviews and 
textbooks cited at the beginning ot this subsection.

The Lagrangian for theories with $N=1$ supersymmetry in four dimenisons 
can be specified fully by three functions of the matter fields $\Phi_i$: 
(i) the superpotential $W$, (ii) the K\"ahler potential $K$, and (iii) 
the gauge 
kinetic function $f$. In addition to constraints from gauge invariance, 
$W$ and $f$ are further constrained to be holomorphic (analytic) 
functions of the fields, while the K\"ahler potential can be any real 
function.  In this 
review, we are concerned with low energy effective theories such 
as the MSSM, and hence consider theories with canonical kinetic terms 
only and confine our attention to the renormalizable couplings.  As 
described 
in Appendix~\ref{backgroundsect}, this indicates a specific (canonical) 
form of $K$ and $f$, and superpotential terms only through dimension 3:
\begin{equation}
W = Y_{ijk} \hat{\Phi}_i \hat{\Phi}_j \hat{\Phi}_k + \mu_{ij} \hat{\Phi}_i 
\hat{\Phi}_j.
\end{equation}
Following the rules to construct the Lagrangian, one can see that 
the trilinear superpotential terms yield Yukawa couplings of the 
form $Y_{ijk}\phi_i\psi_j\psi_k$ and quartic scalar couplings of the form 
$|Y_{ijk}\phi_j\phi_k|^2$. Hence, in supersymmetric 
extensions of the SM the usual Yukawa couplings will be accompanied 
by terms of equal coupling strength involving the scalar partner of one of 
the quark or lepton fields, the remaining quark or lepton field and the 
fermionic partner of the Higgs field. This is an example of a useful 
mnemonic: for each coupling in the original theory, the supersymmetric 
theory includes terms in which any two fields are replaced by their 
superpartners.  

The dimensionful couplings $\mu_{ij}$ give rise to mass terms for all the
components in the chiral supermultiplet.  Such mass terms are of course
only allowed if there are vectorlike pairs in the matter sector. For
example, in supersymmetric extensions of the SM such terms are forbidden
for the SM chiral matter, but are allowed if the model includes a pair of
Higgs doublets with opposite hypercharges, which will turn out to be 
a requirement.  The term involving the electroweak Higgs 
doublets is known as the $\mu$ term; it will be discussed in detail in
Section~\ref{sectmuproblem}.

In the gauge sector, the Lagrangian includes the usual gauge couplings of
the matter fields and kinetic terms for the gauge bosons.  Supersymmetry
also requires a number of additional couplings involving the gauginos and
$D^a$.  The matter fields have interactions with the
gauginos of the form $\sqrt{2} g \phi^* T^a \psi \lambda^a $,
where $T^a$ is the generator of the corresponding gauge symmetry. These
terms can be regarded as the supersymmetric completion of the usual gauge
couplings of the matter fields.  In addition, the Lagrangian includes
kinetic terms for the gauginos of the form $-i \lambda^{a \dagger}
\overline{\sigma}^{\mu} D_{\mu} \lambda^a$, recalling that the generator
in the covariant derivative is written in the adjoint representation.
Finally, there are couplings of the auxiliary field $D^a$. All of these
terms are fixed once the gauge structure and particle content of a model
is specified.

In globally supersymmetric theories, the scalar potential has a specific 
form:
\begin{equation}
\label{globalsusyV}
V(\phi_i)=|F_i|^2+\frac{1}{2}D^aD^a,
\end{equation}
{\it i.e.}, it consists of a sum of {\it F terms} 
and {\it D terms}, which are given by
\begin{eqnarray}
F^*_i&\equiv &W_i=\frac{\partial W}{ \partial \phi^i}\\
D^a&=&-g(\phi_i^*T^a_{ij}\phi_j).
\end{eqnarray}
See also Eq.~(\ref{WtoLrule}) and Eq.~(\ref{dterm1}).  The positive
definite form of Eq.~(\ref{globalsusyV}) has implications for 
supersymmetry breaking.  From the form of the supersymmetry algebra, 
it can be proven that $\langle V \rangle =0$, the global minimum of this 
potential, is a signal of unbroken supersymmetry.  Spontaneous 
supersymmetry breaking is thus characterized by nonvanishing VEVs of $F_i$ 
and/or $D^a$, as discussed further in Section~\ref{lsoftmodelsect}.

Quantum field theories with global supersymmetry provide a natural context
in which to investigate questions within particle physics.  However, in
such models the gravitational sector has been disregarded, even though it
must be included to fully address high energy phenomena.  
Supersymmetrizing the gravitational sector requires that the global
supersymmetry transformations Eq.~(\ref{susytmn}) must be
gauged.\footnote{Recall that the Poincar\'{e} algebra is a subalgebra of
the supersymmetry algebra. Since general relativity arises from gauging
the Poincare spacetime symmetry, within supersymmetry the accompanying
fermionic translations generated by the $Q$s must also be gauged.} For
this reason, local supersymmetry is known as {\it supergravity}, or SUGRA
for short.  Within supergravity theories, the spin 2 graviton is
accompanied by its superpartner, the spin $\frac{3}{2}$ {\it gravitino},
$\widetilde{G}_n$ ($n$ is a spacetime index; the spinor index is
suppressed). The off-shell $N=1$ supergravity multiplet contains a number
of auxiliary fields, which will generally not be of importance for our
purposes within this review.

The most general $N=1$ supergravity Lagrangian \cite{Bagger:1990qh} 
consists of a sum of kinetic terms, gravitational terms, topological 
terms, scalar self-couplings, and fermion interaction terms.  The scalar 
self-couplings and fermion interactions include both renormalizable and 
nonrenormalizable terms.  The theory is specified by the same three 
functions $W$, $K$, and $f$ as in the global case. We describe further 
aspects of this theory in Appendix~\ref{sec:sugra}.  

The supergravity scalar potential is particularly relevant for
phenomenology, because it plays an important role in supersymmetry
breaking.  Following \cite{Bagger:1990qh} (but using slightly different
notation which should be clear from the context\footnote{For simplicity,
in what follows we factor out the dependence on the quantity $e$,
essentially the determinant of the vierbein. In flat space, which is the
situation of interest for most of this review, this quantity is equal to
1.}), the scalar potential is
\begin{equation}
-e^{-1}{\mathcal{L}}_{s}=\frac{1}{2}g^2D_{a}D_{a}+e^Kg^{ij*}(D_iW)(D_jW)^*-3e^KW^*W.
\end{equation}
Note that in supergravity, there is a manifestly nonrenormalizable 
contribution (the last term). The scalar potential is once again a sum of 
D terms and F terms, the analogues of Eq.~(\ref{globalsusyV}) for global 
superymmetry.  The F terms have the generalized form 
$F_i=e^{\frac{K}{2}}g^{ij*}(D_jW)$, in which 
\begin{equation}
D_iW=\frac{\partial W}{\partial \phi_i}+\frac{\partial K}{\partial 
\phi_i}W.
\end{equation}
In the above expressions, we have suppressed the factors of the Planck 
mass; these factors can be restored using dimensional analysis. 
 
\subsection{Introducing the MSSM}
\label{mssmintrosect}
We now present a basic introduction to the minimal supersymmetric 
standard model (MSSM) for those unfamiliar with the details of the model.  
At present we shall focus on the supersymmetric sector; the soft 
supersymmetry-breaking Lagrangian will be introduced in 
Section~\ref{paramsect}.

The MSSM is defined to be the minimal supersymmetric extension of the SM,
and hence is an $SU(3) \times SU(2)_L \times U(1)_Y$ supersymmetric gauge
theory with a general set of soft supersymmetry-breaking terms.  The known
matter and gauge fields of the SM are promoted to superfields in the MSSM:
each known particle has a (presently unobserved) superpartner.  The
superpartners of the SM chiral fermions are the spin zero {\it sfermions},
the {\it squarks} and {\it sleptons}. The superpartners of the gauge
bosons are the spin $1/2$ {\it gauginos}.

The Higgs sector of the MSSM differs from that of the SM (apart from
the presence of superpartners, the spin $1/2$ {\it higgsinos}). The SM 
Higgs sector consists of a single doublet $h$ which couples to all of the 
chiral matter.  In the MSSM, two Higgs doublets $H_u$ and $H_d$, which 
couple at tree level to up and down type chiral fermions separately, 
are required.  The need for two Higgs doublets can be seen from the 
holomorphic property of the superpotential: couplings involving $h^*$, 
necessary in the SM for the up-type quark Yukawa couplings, are not 
allowed by supersymmetry.  
Two Higgs doublets are also required for the model to be anomaly free.  
Since the chiral fermion content of the theory includes the higgsinos,
anomaly constraints require that the Higgs sector be vectorlike, {\it 
i.e.}, that the two Higgs doublets have opposite hypercharges.

With the exception of the Higgs sector, the MSSM particle content, which
is listed in Table~\ref{habertable}, includes only the known SM fields and
their superpartners.  Supersymmetric theories with additional matter
and/or gauge content can of course easily be constructed. We discuss 
several possible extensions of the MSSM in Section~\ref{extensionsect}.

\renewcommand{\arraystretch}{1.3}
\setlength{\tabcolsep}{0.01in}
\begin{table}[htb]
\centering
\caption{The MSSM Particle Spectrum}
\vskip6pt
\begin{tabular}{ccc}
          &             &  \\[-5pt]
Superfield& Bosons & Fermions \\
\hline
\multicolumn{2}{l}{\underline{Gauge}}& \\[3pt]
$\wh G$&  $g$&     $\wt g$ \\
$\wh V^a$&$W^a$&  $\widetilde W^a$ \\
$\wh V\pri$&     $B$&      $\widetilde B$ \\
\hline
\multicolumn{2}{l}{\underline{Matter}} & \\[3pt]
$\displaystyle{\wh L\atop \wh E^c}$&
leptons $\Bigg\{ \displaystyle{\wt L\,=\,(\widetilde\nu,\widetilde 
e^-)_L  \atop
           \displaystyle{\wt E\,=\,\widetilde 
e^+_R\hphantom{(\nu,_L)}}}$\hfill&
            $\displaystyle{ (\nu,e^-)_L \atop  e^c_L}$
\\[16pt]
$\displaystyle{\wh Q\atop \displaystyle{\wh U^c \atop \wh D^c} } $&
  quarks $\left\{\vbox to 27pt{}   \right.  
 \displaystyle{ \wt Q\,=\,(\widetilde u_L,\widetilde d_L)
  \atop \displaystyle{\wt U^c\,=\,\widetilde u^*_R\hphantom{,d_L)^f}
  \atop \displaystyle{\wt D^c\,=\,\widetilde d^*_R\hphantom{,d_L)^f}} } 
}$\hfill&
    $\displaystyle{(u,d)_L \atop\displaystyle{u^c_L \atop d^c_L}}$
\\ \noalign{\vskip8pt}
$\displaystyle{\wh H_d\atop \wh H_u}$&
     Higgs $\Bigg\{ \displaystyle{ H^i_d \atop \displaystyle{H^i_u}}$\hbox 
to 
2.0cm{}&
    $\displaystyle{(\wt H^0_d,\wt H^-_d)_L \atop (\wt H^+_u,\wt H^0_u)_L}$
\\ \noalign{\vskip8pt} \hline \end{tabular} 
\label{habertable}
\end{table}

The renormalizable interactions of the MSSM are encoded as terms of
dimension two and three in the superpotential of the 
theory. The superpotential terms include the Yukawa couplings of the 
quarks and leptons to the Higgs doublets, as well as a mass term which 
couples $H_u$ to $H_d$.

Additional renormalizable superpotential couplings which violate baryon 
number and lepton number are also allowed by gauge invariance, as shown 
explicitly in Section~\ref{RPsect}. Such couplings would 
lead to rapid proton decay, and hence at least certain combinations of 
these terms must be forbidden by imposing additional symmetries on the 
theory. A common, though not absolutely necessary, choice is to impose a 
discrete symmetry known as {\it R-parity}, which forbids all baryon and 
lepton number violation in  the renormalizable superpotential.  R-parity 
and related issues will be discussed in Section~\ref{RPsect}. In this 
review, the definition of the MSSM always includes the assumption of a 
conserved R-parity.  Hence, the MSSM superpotential is 
\begin{equation}
\label{MSSMW}
W=\epsilon_{\alpha 
\beta}[-\hat{H}_u^{\alpha}\hat{Q}_i^{\beta}Y_{u_{ij}}\hat{U}^c_j
+\hat{H}_d^{\alpha}\hat{Q}_i^{\beta}Y_{d_{ij}}\hat{D}^c_j
+\hat{H}_d^{\alpha}\hat{L}_i^{\beta}Y_{e_{ij}}\hat{E}^c_j- 
\mu\hat{H}_d^{\alpha}\hat{H}_u^{\beta}].
\end{equation}
In the above expression, $i$ and $j$ are family indices, while $\alpha$ 
and $\beta$ are $SU(2)_L$ doublet indices (the color indices are suppressed). 
$\epsilon_{\alpha \beta}$ is defined in the standard 
way; see Appendix \ref{conventionsapp}.  

The superpotential of the MSSM dictates all of the supersymmetric
couplings of the theory, aside from the gauge couplings.  The 
superpotential and gauge couplings thus dictate the couplings of the Higgs
potential of the theory.  This would appear to reduce the number of
independent parameters of the MSSM; for example, the tree-level Higgs
quartic couplings are fixed by superysmmetry to be gauge couplings rather
than arbitrary couplings as in the SM.  However, the phenomenological 
requirement of supersymmetry breaking terms in the Lagrangian introduces 
many new parameters, which play crucial roles in the phenomenology of the 
model.  The rest of the review will focus on theoretical and 
phenomenological aspects of the soft supersymmetry-breaking sector of the 
MSSM.

\subsection{The parameters of the MSSM}
\label{paramsect}

At low energies, supersymmetry must be a broken symmetry.  Since this
implies the appearance of supersymmetry-breaking terms in the Lagrangian,
an immediate question is whether such terms spoil supersymmetry's elegant
solution to the hierarchy problem.  As generic quantum field
theories with scalars generally have a hierarchy problem, if all
supersymmetry-breaking terms consistent with other symmetries of the
theory are allowed the dangerous UV divergences may indeed be
reintroduced.
 
Fortunately, such dangerous divergences are not generated to any order in 
perturbation theory if only a certain subset of supersymmetry-breaking 
terms are present in the theory.   Such operators, are said to break 
supersymmetry {\it softly}, and their couplings are collectively denoted 
the {\it soft parameters}.  The part of the Lagrangian which contains 
these terms is 
generically called the soft supersymmetry-breaking Lagrangian \lsoft, or 
simply the soft Lagrangian.  
The soft supersymmetry-breaking operators comprise a consistent truncation of all  
possible operators in that the presence of soft supersymmetry-breaking 
parameters does not regenerate ``hard" supersymmetry-breaking terms at 
higher order.
The complete set of possible soft supersymmetry-breaking parameters was first
classified in the seminal papers 
\cite{Girardello:1981wz,Inoue:1982ej,Inoue:1982pi,Inoue:1983pp}.  The 
classic proof of Girardello and Grisaru \cite{Girardello:1981wz}
will not be repeated here.  The power counting method, which explains why 
certain terms are soft while others are not, is reviewed in 
Appendix~\ref{softapp}.

The soft supersymmetry-breaking Lagrangian is defined to include all
allowed terms that do not introduce quadratic divergences in the theory:
all gauge invariant and Lorentz invariant terms of dimension
two and three ({\it i.e.}, the relevant operators from an effective field
theory viewpoint). The terms of \lsoft can be categorized as follows
(summation convention implied):
\begin{itemize}
\item Soft {\it trilinear scalar} interactions: 
$\frac{1}{3!}\widetilde{A}_{ijk}\phi_i\phi_j\phi_k+{\rm h.c.}$.

\item Soft {\it bilinear scalar} interactions:
$\frac{1}{2}b_{ij}\phi_i\phi_j+{\rm h.c.}$.

\item Soft {\it scalar mass-squares}: $m^2_{ij} \phi^{\dagger}_i\phi_j$.

\item Soft {\it gaugino masses}: 
$\frac{1}{2}M_a\lambda^a\lambda^a+{\rm 
h.c.}$.
\end{itemize}
In the expression above, $a$ labels the gauge group ({\it i.e.}, the 
generator index is suppressed here).
We will not discuss in depth the terms in \lsoft which can be only be soft 
under certain conditions, as described briefly in Appendix~\ref{softapp}.  
Such terms are usually not included since they turn out to be negligible 
in most models of the soft supersymmetry-breaking parameters. 

As stated, our attention will mainly be focused on the MSSM, which is 
defined to be the supersymmetrized 
Standard Model with minimal particle content and the most general set of 
soft supersymmetry-breaking parameters.\footnote{The label MSSM has been 
used in the literature to denote simpler versions of the theory ({\it e.g.} with 
a restricted set of soft supersymmetry-breaking parameters). Here
``minimal'' refers to the particle content, not the parameters.} 
 Of course, the correct theory could be larger than the MSSM. If the
theory is extended, for example by adding an extra singlet scalar or an
additional U(1) symmetry, the associated terms can be added in a
straightforward way; see {\it e.g.} the discussion of the next-to-minimal
supersymmetric standard model (NMSSM) in Section~\ref{nmssm}.  Similarly,
just as it is necessary to add new fields such as right-handed neutrinos
to the SM to incorporate neutrino masses in the SM, such fields and their
superpartners and the associated terms in ${\mathcal{L}}_{soft}$ must be
added to include neutrino masses.  This issue is somewhat model-dependent, 
and will be discussed further in Section~\ref{neutrinosect}.

The matter content and superpotential of the MSSM were presented in 
Table~\ref{habertable} and Eq.~(\ref{MSSMW}) in 
Section~\ref{mssmintrosect}; further details are presented in 
Appendix~\ref{conventionsapp}. 
The soft Lagrangian for the MSSM is presented in Eq.~(\ref{lsoftexpr}), 
which we repeat here:
\begin{eqnarray}
-{\mathcal{L}}_{soft} &=&\frac{1}{2}\left[M_3
\widetilde{g}\widetilde{g}+M_2\widetilde{W}\widetilde{W}+ 
M_1\widetilde{B}\widetilde{B}\right ]\nonumber\\
&+&\epsilon_{\alpha \beta}[-b H^{\alpha}_dH^{\beta}_u-
H^{\alpha}_u\widetilde{Q}^{\beta}_i\widetilde{A}_{u_{ij}}\widetilde{U}^c_j
+H^{\alpha}_d\widetilde{Q}^{\beta}_i\widetilde{A}_{d_{ij}}\widetilde{D}^c_j
+H^{\alpha}_d\widetilde{L}^{\beta}_i\widetilde{A}_{e_{ij}}\widetilde{E}^c_j 
+{\rm h.c.}]\nonumber\\
&+&m_{H_{d}}^{2}|H_{d}|^2+m_{H_{u}}^{2}|H_{u}|^2
+\widetilde{Q}^{\alpha}_i{m^2_Q}_{ij}\widetilde{Q}_j^{{\alpha}*} 
\nonumber\\
&+& \widetilde{L}^{\alpha}_i{m^2_L}_{ij}\widetilde{L}_j^{{\alpha}*}
+\widetilde{U}^{c*}_i{m^2_U}_{ij}\widetilde{U}^c_j
+\widetilde{D}^{c*}_i{m^2_D}_{ij}\widetilde{D}^c_j
+\widetilde{E}^{c*}_i{m^2_E}_{ij}\widetilde{E}^c_j.
\end{eqnarray}
Supersymmetry is broken because these terms contribute explicitly to
masses and interactions of (say) winos or squarks but not to their
superpartners.  The underlying supersymmetry breaking is assumed to be
spontaneous (and presumably take place in a hidden sector, as discussed 
in Section~\ref{lsoftmodelsect}). How supersymmetry breaking is 
transmitted to the superpartners is encoded in the parameters of
${\mathcal{L}}_{soft}$.  All of the quantities in ${\mathcal{L}}_{soft}$ 
receive radiative corrections and thus 
are scale-dependent, satisfying known renormalization group equations.  
The beta functions depend on what new physics is present between the two 
scales.  ${\mathcal{L}}_{soft}$ has the same form at any scale.

The soft parameters clearly have a significant impact on the MSSM mass
spectrum and mixings; the tree-level mass spectrum is presented in
Appendix~\ref{conventionsapp}.  As shown in Eq.~(\ref{pok1}), the mass
matrices of the sfermions are generally not diagonal in the diagonal
fermion basis, with off-diagonal effects dependent on the
soft mass-squares, $\widetilde{A}$ parameters, and the $\mu$ parameter.  The
gauginos and higgsinos with equal electric charges mix, with the charged
superpartners generically denoted as {\it charginos} and the neutral
superpartners as {\it neutralinos}.  The chargino and neutralino mass
matrices depend on the gaugino mass parameters and $\mu$, as shown in
Eq.~(\ref{charginomat}) and Eq.~(\ref{neutralinomat}).  The tree-level
Higgs sector depends on the Higgs soft mass-squares and the $\mu$ and $b$
parameters, as discussed in Section~\ref{EWSBsect}, and many other
parameters filter into the Higgs sector at higher-loop order.  
All of the above quantities also depend nontrivially on 
$\tan\beta$, the ratio of the vacuum expectation values of 
the Higgs doublets ($\tan\beta\equiv \langle H_u \rangle /\langle H_d 
\rangle$). As will become clear
throughout this review, this parameter plays a crucial role in both the
theoretical and phenomenological aspects of the MSSM.

Many of the soft parameters can be complex. The squark and slepton mass
matrices are Hermitian matrices in flavor space, so the diagonal elements
are real while the off-diagonal elements can be complex.  The soft
supersymmetry-breaking trilinear couplings $\widetilde{A}_{u,d,e}$ are general
$3 \times 3$ complex matrices in flavor space. The Yukawa-like $\widetilde{A}$
parameters are often assumed to be proportional to the corresponding
Yukawa matrices.  While this can arise in certain models of the soft
supersymmetry-breaking terms, it is by no means a general feature. In this review, this
proportionality shall not be assumed to be true unless that is explicitly
stated. Symmetries of the theory allow a number of the parameters to be
absorbed or rotated away with field redefinitions. The parameters will be
counted carefully below.

The supersymmetric higgsino mass parameter $\mu$ is also highly relevant 
in the discussion of the constraints on the soft parameters. In 
general, $\mu$ can be a complex parameter, with a phase  $\phi_{\mu}$.  
For the purpose of this review the $\mu$ parameter will be included in the 
general category of the soft parameters, although it is not {\it a 
priori} directly related to supersymmetry  breaking.  The supersymmetric 
interactions of the theory should not include a bare $\mu$ term, because 
the natural scale for $\mu$ would presumably be the high scale at which 
the theory is defined while phenomenology dictates that $\mu$ should have 
the same order of magnitude as the soft terms.  This {\it $\mu$ 
problem} will be discussed in Section~\ref{EWSBsect}.

\setcounter{footnote}{0}
\subsubsection{Parameter counting}
\label{paramcountsect}
Having presented the soft supersymmetry-breaking Lagrangian of the MSSM, we now 
count its physical parameters (see also  
\cite{Dimopoulos:1995ju,Haber:1997if}).

With the exception of $m^2_{H_d}$, $m^2_{H_u}$, and the diagonal
entries of the soft mass-squared parameters of the squarks and sleptons,
every parameter can in principle be complex. The Yukawa couplings of the
SM and the soft supersymmetry-breaking trilinear couplings are each
general complex $3 \times 3$ matrices which involve a total of $54$ real 
parameters and $54$ phases. The soft mass-squared parameters for the
squarks and sleptons are each Hermitian $3 \times 3$ matrices which have
in total $30$ real and $15$ imaginary parameters.  Taking into account the
real soft Higgs mass-squared parameters, complex gaugino masses, $\mu$ and
$b$, the MSSM would appear to have 91 real parameters (masses and mixing 
angles) 
and 74 phases.\footnote{One can also include the complex gravitino mass in 
the parameter count.}

However, a subset of parameters can be eliminated by global rephasings of
the fields and thus are not physical.  In the limit in which the
superpotential and soft supersymmetry-breaking couplings are set to zero, 
the MSSM Lagrangian possesses the global family symmetry
\begin{equation}
\label{u35}
G=U(3)_Q \times U(3)_D \times U(3)_U \times U(3)_L \times U(3)_E.
\end{equation}
As each $U(3)$ can be parameterized by 3 magnitudes and 6 phases, 
$G$ has 15 real parameters and 30 phases.  A subgroup of 
this family symmetry group is left unbroken in the limit that the 
superpotential and soft supersymmetry-breaking interactions are switched on:  
\begin{equation}
G_{{\rm residual}} = U(1)_B \times U(1)_L,
\end{equation}
and hence only 15 magnitudes and 28 phases can be removed 
from the MSSM Lagrangian from such global rephasings of the fields. There 
are two more $U(1)$ global symmetries of the MSSM: $U(1)_R$ and 
$U(1)_{PQ}$, which will be discussed in detail later. Including the rest 
of the SM parameters: the gauge couplings, the QCD $\theta$ 
angle, {\it etc.}, there are
$79$ real parameters and $45$ phases in the MSSM.  For this reason, the
theory has also been labeled the MSSM-124 by Haber \cite{Haber:1997if}. 

Let us look in greater detail at how this elimination of parameters is
usually done.  In the quark/squark sector, global symmetry rotations of
$(U(3)_Q \times U(3)_U \times U(3)_D) /(U(1)_B)$ are used to eliminate 9
real parameters and 17 phases from the Yukawa couplings $Y_{u,d}$, leaving
9 real parameters (the 6 quark masses and 3 CKM angles) and 1 CKM phase.
It is customary to make a further $U(3)_{u_L} \times U(3)_{d_L}$ rotation
on both the quarks and their superpartners.\footnote{This rotation is not
a symmetry of the gauge sector and thus does not further reduce the number
of parameters, but rather introduces the CKM matrix into the charged
current coupling.} In this basis (the super-CKM or SCKM basis), the quark
mass matrices are diagonal but generically the squark mass matrices are
not diagonal because of supersymmetry-breaking effects. Let us first
assume massless neutrinos; the generalization to massive neutrinos will be
discussed in Section~\ref{neutrinosect}.  In the massless neutrino case,
$(U(3)_L \times U(3)_E)/U(1)_L $ symmetry rotations of the lepton/slepton
sector are used to eliminate 6 real parameters and 11 phases, leaving 3
real parameters (the lepton masses) and no phases in $Y_e$. Two phases can
then be removed from the slepton couplings in ${\mathcal{L}}_{soft}$.
These flavor rotations manifestly leave the gaugino mass parameters,
$\mu$, $b$, and the Higgs soft mass-squared parameters invariant.

In the limit that the $\mu$ term and the ${\mathcal{L}}_{soft}$ parameters
are set to zero, the MSSM Lagrangian has two additional global $U(1)$
symmetries, $U(1)_{PQ}$ and $U(1)_R$, which are not a subgroup of
Eq.~(\ref{u35}).  $U(1)_{PQ}$ commutes with supersymmetry; in contrast, 
particles and their respective superpartners have different charges with 
respect to $U(1)_R$. For such symmetries, generically called R-symmetries, 
the charges of the bosonic components of the chiral superfields are 
greater
than the charges of the fermionic components by a fixed amount, typically
normalized to $1/2$. These symmetries do not act on the family indices,
otherwise the Yukawa couplings would not remain invariant. The
corresponding field rephasings thus do not affect the phases 
of the off-diagonal components of either the $m^2$ or the $\widetilde{A}$ 
terms up to an overall phase of the $\widetilde{A}$ terms, as discussed 
below.

These field rephasings do affect the phases of the gaugino mass parameters, 
the phases of $\mu$ and $b$. and the overall phases of 
the $\widetilde{A}$ parameters. The overall $\widetilde{A}$ phases are of 
course not uniquely defined; we'll return to this issue later. 
Global $U(1)_{PQ}$ rotations keep all of the soft trilinear scalar
couplings $\widetilde{A}$ invariant\footnote{The soft 
trilinear couplings involve the same combination of fields as the Yukawa 
couplings; the only difference is that the two fermions are changed to 
their scalar partners, 
which has no effect because $U(1)_{PQ}$ commutes with supersymmetry.} 
while global $U(1)_R$ transformations change 
the phases of the trilinears by a charge 1 rotation.  
$U(1)_{PQ}$ rotates $\mu$ and $b$ by the same
amount and thus has no effect on their relative phase.
$U(1)_R$ can change the relative phase because the charge of $\mu$ 
is greater the the charge of $b$ by 2.\footnote{The relevant terms are the 
higgsino mass term $\mu \widetilde{H}_u \widetilde{H}_d$ and the scalar soft 
bilinear term $bH_u H_d$. The scalar mass terms derived from the $\mu$ 
term are $|\mu|^2  |H_{u,d}|^2$, which are invariant under global phase 
rotations of the Higgs fields.}
$U(1)_{PQ}$ has no effect in the gaugino sector, but $U(1)_R$ 
rotations lead to shifts in the gaugino mass phases.
 
A particular choice of $U(1)_{PQ}$ and $U(1)_R$ charges is
shown in Table~\ref{c_field},
\begin{table}[h]
\begin{center}
\begin{tabular}{|c|c|c|c|c|c|}
\hline
Fields &\, $U(1)_{PQ}$ \, & \multicolumn{2}{c|}{$U(1)_R$} 
&\multicolumn{2}{c|}{$U(1)_{R-PQ}$}\\
\hline
 & &\, boson \,& \, fermion \, & \, boson \, & \, fermion\, \\
\hline
$Q$,$U^c$,$D^c$,$L$,$E^c$ & $-\frac{1}{2}$ & $\frac{1}{2}$ &
$-\frac{1}{2}$ &1&0\\
\hline
$H_u$, $H_d$ & 1 & 1 & 0 &0&$-$1\\
\hline
$V_a$ & 0 & 0& 1&0&1\\
\hline
\end{tabular}
\caption{ \label{c_field} The PQ, R, and R-PQ charge 
assignments of the MSSM fields. }
\end{center}  
\end{table}
in which $V_a=(V_{a\,\mu}, \lambda_a)$ are the vector multiplets of the SM
gauge fields, which include the gauge bosons $V_{a\,\mu}$ and the 
gauginos $\lambda_a$. A useful 
way to keep track of the effect of the global $U(1)$ rotations on the 
phases of the parameters is to assume that the parameters
themselves are actually (VEVs of) fields which transform with respect to
the $U(1)$ symmetries, with charges chosen such that the global $U(1)$s
are symmetries of the full Lagrangian.\footnote{For example, consider a
Lagrangian term $C {\mathcal{O}}$, where ${\mathcal{O}}$ is any given
combination of fields with $U(1)$ charges $c_{\mathcal{O}}$.  Upon a field
rotation ${\mathcal{O}}^{\prime} = e^{i c_{\mathcal{O}} \phi}
{\mathcal{O}}$, the Lagrangian term becomes $C e^{-i c_{\mathcal{O}}\phi }
{\mathcal{O}}^{\prime}$. This is equivalent to assigning the coupling $C$
a $U(1)$ charge $-c_{\mathcal{O}}$ such that the $U(1)$ is a symmetry of
the full Lagrangian.} The classification of the parameters with 
respect to PQ and R was done for the MSSM in 
\cite{Rattazzi:1995gk,Dimopoulos:1995kn}. For completeness, the
the spurion charge assignments for the MSSM parameters under $U(1)_{PQ}$
and $U(1)_R$ are given in Table~\ref{c_spur}.
\begin{table}[h]
\begin{center}
\begin{tabular}{|c|c|c|c|}
\hline
\, Fields \,& \,$U(1)_{PQ}$ \,& \,$U(1)_R$\, &\, $U(1)_{R-PQ} \,$\\
\hline
$\mu$ & $-$2  & 0  &2\\
\hline
$Y_{u,d,e}$ & 0 & 0 &0 \\
\hline
$M_a$ & 0 & $-$2&$-$2\\
\hline
$b$ & $-$2 &$-$2 &0\\
\hline
$\widetilde{A}$ & 0 & $-$2&$-$2 \\
\hline
$m^2$ & 0 & 0 &0\\
\hline
\end{tabular} 
\caption{ \label{c_spur} The PQ, R, and R-PQ charge 
assignments of the MSSM spurions. }
\end{center}
\end{table}
In phenomenological applications, $U(1)_{PQ}$ and $U(1)_R$ rotations are
often used to eliminate certain phases for the sake of simplicity.  The
results must of course be interpreted in terms of the relevant
reparameterization invariant phase combinations.  Reparameterization
invariance can also serve as a useful check of calculations, as the
invariance should be manifest in the final results.

Reparameterization invariant combinations of phases for the MSSM are built
by determining the products of fields and parameters, or equivalently the
linear combinations of phases, for which the total charge sums to zero.
Several obviously invariant combinations include (i) the phases of the
off-diagonal entries of the soft mass-squared parameters, since they are
uncharged under both $U(1)_{PQ}$ and $U(1)_R$, and (ii) the relative
phases of the gaugino masses $\phi_{M_a} - \phi_{M_b}$ 
($a\neq b$) 
and the
relative phases of the $\widetilde{A}$ parameters
$\phi_{\widetilde{A}_{f_{ij}}}-\phi_{\widetilde{A}_{{f\prime}_{i^{\prime}
j^{\prime}}}}$, since they have the same PQ and R charge.  The phases that
{\it are} affected are $\phi_{\mu}$, $\phi_{b}$ and $\phi_{M_a}$, 
and
$\phi_{\widetilde{A}_f}$, the overall phases of the $\widetilde{A}_f$ parameters.
Following \cite{Lebedev:2002wq}, $\phi_{\widetilde{A}_f}$ can be defined in a
basis-independent way as $\phi_{\widetilde{A}_f}\equiv \frac{1}{3}{\rm
Arg}[{\rm Det}(\widetilde{A}_fY^{\dagger}_f)]$ (providing the determinant
exists). Linear combinations of these phases invariant under
reparameterization can be built from the following set of basis vectors:
\begin{eqnarray}
\label{reparaminvbasis}
\phi_{1\,f}&=&\phi_{\mu}+\phi_{\widetilde{A}_f}-\phi_{b}\\
\phi_{2\,{\lambda}}&=&\phi_{\mu}+\phi_{M_a}-\phi_{b}. 
\end{eqnarray}
For example,
$\phi_{M_a}-\phi_{\widetilde{A}_f}=\phi_{2\,{\lambda}}-\phi_{1\,f}$.  
This 
is not to say that all possible invariants will appear in a given
process.  Typically only a few reparameterization 
invariant combinations appear, depending on the details of the 
observable in question.

The previous discussion was based on particular choices of
$U(1)_R$ and $U(1)_{PQ}$.  An alternate choice of 
$U(1)_{R-PQ}$ and $U(1)_{PQ}$ 
is often used in the literature.  The associated charges shown in Tables 
\ref{c_field} and \ref{c_spur}. The R-PQ combination is useful since the Higgs scalars are neutral
under R-PQ, and hence their VEVs leave this combination unbroken.
While $\mu$, $\widetilde{A}_{u,d,e}$ and $M_a$ violate R-PQ,
$Y_{u,d,e}$ and $b$ respect R-PQ.
Since the Higgs fields violate PQ but respect R-PQ,
the PQ symmetry can be used to remove a phase from $b$ in the knowledge
that R-PQ rotations will not put it back.
Further R-PQ rotations can then remove a phase
from $\mu$, $\widetilde{A}_{u,d,e}$,or $M_a$,
after which
both PQ and R-PQ symmetries are exhausted. The Lagrangian can be cast into
a basis where the phase of $b$ is zero and dropped from the invariants
presented above.  One can always choose to work in this basis.  The  
reparameterization invariant combinations used in this review will
be those invariant under R-PQ ({\it e.g.},  
$\phi_{M_a}+\phi_{\mu}$), but one should always keep in mind that 
the full 
invariant must include the phase of $b$ term.  In addition to setting the 
phase of $b$ to zero, it is also common in the literature to use the 
$U(1)_R$ symmetry 
to set another phase to zero; this phase is usually one of the 
$\phi_{M_a}$, but the phase of $\mu$ or an overall $\widetilde{A}$ 
phase of 
could instead be eliminated.  Again, one should keep the full 
reparameterization invariant in mind in such situations.

\subsubsection{The allowed \lsoft parameter space}
\label{validmssmsect}
In the previous subsection, we have seen that the Lagrangian of the 
minimal supersymmetric extension of the SM contains at 
least 105 new parameters in addition to the SM parameters.  These 
parameters include masses, CKM-like mixing angles, and
reparameterization invariant phase combinations. 

The masses, mixings, and couplings of the superpartners and Higgs bosons
depend in complicated ways on the \lsoft parameters as well as on the SM
parameters, as described in detail in Section~\ref{expdisc} and
Appendix~\ref{conventionsapp}.  There are 32 mass eigenstates in the MSSM:
2 charginos, 4 neutralinos, 4 Higgs bosons, 6 charged sleptons, 3
sneutrinos, 6 up-squarks, 6 down-squarks, and the gluino. If it were
possible to measure all the mass eigenstates it would in principle be
possible to determine 32 of the 105 soft parameters.  However, as we will
see, inverting the equations to go from observed mass eigenstates to soft
parameters requires a knowledge of soft phases and flavor-dependent
parameters, or additional experimental information, and hence in practice
it may be difficult or impossible.

This review aims to provide a guide to the allowed regions of the MSSM-124 
parameter space.  Constraints on the 105-dimensional \lsoft parameter
space arise from many phenomenological and theoretical considerations, as
well as direct and indirect experimental bounds.  The restrictions on the 
soft parameters can be loosely classified into two categories: 
\begin{itemize}

\item {\it Constraints from flavor physics}.  

Many of the parameters of the MSSM-124 are present only in flavor-changing
couplings.  Even flavor-conserving MSSM couplings can lead to
flavor-violating effective couplings at higher-loop level.  Such couplings
potentially disrupt the delicate cancellation of flavor-changing neutral
currents (FCNCs) of the SM.  The constraints are particularly stringent
for the parameters associated with the first and second generations of
squarks and sleptons.  This issue, known as the {\it
supersymmetric flavor problem}, will be discussed in 
Section~\ref{CPflsect1}.

\item  {\it Constraints from CP violation}.

The parameters of the MSSM include a number of CP-violating phases, which 
can be classified into two general categories:

\begin{enumerate}
\item Certain phases are present in flavor-conserving as well as
flavor-changing interactions.  These phases
include the phases of the gaugino 
mass parameters $\phi_{M_a}$, the
phases of $\mu$ and $b$, $\phi_{\mu},\phi_{b}$, and
the overall phases of $\widetilde{A}_{u,d,e}$: physical observables 
depend on the reparameterization invariant phase combinations spanned by 
the basis Eq.~(\ref{reparaminvbasis}).   A subset of these phases play a 
role in electroweak baryogenesis, as discussed in Section~\ref{bgensect}.  
However, these phases are also constrained by electric dipole moments 
(EDMs), as discussed in Section~\ref{EDMsect}.

In general, the phases affect many CP-conserving quantities and thus can
be measured, up to some overall signs, in such quantities.  But such
measurements may be model dependent.  There are several ways to
unambiguously demonstrate the existence of soft Lagrangian phases: (1)
detection of EDMs, (2) observation at colliders of explicitly CP-violating
observables such as appropriate triple scalar products of momenta, (3)
observation of CP-violating asymmetries different from the SM expectation
in rare decays such as $b\rightarrow s + \gamma$, or $B\rightarrow \phi
K_s$, (4) observation of production of several neutral Higgs mass
eigenstates at linear colliders in the Z + Higgs channel, and (5) finding
that measurement of parameters such as $\tan\beta$ give different results
when measured different ways assuming phases are zero.  Extended models 
could mimic the last two of these but to do so they will predict other 
states or effects that can be checked.

In summary, the phases, if nonnegligible, not only can have significant
phenomenological implications for CP-violating observables, but also can
have nontrivial consequences for the extraction of the MSSM parameters
from experimental measurements of CP-conserving quantities, since almost
none of the Lagrangian parameters are directly measured
\cite{Brhlik:1998gu}.  The phases will be addressed in the context of
neutralino dark matter in Section~\ref{darkmattersect}, and collider
physics in Section~\ref{expsect}.

\item The remaining phases are present in the off-diagonal entries of the
$\widetilde{A}$ and $m^2$ parameters, and hence occur in flavor-changing
couplings.  In this sense they are analogous to the CKM phase of the SM,
which is most economically expressed in terms of the Jarlskog invariant
\cite{Jarlskog:1985cw}.  Analogous Jarlskog-type invariants have been
constructed for the MSSM \cite{Lebedev:2002wq}.  These phases are
generically constrained by experimental bounds on CP violation in
flavor-changing processes, as discussed in Section~\ref{CPflsect1}.
\end{enumerate}

\item {\it Constraints from EWSB, cosmology, and collider physics}.

The gaugino masses, $\mu$ parameter, and the third family soft mass
parameters play dominant roles in MSSM phenomenology, from electroweak
symmetry breaking to dark matter to collider signatures for the
superpartners and Higgs sector.  Issues related to electroweak symmetry 
breaking will be discussed in Section~\ref{EWSBsect}. Cosmological 
questions such as dark matter and
baryogenesis will be addressed in Section~\ref{darkmattersect} and
Section~\ref{bgensect}. Finally, collider constraints will be
presented in Section~\ref{expsect}.

\end{itemize}
Given the complicated structure of the MSSM-124 parameter space, many of 
the phenomenological analyses of the MSSM assume that the 105 \lsoft 
parameters at electroweak/TeV energies take on simplified forms at
a given (usually high) scale. The next section of the 
review will be
dedicated to a summary of the various possible models of the \lsoft
parameters.  Before discussing the details of various supersymmetry
breaking models it is useful to consider on general grounds a certain
minimal framework for the pattern of \lsoft parameters.  In these classes
of models, the parameters have a minimal flavor structure; {\it i.e.}, all
flavor violation arises from the SM Yukawa couplings.  Many of the
parameters are then flavor-diagonal and may even be universal as well,
drastically reducing the number of independent parameters characteristic
of the MSSM-124.  In such scenarios, the squark and slepton mass-squares
are diagonal in flavor space:
\begin{equation}
\label{diagm2}
{m^2_Q}_{ij}=m_Q^2\delta_{ij};\;{m^2_U}_{ij}=m_U^2\delta_{ij}; 
\;{m^2_D}_{ij}=m_D^2\delta_{ij};\;{m^2_L}_{ij}=m_L^2\delta_{ij};\; 
{m^2_E}_{ij}=m_Q^2\delta_{ij},
\end{equation} 
and the $\widetilde{A}$ terms are proportional to the corresponding Yukawa 
couplings as follows:
\begin{equation} 
\label{AproptoY}
\widetilde{A}_{u_{ij}}=A_u Y_{u_{ij}};\;\widetilde{A}_{d_{ij}}=A_d 
Y_{d_{ij}};\; 
\widetilde{A}_{e_{ij}}=A_e Y_{e_{ij}}.
\end{equation}
Typically this pattern is present at a higher scale, the scale where
the soft parameters are presumably generated.  Therefore, the parameters
must be run to low energy using the renormalization group equations
(RGEs).  The one-loop RGEs for the MSSM-124 are presented in
Appendix~\ref{rgeapp}. For many phenomenological analyses
higher-loop accuracy is needed; see \cite{Martin:1993zk} for the 
full set of two-loop RGEs of the MSSM.

Such scenarios are known as minimal flavor violation (MFV). The squark and
slepton mass matrices are now diagonal in family space, such that their
flavor rotation angles are trivial. There is still LR mixing, but it is
negligibly small for all but third generation squarks and sleptons.  MFV
scenarios also often assume that \lsoft contains no new sources of CP
violation.  While many of the CP-violating phases of the MSSM-124 are
eliminated in minimal flavor violation scenarios by Eq.~(\ref{diagm2}) and
Eq.~(\ref{AproptoY}), the gaugino masses, $\mu$, $b$, and $A_{u,d,e}$
could in principle be complex and subject to the constraints mentioned in
Section~\ref{EDMsect}.

Minimal flavor violation is emphasized here because it is so
commonly used in the literature. It has several practical advantages with
respect to the general MSSM-124.  Simplicity is an obvious virtue; other
advantages will become clear during the course of this review,
particularly after the discussion of CP violation and FCNCs in 
Section~\ref{CPflsect1}. As discussed in the
next section, most attempts so far to build viable models of the \lsoft
parameters involve reproducing the structure of Eq.~(\ref{diagm2})  and
Eq.~(\ref{AproptoY}), or small deviations from it.  Even if this minimal, 
universal structure is assumed to hold at high scales, renormalization 
group evolution to low energies does not typically 
induce unacceptably large departures from this general pattern.

However, such minimal scenarios are {\it not} necessarily expected either
from theoretical or phenomenological considerations.  Despite the
overwhelming focus on this scenario in the literature, minimal
universality should thus not be adhered to blindly, especially in the 
crucial task of learning how to extract the Lagrangian parameters from
observables.

\setcounter{footnote}{0}
\section{Brief overview of models of ${\mathcal{L}}_{soft}$}
\label{lsoftmodelsect}

For phenomenological purposes, the MSSM Lagrangian described in the
previous sections should be viewed simply as a low energy effective
Lagrangian with a number of input parameters; we have seen that the
supersymmetry-breaking sector alone includes at least 105 new parameters.  
While often only subsets of these parameters are relevant for particular
experimental observables, in general the number of parameters is too large
for practical purposes to carry out phenomenological analyses in full
generality.  Furthermore, as outlined in the previous section, a number of
phenomenological constraints indicate that generic points in MSSM-124
parameter space, {\it i.e.}, with all mass parameters of $O({\rm TeV})$,
general flavor mixing angles and phases of $O(1)$, are excluded. 
Acceptable phenomenology does occur for certain regions of the MSSM-124
parameter space; unfortunately, a full map of all the allowed regions of
this parameter space does not exist.  These regions include (but are not
limited to) those clustered about the pattern of soft terms of
Eq.~(\ref{diagm2}) and Eq.~(\ref{AproptoY}).

In a top-down approach, the MSSM parameters are predicted within the
context of an underlying theory, often as functions of fewer basic
parameters.  Specific models can be constructed which approach or
reproduce the minimal/universal scenarios, often further simplifying the
number of independent parameters.  For convenience and practicality,
phenomenological analyses of supersymmetry have always been restricted to
models for the \lsoft parameters which exhibit such drastic
simplifications; as a consequence many results of such analyses are
model-dependent.

In this section, a brief summary of the various classes of models for the
\lsoft parameters is provided.  A proper summary of the various approaches
and models would be a subject for a review in itself.  The following
discussion is meant to familiarize the reader with certain theoretical
frameworks and prototype models which are often used in phenomenological
analyses.

\subsection{TeV scale supersymmetry breaking}
The basic question to be addressed is how to understand the {\it
explicit} soft supersymmetry breaking encoded in the \lsoft parameters as
the result of {\it spontaneous} supersymmetry breaking in a more
fundamental theory. To predict the values of the \lsoft parameters
unambiguously within a more fundamental theory requires a knowledge of the
origin and dynamics of supersymmetry breaking. Despite
significant effort and many model-building attempts, the mechanism of 
spontaneous supersymmetry breaking and how it might be implemented 
consistently within the underlying theory is still largely unknown.  

The most straightforward approach to a theory of \lsoft
is to look at spontaneous breaking of supersymmetry through
the generation of TeV scale F and/or D term VEVs in the MSSM,
or simple extensions of the MSSM. Scenarios of TeV scale
supersymmetry breaking are also called ``visible sector'' supersymmetry breaking,
for reasons which will become apparent in the next subsection.

Remarkably, it is already known that any tree level approach
to TeV scale spontaneous supersymmetry breaking necessarily
leads to an experimentally excluded pattern of bosonic and
fermionic masses assuming the particle content of the MSSM.
Consider a supersymmetric theory with gauge-neutral matter fields 
$\Phi_i$, for 
which the scalar potential $V \propto \sum F_i F_i^{*}$. The potential is 
positive definite and hence the absolute minimum occurs when
$F_i = 0$. The supersymmetric transformation rules imply that 
this absolute minimum  is also supersymmetry preserving.\footnote{To see this 
explicitly, consider the vacuum expectation value of the supersymmetric 
transformation rules of the fermions:  $\langle \delta \psi \rangle = 
\langle i (\sigma^{\mu} \epsilon^{\dagger})
\partial_{\mu} \phi + \epsilon F\rangle$. Lorentz invariance forbids a 
nonzero VEV for the first term but allows a nonzero VEV for the F 
term. If $\langle F \rangle \neq 0$, $<\delta \psi> \neq 0$ 
and supersymmetry is not preserved.} 
It is possible though to construct a scalar
potential in such a way that the $F_i$'s can not be set to zero
simultaneously. This can be achieved using a simple
renormalizable Lagrangian as first shown by O'Raifeartaigh
\cite{O'Raifeartaigh:pr}.
The MSSM coupled directly to such an O'Raifeartaigh sector will
exhibit spontaneous supersymmetry breaking at tree level.
 
Unfortunately this does not lead to a phenomenologically viable pattern of 
supersymmetry-breaking parameters. This can be seen from the following sum 
rule, known as the supertrace relation, for particles of spin $J$ 
\cite{Ferrara:wa,Dimopoulos:1981zb} 
\begin{equation}
\sum m^2_{J=0} - 2 \sum m^2_{J=\frac{1}{2}} + 3 \sum m^2_{J=1} = 0,
\label{eq:supertrace}
\end{equation}
which is valid in the presence of tree level supersymmetry
breaking.
The vanishing of this supertrace is fundamental to tree level soft
supersymmetry breaking, as it is simply the condition that one-loop 
quadratic
divergences cancel.

To understand why this sum rule leads to serious difficulties, consider
the SM particle content and their superpartners. As conservation of 
electric charge, color charge, and global symmetry charges such as 
baryon and lepton number prevents mass mixing between sectors of fields
differing in those quantum numbers, the sum rule holds
separately for each sector. For example, consider the charge
$-\frac{1}{3}$, color red, baryon number $-\frac{1}{3}$ and lepton number
0 sector. The only fermions in this sector are the three generations of
right-handed down type quarks, which contribute to the sum $2 (m_d^2 +
m_s^2 + m_b^2) \sim 2 (5$GeV$)^2$. This implies that in the rest of 
the sum none of the masses of the bosons could be greater than about $7$ 
GeV. Such light bosonic superpartners of quarks are clearly inconsistent 
with experimental searches. 

One can attempt to evade this problem by including D term supersymmetry 
breaking at tree level. For example a Fayet-Iliopoulos term 
\cite{Fayet:jb}
for $U(1)$ hypercharge
can break supersymmetry via a D term VEV. The MSSM mass splittings
are then determined by the known SM hypercharge assignments, but one
again fails to obtain a viable spectrum. One is then led to extensions
of the MSSM which have additional $U(1)$ gauge symmetries. To cancel
anomalies, this generally also requires the addition of extra chiral
superfields which carry SM quantum numbers. In any such
model, the effect on the supertrace formula (\ref{eq:supertrace})
is to replace the right hand side by D term contributions proportional
to traces over the new $U(1)$ charges. However these traces must all 
vanish, as otherwise they imply mixed gravitational-gauge anomalies,
and produce a one-loop quadratically divergent contribution to the
corresponding Fayet-Iliopoulos parameter \cite{Fischler:1981zk}. 
Thus one expects that
all such models have difficulty generating sufficiently large
superpartner masses.

Indeed, the best existing models \cite{Hall:1982vu,Oshimo:rn}
of tree level supersymmetry breaking
in an extended MSSM fail to obtain superpartner spectra consistent
with current experimental lower bounds. Thus TeV scale
supersymmetry breaking would appear to be ruled out by experiment.
Like most ``no-go'' results,
this one should be taken with a grain of salt.
The supertrace formula is only valid at tree level,
and assumes minimal (thus renomalizable) kinetic terms.
It may be possible to get viable spectra
from models similar to \cite{Hall:1982vu,Oshimo:rn}
by including loop effects and raising somewhat
the scale of supersymmetry breaking, from TeV to $\sim 10$ TeV \cite{Kumar:tbp}.
Or one can consider models in which the MSSM is
enhanced by new strong interactions and new mass scales,
such that the effective low energy Lagrangian for the
MSSM fields has nonvanishing supertrace. This is
the route taken in models of direct gauge mediation, discussed
below, but these already require raising the scale
of supersymmetry breaking to at least $\sim 100$ TeV \cite{Cheng:1998nb}.

\subsection{The hidden sector framework}
The negative results of the previous subsection
are a strong motivation to consider alternatives to TeV scale spontaneous
supersymmetry breaking in a renormalizable Lagrangian. As first noted by
\cite{Polonyi:pj,Nilles:1982ik,Nilles:1982xx,Nilles:1982dy}, a resolution
of this issue leads one to assume that the theory can be split
into at least two sectors with no direct renormalizable couplings between 
them:
\begin{itemize}
\item The {\it observable} or {\it visible sector}, which contains the SM 
fields and their 
superpartners.

\item The {\it  hidden sector}, in which supersymmetry is spontaneously 
broken by a dynamical mechanism, such as gaugino condensation.  

\end{itemize}

Within this framework, supersymmetry breaking is communicated from the
hidden sector where it originates to the observable sector via suppressed
interactions (loop-suppressed or nonrenormalizable 
operators) involving a third set of fields, the mediator or messenger 
fields. The
result is the effective soft supersymmetry breaking Lagrangian, \lsoft, in
the observable sector. Though somewhat {\it ad hoc}, this approach is 
successful in that the sum rule (\ref{eq:supertrace}) can be avoided, and
it can be easily realized in a wide variety of models.
Since the mediator interactions which generate
\lsoft are suppressed, the hidden sector framework implies that
the fundamental scale of supersymmetry breaking $M_S$, as exemplified by 
the F and/or D term VEVs, is hierarchically larger than the TeV scale.
Indeed, as we will see later, $M_S$
may be related to other postulated heavy mass
scales, such as the Majorana neutrino mass scale, the GUT scale,
or scales in extra-dimensional braneworlds.

Because both $M_S$ and the scales associated with the mediator
interactions are much larger than the TeV scale, renormalization group
analysis is necessary in order to obtain the low energy values of the
\lsoft parameters. Specific mechanisms for how supersymmetry breaking is
mediated between the hidden and observable sectors imply specific energy
scales at which the soft terms are generated. These generated values are
then used to compute the values at observable energy scales, using the
scale dependence of the \lsoft parameters as dictated by their RGEs.

The two-loop MSSM RGEs are presented in 
\cite{Martin:1993zk}, in which
the two-loop beta functions for the soft parameters were derived. We 
refer the reader to this paper and the references therein for earlier work 
on the beta functions of the supersymmetric sector such as the gauge 
couplings and Yukawa couplings.  While the one-loop RGEs are in general 
not sufficient for detailed phenomenological analyses, they encapsulate 
much of the essential physics.  Hence,  the complete set of 
one-loop renormalization group equations is presented for reference in 
Appendix~\ref{rgeapp}.  There have been many phenomenological analyses of 
the MSSM soft parameters. Classic
studies include
\cite{Arason:eb,Barger:1993vu,Castano:1993ri,Barger:1993gh}. In this
review, we will not present a complete RG analysis of the soft
parameters in different scenarios.  This type of study has evolved
into a large industry in recent years. Rather, we will explain the
necessary details of RG running when necessary and refer further
detail to the references.

\subsection{A taxonomy of hidden sector models}
There is a bewildering variety of phenomenologically viable
hidden sector models already on the market, many developed
in just the past few years. To organize our thinking, we need a 
reasonable taxonomy for these models. What constitutes a reasonable
taxonomy depends entirely on what you care about,
which in our case is the different patterns of \lsoft parameters which
are the outputs of these models. Thus we need to understand
what characteristics of hidden sector models are most important
in determining the resultant patterns of \lsoft parameters.

As it turns out, the pattern of MSSM soft terms depends
most crucially upon
\begin{itemize}
\item What is the mediation mechanism of supersymmetry breaking.
\item Which fields get the largest F and/or D term VEVs.
\item What are the dominant effects producing the couplings between
these VEVs and the MSSM fields: tree level, one-loop, one-loop anomaly, 
two-loop, nonperturbative, Planck scale. 
\end{itemize}

Surprisingly, the pattern of the soft terms usually turns out to
be relatively insensitive to the exact mechanism of the supersymmetry
breaking initiated in the hidden sector. 
While this
is good news in that our ignorance of the origin of supersymmetry breaking does not
prevent us from doing phenomenological analyses of theories such as the
MSSM with softly broken supersymmetry, it is unfortunate that
it becomes more difficult to infer the mechanism of supersymmetry breaking
from data. 

Many generic features of the soft terms are determined by the
basic mechanism by which
supersymmetry breaking is mediated to the observable sector.
The known scenarios for the mediation of
supersymmetry breaking are {\it gravity mediation}, {\it gauge
mediation}, and {\it bulk mediation}. These are the highest
level classifications in our taxonomy. Simply put, in gravity mediation
the soft parameters arise due to couplings which vanish as
$M_{Pl} \rightarrow \infty$. In gauge mediation, the soft parameters
arise from loop diagrams involving new messenger fields with SM
quantum numbers.
In bulk mediation, the hidden and observable
sectors reside on different branes separated in extra dimensions,
and supersymmetry breaking is mediated by fields which propagate in between
them, ``in the bulk.''

Even this highest level of our taxonomy is not completely clean.
For example, since gravity is a bulk field, some subset of gravity mediation
models are also bulk mediation models; these are among the ``sequestered''
supergravity models discussed below. Another example is
models of ``direct'' gauge mediation, which could
as well be classified as visible sector supersymmetry breaking models, with
their additional dynamics allowing them to circumvent the no-go
results reviewed earlier.

\subsection{Gravity mediated supersymmetry breaking}
\label{grav_susyb_sect}
As gravitational interactions are shared by all particles,
gravity is a leading candidate for the mediation of supersymmetry 
breaking. It is quite natural to imagine gravity (and whatever 
Planck-suppressed effects accompany gravity) to be
the only interaction shared by both the hidden and the observable sector.
Such a situation can be naturally addressed within $N=1$ supergravity, 
which is a nonrenormalizable supersymmetric effective field theory of 
gravity coupling to matter obtained by gauging global supersymmetry.  
Supergravity was already introduced in this review in 
subsection~\ref{briefsusyintro} and further details are presented in
Appendix~\ref{sec:sugra}. All gravity mediated models are based
on the formalism of $N=1$ supergravity, sometimes with additional stringy
or higher dimensional refinements. Note that gravity mediation
does not refer to interactions involving graviton exchange, but rather to
supergravity interactions dictated by the necessity,
in the presence of gravity, of promoting
global supersymmetry to local supersymmetry.

Within the framework of $N=1$ supergravity, local supersymmetry is assumed to 
be spontaneously broken in the hidden sector and mediated to the observable
sector by Planck-suppressed nonrenormalizable contact terms. These contact
terms couple hidden sector fields to visible sector fields; their
existence is required by local supersymmetry and their form is {\it almost}
completely fixed by symmetry considerations alone. These powerful symmetry
considerations are what allow us to make predictive statements from
nonrenormalizable theories of Planck scale physics.  

The mediating contact terms can be regarded as couplings of the
visible sector fields to F term VEVS of supergravity auxiliary
fields.
Since the supergravity interactions are 
Planck-suppressed, on dimensional grounds the soft parameters generated in 
this way are of order 
\begin{equation}
m\sim \frac{F}{M_{Pl}}.
\end{equation}
For $m \sim O({\rm TeV})$, the scale of spontaneous supersymmetry breaking 
$M_S\sim \sqrt{F}$ is $10^{11-13}$ GeV.
This dimensional analysis is modified in the case of
dynamical breakdown of supersymmetry 
via gaugino condensation in the hidden
sector \cite{Ferrara:1982qs}. A gaugino condensate 
$\langle\lambda^a\lambda^a\rangle \sim \Lambda ^3$ is not itself
an F term, but can appear in the F terms of matter superfields
due to nonrenormalizable couplings allowed by supergravity.
The resulting F term VEVs are of order $\Lambda^3/M_{Pl}$,
and thus generate soft terms of order $\Lambda^3/M_{Pl}^2$.
In this case TeV soft terms implies that the gaugino
condensation scale $\Lambda$ should be $10^{13-15}$ GeV.   

Goldstone's theorem dictates that if a global symmetry is
spontaneously broken, there will be a massless (Goldstone) particle with 
the same spin as the broken symmetry generator. For spontaneously broken 
supersymmetry, this implies the presence of a massless fermion, since the 
supersymmetry generators are spinors. This massless
fermion is called the Goldstino $\widetilde{G}$. For spontaneously broken 
local or gauge  symmetries, the Higgs mechanism states that the massless 
Goldstone particle will be eaten to become the longitudinal component of 
the corresponding massive gauge field. For 
spontaneous local supersymmetry breaking
in supergravity, the supersymmetric version of the Higgs mechanism (the 
superHiggs mechanism) implies that the Goldstino will be eaten by the
gravitino (the spin 3/2 partner of the spin 2 graviton), such that the
gravitino becomes massive, with 
\begin{equation}
m_{\widetilde{G}} \sim \frac{M_S^2}{M_{Pl}}.
\end{equation}

In gravity mediated supersymmetry breaking, the gravitino mass
$m_{\widetilde{G}}$ generically sets the overall scale for all of the soft
supersymmetry breaking mass parameters. In fact, the supertrace in
(\ref{eq:supertrace}) does not vanish for gravity mediated
supersymmetry breaking, instead it is positive and proportional to
$m_{\widetilde{G}}^2$. This implies that on the average bosons are
heavier than fermions, a result which is certainly more in concert with
experimental observations than (\ref{eq:supertrace}).

As previously discussed, the Lagrangian of $N=1$ supergravity, shown 
explicitly in Appendix~B, is completely fixed by symmetries up to the 
specification of three functionals of the matter superfields: the 
K\"{a}hler potential $K$, the superpotential $W$, and the gauge kinetic 
functions $f_a$, where $a$ labels the gauge groups.  

At tree level, the soft breaking parameters can be computed   
directly from the supergravity Lagrangian 
\cite{Hall:iz,Soni:1983rm,Lykken:th}; 
this is explained in more detail in Appendix~B.
The details of the resulting soft supersymmetry breaking terms for the
observable sector will of course depend crucially on the assumed form of
the functionals given above and their dependence on the F and D term
VEVs that break supersymmetry. In all cases what is determined are the
high energy values of the soft parameters, and an RGE analysis is necessary
to run these values down to lower energies. The high energy scale is
either the Planck scale, the string scale\footnote{Estimates of the
string scale range from a few times $10^{17}$ GeV down to as low
as a few TeV \cite{Lykken:1996fj}. Models with an intermediate string scale
$10^{11}$ GeV can still be accommodated by the supergravity framework
discussed here \cite{Allanach:2001qe}.},  
or the GUT scale, depending
upon how one is imagining matching the effective $N=1$ supergravity
Lagrangian onto a more fundamental ultraviolet theory.

As explained in Appendix B, the $N=1$ supergravity Lagrangian has a
tree level invariance under K\"{a}hler-Weyl transformations. When
supersymmetry is broken this invariance can be used to express $K$ and $W$
in terms of a single functional $G$:
\begin{equation}
G = {K\over M_{Pl}^2} + \ln{W\over M_{Pl}^3} + \ln{W^*\over M_{Pl}^3} \,.
\end{equation}
The choice of the functional $G$ will determine, among other things,
the pattern of soft scalar masses, the trilinear $A$ terms, and the
bilinear $B$ term.
$G$ can also be chosen in a way (the Giudice-Masiero
mechanism) that naturally gives a value for the $\mu$
parameter of order $m_{\widetilde{G}}$,
and $G$ can be fine-tuned to make the cosmological constant
vanish after supersymmetry breaking.

The gaugino masses are determined by the gauge kinetic terms
$f_a$. At the renormalizable level the $f_a$ are just
constants
\begin{equation}
f_a = {4\pi\over g_a^2} + {i\theta_a\over 2\pi}\; .
\end{equation}
However these functionals may also include tree-level
(Planck-suppressed) couplings to F term VEVs of messenger superfields, 
which if present imply tree level gaugino masses of order $m_{\widetilde{G}}$.
Gauge invariance requires that
these messenger superfields must be singlets under the SM gauge group.
More generally in a GUT framework these messenger fields
must transform in a representation of the GUT gauge group that
is contained in the tensor product of two adjoints \cite{Anderson:1996bg}.

\subsection{Taxonomy of gravity mediation models}
\label{grav_models_sect}
From the above discussion it would seem that the obvious way to
make hidden sector models with gravity mediation is by theoretically
motivated choices of the functionals $K$, $W$, and $f_a$.
However, to understand the underlying physics, it is better
to approach this model building in two stages. 

Consider first the
limit in which all of the supergravity fields are {\it turned off}.
Let $K^0$, $W^0$ and $f_a^0$ denote the K\"{a}hler potential,
the superpotential and the gauge kinetic functions in this limit.
At the renormalizable level $K^0$ and $W^0$ are just bilinear
and trilinear polynomials of the superfields, while the
$f_a^0$ are just constants. The hypothesis of the hidden sector
places a strong constraint on the form of $K^0$ and $W^0$:
\begin{eqnarray}
K^0(\Phi^{\dagger},\Phi) &= K^0_{vis} + K^0_{hid} \; ,\\
W^0(\Phi) &= W^0_{vis} + W^0_{hid} \; ,
\end{eqnarray}
where  $K^0_{vis}$,$W^0_{vis}$ are functionals only of the visible
sector fields, while $K^0_{hid}$,$W^0_{hid}$ are functionals
only of the hidden sector fields. 

We expect that $K^0$, $W^0$ and $f_a^0$
also contain explicit nonrenormalizable
couplings, suppressed by powers of $M_{Pl}$. These Planck
suppressed couplings are determined, in principle, by
matching this effective Lagrangian onto whatever is the more
fundamental Planck scale theory ({\it e.g.} string theory). The
hypothesis of the hidden sector does {\it not} imply the absence
of nonrenormalizable couplings which contain both visible and
hidden sector fields. In general such mixed couplings will be present,
and they represent supersymmetry breaking mediated not by supergravity 
{\it per se}, but
rather by other Planck scale physics ({\it e.g.} string mode exchange or
couplings dictated by stringy symmetries).

Thus an essential part of building gravity mediation models
is the specification of these explicit Planck suppressed couplings
between the visible and hidden sectors. This is done either
by deriving these couplings from a particular stringy scenario,
or just by postulating some simple form. Several classes of
gravity mediation models are distinguished by this specification:
\begin{itemize}
\item {\bf Dilaton dominated supersymmetry breaking models:} The dilaton
superfield is inevitable in string theory, and the dilaton
dependence of $K^0$, $W^0$ and $f_a^0$ for weakly coupled
strings is completely specified at the
perturbative level \cite{Brignole:1993dj,Brignole:1995fb,Binetruy:2000md}. 
Other considerations, e.g. string dualities and the
dilaton ``runaway'' problem, give us important information
about nonperturbative couplings involving the
dilaton \cite{Binetruy:1996xjb,Nelson:2003tk}.
Hidden sector gaugino condensation automatically generates
an F term VEV for the dilaton. Thus if this dilaton F
term turns out to be the dominant contribution to visible
sector supersymmetry breaking, we obtain a well-motivated scenario
for generating \lsoft that has essentially no free parameters
besides $m_{\widetilde{G}}$.
\item {\bf Moduli dominated supersymmetry breaking models:} String
theory also contains many other (too many other) moduli superfields,
associated with the various possibilities for string compactifications.
In some cases the dependence of $K^0$, $W^0$ and $f_a^0$ on
other moduli can be constrained almost as well as for the dilaton,
and one can make strong arguments that these moduli obtain F term
VEVS, which may be the dominant contribution to visible sector
soft terms. Thus again one obtains well-motivated scenarios
for generating \lsoft that have very few free parameters. It is
also popular to consider scenarios where a combination of dilaton
and moduli F term VEVs dominate, with ``goldstino angles''
parametrizing the relative
contributions \cite{Brignole:1997dp,Allanach:2001qe}.
\item {\bf Sequestered models:}
The simplest assumption about explicit nonrenormalizable couplings
--- in the limit that supergravity is turned off ---
is to postulate that all Planck suppressed
mixed couplings are absent. Such models are called
{\it sequestered}. In the general context of gravity mediation
this choice is poorly motivated. We will see later, however, that
in the context of bulk mediation sequestered models are very natural,
if we imagine that the visible and hidden sectors reside on different
branes \cite{Randall:1998uk}.
\end{itemize}

Now let us turn supergravity back on, and ask in more detail
how supergravity itself communicates supersymmetry breaking in the hidden sector
to visible sector fields. The off-shell $N=1$ supergravity multiplet
only contains one scalar field: a complex auxiliary field $u(x)$.
Thus since we are attempting to communicate supersymmetry breaking (at leading
order in $1/M_{Pl}^2$) with supergravity messengers,
it is not surprising that this occurs entirely via couplings of the
visible sector fields to $u(x)$, which has a nonzero VEV induced
by hidden sector supersymmetry breaking. A covariant approach to studying these
couplings is to introduce a ``spurion'' chiral superfield $\phi$,
defined as
\begin{equation}
\phi = 1 + \theta^2 F_{\phi} = 1 + \theta^2 u/3 \; .
\end{equation}
The couplings of $\phi$ then determine in an obvious way the soft
terms induced in the visible sector.

As already noted, couplings of $\phi$ to the visible sector
are required by local supersymmetry. In fact these couplings are
modifications (replacements) for the couplings that we had input
with supergravity turned off. Remarkably these modified couplings
are determined from the original couplings very simply,
by the broken super-Weyl invariance of $N=1$ supergravity. 
The rule is that $\phi$ appears only in couplings that were
not scale invariant, and that $\phi$ appears to the appropriate
power such that the contribution from its canonical scale dimension
renders the modified couplings scale invariant (we are ignoring
some complications here but this is the basic idea). Thus for
example \cite{Rattazzi:1999qg} if we had chosen
\begin{equation}
W^0(C) = m_1C^2 + \lambda C^3 + {1\over m_2}C^4
\end{equation}
as the superpotential for a visible sector chiral superfield $C$
with supergravity turned off, then with supergravity turned on
we obtain:
\begin{equation}
W(C) = m_1\phi C^2 + \lambda C^3 + {1\over m_2\phi}C^4
= \phi^3 W^0(C/\phi) \; .
\end{equation}
This is a powerful result. It implies that, at tree level, supergravity
{\it per se} does not generate any soft terms for a scale invariant
visible sector. Since the renormalizable couplings of the MSSM are
all scale invariant with exception of the $\mu$ term, only the $B$
bilinear soft term arises from tree-level supergravity couplings
to a renormalizable MSSM. All of the other soft terms can arise
only through loop-induced MSSM supergravity couplings, or through
nonrenormalizable (and scale noninvariant) MSSM couplings.

Let us now ask
what is the condition to have a sequestered model once supergravity
is turned on, {\it i.e.} what form is required for $K$, $W$ and $f_a$?
Since  $W(C) = \phi^3 W^0(C/\phi)$, we could just as well have
written $W = W_{vis} + W_{hid}$ as the condition for a sequestered
superpotential in supergravity. The same comment applies for the
gauge kinetic functions $f_a$. However, things are more complicated
for the supergravity K\"{a}hler potential $K$, which has a nonlinear
relation to the input K\"{a}hler potential $K^0$:
\begin{equation}
K(C,h) = -3M_{Pl}^2\ln \left( 
1 - {\phi^{\dagger}\phi K^0(C/\phi,h/\phi)\over 3M_{Pl}^2} \right) \; ,
\label{eq:kahlerrel}
\end{equation}
where $C$ and $h$ denote visible sector and hidden sector superfields,
and we have suppressed complications involving derivatives. Note
that, expanding in powers of $1/M_{Pl}^2$ and suppressing the $\phi$
dependence:
\begin{equation}
K(C,h) = K^0 + {\cal O}\left( {(K^0)^2\over M_{Pl}^2}\right) \; .
\end{equation}
Thus a sequestered $K^0$ does {\it not} imply that $K$ is of the
form $K= K_v(C) + K_h(h)$, nor vice-versa.
Instead we see from (\ref{eq:kahlerrel}) that sequestering
implies a supergravity K\"{a}hler potential with the following
special form:
\begin{equation}
K(C,h) = -3M_{Pl}^2 \ln \left[ -{K_v(C)\over 3M_{Pl}^2} - {K_h(h)\over 
3M_{Pl}^2}
\right] \; .
\label{eq:sequesterK}
\end{equation}

Several classes of
gravity mediation models are distinguished by these considerations:

\subsubsection{Anomaly mediation}
The renormalizable couplings of the MSSM are
all scale invariant at tree level with the exception of the $\mu$ term.
However at the loop level all of the couplings run, and this
renomalization scale dependence represents an anomaly in the scaling
symmetry. Thus at the loop level we induce soft-term-generating
supergravity couplings from all of the couplings of the MSSM.
Furthermore the soft terms generated by these effects are computable
in terms of the beta functions and anomalous dimensions of the
MSSM sector. If we turn off {\it all} of the nonrenormalizable
visible sector and mixed couplings in $K^0$, $W^0$ and $f_a^0$,
then this anomaly mediation will be the dominant (only) source
of \lsoft \cite{Randall:1998uk,Giudice:1998xp}.

The soft masses in a pure AMSB scenario can be obtained using either the 
spurion technique (see {\it e.g.} \cite{Pomarol:1999ie}) or by carefully 
regulating the supersymmetric Lagrangian (see {\it e.g.}  
\cite{Giudice:1998xp,Boyda:2001nh,Binetruy:2000md}).  In the minimal 
realization of AMSB, the soft parameters are given by
\begin{eqnarray}
\label{amsbparams}
M_a &=& \frac{\beta_{g_a}}{g_a} m_{3/2}, \nonumber \\
m^2_{\widetilde{f}} &=& -\frac{1}{4} \left( \frac{\partial \gamma }{\partial
g } \beta_g + \frac{\partial \gamma }{\partial y } \beta_y \right)
m_{3/2}^2 \nonumber \\
A_y &=& - \frac{\beta_y}{y} m_{3/2},
\end{eqnarray}
in which $y$ collectively denotes the Yukawa couplings. The
$\beta$-functions and anomalous dimensions $\gamma$ are functions of 
the gauge couplings and superpotential parameters.  Typically soft 
supersymmetry breaking masses generated this way are of order the 
gravitino mass suppressed by a loop factor,  
\begin{equation}
m\sim \frac{m_{3/2}}{16\pi^2},
\end{equation}
which implies that for soft masses of order a TeV, the gravitino mass 
should be about two orders of magnitude larger. 

An interesting feature of Eq.~(\ref{amsbparams}) is that the
form of the soft parameters is scale independent, provided the appropriate 
running parameter is used in the computation of $\beta$ and $\gamma$. The 
UV insensitivity reflects the elegant solution of the flavor problem 
within anomaly mediation: the soft masses are independent of high energy 
flavor violating effects.

The soft parameters in anomaly mediation have distinctive phenomenological
implications.  The main feature is that the gaugino masses are in the
ratio: \begin{equation} M_1:M_2:M_3=2.8:1:7.1 \end{equation} such that the
LSP is the neutral wino, which is only slightly lighter than the charged
wino (by a few hundred MeV). This leads to a long lived lightest chargino
with a distinctive signature
\cite{Feng:1999fu,Gherghetta:1999sw,Gunion:1999jr}. The wino LSP also has
interesting implications for dark matter (see {\it e.g.} 
\cite{Moroi:1999zb}).

Unfortunately, there is also an unattractive phenomenological prediction
of the AMSB soft parameters of Eq.~(\ref{amsbparams}).  The problem
is that the slepton mass squareds turn out to be negative, which is
clearly unacceptable (this leads to charge breaking minima, as discussed
in Section~\ref{CCBsect}). The slepton mass problem has many proposed
solutions, of which the simplest \cite{Gherghetta:1999sw} is to add
a common $m_0^2$ to the scalar mass-squares. However, one can argue that
such a phenomenological solution undermines the elegant solution to the
flavor problem in the flavor problem, because there is no fundamental
reason to assume that the additional physics responsible for generating
the $m_0^2$ contribution is flavor blind.  Other solutions include
``deflected" anomaly mediation \cite{Pomarol:1999ie,Rattazzi:1999qg},
coupling additional Higgs doublets to the leptons \cite{Chacko:1999am},
and combining this mechanism with D term supersymmetry breaking
\cite{Jack:2000cd,Jack:2002pn,Murakami:2003pb}, among others.

\subsubsection{No-scale models} 
No-scale models 
\cite{Cremmer:1983bf,Ellis:1983sf,Ellis:1984kd,Lahanas:1986uc}
are a special case
of the sequestered models discussed above. Let us suppose
that the hidden sector includes
a singlet (modulus) superfield $T$. $T$ does not
appear in the superpotential, but hidden sector
gaugino condensation produces a VEV for the superpotential,
breaking supersymmetry.
We further assume that
that the supergravity K\"{a}hler potential is of the sequestered
form (\ref{eq:sequesterK}) with
\begin{equation}
K(C,h,T) = -3\ln \left[ T + T^{\dagger} -C^{\dagger}C - K_h(h) \right] \; ,
\end{equation}
where we have suppressed factors of $M_{Pl}$.
In this sort of model the (high scale) values of the soft scalar
masses and the trilinear $A$ terms all vanish at
tree level. The cosmological constant also vanishes
automatically at tree level. Interestingly, the anomaly mediated
contributions to the gaugino masses also vanish
in this model \cite{Bagger:1999rd},
but we can generate gaugino masses at tree level through
$T$ dependent (nonrenormalizable) gauge kinetic functions.
Obviously no-scale models have the virtue of a small number of free 
parameters.
It has been argued that the strongly coupled heterotic string
produces a no-scale effective theory \cite{Banks:1996ss}.

\subsubsection{Minimal supergravity} 
This model is obtained by
assuming universal gauge kinetic functions for the three 
SM gauge groups, with tree level gaugino mass generation, and
by assuming that the supergravity K\"{a}hler potential has
the ``canonical'' form:
\begin{equation}
K(\Phi_i) = \sum_i \vert \Phi_i \vert^2 \; ,
\end{equation}
where the label $i$ runs over all the MSSM chiral superfields
and at least those hidden superfields which participate in supersymmetry 
breaking. The assumption of a canonical
K\"{a}hler potential produces (at the high scale) universal soft
scalar masses, and a common overall soft trilinear parameter \cite{Hall:iz}.
The resulting model of the \lsoft parameters is often 
labeled as the {\it minimal supergravity}
(mSUGRA) model \cite{Nilles:1983ge}. A subset of the mSUGRA
parameter space gives low energy models that satisfy the
basic phenomenological requirements ({\it e.g.} electroweak symmetry 
breaking) incorporated into what is known as the
constrained MSSM (CMSSM) \cite{Kane:1993td}.  
The CMSSM is by far the most popular scenario for \lsoft amongst
phenomenologists and experimenters; more phenomenological analyses
have been performed for mSUGRA/CMSSM than for all other scenarios combined.
 
The complete list 
of mSUGRA soft parameters is:
\begin{itemize} 
\item a common gaugino mass $m_{1/2}$

\item  a common soft scalar mass $m_0$

\item a common soft trilinear parameter $A_0$ 
($\widetilde{A}_{ij}=A_0Y_{ij}$)

\item a bilinear term $b_0$

\end{itemize}
These parameters plus the $\mu$ term
are often traded for the mass of the $Z$ boson 
$m_Z$, $\tan\beta$, and the sign of $\mu$ relative to $m_{1/2}$ or $A_0$ 
by imposing consistent radiative electroweak symmetry
breaking, as will be discussed in Section~\ref{EWSBsect}. The 
origin of $\mu$ and $b$ is quite model-dependent,
and hence it is can be useful to trade their magnitudes for $m_Z$ and
$\tan\beta$ to implement the phenomenologically desirable radiative
electroweak symmetry breaking mechanism.  This, however, does not
constrain the phases of the parameters, or the overall signs (if the
parameters are real).  The phase of $b$ can always be
consistently rotated to zero using the PQ symmetry, while the phase of
$\mu$ relative to the other soft parameters is undetermined. These issues 
will be discussed in Section~\ref{EWSBsect}.
In general the
PQ and R symmetries allow only two irremovable phases.  The two
reparameterization invariant combinations are often written as ${\rm
Arg}(A_0^* m_{1/2})$ and ${\rm Arg}(A_0 B^*)$.

The alert reader will have already objected that the assumption
of a canonical supergravity K\"{a}hler potential has very poor
theoretical motivation, since from (\ref{eq:kahlerrel})
we see that this assumption requires that, with supergravity
turned off, we have a conspiracy between a noncanonical
K\"{a}hler potential and explicit Planck suppressed couplings.
However it was shown in \cite{Hall:iz} that the CMSSM will
also arise from the MSSM if we assume that the K\"{a}hler potential
is canonical with supergravity turned off, or more generally from
the entire $U(N)$ symmetric class of K\"{a}hler potentials which are
functionals only of $\sum_{i=1}^{N} \vert \Phi_i \vert^2$.
This is a stronger result, but this $U(N)$ symmetry is certainly
not respected by the superpotential, and is generally violated in
string-derived models \cite{Choi:1997cm,Kobayashi:1994eh}. 

By the same token string-derived models
generally violate the assumption of universal
gaugino masses \cite{Kane:2002qp}.
One can attempt to impose gaugino mass universality at the high
scale via grand unification, but in a real model GUT threshold
effects will typically give significant departures from \lsoft
universality for the effective theory below the
GUT breaking scale \cite{Polonsky:1994rz}.

\subsection{Gauge mediated supersymmetry breaking}
Theories in which supersymmetry breaking is mediated by gauge interactions
provide an important alternative framework to gravity
mediation for constructing models of the soft supersymmetry breaking
parameters.  The canonical models were first put forth in the older works
of \cite{Dine:za,Dimopoulos:1981au,Nappi:1982hm,Alvarez-Gaume:1981wy} but
interest was renewed in the scenario by models of Dine, Nelson and
collaborators \cite{Dine:1993yw,Dine:1994vc,Dine:1995ag}.  

The ingredients of gauge mediated supersymmetry breaking (GMSB) in its
most basic implementation are as follows.  As usual, there is the 
observable sector and the hidden sector, where as usual supersymmetry is 
assumed to be broken dynamically such that nonzero F component VEVs of 
the hidden sector fields are generated. In addition, there is a {\it 
messenger} sector with messenger fields $S_i$. The messenger fields couple 
to the goldstino field of the hidden sector, which generates nonzero $F_S$ 
terms. The $S_i$ also couple to the MSSM gauge bosons and gauginos and 
are typically assumed to be complete
multiplets under a given GUT group to preserve successful gauge coupling
unification.  Supersymmetry breaking is then communicated to the observable sector
through radiative corrections involving messenger field loops to the
propagators of the observable sector fields. 
On purely dimensional grounds, it can be inferred that the soft mass
spectrum resulting from this scenario is 
\begin{equation} 
M_a \sim \frac{g_a^2}{(4 \pi)^2}\frac{F_S}{M_S}, 
\end{equation} 
where $M_S$ is a typical mass scale associated with the messenger sector
and $g$ is an $O(1)$ gauge coupling. To estimate the sizes of $F_S$ and
$M_S$ which yield phenomenologically desirable soft supersymmetry breaking
mass parameters of $\sim O({\rm TeV})$: if $F_S \sim
M_S^2$, $M_S \sim 10^5$ GeV. For larger values of $F_S$ such as $F_S \sim
10^{14}$ GeV$^2$, $M_S \sim 10^9$ GeV. Therefore, $M_S$ is generally much
smaller in gauge mediated models than it is in gravity mediated scenarios
(even when $\sqrt{F_S} \ll M_S $). In models of ``direct'' gauge mediation,
where the messenger fields carry the quantum numbers of the gauge
fields that break supersymmetry, $M_S$ can be as low as 
100 to 1000 TeV \cite{Murayama:1997pb,Cheng:1998nb,Agashe:1998wm}.

The gauge mediation framework has certain advantages on theoretical and
phenomenological grounds.  A major success of gauge mediation is that
gaugino masses are generated at one-loop order, while scalar mass-{\it
squares} are generated at two-loop order. Generically, they are of the
form 
\begin{equation}
m^2_{\widetilde{f}} \sim \frac{g^4}{(16 \pi^2)^2} \frac{F_S^2}{M_S^2},
\end{equation}  
where we include the two-loop suppression factor explicitly. Hence,
gaugino and scalar masses are comparable in magnitude.  

In contrast, the soft trilinear $\widetilde{A}$ terms arise at two-loop order and
are negligible.\footnote{The issue of how $\mu$ and $b$ are generated is
more complicated; see Section~\ref{EWSBsect}.}
This underlies one of the advantages of the framework in 
that it is not necessary to work hard to achieve minimal flavor violation.
As gauge interactions are flavor-blind, the soft mass-squares are
automatically flavor diagonal as in Eq.~(\ref{diagm2}); the $\widetilde{A}$
terms are generated by RG evolution and thus are automatically of the form 
given in Eq.~(\ref{AproptoY}).

Since any fundamental theory must contain gravity, we must consider the
coupling of the present scenario to a supersymmetric generalization of
gravity, usually assumed to be 4-dimensional $N=1$ supergravity.  Given the
typical sizes of $F_S$ and $M_S$, gauge mediation provides the dominant
contribution to the \lsoft parameters. One main consequence of coupling
this supersymmetry breaking scenario to supergravity is that it will also
break local supersymmetry. However, due to the low value of $M_S$, the 
gravitino mass will be very light ($m_{\widetilde{G}} \sim M_S^2/M_{Pl}$) and 
is invariably the LSP 
within GMSB, leading to distinctive phenomenological signatures. Aspects 
of the phenomenology of gauge-mediated models are presented in 
Section~\ref{expsect}; see 
\cite{Dimopoulos:1996va,Dimopoulos:1996yq,Bagger:1996bt,Ambrosanio:1996jn}
and the review \cite{Giudice:1998bp} for details.

\subsubsection{Minimal gauge mediation}
Using these building blocks, there are many possibilities for
model building in the gauge mediation framework, {\it e.g.} by varying the
matter content and couplings of the messenger sector and the scale 
$\Lambda=F_S/M_S$. In this review, 
the examples we will consider will be minimal GMSB models (MGM), which are 
utilized in many phenomenological analyses \cite{Culbertson:2000am}.
In such models, the 
messenger sector is assumed to consist of $N_5$ complete vectorlike pairs 
of $SU(5)$ GUT 5-plets. The use of complete $SU(5)$ multiplets
preserves gauge coupling unification, and $N_5$ can be as large as
5 to 10 (depending on $M_S$) without spoiling perturbativity of the
theory up to the GUT scale.
In addition, once again the $\mu$ and $b$ 
terms are traded for $m_Z$, $\tan\beta$, and the sign of $\mu$ relative to 
the gaugino masses. The soft masses are given by:
\begin{eqnarray}
M_3&=&\frac{\alpha_s}{4\pi}\Lambda N_5\\
m^2_{\widetilde{e}_L}&=&\frac{3\alpha_2^2}{32\pi^2}\Lambda^2N_5+ 
\frac{3\alpha_1^2}{160\pi^2}\Lambda^2N_5\\
m^2_{\widetilde{e}_R}&=&\frac{3\alpha_2^2}{32\pi^2}\Lambda^2N_5+
\frac{3\alpha_1^2}{160\pi^2}\Lambda^2N_5\\
m^2_{\widetilde{u}_L}&=&\frac{\alpha_s^2}{6\pi^2}\Lambda^2N_5 
\frac{3\alpha_2^2}{32\pi^2}\Lambda^2N_5+
\frac{\alpha_1^2}{480\pi^2}\Lambda^2N_5\\
m^2_{\widetilde{u}_R}&=&\frac{\alpha_s^2}{6\pi^2}\Lambda^2N_5+
\frac{\alpha_1^2}{30\pi^2}\Lambda^2N_5.
\end{eqnarray}
Thus it appears that for minimal gauge mediation \lsoft
is determined by only three parameters ($\Lambda$,
$N_5$, $\tan\beta$) together with the sign of $\mu$. This is
not quite true, as the low energy spectrum obtained by RGE
running depends significantly on the starting point of the
RGE, {\it i.e.} on the high energy messenger scale $M_S$.

\subsubsection{The NLSP}
Since the gravitino is always the LSP in gauge mediation models,
superpartner decay chains terminate with the decay of
the next-to-lightest-superpartner (NLSP) into the goldstino component
of the gravitino. The decay length of the NLSP is given
by the formula \cite{Culbertson:2000am}:
\begin{equation}
c\tau(\widetilde{X}\rightarrow X\widetilde{G}) \simeq {\rm 100\ }\mu{\rm m}\,
\left({{\rm 100 GeV}\over m_{\widetilde{X}}}\right)^5
\left({\sqrt{F}\over {\rm 100 TeV}}\right)^4
\left(1-{m_X^2\over m_{\widetilde{X}^2}} \right)^{-4}
\, .
\end{equation}
Note that this decay length depends on the {\it instrinsic} supersymmetry
breaking scale $\sqrt{F}$, which may be larger than the effective supersymmetry
breaking scale $\sqrt{F_S}$ communicated by the messenger sector.
Thus this introduces another phenomenologically relevant parameter
$C_G \equiv F/F_S$. The NLSP decay length is of great importance,
since for $\sqrt{F}$ greater than about 1000 TeV, the NLSP will decay
outside a conventional collider detector.

In gravity mediated models, the identity of the LSP varies according
to models and parameters, but if R-parity is conserved models with
a neutralino LSP are strongly favored phenomenologically. For gauge
mediation there is no analogous phenomenological preference for a
neutralino NLSP. The lightest stau $\widetilde{\tau}$ is an
equally plausible candidate for the NLSP, and it is even possible to
construct models with a gluino NLSP. Furthermore it is not unlikely
in gauge mediation models to encounter ``co-NLSPs'', {\it e.g.} a nearly
degenerate lightest neutralino and lightest stau. 

In any taxonomy of gauge mediation models, it is crucial to make a
clear link between the underlying model parameters and
the identity of the NLSP or co-NLSPs. The identity and decay
length of the NLSP determines whether supersymmetry collider events
are characterized by hard photon pairs, leptons, Higgs, or 
exotic charged tracks.
The interested reader should consult the excellent reviews
\cite{Giudice:1998bp,Kolda:1997wt,Culbertson:2000am} 
for details of GMSB model building and the associated phenomenology.

\subsection{Bulk mediation}
\label{othermechsect}
Several supersymmetry breaking and mediation mechanisms are inspired by
brane-world constructions in which there are two 4D branes separated by a
single extra dimension.  In this review we do not generally consider extra
dimensional scenarios, but we do often mention string theory as a
candidate primary theory. String theories are generally formulated in
larger numbers of dimensions, with the
extra dimensions being either compactified with a small radius of
compactification, or warped in such as way as to make them consistent with
the apparent 4D description with which we are familiar. The discovery of
branes opens up the possibility that different sectors of the theory live
in different places, for example on either one of the two branes or in the
bulk, in the example of two 4D branes separated by a single extra
dimension mentioned above. Such a set-up is motivated by the Horava-Witten
construction for example \cite{Horava:1995qa}. In such scenarios, it is
possible to envisage supersymmetry breaking occuring on one of the branes
(the hidden brane), and part or all of the 
MSSM living on the other brane (the visible brane). As
already mentioned, this geometrical picture
of sequestering was first actively pursued
by \cite{Randall:1998uk} in the context of anomaly mediation.
The precise way
that the supersymmetry breaking is mediated to the brane in which we live
has given rise to several different scenarios in addition to anomaly
mediation.

\subsubsection{Gaugino mediation}
A now classic example within this context is {\it gaugino mediation}
($\widetilde{\rm g}$MSB)
\cite{Kaplan:1999ac,Chacko:1999mi}, which is similar to the anomaly
mediation scenario with the exception that the gauginos are now allowed to
propagate in the bulk and hence can have direct couplings to the
supersymmetry breaking on the hidden brane.
Therefore, their soft masses are $\propto
F/M$, where $F$ is supersymmetry breaking order parameter and $M$ is the
scale that characterizes the coupling between gaugino and the hidden
sector (since the coupling is usually of the form of a nonrenormalizable
term suppressed by $M$). With proper choice of $F$ and $M$, the gaugino
mass in this scenario can be chosen to be similar to any of the other
supersymmetry breaking mediation scenarios.  The soft scalar masses are
generated from loop diagrams in which gauginos propagate between the
visible sector brane and the supersymmetry breaking brane.  They are then
suppressed compared to the gaugino mass by a loop factor $m^2_{\widetilde{f}}
\sim M_{\lambda}/(16 \pi^2)$, but receive positive  
flavor-diagonal contributions proportional to the gaugino masses through
RG running. The flavor problem is thus alleviated in this scenario in a
way similar to gauge mediation.  There are a number of variations on this
basic theme (see {\it e.g.} 
\cite{Schmaltz:2000gy,Kaplan:2000jz,King:2000ir,Chacko:2001km}, 
among others).

\subsubsection{Radion mediation}
Brane scenarios generically have moduli fields called
radions related to the brane separations; with supersymmetry these become chiral
superfields that live in the bulk. Formally, this is no different
than the other string moduli superfields which we discussed in
the context of gravity mediation.
When gauge boson superfields
also live in the bulk, as in the $\widetilde{\rm g}$MSB models just
discussed, the radion superfield appears linearly in the gauge
kinetic terms. This means that an F term VEV for the radion
will generate tree level gaugino masses. This mechanism, called
radion mediated supersymmetry breaking (RMSB), is larger than the
contribution to gaugino masses from anomaly mediation, and can
thus dominate when the direct hidden sector gaugino couplings of
$\widetilde{\rm g}$MSB are absent. Nonuniversal gaugino masses result
from the sum of the RMSB and anomaly mediated contributions. 
In explicit models of radion mediation, the F term radion VEV
is generated by the dynamics which stabilizes the radion scalar
VEV \cite{Chacko:2000fn,Kobayashi:2000ak,Agashe:2000nk,Kaplan:2001cg}.

\subsection{D term breaking} 
In the models discussed so far the possibility of significant
D term contributions to the soft parameters was mostly ignored.
However, D term contributions to scalar soft masses arise
generically whenever a gauge group is spontaneously broken with
a reduction in rank. In extensions of the MSSM to GUTs or strings,
we introduce additional $U(1)$ factors which are certainly candidates
for D term contributions to \lsoft. These contributions depend
on the charges of the MSSM fields under these extra $U(1)$s, and
thus typically generate nonuniversal contributions to the
soft scalar masses. A general analysis for extra $U(1)$s which
are contained in $E_6$ can be found in \cite{Kolda:1995iw}.

\subsubsection{Anomalous $U(1)$ mediated supersymmetry breaking}
D term supersymmetry breaking using anomalous $U(1)$'s is also an 
interesting framework for generating models of the soft parameters.
This mechanism is inspired by string constructions in which there   
are many extra $U(1)$ gauge groups, at least one of which is an anomalous 
$U(1)$ gauge group with anomalies cancelled by a Green-Schwartz (GS) 
mechanism.  As the GS mechanism requires both the hidden sector and the 
observable fields transform nontrivially under the $U(1)$, this $U(1)$
is a natural candidate for transmitting the supersymmetry breaking from 
the hidden to the observable sector, as was first pointed out in
\cite{Dvali:1996rj,Binetruy:1996uv}.
For example in the model in \cite{Dvali:1996rj}, a pair of chiral  
superfields $\phi^-$ and $\phi^+$ are introduced with charges equal to
$-1$ and $+1$ respectively under the $U(1)$. Observable matter
superfields $Q_i$ carry charges $q_i$ resulting in the D term  
\begin{equation}
\frac{g^2}{2}D^2=\frac{g^2}{2}\left( \sum_iq_i|Q_i|^2+|\phi^+|^2    
-|\phi^-|^2+\xi\right)^2,
\label{Dterm}
\end{equation}
where
\begin{equation}
\xi = \frac{g^2TrQ}{192\pi^2}M_{Pl}^2.
\end{equation}
If Eq.~(\ref{Dterm}) is the only term in the potential then supersymmetry
will not be broken since the D term is zero at the minimum.
However by including a mass term $W=m\phi^+\phi^-$ supersymmetry
is broken at the global minimum with both F terms and D terms acquiring
vacuum expectation values, and this results in scalar mass
contributions of order \cite{Dvali:1996rj},
\begin{equation}
m_Q^2\approx \frac{<F_{\phi^+}>^2}{M_{Pl}^2}.
\label{scalarmass}
\end{equation}
From this basic starting point, various models have been constructed
with different phenomenologies, 
for example \cite{Mohapatra:1996in,Irges:1998ax}.

\subsection{Why so many models?}
This brief overview of models serves to illustrate the enormous
variety of interesting scenarios and powerful ideas which have been
developed to make models of supersymmetry breaking and its mediation to
the MSSM. It is particularly impressive that, fully twenty years after
the onset of serious supersymmetry model building, new ideas are still
surfacing.

Many concrete and detailed models have been proposed which can
be considered phenomenologically viable. However if one combines the
now rather stringent phenomenological constraints,
with our theoretical bias towards simple and robust models, it
must be admitted that no existing approach has yet emerged as
compelling. This is clearly a fruitful area for further
theoretical study, and future progress will be greatly
aided and accelerated by experimental guidance.

\setcounter{footnote}{0}
\section{Constraints on \lsoft from electroweak symmetry breaking} 
\subsection{Radiative electroweak symmetry breaking}
\label{EWSBsect}
Arguably the most important success of supersymmetry is that it can
provide a natural mechanism for understanding Higgs physics and
electroweak symmetry breaking
\cite{Ibanez:fr,Alvarez-Gaume:1981wy,Inoue:1982ej,Inoue:1982pi,Inoue:1983pp}.  
While the basic physics here is nearly two decades old, it is less
familiar to many particle physicists today than it should be.  Therefore,
this subsection is devoted to a basic explanation of this mechanism. The
main result is that this mechanism requires basic correlations among the
Higgs soft supersymmetry-breaking parameters and the supersymmetric Higgs 
mass parameter $\mu$, which leads naturally into a discussion of the $\mu$ 
problem of the MSSM.

Let us begin by considering the Higgs potential in the MSSM (for 
further details and more explicit notation, see the Appendix). Anomaly
conditions, or equivalently the requirement that the superpotential is 
holomorphic and has both up-type and down-type quark Yukawa couplings,
require two electroweak Higgs doublets
\begin{eqnarray}
H_{d}=\left(\begin{array}{c c} H_{d}^0\\H_d^-\end{array}\right),\;\;
H_{u}=\left(\begin{array}{c c} H_u^+\\H_u^0\end{array}\right),
\end{eqnarray}
with hypercharges $\mp 1/2$.
The tree-level scalar potential for the two Higgs doublets is a sum of
F terms, D terms, and soft supersymmetry-breaking terms:
\begin{eqnarray}
V_{Higgs}&=&(|\mu|^2+m^2_{H_u})|H_u^a|^2+(|\mu|^2+m^2_{H_d})|H_d^a|^2\nonumber
\\&+&\frac{1}{8}(g^2+{g'}^{2})(|H_u^a|^2-|H_d^a|^2)^2
+\frac{1}{2}g^2|H_u^aH_d^{a*}|^2\nonumber\\&-&
(\epsilon_{ab}b H^a_dH^b_u+{\rm h.c.}),
\end{eqnarray}
in which $g\equiv g_2$ is the $SU(2)_L$ gauge coupling and $g'$ is the 
hypercharge gauge coupling.  Electroweak symmetry 
breaking requires that the parameters of this potential must take
on correlated values, such that the potential is minimized with
nonzero VEVs for the neutral components of the Higgs doublets:
\begin{eqnarray}
\langle H_{d} \rangle=\left(\begin{array}{c c}
v_d\\0\end{array}\right),\;\;
\langle H_{u}\rangle=\left(\begin{array}{c c}
0\\v_u\end{array}\right),
\end{eqnarray}
in which $v_d^2+v_u^2=v^2$, $v=174$ GeV, and $\tan \beta = v_u/v_d$.
It is always possible by $SU(2)_L$ gauge transformations to set the vacuum
expectation values of the charged Higgs components to zero.  Furthermore,
we can see that in this tree-level potential it is always possible to
choose global phases of the Higgs fields to eliminate any complex phase in
the $b$ parameter, such that $v_{u,d}$ can be chosen real and positive.
CP symmetry is thus not broken at tree level and the Higgs mass
eigenstates have definite CP quantum numbers. As the two Higgs doublets 
each contain 4 real degrees of freedom and 3 generators are broken when 
$SU(2)_L\times U(1)_Y \rightarrow U(1)_{EM}$, there are 5 physical Higgs 
bosons.  The physical spectrum of Higgs bosons includes 3 neutral Higgs 
bosons (the CP-even $h$, $H$ and CP-odd $A$) and 1 charged Higgs 
($H^{\pm}$).  See {\it e.g.} the review \cite{Carena:2002es} for further 
details 
of the Higgs mass spectrum at tree-level and higher-loop order.  

After replacing the Higgs doublets in the potential by their VEVs, the 
potential takes the form
\begin{eqnarray}
V_{Higgs}=(|\mu|^2+m^2_{H_u})v_u^2+(|\mu|^2+m^2_{H_d})v_d^2-2b v_d
v_u+\frac{1}{8}(g^2+{g'}^{2})(v_u^2-v_d^2)^2.   
\end{eqnarray}
As a brief digression let us consider the conditions on the potential in
the unphysical limit of unbroken supersymmetry but
broken gauge symmetry.  If the soft supersymmetry-breaking terms
$m^2_{H_u}$, $m^2_{H_d}$, and $b$ are zero, the potential is given by
\begin{equation}
V^{SUSY}_{Higgs}=|\mu|^2(v_d^2+v_u^2)+\frac{1}{8}(g^2+g^{'2})(v_u^2-v_d^2)^2,
\end{equation}
which is a positive definite quantity.  This potential is minimized
for nonzero $v_{u,d}$ if and only if $\mu=0$ and $\tan\beta\equiv
v_u/v_d=1$; hence, unbroken supersymmetry but broken
gauge symmetry is possible only in this limit.  Of course, the unbroken 
supersymmetry
limit is unphysical; furthermore, $\mu=0$ and $\tan\beta=1$ have
both been excluded experimentally by direct and indirect searches at
colliders such as LEP. Nevertheless, this limit will prove instructive 
later on when considering certain loop-suppressed processes 
such as magnetic dipole transitions, where the SM and superpartner
contributions cancel \cite{Ferrara:1974wb}.

Let us now consider the phenomenologically viable situation in which the 
soft terms and $\mu$ are nonzero.  The minimum of the potential must break  
$SU(2)_L\times U(1)_Y$; {\it i.e.}, the minimum of the potential should 
not occur for $v_{u,d}=0$. This leads to the condition
\begin{equation}
\label{brk1}
(|\mu|^2+m^2_{H_d})(|\mu|^2+m^2_{H_u})<b^2.
\end{equation}
The potential must be also bounded from below along D flat directions 
({\it i.e.}, with vanishing D terms), yielding the constraint 
\begin{equation}
\label{brk2}
2|\mu|^2+m^2_{H_d}+m^2_{H_u}\geq 2|b|.
\end{equation}
The minimization conditions for this potential are as follows:
\begin{eqnarray}
|\mu|^2+m^2_{H_d}&=&b \tan\beta-\frac{m_Z^2}{2}\cos 2\beta\\
|\mu|^2+m^2_{H_u}&=&b \cot\beta+\frac{m_Z^2}{2}\cos 2\beta.
\label{min}
\end{eqnarray}
The minimization conditions demonstrate explicitly that the soft
parameters $m^2_{H_u}$, $m^2_{H_d}$, $b$ and the supersymmetric parameter 
$\mu$ all must be of approximately the same order of magnitude as $m_Z$ 
for the electroweak symmetry breaking to occur in a natural manner, 
{\it i.e.}
without requiring large cancellations. Here we
mean technically natural in the 't Hooft sense in that there is no
symmetry in the effective theory at the electroweak scale to protect this
cancellation, and the cancellations in the loop corrections to the masses, 
if the particle/sparticle mass differences are not of order the 
electroweak/TeV scale.

The minimization conditions for an $SU(2)_L\times U(1)_Y$ breaking vacuum
suggest that one or both of the Higgs doublets has a
negative mass-squared at $v_d=v_u=0$, like the negative mass-squared in
the SM.  In a single Higgs doublet model, the
usual condition is that the mass-squared parameter is negative.  However,  
the requirements are more subtle in two Higgs doublet models, in which 
the condition $m^2_{H_u}<0$ is neither necessary
nor sufficient (although it helps). 

Nevertheless, a celebrated features of the MSSM is that the up-type Higgs
soft mass-squared parameter does get driven negative via renormalization
group running due to the large top quark Yukawa coupling
\cite{Ibanez:fr,Alvarez-Gaume:1981wy,Inoue:1982ej,Inoue:1982pi,Inoue:1983pp}.
This can be seen upon an inspection of the renormalization group equations
for the relevant soft parameters. For this purpose, it suffices to retain
only the third family contributions in the approximation of
Eq.~(\ref{thirdfamilyYs}), as presented in
Eq.~(\ref{thirdfamilyYrges})--Eq.~(\ref{thirdfamilym2rges}) of 
Appendix~\ref{rgeapp}.  Retaining only the top quark Yukawa coupling, one 
can see that the $m^2_{H_u}$ parameter is driven down 
by the large top Yukawa terms as one runs down from the high scale to the 
low scale.  In the large $\tan\beta$ regime in which the bottom and tau 
Yukawas are also large,  there is a similar effect for $m^2_{H_d}$, as 
will be discussed later.  Other masses such as the stop mass-squared 
parameters also are driven down by the Yukawa terms; however, they also 
receive large positive contributions from gluino loops, so they don't 
usually run negative, although they can.  
Therefore, the Higgs soft mass-squared parameters can be 
driven to negative  values near the electroweak scale due
to perturbative logarithmic running.\footnote{Note however that
electroweak symmetry breaking is possible even if $m^2_{H_u}$ is positive
as long as $b$ is large enough.}

\subsection{The $\mu$ problem}
\label{sectmuproblem}
Electroweak symmetry breaking can thus take place in a natural  
way in the MSSM via a radiative mechanism by which the soft mass-squared
parameter of the up-type Higgs doublet (and also that of the down-type
Higgs when  $\tan\beta$ is large) approaches or becomes zero,  
provided that $\mu$ and $b$ are nonzero and take values roughly of the   
same order as $m_Z$.  To see this correlation let us demonstrate it
explicitly for the $\mu$ parameter. Rewriting the minimization conditions
yields the following expression: 
\begin{equation}
\label{mucond}
\mu^2=\frac{m^2_{H_d}-m^2_{H_u}\tan^2\beta}{\tan^2\beta
-1}-\frac{1}{2}m_Z^2.
\end{equation}
This correlation leads to a puzzle. Just as we are ignorant of the origin
and dynamical mechanism of supersymmetry breaking, we do not know 
why the supersymmetric mass parameter $\mu$ should be of the order of the
electroweak scale, and of the same order as the supersymmetry {\it
breaking} parameters (or else there would be a chargino lighter than the W 
boson, which has been excluded experimentally).  Given that $\mu$ is a
superpotential parameter one might expect $\mu \sim O(M_X)$, where $M_X$
is a high scale, {\it e.g.} the unification or GUT scale.  If this were
true, the hierarchy problem is clearly restored.  This puzzle, known as 
the {\it $\mu$ problem}, was first pointed out in \cite{Kim:1983dt}.

Operationally, one can trade the unknown input values
of $\mu$ and $b$ for $m_Z$ and $\tan\beta$; however, this does
not constrain the phase or sign of the $\mu$ parameter relative to the
other soft supersymmetry-breaking terms. In practice, this is the 
standard approach for most phenomenological analyses of the MSSM, in which 
$\tan\beta$ is typically taken to be an input parameter.  

However, one can view the $\mu$ problem in another way. The small value of 
the $\mu$ parameter relative to the fundamental scale suggests that the 
$\mu$ term is not a fundamental parameter, but rather parameterizes more
fundamental physics associated with the breakdown of supersymmetry at
scales higher than the electroweak scale.  In this way understanding the
size of $\mu$ might lead to new insight about the origin of
supersymmetry breaking. 

The ways in which $\mu$ and $b$ are generated are highly model
dependent.\footnote{An optimist would argue that this model dependence can
be viewed as a positive feature, since then data may point to how $\mu$
and $b$ are actually generated, rather than having to decide from purely
theoretical arguments.} Let us consider a few standard examples --- these
by no means exhaust the possible models. The interested reader should
consult the excellent review \cite{Polonsky:1999qd} for further details
and a more complete classification.

\begin{itemize}

\item The $\mu$ 
term can be generated from a
renormalizable superpotential coupling 
\begin{equation}
W=\hat{N} \hat{H}_u \hat{H}_d,
\end{equation}
which occurs for example in the NMSSM, as dicussed in  
Section~\ref{nmssm}.  
This renormalizable superpotential leads to the generation of $\mu\sim 
\langle N \rangle$, and the $b$ term can be due to the associated soft 
trilinear coupling $A_N$.  The VEV of $N$ can be triggered
in ways similar to the usual radiative breaking mechanism in the MSSM, for
example if the $N$ field couples to heavy exotic particles with large
Yukawa couplings.  $N$ can either be a total singlet with respect to any
gauge group, as in the NMSSM, or a SM singlet charged with
respect to an additional gauged $U(1)'$ (see {\it 
e.g.} \cite{Cvetic:1997ky}).

\item Another possibility which can naturally occur within the 
supergravity
framework is the Giudice-Masiero mechanism \cite{Giudice:1988yz}, 
which uses K\"ahler potential couplings that mix the
up and down-type Higgs:
\begin{equation}
K_{GM}\propto \hat{H}_u\hat{H}_d+{\rm h.c.}.
\end{equation}
This term becomes an effective superpotential term after supersymmetry 
breaking. The $\mu$ and $b$ terms are naturally of a similar
order of magnitude as the gravitino mass, which sets the scale for the
soft supersymmetry-breaking terms.
\end{itemize}

The examples described here both naturally fit in with the supergravity
mediation scheme for supersymmetry breaking. There are several other
possible mediation schemes, such as gauge mediation, which have lower
mediation scales and a different hierarchy between the VEVs of the hidden
sector fields and the supersymmetry-breaking F terms.  Within these other
schemes other possible operators can be used to obtain $\mu$ and $b$ with
correct orders of magnitude. However, in gauge mediation it takes a
certain amount of work to arrange that $\mu$ and $b$ are not generated at
the same loop order, which would be problematic for viable phenomenology
(see {\it e.g.} \cite{Giudice:1998bp,Kolda:1997wt} for further
discussions).

\subsection{The ubiquitous $\tan \beta$}
\label{largetanbetasect}
An important quantity in relating supersymmetry to the
real world is $\tan\beta \equiv v_u/v_d$.  ${\rm Tan}\beta$ does not 
exist in the high scale theory, since it is the ratio of the vacuum 
expectation 
values for the two Higgs doublets. The 
VEV's become nonzero at the electroweak phase transition at a few hundred 
GeV as the universe cools; above that scale the electroweak symmetry is 
unbroken.  
Thus $\tan \beta $ has an unusual status in the theory because it does not
appear in the superpotential or ${\mathcal{L}}_{soft},$ yet it enters
significantly in almost every experimental prediction.  It is often 
used as an input parameter in phenomenological analyses of the MSSM, 
typically under 
the assumption of perturbative radiative electroweak symmetry breaking.  
As discussed in 
Section~\ref{EWSBsect}, the tree-level minimization conditions of the 
Higgs potential allow $b$ and $\mu$ to be eliminated in favor $\tan\beta$ 
and the Z mass up to a phase ambiguity.   It is then possible to calculate 
$\tan\beta $ within the framework of the high 
energy theory, which should predict the source of $b$ and $\mu$. The 
result of course will depend on a number of soft parameters.

There is information available about $\tan \beta $ from both theory and
phenomenology.  Bounds on the possible range of $\tan\beta$ can be
obtained under the plausible assumption that the theory stays perturbative
at energies up to the unification scale; recall the evidence for this
includes gauge coupling unification and successful radiative electroweak
symmetry breaking.  As $\tan\beta$ relates the Yukawa couplings to the
masses, $\tan \beta $ cannot be too small or too large because the Yukawa
couplings should be bounded.  This gives a lower limit of about 1 and an
upper limit of about 60. These limits will not be discussed in detail
since phenomenological information is anticipated to improve on them in
the near future.

An additional constraint arises from the upper bound on
the lightest Higgs mass, which at tree level is given by 
\begin{equation}
\label{treehiggsbound}
m_{h^0}\lesssim m_Z \left| \cos 2\beta \right|.
\end{equation}
It has been known for more than a decade that there are large loop 
corrections to this tree-level bound (see {\it e.g.} \cite{Carena:2002es} 
for a review).  At very low values of 
$\tan\beta$, large loop corrections are needed, which makes it more 
difficult for such low $\tan\beta$ values to be consistent with LEP 
Higgs mass bounds. Indeed,  
the absence of a Higgs boson lighter than about 110 GeV implies 
$\left| \cos 2\beta \right| $ is very near unity, which implies $\tan 
\beta $ is larger than about 4.\footnote{To do this precisely one should 
allow for CP-violating effects which can
lower the limit; see Section~\ref{expsect}.} 

There are other hints of a lower limit of a few 
on $\tan \beta $ --- the precision data from LEP, SLC, and the Tevatron 
is described a little better \cite{Cho:2001nf,Altarelli:2001wx} if there 
are light superpartners and in particular if sneutrinos are significantly 
lighter than charged sleptons. 
Their masses-squared are separated by the $SU(2)$ D term $\left| \cos 
2\beta \right|  m_W^{2}$, so again the implication is that $\left| \cos 
2\beta \right| $ is near unity.\footnote{Also, as described in 
Section~\ref{CPflsect1}, the recent data for the muon anomalous magnetic 
moment may show a deviation from the SM. If so, and if the effect is 
indeed due to supersymmetry, the supersymmetry contribution needs to be a 
few times the electroweak contribution. This is reasonable if $\tan\beta$ 
is greater than about 3, since the supersymmetry contribution grows with 
$\tan\beta$.}
In general, deducing upper limits on 
$\tan \beta $ is more involved 
because at larger $\tan \beta $ it is necessary to include effects of 
$\tan \beta $ itself on masses and other quantities that enter into 
estimating the limits.

On the theoretical side, there has long been a bias toward having $\tan
\beta $ near unity for several reasons.  First, in the supersymmetric
limit the Higgs potential is minimized when $\tan \beta =1$, as shown in
Section~\ref{EWSBsect}. Second, if the parameters of the Higgs potential
are comparable in size, it is natural for the Higgs fields to have VEVs of
similar magnitudes.  One argument in the opposite direction is that the
attractive idea that the $t$, $b$, and $\tau$ Yukawa couplings unify at a
high scale requires large $\tan \beta$
\cite{Ananthanarayan:xp,Langacker:1993xb,Kelley:1991aj,Carena:1993ag,Allanach:1995ji,Carena:1994bv,Hall:1993gn,Rattazzi:1995gk,Olechowski:1994gm,Baer:1999mc}.  
Precisely how large is subtle, since one must include running effects on
masses and higher order effects.

It was noticed quite some time ago that 
radiative electroweak symmetry breaking without fine tuning can be more 
difficult to achieve in the very large $\tan\beta$ limit 
\cite{Nelson:1993vc,Rattazzi:1995gk}. 
To see this, rewrite the minimization conditions as follows:
\begin{eqnarray}
m^2_{H_u}\tan^2\beta-m^2_{H_d}&=&- 
\left (|\mu|^2+\frac{m_Z^2}{2} \right )(\tan^2\beta-1)\\
\frac{2b}{\sin 2\beta}&=&2|\mu|^2+m^2_{H_d}+m^2_{H_u}=m^2_A,
\end{eqnarray}
in which $m^2_A$ is the mass of the CP-odd Higgs boson. In the 
large  $\tan\beta$ limit, 
\begin{eqnarray}
|\mu|^2&=&-m^2_{H_u}-\frac{1}{2}m_Z^2+O\left (\frac{1}{\tan^2\beta}\right 
)\\
b&=&\frac{1}{\tan\beta}(m^2_{H_d}-m^2_{H_u}-m_Z^2)+O\left
(\frac{1}{\tan^3\beta}\right ).
\end{eqnarray}
This shows that there must be a hierarchy among the soft parameters:
\begin{equation}
\label{hierestimate}
b\lesssim 
m_W^{2}/\tan \beta,
\end{equation}
while one would expect $b$ to be the size of a typical soft mass-squared.  
More precisely, 
\begin{equation}
\frac{1}{\tan\beta}=\frac{b}{2|\mu|^2+m^2_{H_d}+m^2_{H_u}}.
\end{equation}
This hierarchy does not appear to be explained by any approximate symmetry 
in two Higgs doublet models such as the MSSM (and even in singlet-extended 
models such as the NMSSM), as the most obvious symmetries that can 
do the job ({\it e.g.} the $U(1)_{PQ}$ and $U(1)_R$ symmetries of the MSSM) 
result in a light
chargino with mass $\ll m_Z$, which is ruled out experimentally 
\cite{Nelson:1993vc,Rattazzi:1995gk}. For example, $\mu$ is typically much 
lighter than
$m_Z$ in the $U(1)_{PQ}$ scenario, while the soft parameters $B\equiv
b/\mu \sim M_2\sim A$ are typically much lighter than $m_Z$ for $U(1)_R$
\cite{Rattazzi:1995gk}.  Either scenario predicts a chargino lighter than
the current LEP limits. It appears to be necessary to 
take the scale of (at least a subset of) the soft parameters larger than 
the electroweak scale by a multiplicative factor of $\sqrt{\tan\beta}$, which is not favored by naturalness
arguments.

Clearly the issue of how to achieve the hierarchy of
Eq.~(\ref{hierestimate}) must be addressed in model-building. Such a
hierarchy is not in general favored within the simplest SUGRA scenarios,
in which $\mu^2 \sim b$ unless specific cancellations occur, although it
can be achieved within GMSB (see {\it e.g.}
\cite{Babu:1996jf,Dine:1996xk}).  
Strictly speaking, the constraints here apply to the values of the
parameters at the electroweak scale. Since $\mu$ is a superpotential
parameter and hence only receives wavefunction renormalization, its
running is mild. However, $b$ is a soft supersymmetry-breaking parameter
which can receive large corrections not proportional to its initial value.
In carefully chosen scenarios, $b$ and $\mu$ could start with similar 
values but run to very different values at low energy. If there is no 
compelling theoretical motivation for such a scenario, though, a certain 
degree of fine-tuning is inherently present.

Radiative electroweak symmetry breaking with large $\tan\beta$ is also 
complicated by the similar running of the soft 
mass-squared parameters of the two Higgs doublets when the $t$ 
and $b$ quark Yukawa couplings are comparable \cite{Rattazzi:1995gk}. 
The key point is Eq.~(\ref{brk1})  and Eq.~(\ref{brk2}) cannot be
satisfied if $m^2_{H_u}=m^2_{H_d}$, indicating the need for violation of
the custodial $u \leftrightarrow d$ symmetry.  In principle, this breaking 
can be provided by the hierarchy between the $t$ and $b$ Yukawa couplings, 
with the heavy top Yukawa coupling driving $m_{H_u}^2$ negative.  However,  
this is not possible in the large $\tan\beta$ regime because the Yukawas 
are comparable.  Both $m_{H_u}^2$ and $m_{H_d}^2$ will run to 
negative and comparable values if their initial values are similar, which 
is generally problematic for electroweak symmetry breaking. This is
particularly an issue for GUT models in which the two electroweak
Higgs doublets reside in a single GUT multiplet as the initial values of
their soft mass-squared parameters are equal.

However, this problem can be alleviated via the well-known mechanism of
splitting the scalar masses using additional D term contributions 
\cite{Drees:1986vd,Hagelin:1989ta,Kawamura:yv,Cheng:1994bi,Kolda:1995iw}.
Whenever a gauged $U(1)$ symmetry is broken, contributions to soft scalar
mass squareds can result via the D terms, which can change
the superpartner spectrum in a significant way. The typical structure of a
D term is
\begin{equation}
D^a=\phi^*_i T^a_{ij} \phi_j
\end{equation}
where $T^a_{ij}$ is a gauge group generator and
$\phi_i$ is a scalar component of a chiral superfield which
transforms under the gauge group.
The contribution to the soft potential is then of the form
\begin{equation}
\Delta V= \frac{1}{2}g^2\sum_a D^aD^a
\end{equation}
where $g$ is a gauge coupling associated with the gauge group
under which $T^a$ is a generator. For commuting gauge groups
the potential is constructed by summing over the terms
for each gauge group.

For a $U(1)$ gauge group, such D terms were first discussed by Fayet and
Illiopoulos \cite{Fayet:jb}. These D terms can lead to contributions to
soft masses when Higgs fields develop VEVs which break the $U(1)$. Such
contributions to the masses of the squarks and sleptons are already
present in the MSSM due to the breaking of the electroweak symmetry,
contributing essentially $m_Z^2(T_3-Q\sin^2 \theta_W)$ for each, which is
relatively small.  However, further $U(1)$ gauge groups could exist as
additional commuting Abelian gauge groups, or corresponding to diagonal
generators of non-Abelian gauge groups which are broken; these could lead
to additional contributions to the soft scalar masses while leaving the
other soft parameters unchanged.

In supersymmetric GUT models, the GUT symmetry breaking can have 
consequences for low energy phenomenology via such D term contributions to 
the scalar masses if the SM particles are charged under the resulting 
$U(1)$ symmetries. This has been studied within supersymmetric GUT
frameworks such as $SO(10)$ and Pati-Salam $SU(4)\times SU(2)_R\times
SU(2)_L$ 
\cite{King:2000vp,Baer:1999mc,Kawamura:1993uf,Kawamura:1994ys,Faraggi:1991bb}. For
example, within the Pati-Salam model the D term corrections must be 
included because they leave an imprint in the scalar masses of
the charges carried by the broken GUT generator (these charges determine
the coefficients of the $g^2$ terms above). Therefore the analysis of the
sparticle spectra \cite{Faraggi:1991bb} might reveal the nature of the GUT
symmetry breaking pattern.  In addition, they split the soft Higgs masses 
by
\begin{equation}
m^2_{H_u} - m^2_{H_d} \sim -4 g^2_X D, 
\end{equation}
where $g_X$ is the gauge coupling constant defined at GUT
scale. The positive D term thus facilitates radiative electroweak 
symmetry breaking, particularly for large 
$\tan\beta$.  Such
results are expected to be quite generic and apply in string theory for
example where the symmetry breaking is more obscure. In general whenever
there is a D flat direction which may be lifted by soft supersymmetry-breaking
terms, there will be D term contributions to soft masses. Thus any
discussion of soft squark and slepton masses must include an examination
of the presence of D terms, which can give significant contributions to
the soft mass matrices. The D terms always lead to additional
soft mass squared contributions which are always real. The possible
presence of such terms is one reason why assuming degenerate scalar masses
for phenomenological studies may be unwise.

\subsection{Charge and color breaking minima}
\label{CCBsect}
In the SM, the quartic coupling $\lambda$ of the Higgs potential must be 
positive, or else the Higgs potential has no minimum and the resulting 
field theory is ill defined.  In the MSSM, the quartic scalar couplings 
arise from D terms, which are positive semi-definite by definition but can 
be zero along certain directions in field space.  For example, the Higgs 
scalar potential projected along the neutral components
\begin{eqnarray}
V_{Higgs}&=&(|\mu|^2+m^2_{H_u})|H_u^0|^2+(|\mu|^2+m^2_{H_d})|H_d^0|^2\nonumber
\\&+&\frac{1}{8}(g^2+{g'}^{2})(|H_u^0|^2-|H_d^0|^2)^2-b H^0_dH^0_u+{\rm h.c.})
\end{eqnarray}
has D terms which vanish if $\langle H^0_u \rangle=\langle 
H^0_d\rangle$; technically the conditions for such vanishing D terms are 
known as D flatness conditions. Along this D flat direction in field 
space, the 
Higgs VEVs can be too large and hence unphysical.\footnote{Note that since 
the D term involves quartic Higgs VEVs, it would dominate in the large VEV 
limit. Therefore, since the D term is positive (if it is nonzero), it 
would prohibit the Higgs VEVs from ever becoming large.} The 
quadratic terms, which determine the shape of the potential, must be
positive or else the Higgs potential becomes unbounded from below (UFB).  
More precisely, the condition to avoid a tree-level UFB potential is:
\begin{equation}
m_{H_u}^2+m_{H_d}^2+2|\mu |^2-2b>0,
\end{equation}
which must be satisfied for all scales between $M_{GUT}$ and $m_Z$. 
Once radiative corrections are included the potential is no longer 
strictly UFB; perhaps then the problem should be called ``the problem of 
large unphysical minima'' since the potential will develop a deep 
unphysical minimum at a large Higgs VEV. Typically the tunneling 
transition rate
from the physical Higgs VEV to a large unphysical Higgs VEV is so slow as 
to not yet have happened. The problem then is a cosmological
one, namely why would the universe end up in our shallow, observed minimum 
when there is a much deeper, but unphysical, one available?
For this reason the UFB constraint should perhaps be regarded as a
theoretical cosmological constraint rather than a collider constraint.

The MSSM differs from the SM in that the full scalar potential is not 
just the potential of the Higgs doublets, but also includes the 
potential of the squarks and sleptons, any of which could acquire a 
phenomenologically disastrous VEV if certain conditions are not met.  For 
example, there is 
a D flat direction in which 
$\widetilde{U}^c$, $H_u$, and the $\widetilde{U}_L$ component of $\widetilde{Q}$ 
all have equal VEVs. However, unlike the Higgs doublet 
case, this direction also has a cubic contribution in the potential, the 
soft supersymmetry-breaking trilinear term 
$H_u\widetilde{Q}\widetilde{A}_u\widetilde{U}^c$. 
If this trilinear term gives a negative contribution to the
potential, then a very deep CCB minimum appears unless the 
following constraint is satisfied \cite{Frere:ag}:
\begin{equation}
|\widetilde{A}_u|^2\leq 3(m_Q^2+m_U^2+m_{H_u}^2+|\mu|^2).
\end{equation}
There are similar constraints for all the trilinear terms, including
off-diagonal flavor changing ones \cite{Casas:1996de}.
The CCB minima are those which lead to a deeper minimum than the physical 
one, even at tree-level.

The presence of squarks and sleptons also allows new UFB
problems with the full scalar potential, analogous to the Higgs UFB 
problems discussed above \cite{Komatsu:1988mt}.
As before the UFB potential at tree-level becomes converted
into a large deep minimum once radiative corrections are included,
and so strictly speaking the UFB vacua involving squarks and
sleptons are really further examples of CCB vacua.
Many dangerous CCB minima of both types were subsequently 
classified and studied in detail for different physical situations
\cite{Casas:1995pd,Strumia:1996pr,Abel:1998cc,Abel:1998wr}.
All the dangerous directions have the feature that they are
both D flat and F flat, where the F flatness conditions are defined to be 
$\langle F_i \rangle \equiv \langle \partial W/\partial \phi_i\rangle=0$ 
for all fields $\phi_i$ in the model. 

A particularly dangerous set of flat directions involve the Higgs VEV 
$H_u$, since the mass squared $m_{H_u}^2$ is naturally negative as it runs 
below the GUT scale. For example consider the flat direction
characterized by $L_iQ_3D_3^c$ and $L_iH_u$, where we
have used the correspondence between flat directions
and holomorphic gauge invariant polynomials of chiral superfields
\cite{Gherghetta:1995dv}. The dangerous flat direction occurs
when the VEV of the $\widetilde{D}_3$ component of $\widetilde{Q}_3$
equals that of $\widetilde{D}^c_3$ and in addition the
VEVs of $H_u$ and a slepton doublet $\widetilde{L}_i$ are
related by \cite{Komatsu:1988mt}
\begin{equation}
|\widetilde{L}_i|^2=|H_u|^2+|\widetilde{D}_3|^2.
\end{equation}
This leads to the constraint \cite{Komatsu:1988mt}
\begin{equation}
m_{H_u}^2+m_{L_i}^2 >0,
\end{equation}
which must be satisfied over the whole range between $M_{GUT}$
and $m_Z$. Since $m_{H_u}^2$ runs negative this condition
can easily be violated.
This constraint is only approximate; the full constraint has been 
subsequently studied in detail \cite{Abel:1998cc,Abel:1998wr}, where other 
equally dangerous flat directions $L_iL_jE_k^c$ and $L_iH_u$ were also 
considered.  

The requirement of no CCB minima arising from the dangerous directions 
leads to severe conditions on the parameter space of the constrained MSSM. 
Generally the CCB constraints prefer models where $m_0$ is high and 
$m_{1/2}$ is low \cite{Abel:1998wr}. For minimal
models based on dilaton-dominated supersymmetry breaking, for example, the
CCB requirements rule out the entire experimentally allowed parameter
space. Other nonuniversal models must be studied case by case. 
However, we repeat that the CCB constraints should properly be regarded as
cosmological constraints rather than particle physics constraints. For
this reason, it is not certain how seriously these constraints should be
taken in phenomenological analyses.

\subsection{Upper limits on superpartner masses and fine-tuning}
\label{finetunesect}
There are several arguments which have been used to suggest that at least 
a subset of the superpartners will be light.  In this section, we briefly 
discuss these arguments and discuss issues of fine-tuning in the context 
of the MSSM.

Superpartners get mass from both the Higgs mechanism and supersymmetry
breaking, the latter entering through the soft masses.  Generically, the 
superpartner masses are dominantly due to the soft masses (and
$\mu$ and $\tan \beta$) and not electroweak symmetry breaking effects. For 
example, in the chargino mass matrix the off-diagonal elements are 
electroweak symmetry breaking effects and the
diagonal elements come from ${\mathcal{L}}_{soft}$.  The electroweak
contributions are typically of order $m_{W}$ or less.  If the soft masses
are large, the superpartner masses will generally be large. Whether there
are upper limits on superpartner masses is of interest because
superpartners have not yet been observed directly, and because such
considerations are of crucial importance in the planning and construction
of future colliders.

Perhaps the most compelling argument in favor of light superpartners
comes from the hierarchy problem, which remains the basic motivation for 
low energy supersymmetry.
From a bottom-up perspective, the hierarchy problem is encountered
in the Standard Model as one-loop radiative corrections to the Higgs
mass parameter $m_H^2$ in the Higgs potential. Since the top quark is 
heavy, the dominant one-loop correction arises from top loops:  
\begin{equation}
\delta m_H^2 ({\rm top \, loop})
=-(900\,{\rm GeV})^2\left(\frac{\Lambda}{3\,{\rm TeV}}\right)^2
\label{oneloop}
\end{equation}
where $\Lambda$ is a cutoff scale.
In the SM, electroweak symmetry breaking requires
\begin{equation}
m_H^2+\delta m_H^2=-\lambda (246\,{\rm GeV})^2
\label{sm}
\end{equation}
where $\lambda$ is the quartic Higgs coupling.
By comparing Eq.~(\ref{oneloop}) to Eq.~(\ref{sm}) it is clear that
fine-tuning of the unrenormalized parameter $m_H^2$
is required if $\Lambda \gg 1$TeV.
Loops involving stop squarks, whose couplings to the
Higgs are equal to the top couplings by virtue of supersymmetry,
give opposite sign contributions which cancel the leading quadratic
divergence, leaving only a subleading logarithmic divergence.
The condition of no fine tuning then apparently implies that
the stop masses, identified with the cutoff $\Lambda$ in 
Eq.~(\ref{oneloop}),
should be not much larger than the TeV scale.
According to similar arguments the other superpartners would have higher
upper mass limits since the top quark is the heaviest known particle.

From a top-down perspective the requirement that the MSSM gives
radiative electroweak symmetry breaking without fine-tuning can
again give upper limits on superpartner masses. A very attractive
feature of the MSSM is that the effective Higgs mass parameters
$m_{H_{u}}^{2}+|\mu|^2$ and $m_{H_{d}}^{2}+|\mu|^2$ can both
start out positive and equal at the high energy scale, then
when they are run down to low energy using the RG equations
$m_{H_{u}}^{2}$ can get driven negative due to the effects of 
top quark loops, resulting in electroweak symmetry breaking as  
discussed in Section~\ref{EWSBsect}. This radiative breaking mechanism
requires a sufficiently heavy top quark in order to work. However, 
$m_{H_{u}}^{2}$ is typically driven much more negative than 
$-m_Z^2$, depending on the sizes of the superpartner masses.
According to the minimization conditions in Eq.~(\ref{min}), this
effect can be compensated by choosing the value of $|\mu|^2$
(which does not run very strongly) to cancel against the
excess negative low energy value of $m_{H_{u}}^{2}$, but at the
in expense of a certain amount of fine-tuning. The resulting fine-tuning
was first studied by 
\cite{Ellis:1985yc,Barbieri:1987fn,Dimopoulos:1995mi}.
The price of such fine-tuning imposed by the failure to
find superpartners at LEP was subsequently discussed in
\cite{Chankowski:1997zh,Chankowski:1998xv,Chankowski:1998za,Barbieri:1998uv}.

Generically, for a given fixed top quark mass, the larger the high energy
soft masses the more negative $m_{H_{u}}^{2}$ is driven at    
low energies and the greater the fine-tuning. 
In many cases, the soft mass parameter ultimately most responsible for 
driving $m_{H_{u}}^{2}$ negative is the gluino mass $M_3$ 
\cite{Kane:1998im,Bastero-Gil:1999gu}.\footnote{A counterexample is the 
``focus-point'' regime \cite{Feng:1999zg} of {\it e.g.} mSUGRA models, in 
which the scalar masses are much larger than the gaugino masses; in this 
case the stop masses control the RG running.} This 
has the effect of increasing the stop soft masses, and since the RGEs for
the up-type Higgs and the stop soft masses are strongly coupled
due to the large top Yukawa coupling, $m_{H_{u}}^{2}$ is driven
more negative in response. The requirement of a large Higgs boson mass
is indirectly responsible for fine-tuning,
since in the MSSM it must derive all of its mass in excess of
$m_Z$ from radiative corrections, and these dominantly originate
from the stop sector. Therefore the
more the Higgs mass exceeds $m_Z$, the heavier the
low energy soft mass parameters associated with the stop sector must be,
and the more negative $m_{H_{u}}^{2}$ becomes.
Since the Higgs mass only receives radiative corrections logarithmically,
this implies that fine-tuning increases exponentially with the
Higgs boson mass.
If the Higgs boson mass can exceed $m_Z$ at tree-level as in the
NMSSM then the fine-tuning arising from the Higgs boson mass will
be significantly decreased \cite{Bastero-Gil:2000bw}.

One can argue that there are essentially no instances in physics where 
large fine-tuning occurs or is acceptable once there is a theory, so it 
is appropriate to impose such a condition.  On the other hand, imposing a 
numerical value to quantify fine-tuning and using it to obtain
upper limits on superpartners is fraught with difficulties.
Even the question of how to define a measure of the fine-tuning   
associated with the radiative breaking mechanism is not settled. 
Several analyses 
\cite{Anderson:1994dz,Anderson:1994tr,Anderson:1995cp,Anderson:1996ew} 
dispute the relevance of the definition of fine-tuning in terms
of a sensitivity parameter on which all of the discussion
above is based. They argue that one must take into account the
normalization of any naturalness measure, and claim that this results
in significantly reduced fine-tuning.

What appears as fine-tuning is of course theory-dependent.
The usual example is the precise equality of the electric charges of
the proton and the electron, so atoms are neutral to a part in about 
10$^{20}$. If electric charge is quantized that is reasonable, if not it 
requires a huge
fine-tuning.  So one expects any acceptable theory to imply quantization
of electric charge.  Similarly, one should judge the fine-tuning
of the soft masses in the presence of a theory that can relate the
parameters.  Even then, constraints remain because parameters generally
have different physical origins and run differently from the 
high or unification scale where the theory is defined to the electroweak 
scale. \emph{If }supersymmetry is indeed the explanation for electroweak 
symmetry
breaking, then it is appropriate to impose reasonable fine-tuning
constraints on the soft parameters. These issues and possible ways to 
evade constraints have recently been reexamined in \cite{Kane:2002ap}.

There are other arguments \cite{Ellis:1985yc,Roszkowski:1991qp} that 
certain
superpartners, most likely sleptons, should be light or the lightest 
supersymmetric particle (LSP) would
annihilate too poorly and the large number of LSPs left would overclose
the universe.  This assumes the LSP is the dark matter, which is an
extra, although likely, assumption.  There can also be loopholes
\cite{Ellis:1985yc,Roszkowski:1991qp} from annihilation through a
resonance or along particular directions in parameter space.  
A third argument is that electroweak baryogenesis requires charginos and 
stops to be lighter than about $m_{top}$ and Higgs bosons to be fairly 
light.  Of course, this assumes the baryon asymmetry is indeed produced 
this way; see Section~\ref{bgensect}.  
Finally, one of the stop masses is typically lighter than those of the 
first two generations of squarks for two reasons: (i) the stop soft 
mass-squared parameters are driven down by RG running much like 
$m^2_{H_u}$, and (ii) 
they can have large LR mixing, 
which further pushes down the mass of the lighter stop (for 
large $\tan\beta$, the sbottom and stau soft mass-squares are also reduced
substantially). These arguments reinforce the expectation that some
superpartners are light and perhaps in the Tevatron domain, but none are 
definitive.

\setcounter{footnote}{0}
\section{CP violation and flavor --- origin and connections to 
${\mathcal{L}}_{soft}$}
\label{CPflsect1}
The flavor problem of the SM quarks and leptons is among the most 
intriguing issues in high energy physics. The SM flavor problem can be 
summarized by the following questions: (i) why are there three standard 
families of quarks and leptons, not more or less, and (ii) what is the 
origin of their hierarchical masses and mixing angles. In the SM, this 
can be rephrased as follows: what is the theoretical explanation of the 
quark and lepton Yukawa matrices?

The origin of CP violation is also a mystery. CP violation was observed in
the kaon system in the 1960's \cite{Christenson:fg}, and more recently in
the B system \cite{Aubert:2002ic,Abe:2002px}.  CP violation is also a
necessary ingredient for baryogenesis \cite{Sakharov:dj}, as discussed in
Section ~\ref{bgensect}. Whether the observed CP violation in the neutral
meson systems is related to the CP violation that affects the baryon
asymmetry is an open question (see {\it e.g.} \cite{Chang:fx}). However,
other CP-violating observables, most notably the fermion electric dipole
moments (EDMs), have not been observed experimentally.

The three-family SM provides a well-known source of CP violation in the
quark sector\footnote{We defer the discussion of phases in the lepton
sector to Section~\ref{neutrinosect}, in which we discuss the minimally
extended MSSM including right-handed neutrinos.} through a single phase in
the CKM matrix \cite{Kobayashi:fv}. The CKM phase does not lead to
observable EDMs\footnote{EDMs are flavor-conserving, while the CKM phase
is associated with flavor-changing couplings. Hence, the first
nonvanishing contribution to the EDMs occurs at three-loop order and is
highly suppressed \cite{Shabalin:rs}.} and there is emerging, but not
definitive, evidence that the CKM phase is the dominant or only source
of CP violation in the neutral meson systems. However, the strength of CP
violation, which is proportional to the Jarlskog invariant 
\cite{Jarlskog:1985cw}, is insufficient for electroweak baryogenesis, as 
discussed in Section~\ref{bgensect}.  The EDM problem is also not 
solved because the QCD \( \theta \)  parameter generically overproduces 
the neutron EDM by many orders of magnitude. This {\it strong CP problem} 
will be addressed in Section~\ref{strongCPsect}.

Aside from the caveats mentioned above regarding the origin of the baryon
asymmetry and the resolution to the strong CP problem (which both have
possible solutions discussed in this review), the key to understanding the 
SM flavor and CP problems is to understand the origin of the Yukawa
couplings of the quarks and leptons.  However, the SM is an effective
theory which does not provide a framework in which to address the origin 
of CP violation and flavor. These questions must be reserved for a more 
fundamental underlying theory.  As the MSSM is itself an effective theory, 
making the theory supersymmetric simply transports the problem of the 
Yukawa matrices from the Lagrangian to the effective low energy 
superpotential of the MSSM.

However, supersymmetry {\it breaking} introduces new flavor and CP
questions because there are many new sources of complex flavor-changing
couplings and complex flavor-conserving couplings due to the structure of
\lsoft.  These questions can be summarized as follows:
\begin{itemize} 

\item The complex flavor-conserving couplings of
\lsoft can overproduce the electric dipole moments (see {\it e.g.}
\cite{Dugan:1984qf}). This is commonly known as the {\it supersymmetric CP
problem}; it will be addressed in Section~\ref{EDMsect}.

\item These new sources of flavor and CP violation can also disrupt the 
delicate mechanism which suppresses FCNCs to acceptably low levels in the 
SM (the GIM mechanism \cite{Glashow:gm}).  If the off-diagonal elements
of the squark or slepton soft parameters are of order the typical squark
or slepton masses, then generically there would be large flavor-mixing
effects \cite{Donoghue:1983mx}, because the rotations that diagonalize the quarks and charged
leptons need not diagonalize the squarks and sleptons. FCNCs thus 
significantly constrain the \lsoft parameter space.  This is commonly
known as the {\it supersymmetric flavor problem}, which will be discussed
in Section~\ref{fcncex}.

\end{itemize}
\subsection{Constraints on \lsoft from FCNCs}
\label{fcncex}
\subsubsection{FCNCs and the mass insertion approximation}
The explanation for the suppression of FCNCs is a great success of the SM. 
The tree level couplings of the fermions to the neutral gauge bosons 
do not change flavor because the fermions
are rotated from gauge to mass eigenstates by unitary diagonalization
matrices. In addition, the higher order contributions from charged
currents at one-loop vanish in the limit of degenerate fermion masses: 
this is the GIM mechanism. 

For example, consider \( K^{0}-\overline{K}^{0} \) mixing in the SM, which 
proceeds via the box diagram involving \( W \) bosons and
up-type quarks \( u,c,t \). The GIM mechanism dictates that
the amplitude is suppressed (in addition to the loop suppression) by small
fermion mass differences. The leading contribution is \( \sim
(m^{2}_{c}-m^{2}_{u})/M^{2}_{W} \); other contributions are further
Cabibbo-suppressed.

In the MSSM, there are many additional flavor-changing couplings which can
contribute to FCNCs at one loop. Consider for example the implications for 
the \( K^{0}-\overline{K}^{0} \) 
mixing example given above. In addition to the \( W \) box diagram, there 
are now diagrams with \( \widetilde{W} \)s and up-type squarks \( 
\widetilde{u},\widetilde{c},\widetilde{t} \),
which are proportional to {\it sfermion} mass differences, {\it e.g.} \(
(m_{\widetilde{c}}^{2}-m_{\widetilde{u}}^{2})/\widetilde{m}^{2} \), in 
which $\widetilde{m}$ denotes a typical soft mass.  Therefore, the 
superpartner loop contributions in general involve an unsuppressed factor
of order unity unless there is an approximate degeneracy of the squarks;  
of course, the overall magnitude of the diagram may be smaller because the
superpartners in the loop are typically heavier than \( m_W \).  If there 
is an approximate squark degeneracy, this type of 
contribution is not a serious problem; {\it i.e.}, there is a 
``super-GIM mechanism.''

The supersymmetrized charged current interactions contribute to FCNCs even 
if \lsoft is flavor diagonal. If \lsoft has nontrivial flavor 
structure at low energies, then there are additional contributions to 
FCNC which arise from supersymmetrizing the fermion couplings
to the neutral gauge bosons. The resulting fermion-sfermion-gaugino 
couplings, such as the quark-squark-gluino couplings and the 
quark-squark-neutralino couplings, are generically not flavor diagonal. 
This is because the squark mass matrices are typically not diagonal in the 
basis in which the quarks are diagonal, as shown explicitly in 
Section~\ref{conventionsapp}. 
In this case, gluino and neutralino loops can also contribute to FCNCs at  
one-loop order.\footnote{Diagrams 
involving charged Higgs bosons are also present.  The couplings of the 
charged Higgs to quarks obey the CKM hierarchy, and hence their 
interactions cannot probe genuine supersymmetry flavor-violating effects 
such as those involving the gluinos and neutralinos.} Hence, in 
generic supersymmetric models there is an 
explicit failure of the supersymmetric version of the GIM mechanism.

The amplitudes for such flavor-changing and CP-violating processes 
of course depend on various entries of the $6\times 6$ sfermion 
diagonalization matrices, given explicitly in Eq.~(\ref{diagUSCKM}) and 
Eq.~(\ref{diagDSCKM}). These matrices are related in complicated ways to 
the original parameters of \lsoft expressed in the SCKM basis.  Rather 
than working with the explicit diagonalization matrices, it 
is often useful to recall that the size of the flavor-violating 
effects can be related to the off-diagonal elements of the sfermion mass 
matrices.  If these off-diagonal entries are small compared to the 
diagonal ones, it is illustrative to use the {\it mass insertion 
approximation}, in which the sfermion diagonalization matrices can be 
expressed as a perturbation expansion in the off-diagonal entries of the 
sfermion mass matrices normalized by a common sfermion mass 
\cite{Hagelin:1992tc}.\footnote{For those unfamiliar with the mass 
insertion approximation, we present a simple two-family example in 
Appendix~\ref{FCNCexapp}.} 

Explicitly, consider the full $6\times 6$ sfermion mass matrices expressed 
in the SCKM basis, as presented in Eqs.~(\ref{pok1}). The 
diagonal terms are denoted 
as $(m^{2}_{{\rm AA}})_{ii}$, in which AA can be LL or RR,
and $i=1,2,3$ is a 
family index.  For notational simplicity, here we have suppressed the 
sfermion flavor index (for up-type squarks, 
down-type squarks, charged sleptons, and sneutrinos).  
The off-diagonal terms in the sfermion mass matrices are
$(\Delta_{{\rm AB}})_{ij}$, where AB is LL, RR, LR, or RL (see 
Eq.~(\ref{pok1a})). For example, $m^2_{LL}$ may be written as
\begin{equation}
\label{pok2}
m^2_{LL} = \left(
\begin{array}{ccc}
  (m^{2}_{LL})_{11}    & (\Delta_{LL})_{12} &
(\Delta_{LL})_{13}
\vspace{0.2cm} \\
(\Delta_{LL})_{21} &    (m^{2}_{LL})_{22}  & (\Delta_{LL})_{23}
\vspace{0.2cm} \\
(\Delta_{LL})_{31} & (\Delta_{LL})_{32} &   (m^{2}_{LL})_{33}
\end{array}\right),
\end{equation}
and analogously for all the other matrices. Hermiticity dictates that
$(\Delta_{LL})_{ij} = (\Delta^{*}_{LL})_{ji}$ and
$(\Delta_{RR})_{ij}= (\Delta^{*}_{RR})_{ji}$, as well as 
$(\Delta_{LR})_{ij}=(\Delta^{*}_{RL})_{ji}$.

FCNC constraints translate most naturally into bounds on the mass 
insertion parameters, which are defined to be the $\Delta$s normalized by 
a common soft mass.  For example, the mass insertion parameters can be  
defined as follows:
\begin{equation} 
\label{deltadef}
(\delta_{AB})_{ij}=
\frac{(\Delta_{AB})_{ij}}{\sqrt{(m^{2}_{AA})_{ii}(m^{2}_{BB})_{jj}}}.
\end{equation}
The choice of the denominator is of course not unique, as any mass scale 
which characterizes the diagonal terms would suffice. Arguments for the 
choice of this denominator were first presented in \cite{Gabbiani:1996hi}. 

In the above expressions, the LL and RR mass insertion
parameters involve the soft mass-squared parameters $m^2_Q$ 
and $m^2_U$ rotated by the left-handed and right-handed quark 
diagonalization matrices, respectively.  The LR and RL mass insertion 
parameters involve
linear combinations of $\widetilde{A}$ and $\mu$, rotated by the same
combination of matrices which diagonalize the Yukawas.  The LR and RL 
blocks are generated only after electroweak breaking, and consequently   
their size is typically the geometric mean of the electroweak scale and 
the scale of the soft supersymmetry-breaking parameters. On the other 
hand, only the diagonal entries of the LL and RR blocks are influenced by 
electroweak breaking; the flavor-violating entries originate solely from 
supersymmetry breaking. In addition, while the LL and RR parameters are 
invariant under 
$U(1)_{PQ}$ and $U(1)_{R}$, the LR and RL parameters are not R invariant 
(they have R charge $\pm 2$ according to our conventions in 
Table~\ref{c_spur}). Physical observables are either functions of the 
absolute squares of LR/RL quantities or of the LR/RL quantities  
multiplied by the appropriate R-charged soft parameters.

In the next section we briefly discuss connections between data and the
flavor-dependent soft parameters.  There has been a tremendous amount of
work studying the implications of FCNCs for various supersymmetric models,
and it is beyond the scope of this review to cover all models or discuss
each process in detail.  A number of excellent reviews exist 
which provide a comprehensive approach to this subject
\cite{Buras:2003wd,Buras:vd,Buras:1999tb,Buras:1998ra,Nir:2002gu,Nir:2001ge,Nir:1998mq,Nir:1998pg} 
for those who want more detail in this area.

\setcounter{footnote}{0}
\subsubsection{Constraints from FC processes} 
The absence of flavor-changing decays for many systems puts 
strong constraints on certain combinations of the soft parameters. 
There are various observables which are and/or will be under experimental 
investigation at various meson factories. A partial list would include
the mass differences and CP-violating mixings of beauty, charm and strange 
mesons as well as rare decays such as $b\rightarrow s \gamma$. In the 
presentation that follows, the experimental bounds are all taken from the 
Particle Data Group Collaboration \cite{Hagiwara:fs} unless otherwise 
indicated. 

As the FCNC constraints generically require that the off-diagonal entries
of the sfermion mass matrices in the SCKM basis are suppressed to some
degree, it is standard to express the constraints in the context of the
mass insertion parameters defined in the previous subsection. Before
discussing specific constraints, we emphasize that many of the constraints
on the flavor-changing parameters in the literature have been evaluated
with simplified assumptions.  In general, these assumptions need not apply
and nontrivial cancellations may occur which can relax certain
constraints.  We depict several examples of FCNC observables, including
the SM predictions and their sensitivities to the MSSM parameters, for 
both the hadronic (Table~\ref{QFV_tab}) and leptonic (Table~\ref{LFV_tab})
sectors.

\begin{table}[tbp]
\begin{center}
\begin{tabular}{|c|c|c|}
\hline
&  & \\
Observable & SM Prediction&$\;$ MSSM Flavor Content $\;$
\\ 
  & &\\
\hline
  & &\\
$\Delta m_K $& $\sim (V^*_{cs}V_{cd})^2$ 
& $(\delta_{AB})_{12}$ \\
& &  \\
$\epsilon$ & 
$\sim$Im$(V_{ts}^* 
V_{td})$Re$(V_{cs}^* V_{cd})$&
$(\delta_{AB})_{12}$\\
 & & \\
$\epsilon^{'}/\epsilon$& 
$\sim$Im$(V_{ts}^* V_{td})$&
$(\delta_{AB})_{12}$\\
& & \\

$b\rightarrow s \gamma$& 
$\sim V_{tb} V_{ts}^*$
&$(\delta_{AB})_{23}$\\
& & \\

$A_{CP}(b\rightarrow~s\gamma)$& 
$\sim 
\alpha_s(m_b) \frac{V_{ub}}{V_{cb}}\frac{m_c^2}{m_b^2}$ & 
$(\delta_{AB})_{23}$\\
 & & \\

$\Delta m_{B_d}$ & 
$\sim (V_{td}^* 
V_{tb})^2$&
$(\delta_{AB})_{13}$\\
 & & \\

$\Delta m_{B_s}$ & 
$\sim (V_{ts}^*V_{tb})^2$&
$(\delta_{AB})_{23}$\\
 & & \\

$A_{CP}(B\rightarrow~\psi K_S)$& $=\sin 
2\beta$&$(\delta_{AB})_{13}$\\
 & &\\

$A_{CP}(B\rightarrow~\phi K_S)$ &$=\sin 
2\beta$&
$(\delta_{AB})_{23}$ \\

 & & \\

\hline
\end{tabular}
\caption{A partial list of flavor-violating observables in the quark
sector and their relation to SM and MSSM parameters. The $\delta$s are the 
mass insertion parameters for the up- and down-type squark sectors, with 
AB denoting LL, LR, RL, or RR. \label{QFV_tab}} 
\end{center}
\end{table}

A model-independent parameterization of such new FCNC effects based on the
mass insertion approximation, with a leading order linear mass insertion,
has been used to set limits on the off-diagonal mass parameters 
\cite{Gabbiani:1996hi,Misiak:1997ei}. The full panoply of FCNC constraints 
on the off-diagonal masses include those which arise from $\Delta m_K$,
$\Delta m_B$, $\Delta m_D$, $\epsilon$, $\epsilon'/\epsilon$,
$b\rightarrow s \gamma$, $\mu\rightarrow e \gamma$, and the electric 
dipole moments $d_n$ and $d_e$ (these will be discussed in 
Section~\ref{EDMsect}).  In 
much of the analysis of \cite{Gabbiani:1996hi}, the
gluino-mediated loops are the dominant source of FCNC; {\it i.e.}, the 
chargino contributions, which can be significant, are not included. In 
general, the bounds are derived assuming that single mass insertion 
parameters saturate the FCNC constraints.

The strongest FCNC constraints by far arise from the kaon system, imposing
very severe limits on mixing of the first and second generation squarks.
The kaon system suffers from large hadronic uncertainties, and hence care 
must be taken in the interpretation of the results both within the SM and 
supersymmetry. The relevant observables include:
\begin{itemize}

\item {\bf $\Delta m_K= m_{K_L}-m_{K_S}$:} The experimental bound 
quoted by the PDG is $\Delta m_K=3.490\pm 0.006 \times 10^{-12}$ MeV 
\cite{Hagiwara:fs}. The leading SM contribution is $\sim 
(V_{cs}^*V_{cd})^2$. The most significant MSSM contributions typically are 
those involving gluinos and down-type squarks, and charginos and up-type 
quarks.  As shown in the table, the results are sensitive to the 12 
entries of the 
LL, LR, and RR subblocks of the squark mass matrices in the SCKM basis.  
There are also neutralino--down-type squark and charged Higgs--up-type 
quark diagrams, but they tend to be numerically less significant 
in most regions of parameter space. 

\item {\bf $\epsilon$:} 
This parameter measures the CP violation 
due to mixing of short- and long-lived kaons and is used to fix the 
unitarity triangle.  The experimental value is $\epsilon=2.28 \times 10^{-3}$. 
In the SM, $\epsilon \sim$Im$(V_{ts}^* V_{td})$Re$(V_{cs}^* V_{cd})$.  
Roughly, the MSSM contributions are due to the imaginary part of the 
amplitude of the diagrams which contribute to $\Delta m_{K}$.

\item {\bf $\epsilon'$:} 
This parameter measures the CP violation due to decay in the K 
system; the experimental world average is 
$\epsilon'/\epsilon=(16.6\pm 1.6)\times 10^{-4}$. The SM 
contributions include $W-q$ penguin diagrams 
$\sim {\rm Im}(V_{ts}^* V_{td})$. The supersymmetric contributions include 
box and penguin
diagrams also involving gluinos and charginos, which probe similar \lsoft
parameters as $\epsilon$, However, $\epsilon'$ is particularly sensitive 
to the 12 
entry of the LR blocks of the squark mass matrices.  This quantity suffers 
from large hadronic uncertainties. 
\end{itemize}

In the kaon system, $K^0-\overline{K}^0$ mixing constraints allow for
limits to be placed on the real parts ${\rm Re}(\delta^d_{12})_{LL}< {\rm
few}.10^{-2}$ and ${\rm Re}(\delta^d_{12})_{LR}\sim {\rm 
few}.10^{-3}$.\footnote{The constraints on the mass insertions depend of 
course on the magnitudes of the soft parameters: the bounds mentioned here 
assume 
$m_{\widetilde{g}} \sim m_{\widetilde{q}}\sim 500\ {\rm GeV}$ and that the 
gluino--squark diagrams are the dominant ones.} 
The $\epsilon$ parameter provides an extremely stringent 
constraint on supersymmetric models (and any new flavor-violating physics 
in which the SM GIM mechanism is violated), because a {\it generic} \lsoft 
with superpartner masses of order the electroweak scale, diagonal and
off-diagonal squark masses of similar orders of magnitude in the SCKM
basis, and off-diagonal phases of $O(1)$ overproduces $\epsilon$ by
seven orders of magnitude.  The direct CP-violating parameter
$\epsilon'/\epsilon$ also leads to strong constraints, in particular on
the imaginary part ${\rm Im}(\delta^d_{12})_{LR}\sim {\rm few}.10^{-5}$.
$\epsilon'/\epsilon$ in particular suffers from large hadronic
uncertainties, such that it is not absolutely clear whether the SM
prediction is in agreement with the experimental result, although they are
consistent.  Many authors have speculated whether or not supersymmetry
could provide the dominant contribution to $\epsilon'/\epsilon$
\cite{Masiero:1999ub,Baek:1999jq,Baek:2001kc,Khalil:1999zn,Babu:1999xf,Khalil:1999ym,Kagan:1999iq,Khalil:2000ci,Eyal:1999gk}.

The B system also yields constraints on the allowed forms of 
the \lsoft parameters, and is theoretically relatively clean 
in comparison to the kaon system. For a recent review, see {\it e.g.} 
\cite{Neubert:2002ku}.\footnote{The present experimental and theoretical 
situations for the inclusive B decays are summarized in the recent review 
\cite{Hurth:2003vb}.}
The relevant observables include:
\begin{itemize}

\item {\bf $BR(b\rightarrow s \gamma$)} and 
{\it $A_{CP}(b\rightarrow~s\gamma)$:}
It has been known for quite some time that $b\rightarrow s \gamma$  
provides important tests of supersymmetry 
\cite{Bertolini:1986th,Bertolini:1990if}. 
The leading SM contribution to the branching ratio appears at one loop 
level, with the characteristic Cabibbo suppression. Supersymmetry 
contributions also arise at one loop, and are generically comparable to 
or larger than 
the SM contributions if no mechanisms for suppressing the new sources of 
flavor violation exist. The current experimental weighted average of the 
inclusive $B \rightarrow X_s \gamma$ branching
ratio \cite{aleph,belle,cleo} is
$BR(B\rightarrow X_s \gamma)_{exp}=
(3.23 \pm 0.41) \times 10^{-4}$, which
is in rough agreement with the SM theoretical prediction (at NLO in QCD)
$BR(B\rightarrow X_s \gamma)_{SM} = (3.73 \pm 0.30) \times 10^{-4}$; see
{\it e.g.} \cite{Gambino:2001ew}.

The general agreement between the SM theoretical prediction and the   
experimental results for $b\rightarrow s \gamma$ have provided useful
guidelines for constraining the MSSM parameter space.  Superpartners and
charged Higgs loops generically contribute to $b\rightarrow s \gamma$, at 
a level competitive with the SM, with contributions that depend strongly 
on the parameters of \lsoft, as well as $\mu$ and $\tan\beta$.  This 
process has been most often studied in the MFV scenario at 
LO \cite{Bertolini:1990if,Baer:1997jq}, in certain limits at NLO 
\cite{Ciuchini:1997xe}, and including large $\tan\beta$ enhanced two-loop 
supersymmetry contributions \cite{Degrassi:2000qf,Demir:2001yz}, and 
all-order resummation of $\tan\beta$ enhanced QCD corrections 
\cite{Carena:2000uj}.  

In MFV scenarios, $b\rightarrow s \gamma$ receives contributions from 
charged Higgs and chargino exchange diagrams. The charged Higgs diagram 
has the same sign as the $W$ boson contribution, which already saturates 
the experimental result. Therefore, the chargino and charged Higgs 
contributions must interfere destructively if the charged Higgs, 
charginos, and stops have masses near their present experimental lower 
bounds. In mSUGRA parameter space, this cancellation occurs for a 
particular ``sign of $\mu$'' --- more precisely, when the $\mu$ 
parameter and the stop trilinear couplings are of opposite sign.

If new sources of flavor violation exist in \lsoft, there are additional 
contributions to $b\rightarrow s \gamma$ involving the exchange of 
down-type squarks together with gluinos or neutralinos. 
Depending on the magnitude of the flavor violation in the down squark 
sector, the charged Higgs and chargino contributions can become 
subleading. In particular, in the presence of a chirality-flipping 
mixing between the $\widetilde{b}$ and $\widetilde{s}$ squarks, the gluino 
exchange diagram contributes to the dipole coefficient 
\begin{equation}
\sim \left ( \frac{m_W}{m_{\widetilde{q}}}\right )^{2} 
\frac{m_{\widetilde{g}}}{m_b}\frac{\alpha_s}{\alpha} 
\frac{(\delta_{23}^{d})_{LR}}{V_{tb}V_{ts}^{\star}}, 
\end{equation}
which becomes quite large unless the supersymmetry 
breaking scale is high enough or flavor violation is shut off. The 
present contraints from the experimental knowledge of $b\rightarrow s 
\gamma$ rate is $(\delta_{23}^{d})_{LR}\sim {\cal{O}}(10^{-2})$ when the 
strange quark mass effects are neglected \cite{Gabbiani:1996hi}. As an 
alternative view, one can consider the scenario discussed in 
\cite{Everett:2001yy}, where it was found that the
amplitudes involving the right-handed $b$ quark can cancel with   
the SM, charged Higgs, and chargino contributions, and the
present bounds on the branching ratio can be saturated via
amplitudes involving right-handed $s$ quarks with a much larger 
$(\delta_{23}^{d})_{LR}$.

The CP asymmetry of the $b\rightarrow s\gamma$ is an excellent probe of
new physics, as the SM contribution is less than 1\% 
\cite{Kagan:1998bh}.
The current experimental bounds on this quantity are $-0.3 < A_{CP} <
0.14$, which are consistent with zero but also may allow non-SM effects.
Supersymmetry contributions could in general be quite a bit larger than
the SM prediction due to the additional CP-violating \lsoft phases.

\item {\bf $A_{CP}(B\rightarrow~\psi K_S)$:}
This observable is the ``golden mode" for the study of CP violation in the 
B system, as it is theoretically very clean and provides a measurement of 
the angle $\beta=~Arg\left[-\frac{V_{cd} 
V_{cb}^*}{V_{td}V_{tb}^*}\right]$ 
of the unitarity triangle ($A_{CP}(B\rightarrow~\psi 
K_S)\propto \sin 2\beta$).
There has been experimental observation of an $O(1)$ CP 
asymmetry in this decay.  The experimental world average 
\cite{Nir:2002gu} is
\begin{equation}
\sin 2\beta=0.734\pm 0.054,
\end{equation}
which has provided the first conclusive evidence
supporting the Kobayashi-Maskawa picture of CP violation in the 
SM.\footnote{Recall 
the SM picture of CP violation provides an elegant  
explanation for the size of $\epsilon$, but the theoretical
uncertainties in $\epsilon'/\epsilon$ do not allow for corroborating
evidence from that observable.} 
It is difficult (though not impossible, see {\it e.g.} \cite{Brhlik:1999hs}) to 
have such $O(1)$ effects in the B decays if the phases of \lsoft are the 
dominant source of CP violation.  There is a tree-level SM 
contribution to the decay amplitude, such that supersymmetric 
contributions are negligible and supersymmetry can only influence the CP 
asymmetry of the B decays through $B-\overline{B}$ mixing. 

\item {\bf $A_{CP}(B\rightarrow~\phi K_S)$:}
Recently the CP asymmetries for this exclusive process have been reported.  
In the SM the time-dependent CP asymmetry should arise only from
$B_d-\overline{B}_d$ mixing, as for the analogous CP asymmetry of $\psi K_S$, 
and should be essentially equal to $\sin 2\beta$.  The reported asymmetry 
is $2.7\sigma$ away from this value, although the error bars are large.  
Several recent analyses have studied 
this situation, both in model-dependent and model-independent analyses
\cite{Khalil:2002fm,Kane:2002sp,Ciuchini:2002uv,Harnik:2002vs,Chiang:2003jn}.\footnote{There
are many possible scenarios here.  For example, one scenario 
\cite{Kane:2002sp} uses the gluino diagram with the $(\delta^d_{RL})_{23}$ 
insertion that also gives a satisfactory description of $b\rightarrow s 
\gamma$ \cite{Everett:2001yy}.}

\item {\bf $\Delta m_{B_d}$:}
This quantity measures the mass mixing in the $B_d$ meson system; its 
experimental value is $\Delta m_{B_d}=3.22 \times 10^{-10}$ MeV. In the 
SM this is dominated by the $W-t$ box diagram $\sim (V_{td}^* V_{tb})^2$.  
It is used as a constraint to fix the unitarity triangle and also provides 
constraints on the MSSM flavor violating parameters, especially 
$(\delta_{AB})_{13}$.

\item {\bf $\Delta m_{B_s}$:}
Mass mixing in the $B_s$ meson system is also dominated by the SM $W-t$ 
box diagram $\sim (V_{ts}^* V_{tb})^2$.   In the MSSM, it has similar 
dependence on the mass insertion  parameters with 
$(\delta_{AB})_{13}\leftrightarrow 
(\delta_{AB})_{23}$. The current experimental bound is $\Delta 
m_{B_s}>8.62 \times 10^{-9}$ MeV.  Forthcoming experiments at the b 
factories and the LHC should provide detailed measurements of the $B_s$ 
system.
\end{itemize}
Typical bounds on the $\delta_{13,23}$ parameters from the B systems are 
less stringent than the analogous bounds in the K system 
\cite{Gabbiani:1996hi,Misiak:1997ei}. The lone exception is $b\rightarrow 
s \gamma$, which generically provides significant constraints on the 
\lsoft parameter space.

\begin{table}[tbp]
\begin{center}
\begin{tabular}{|c|c|}   
\hline
& \\   
Observable & $\;$ MSSM Flavor Content $\;$\\
& \\
\hline
 &\\

$\mu\rightarrow e \gamma$& $(\delta_{AB})_{12}$ \\
 &\\

$\tau\rightarrow \mu \gamma$
&$(\delta_{AB})_{23}$\\
 & \\

$\tau\rightarrow e \gamma$&
$(\delta_{AB})_{13}$\\
  &\\   
\hline
\end{tabular}
\caption{A partial list of lepton flavor-violating observables and
their relation to MSSM parameters. The $\delta$s should be understood as 
those arising from the slepton sector. In each case the SM 
contribution is identically zero in the absence of 
right-handed neutrinos due to the conservation of individual lepton 
numbers $L_e$, $L_{\mu}$, and $L_{\tau}$. \label{LFV_tab}} \end{center}
\end{table}

In the leptonic sector, the off-diagonal slepton masses give rise to 
flavor violating processes such as $\mu\rightarrow e \gamma$, 
$\tau\rightarrow \mu \gamma$, $\tau\rightarrow e \gamma$, $\tau 
\rightarrow \mu \mu \mu$.
Therefore, lepton 
flavor violating (LFV) processes in principle will also give rise to 
signals/constraints of the mass parameters in the lepton sector of the 
MSSM; see {\it e.g.} 
\cite{Borzumati:1986qx,Hisano:1995cp,Gabbiani:1996hi}. 
A brief list of 
such observables is given in Table~\ref{LFV_tab}. 

The experimental prospects for improving
the limits or actually measuring LFV processes are very promising.
The 90\% C.L. limits of
${\rm BR}(\tau \rightarrow \mu \gamma)< 1.1\times 10^{-6}$
\cite{Ahmed:1999gh}
and ${\rm BR}(\mu \rightarrow e \gamma)< 1.2\times 10^{-11}$
\cite{Brooks:1999pu}
are particularly stringent in constraining supersymmetric models.
These limits will be lowered in the next 2-3 years
as the present B factories,
inevitably producing tau leptons along with the b quarks, will collect
15-20 times more data and as the new
$\mu\to e\gamma$ experiment at PSI probes the branching ratio
down to $10^{-14}$
\cite{BABAR_tmg:2002,meg_at_PSI:2002}.\\

We close this subsection by pointing out that in the large $\tan\beta$
regime, the above FCNC constraints must be reevaluated for a number of
reasons.  One important effect is that certain diagrams discussed in the
general considerations above are $\tan\beta$-enhanced.  However, it has
recently been realized that additional contributions to FCNC mediated by
Higgs bosons emerge in the large $\tan\beta$ limit.

The essential physics is as follows.  At tree level, the MSSM is a two 
Higgs doublet model in which the up-type and down-type quarks couple to 
different Higgs bosons.  This class of two Higgs doublet models is 
free of tree-level FCNCs, as shown by \cite{Glashow:1976nt}.
This property of the quark-Higgs Yukawa couplings is enforced 
by the analyticity requirement of the superpotential in supersymmetric 
theories.  However, since supersymmetry is softly broken, one should expect that this property 
does not hold at higher orders in perturbation theory.  Indeed, there 
are new effective flavor-changing couplings which arise from large loop 
corrections to the couplings of Higgs bosons to down-type quarks and leptons 
\cite{Hall:1993gn,Carena:1994bv}. 

This effect in the MSSM at large $\tan\beta$ was 
pointed out for the quarks in \cite{Babu:1999hn,Hamzaoui:1998nu} and for 
the leptons in \cite{Babu:2002et}; the CKM matrix also receives finite 
radiative corrections, as discussed in \cite{Blazek:1995nv}. The 
Higgs-mediated FCNC contributions also have a unique feature: they do not 
decouple when the superpartner masses are much larger than the electroweak 
scale, provided that the Higgs sector remains light.

Higgs-mediated effects have been discussed for various FCNC processes
including $B\rightarrow X_s \gamma$ \cite{Degrassi:2000qf,Carena:2000uj},
leptonic and semileptonic $B$ decays
\cite{Huang:2000sm,Chankowski:2000ng,Bobeth:2001sq,Isidori:2001fv,Bobeth:2001jm,Arnowitt:2002cq,Demir:2002cj,Bobeth:2002ch,Mizukoshi:2002gs,Baek:2002wm}
as well as $B^0$--$\overline{B^0}$ mixing \cite{Buras:2001mb} either
individually or combined \cite{Buras:2002wq,Buras:2002vd}.  See also {\it
e.g.} \cite{D'Ambrosio:2002ex} for a recent analysis using an effective
field theory approach. For example, the branching ratio of $B_s\rightarrow
\mu^+\mu^-$ decay, which is $O(10^{-9})$ in the SM, is enhanced by
Higgs-mediated effects to $O(10^{-6})$ or larger for $\tan\beta\geq 50$
and $m_A\sim m_t$, in which $m_A$ denotes the usual pseudoscalar mass 
parameter.  Future measurements at the Tevatron and LHC will be able to 
determine whether such nonstandard effects in $B_s\rightarrow \mu^+\mu^-$ 
are present.

Higgs-mediated FCNC processes in the presence of both supersymmetric CP
and flavor violation lead to a host of interesting phenomena
\cite{Curiel:2002pf,Demir:2003bv}. For example, the CP asymmetry of
$B\rightarrow X_s \gamma$ can be enhanced by such large $\tan\beta$
effects \cite{Demir:2001yz}.  The Higgs-mediated amplitudes can compete,
for instance, with the box diagram contributions to
$B^0$--$\overline{B^0}$ mixing and their interference can either relax or
strengthen existing bounds on various mass
insertions.\footnote{Furthermore, for large values of $\tan\beta$, the
Yukawa couplings of all down quarks assume universal size whereby leading
to experimentally testable signatures for Higgs decays for both
flavor-changing and flavor-conserving channels.} Supersymmetric flavor
violation effects are also important for Higgs couplings to leptons,
though various effects, such as the enhancement of light quark Yukawas,
are typically milder due to the absence of supersymmetry QCD corrections.

\subsubsection{Implications for model building}
Given the tightness of the FCNC constraints, it is apparent that to good
approximation supersymmetry must realize a super-GIM mechanism, thereby 
restricting the class of viable models of \lsoft.  
One way to avoid the FCNC constraints is to assume that at least a subset
of the soft scalar masses are multi-TeV such that flavor-violating effects
decouple.  The heavy-superpartner approach is in contrast to the
philosophy that the scale of the soft supersymmetry-breaking parameters is related to
the origin of the electroweak scale, although models can be constructed in
which the third family sparticles, which have the strongest effects on
radiative electroweak symmetry breaking, are relatively light
\cite{Cohen:1996vb}. This may be a viable possibility, although two-loop
effects may spoil the decoupling \cite{Arkani-Hamed:1997ab}. In this
review, we are mainly interested in light superpartners, and thus do not
discuss this scenario further.

Much effort has gone into constructing models of ${\mathcal{L}}_{soft}$ 
that guarantee without tuning the absence of FCNC.  With light 
superpartners, there are two general approaches: (i) universality, which 
assumes that the soft masses are universal and flavor diagonal, (ii) 
alignment, which assumes that the soft masses have a structure that allows 
the quark and squark masses to be simultaneously diagonalizable. 
The super-GIM mechanism arises in the universal, flavor-diagonal scenario 
since the squark and slepton mass matrices are all proportional 
to the unit matrix in flavor space. When the Yukawa couplings 
are rotated to the diagonal mass basis no 
off-diagonal soft masses are generated and the diagonal masses are 
approximately degenerate.
The super-GIM mechanism also arises in the alignment mechanism: if the 
soft mass matrices and trilinears are diagonalized by exactly the same 
rotations that diagonalize the Yukawa matrices 
\cite{Nir:1993mx,Hall:1990ac}. For example if there is
a non-Abelian family symmetry in some supergravity mediation model, at
leading order the soft matrices are diagonal and the operators which
generate the Yukawa matrices will also generate soft mass matrices tending
to align the Yukawa and soft matrices, with the approximate degeneracy of
the diagonal masses enforced by the family symmetry \cite{Dine:1993np}.

Supergravity-mediated supersymmetry-breaking models do not typically
possess a super-GIM mechanism. In other words, the off-diagonal elements
of the soft mass matrices can generally be nonzero. In addition, the 
diagonal elements of the soft mass matrices may not be accurately 
degenerate. The off-diagonal soft masses at low energies arise
both because of explicit flavor-dependence of supersymmetry breaking at
the high energy scale and the effects of RG running due to the effects of
Yukawa matrices in going down to low energies.  In nonminimal supergravity
models, there is also generically an explicit failure of the alignment
mechanism because the trilinear couplings are generically not proportional
to the corresponding Yukawa couplings; see {\it e.g.} \cite{Kobayashi:2000br} 
for further discussions.\footnote{This feature can have implications for 
EDM constraints, as discussed in Section~\ref{EDMsect}.}

Approaches for which the only source of flavor violation arises
in the Yukawa couplings, such as gauge-mediated supersymmetry-breaking
scenarios or MFV scenarios in minimal supergravity, pass the FCNC
constraints, although $b\rightarrow s\gamma$ provides substantial
constraints on the allowed parameter space. 
Several approaches, such as
the alignment and decoupling mechanisms mentioned previously, can (in
their simplest implementations) be insufficient for the strong FCNC bounds
from the K system, although models can certainly be built which pass the
tests. The approximate CP approach \cite{Eyal:1998bk}, in which all phases
(including the CKM phase) are assumed to be small, has been disfavored
from the observation of large CP-violating effects in the B system.
However, having no new flavor-violating effects in the parameters of
\lsoft is not necessarily the only option; nonuniversality is in
particular more tolerable for the soft supersymmetry-breaking parameters 
of the third generation.  \\

Let us conclude this section by considering the following natural question
in this context: how is the theoretical origin of the soft mass matrices
related to that of the Yukawa matrices? Different mechanisms for
supersymmetry breaking and mediation illustrate the different
possibilities for both the scale at which the soft masses are generated
and the flavor dependence of the soft masses at that scale. In this review
we assume that the Yukawa matrices are generated at a high scale at or
close to the string scale. By contrast supersymmetry breaking may occur at
either a high scale, as in gravity mediation, or a lower scale, as in
gauge mediation. In addition the soft mass matrices may have flavor
dependence, as is generically true in gravity mediation, or they may be 
flavor diagonal, such in gauge and anomaly mediation.  
It is also possible that the gravity mediated models predict flavor 
diagonal soft mass parameters at the high energy scale, such as in
mSUGRA or the dilaton-dominated scenario in string-motivated supergravity.  
In such MFV scenarios, the Yukawa couplings are the only source of flavor
violation in the theory and their effects are filtered to the soft masses
through RG evolution.  An inspection of the RGEs for the soft mass
parameters (see Appendix~\ref{rgeapp}) demonstrates that the
flavor-violating effects of the Yukawa couplings leads to low energy soft
mass matrices which exhibit some degree of flavor dependence.

From a purely bottom-up perspective the soft parameters and Yukawa
structure are intimately linked and cannot be untangled solely from
experimental information.  Nevertheless, if one is willing to make
theoretical assumptions about the form of the soft supersymmetry-breaking
parameters, the observed flavor dependence of the low energy soft masses
could provide a window into the structure of the Yukawa matrices that
would not be possible to obtain from the observed low energy masses and
mixing angles alone.  However, experimental data alone cannot confirm that
the measured soft parameters are consistent with such theoretical
assumptions. This is because the observable quantities not only involve 
the soft parameters, but also the individual left-handed and right-handed 
quark rotation matrices, of which only a subset of parameters can be 
measured independently --- the masses, CKM 
entries, and Jarlskog-type invariants. Therefore, additional theoretical 
input is required in order to learn any further details of the Yukawa
couplings. The issue can be summarized as follows: the observable flavor
structure of the sfermion sector depends on {\it two} unknown mechanisms
which presumably have their resolution in high scale physics:  the origin
of the fermion mass hierarchy (the usual flavor problem of the fermion
sector), and the supersymmetry-breaking/mediation mechanisms.

\subsection{Dipole moment constraints}
\setcounter{footnote}{0}
\subsubsection{$g_{\mu}-2$}
\qquad Recently, precise measurements of the anomalous magnetic moment of
the muon, $a_{\mu}= (g_{\mu}-2)/2,$ were reported \cite{Bennett:2002jb}.  
In a supersymmetric world the entire anomalous magnetic moment of any
fermion vanishes if supersymmetry is unbroken \cite{Ferrara:1974wb}, so
magnetic moments have long been expected to be very sensitive to the
presence of low energy supersymmetry, and particularly to supersymmetry
breaking
\cite{Barbieri:aj,Kosower:1983yw,Yuan:ww,Arzt:1992wz,Lopez:1993vi,Chattopadhyay:1995ae,Moroi:1995yh,Carena:1996qa}.  
The theoretical analysis can be done in a very general and
model-independent manner, and illustrates nicely how one can draw
significant conclusions about the MSSM parameter space from this process.
We describe the situation here both because the effect may be a
measurement of physics beyond the SM, and to illustrate the connections of
$g_{\mu}-2$ to the soft parameters.

\qquad Knowing whether the $g_{\mu}-2$ data indicates a deviation from the
SM depends on knowing the SM theory prediction.  The SM prediction 
is difficult to ascertain, though, because the SM contributions to 
$g_{\mu}-2$ include nonperturbative QCD effects (such as
the hadronic vacuum polarization) which are not calculable from first
principles. Such effects are calculated using data to replace the
nonperturbative parts.  Recent calculations
\cite{Davier:2002dy,Hagiwara:2002ma} use two methods to carry out this
procedure.  If the method using data from low energy $e^{+}e^{-}$
collisions is used, experiment and theory differ by about $3\sigma$
\cite{Davier:2002dy,Hagiwara:2002ma}.  Of course, standard deviations 
from a calculable number are more significant than those in one bin of a 
histogram where any of a number of bins could fluctuate, so $3\sigma$ is a 
very significant deviation.  However, an alternative method using 
information from $\tau$ decays leads to a deviation less than $1 \sigma$ 
\cite{Davier:2002dy}, while it should in principle give the same result. 
Until this discrepancy is understood, it cannot be concluded that there 
is a significant deviation from the
SM.\footnote{It can be argued, though, that considerable theoretical
extrapolation is needed for the $\tau$ decays method (for a detailed
critique see \cite{Melnikov:2001uw}), such that the $\tau$ discrepancy may
not be relevant.} If the deviation is real then the supersymmetric
contribution needs to be about a few times the electroweak SM
contribution.

\qquad The SM deviations of $g_{\mu}$ from 2 arise from the triangle loop
with an internal muon and photon or Z, and the associated loop with W and
$\nu_{\mu}$. The superpartner loops are just those that arise from
$\mu\rightarrow \widetilde{\mu},$ $W\rightarrow$ chargino,
$\nu\rightarrow\widetilde{\nu},$ and $\gamma$ and $Z\rightarrow$ neutralinos.  
11 MSSM parameters can enter (10 from \lsoft and $\tan\beta$): the soft
parameters are $M_{1},$ $M_{2},$ $\mu,$ $A_{\mu},$ $m_{\widetilde{\mu}_{L}},$
$m_{\widetilde{\mu}_{R}},$ $m_{\widetilde{\nu}},\phi_{M_2}+\phi_{\mu},$
$\phi_{M_1}+\phi_{\mu},$ and $\phi_{A}+\phi_{\mu}$.  Although in the
supersymmetric limit $g_{\mu}-2$ vanishes because there is an exact
cancellation between the SM and superpartner loops, when supersymmetry is
broken the cancellation is far from complete. Depending on the soft
parameters, they can even contribute with the same sign. Since the
experimental result is larger than the SM, this is indeed what is
required.

For large $\tan\beta$, the chargino
diagram dominates over the neutralino diagram over most of the parameter 
space \cite{Lopez:1993vi,Chattopadhyay:1995ae,Moroi:1995yh,Carena:1996qa},
and is linear in $\tan\beta$. This effect can be seen most easily in the 
mass insertion approximation, where the main contribution arises from the 
chargino diagram in which the required chirality flip takes place via
gaugino-higgsino mixing rather than by an explicit mass insertion on 
the external muon  
\cite{Lopez:1993vi,Chattopadhyay:1995ae,Moroi:1995yh,Carena:1996qa}. 
Assuming the superpartners are all approximately 
degenerate with masses given by $\widetilde{m}$, in this case the leading 
chargino contribution  is of the order
\begin{equation}
\label{gmu-2}
a_{\mu}^{susy}/a_{\mu}^{SM}\approx\left( \frac{100\,{\rm GeV}}{\widetilde{m}}%
\right) ^{2}\tan\beta\cos(\phi_{M_2}+\phi_{\mu}).
\end{equation}
The chargino sector phase which enters in this leading 
contribution\footnote{The phase 
dependence is of course more complicated when considering all 
contributions; see {\it e.g.} \cite{Ibrahim:2001ym,Ibrahim:1999aj}.} is 
constrained by electric dipole 
moment constraints, as discussed in 
Section~\ref{EDMsect}.  In models such as minimal supergravity where the 
gaugino masses and $\mu$ are assumed to be real, the cosine then reduces 
to the ``sign of $\mu$'' in models where the gaugino masses can be taken 
to be positive without loss of generality.

There have been many analyses of the phenomenological implications for the 
MSSM parameters from the $g_{\mu} -2$ measurement since the 
initial report of the data, {\it e.g.} 
\cite{Czarnecki:2001pv,Feng:2001tr,Everett:2001tq,Chattopadhyay:2001vx,Arnowitt:2001be,Baer:2001kn,Carvalho:2001ex,Komine:2001hy,Choi:2001pz,Ellis:2001yu,Hisano:2001qz,Ma:2001mr}
(among others).  One obvious question addressed in a number of these
analyses is if an upper limit on superpartner masses could be deduced
assuming there is such a deviation; in looking for such an upper limit one 
can of course drop the phase dependence. Once the situation with the 
vacuum polarization is settled, if there is indeed
a real contribution beyond the SM it will be possible to determine useful
upper limits on some superpartner masses as a function of $\tan\beta.$ If
$\tan\beta$ can be measured other ways then $g_{\mu}-2$ will provide a
strong constraint on superpartner masses.
Even if there is no effect beyond the SM, the existence of a measurement
and the SM theory prediction put a limit on how large a supersymmetry
contribution could be (see {\it e.g.} \cite{Martin:2002eu}). A significant
region of supersymmetry parameter space can be excluded in this way, a
region that is not probed by previous experiments.  More extensive recent
analyses of the data have also been carried out by
\cite{Chattopadhyay:2002jx,Byrne:2002cw}.  The measurement can of course
also provide important constraints on models of \lsoft, such as mSUGRA and
gauge mediation; for examples of the effects on mSUGRA parameter space see
{\it e.g.} \cite{Baer:2002ay,Ellis:2001yu}.

\qquad Further data will reduce the experimental errors during 2003.  
Additional experimental data on $e^{+}e^{-}$ collisions will further test 
that the current values are correct, and somewhat reduce errors.  Further 
theoretical work should lead to an understanding of the discrepancy 
between the $e^{+}e^{-}$ and the $\tau$ vacuum polarization results.  
Sometime in 2004 the situation with $g_{\mu}-2$ should be clear.  If there 
is indeed a significant difference between the SM prediction and the data, 
it may be the first signal of physics beyond the SM that has to be 
accounted for by particles with masses of order the electroweak scale.

\subsubsection{CP violation and electric dipole moments}
\label{EDMsect}
In the SM, the only source of CP violation is present in the CKM matrix
and thus CP violation is intimately tied to flavor physics.  In the MSSM,
however, CP-violating phases within supersymmetric models can occur in both
flavor-conserving and flavor-changing couplings.  The phases of the 
flavor-conserving couplings, which have no analogue in 
the SM, are of particular interest because they can have significant 
phenomenological implications which can be studied without knowledge of 
the origin of intergenerational mixing.  In the MSSM, these phases are 
given by reparameterization invariant combinations of the phases of the 
gaugino mass parameters, the trilinear couplings, and the $\mu$ 
and $b\equiv \mu B$ parameters.  
A useful basis of the reparameterization invariant phase combinations is 
given in Eq.~(\ref{reparaminvbasis}): 
$\phi_{1\,f}=\phi_{\mu}+\phi_{A_f}-\phi_{b}$ and
$\phi_{2\,a}=\phi_{\mu}+\phi_{M_a}-\phi_{b}$, as 
previously discussed in 
Section~\ref{paramsect}. 

The presence of these phases leads to 
what traditionally has been called the 
supersymmetric CP problem: the fermion electric dipole moments (EDMs) 
receive one-loop contributions due to superpartner exchange which 
for generic phases can exceed the experimental bounds. Early references 
include 
\cite{Ellis:1982tk,Buchmuller:1982ye,Polchinski:1983zd,delAguila:kc,Franco:1983xm,Gerard:1984bg,Gerard:1984vb,Dugan:1984qf,Sanda:ib} 
and slightly later references include 
\cite{Arnowitt:1990eh,Arnowitt:je,Nath:dn,Kizukuri:mb,Kizukuri:nj,Fischler:ha}.
Using the rough estimate of the one-loop EDMs for {\it e.g.} the neutron 
\cite{Fischler:ha}
\begin{equation}
d_n\approx 2 
\left (\frac{100\,{\rm GeV}}{\widetilde{m}}\right )^2 10^{-23}
\sin \phi,
\end{equation}
in which $\widetilde{m}$ denotes a general soft supersymmetry-breaking mass and $\phi$ can be any of 
the reparameterization invariant phase combinations in 
Eq.~(\ref{reparaminvbasis}),
the bounds for the electron
\cite{Abdullah:nh,Commins:gv} and neutron \cite{Smith:ke,Altarev:cf} EDMs
\begin{eqnarray}
|d_e| &<& 4.3 \times 10^{-27} {\ \rm e-cm} {\ \rm (95\% \, c.l.)}\\  
|d_n| &<& 6.3 \times 10^{-26} {\ \rm e-cm} {\ \rm (90\% \, c.l.)}
\end{eqnarray}
individually constrain the phases to be ${\cal 
O}(10^{-2})$ for sparticle masses consistent with naturalness.  
Such constraints can be expressed as bounds on the 
imaginary parts of the $(\delta^{u,d,e}_{LR})_{11}$ parameters 
\cite{Gabbiani:1996hi}, keeping in mind that by $U(1)_R$ invariance the 
bounds should include the phases of the gaugino masses or $\mu$.

Such small phases have a negligible impact on collider phenomenology, 
although they may still be relevant in the context of baryogenesis, {\it 
e.g.} perhaps in the Affleck-Dine baryogenesis scenarios discussed in 
Section~\ref{bgensect}. Hence, they  have typically been neglected in 
phenomenological analyses. However, recent studies have shown that EDM 
bounds can be satisfied without requiring all reparameterization invariant 
phase combinations to be small, if either 
\begin{itemize}
\item The sparticles of the first and second families have multi-TeV 
masses \cite{Cohen:1996vb}.

\item Certain cancellations exist between the various one-loop diagrams
which contribute to EDMs
\cite{Ibrahim:1998je,Ibrahim:1997gj,Brhlik:1998zn,Falk:1998pu,Pokorski:1999hz,Bartl:1999bc}
(see also \cite{Falk:1995fk,Falk:1996ni,Ibrahim:1997nc}). These 
cancellations are accidental cancellations and are not due to a 
fundamental low energy
symmetry. In a purely low energy context, such cancellations can be
interpreted as fine-tuning.  As discussed below, the question of whether
phases are large and cancellations occur in this manner is arguably most
interesting in the context of model-building.  For example, 
string-motivated supergravity models can be constructed with large phases
which evade the electron and neutron EDM bounds (see e.g
\cite{Brhlik:1999ub,Brhlik:1999pw,Accomando:1999zf,Ibrahim:1999af}); however,
these models often do not pass the mercury EDM constraint 
\cite{Abel:2001mc}, as discussed below.

\end{itemize}
In each of these scenarios, the EDM bounds are more difficult to satisfy 
when $\tan\beta$ is relatively large. First, cancellations in the 
one-loop EDMs more difficult to achieve; see {\it e.g.} 
\cite{Pokorski:1999hz} for a clear presentation of these difficulties.  
Second, certain two-loop contributions are then enhanced 
\cite{Chang:1998uc,Pilaftsis:1999td,Pilaftsis:2002fe,Lebedev:2002ne} which 
do not decouple when the sfermions are heavy.\footnote{For example, in the 
large $\tan\beta$ regime the atomic EDMs receive large contributions 
from Higgs-mediated semileptonic four-fermion operators 
\cite{Barr:cm,Lebedev:2002ne}. The thallium EDM is 
highly sensitive to such contributions: existing bounds are 
violated for $\tan\beta\geq 10$ when $\phi_{\mu}\sim {\cal{O}}(1)$ and 
$M_{A}\sim 200\ {\rm GeV}$. On the other hand, the two-loop electron EDM 
has an important impact on the thallium EDM in that it can partially 
cancel the contributions of the four-fermion operators 
\cite{Pilaftsis:2002fe}.}  

Within each of these scenarios there also are particularly strong
constraints arising from the atomic EDMs such as the mercury EDM 
\cite{Falk:1999tm}, which appear
to rule out many of the ``cancellation" models constructed so far
\cite{Abel:2001vy,Abel:2001mc,Barger:2001nu}. However, there are
unavoidable theoretical uncertainties involved in the determination of the
hadronic EDMs and the atomic EDMs (see {\it e.g.}  
\cite{Pilaftsis:2002fe,Demir:2002gg} for discussions). These uncertainties
are arguably problematic for the mercury EDM (its measurement is reported 
in \cite{Romalis:2000mg}), which yields the strongest 
constraints on the phases.  
For this reason, there are disagreements in the literature 
over how to include this bound and various ranges in the subsequent limits 
on the \lsoft phases. Including all atomic EDM bounds and allowing for EDM 
cancellations, a general low energy 
analysis of the MSSM parameter space leads to a general upper bound of 
$\sim \pi/(5\tan\beta)$ on the reparameterization invariant phase present 
in the chargino sector ($\phi_{\mu}+\phi_{M_2}-\phi_b \equiv \phi_{22}$ in 
our notation), while the other phases are comparatively unconstrained
\cite{Barger:2001nu}; stronger bounds on 
this phase of $O(10^{-2})$ are presented in \cite{Abel:2001vy}, due 
to differences in implementing the mercury EDM constraint. In the language 
used in many EDM papers --- particularly in the mSUGRA analyses --- in 
which 
the phase of $M_2$ is set to zero using $U(1)_R$, this constraint 
thus applies to the ``phase of $\mu$''.  The above bounds on 
$(\phi_{\mu}+\phi_{M_2})$ are quite conservative in that they assume the 
superpartner masses can be of order TeV and that cancellations 
can occur; the bound is $\leq O(10^{-2})$ if the superpartner masses are 
of order $m_Z$.

Recently, it was pointed out \cite{Abel:2001cv} that even if the
supersymmetry-breaking terms conserve CP, {\it e.g.} in a 
high scale supergravity theory
where they are defined, the Yukawa coupling phases required to
achieve a significant CKM phase can filter into the $(\delta_{LR})_{11}$
parameters and overproduce the EDMs. This can occur in
supergravity models because the $\widetilde{A}$ parameters typically do not
have a simple proportionality to the Yukawa couplings and are not 
diagonal in the diagonal quark (SCKM) basis.  More precisely, the 
structure of the $\widetilde{A}$ parameters in supergravity
models leads to contributions to the LR and RL subblocks which are not
suppressed by the corresponding fermion masses in the SCKM basis
\cite{Kobayashi:2000br,Abel:2001cv}.  These contributions are proportional
to derivatives of the Yukawa couplings with respect to the fields which
break supersymmetry, and hence are relevant in scenarios with models for the Yukawa
couplings such as string models, or models using the Froggatt-Nielsen (FN)
mechanism \cite{Froggatt:1978nt}.\footnote{It was pointed out in
\cite{Abel:2001cv} that in supergravity the FN fields necessarily
participate in supersymmetry breaking and thus contribute to the soft trilinear
couplings.  Such FN scenarios in supergravity were subsequently analyzed
in \cite{Ross:2002mr}, with the conclusion that such contributions are
indeed relevant but do not typically exceeed the phenomenological
constraints.} A further observation is that if the $\widetilde{A}$ terms are
Hermitian, the corresponding diagonal entries of the LR and RL subblocks
are then real, alleviating EDM constraints \cite{Abel:2000hn}.  However,
this approach appears to be difficult to implement in models.

Phenomenologically, the question of whether the phases are large must be 
addressed because if the superpartner masses are relatively
light, large phases can have very significant effects \cite{Brhlik:1998gu} 
on a variety of interesting phenomena --- they generate CP violation, they
affect the baryon asymmetry of the universe, the relic density and
detectability of cold dark matter, rare decays, implications of the Higgs
sector, and superpartner masses, cross sections, and branching ratios.  
The patterns of the phases and whether they are measured to be large or
small, may provide a link to the nature of the high energy
theory. Certainly whether the phases are large or small affects how to
extract the Lagrangian parameters from experimental measurements. For
certain particle physics and cosmology phenomena one can be badly misled
if phases are large but are not included in the analysis.

The nonobservation of electric dipole moments provide interesting
constraints on the MSSM phases. One could of course set all the soft
phases to zero, which may suggest that a presently unknown symmetry of the
high scale theory existed. Alternatively, it could happen that the high
scale theory had a structure that led to apparent cancellations in the low
energy effective theory for the phase combinations that are significant
for EDMs. The contributions to EDMs do allow the cancellation
interpretation, but probably only if $\tan \beta$ is not too large and
only if nonzero EDMs appear with the next round of experimental
improvements.

This apparent smallness of the soft phases is referred to as
the supersymmetry  CP problem. The point is somewhat subtle and
sometimes misunderstood. Consider the quark CKM phase. No one would
argue that it is calculable theoretically yet, since we do not
understand the origin of the superpotential Yukawas. The situation is
the same for the supersymmetry soft phases. They are also not
calculable yet. But no 
experiment strongly constrains the CKM phase yet, while the EDMs do
constrain certain combinations of soft phases weighted by soft masses   
and functions of $\tan \beta$. The existence of this constraint that
is not automatically satisfied is the supersymmetric CP problem.  These
arguments refer to the electroweak phase structure and all assume that the 
strong CP problem in the presence of supersymmetry has been addressed.  We 
review the strong CP problem separately in the following section.

\subsubsection{The strong CP problem} 
\label{strongCPsect} 
The strong CP problem (see
\cite{Cheng:1987gp,Kim:ax} for excellent general reviews) of the SM is 
that the unobserved neutron EDM 
forces a dimensionless coefficient \( \theta \) multiplying a CP-violating 
term of the SM QCD Lagrangian to be less than \( 10^{-10}\)
\cite{Baluni:1978rf}, when there is no symmetry reason for such a small
number. More precisely, the term responsible for the problem is the 
following CP-odd term: 
\begin{equation} 
\label{eq:thetaterm} 
\delta {\mathcal{L}}_{SCPV}=\frac{\theta }{64\pi ^{2}}\epsilon _{\mu \nu 
\rho \sigma }G_{a}^{\mu \nu }G_{a}^{\rho \sigma }, 
\end{equation}
where \( G^{\rho \sigma }_{a} \) is the field strength of the
\( SU(3)_{C} \) gluons.  The total derivative nature of
Eq.~(\ref{eq:thetaterm}) would make it unphysical in the absence of
instantons.  For example, an analogous term for the \( U(1)_{Y} \)
sector, where the vacuum manifold is topologically trivial, is
unphysical.

Even without any other source of CP violation, this term leads to the
effective CP-violating operator in the context of chiral perturbation
theory \cite{Crewther:1979pi,Baluni:1978rf}: 
\begin{equation} 
{\mathcal{L}}_{CPV}=\frac{-\theta
}{f_{\pi }}\frac{m_{u}m_{d}(M_{\Xi
}-M_{N})}{(m_{u}+m_{d})(2m_{s}-m_{u}-m_{d})}\vec{\pi }\cdot
\overline{N}\vec{\tau }N, 
\end{equation}
in which \( \vec{\pi } \) is the pion isotriplet, \( N \) is the nucleon
field, \( f_{\pi }=93 \) MeV is the measured pion decay constant, and \( 
\{M_{\Xi },M_{N}\} \) and \( \{m_{s},m_{u},m_{d}\} \)
are the measured baryon and quark masses, respectively.
This leads to an NEDM of
\begin{equation}
D_{n}\approx 10^{-16}\theta \textrm{ e}-\textrm{cm},
\end{equation}
which when compared to the experimental bound leads to the 
unnaturally small \( \theta <10^{-10} \). In this section we briefly 
describe connections of the strong CP problem to supersymmetry and the 
soft supersymmetry-breaking Lagrangian.  In particular, we are not 
surveying the many published appraoches to solving the strong CP problem, 
though we will mention the three main categories.

Because \( \theta  \) transforms nontrivially under the chiral redefinitions
of fermions charged under \( SU(3)_{c} \) due to the chiral
anomaly, \( \theta  \) by itself is not a physically meaningful
parameter. In the SM, the quarks are the only fermions charged under \( 
SU(3)_c \) whose transformations can induce transformations in \( \theta 
\). For example, under the chiral rotations of the first generation up
quarks \begin{equation}
Q_{u}\rightarrow e^{i\alpha }Q_{u}\, \, \, \, 
U^{c}\rightarrow e^{i\alpha }U^{c}, 
\end{equation}
 \( \theta  \) undergoes transformations\begin{equation}
\theta \rightarrow \theta +2\alpha , 
\end{equation}
because of the noninvariance (anomaly) of the measure of the path
integral. This is the key nontrivial property of the \( \theta  \)
term. Denoting the mass matrices for the up-type and 
down-type quarks as \( M_{u,d} \), respectively, the physically meaningful 
parameter is
\begin{equation}
\label{eq:sminv}
\overline{\theta }=\theta -\textrm{Arg}[\textrm{Det}[Y_{u}Y_{d}]],
\end{equation}
which is invariant under \( U(3)_{Q}\times U(3)_{U}\times U(3)_{D} \) 
global quark field redefinitions.

In the SM, the leading divergent radiative corrections to \(
\overline{\theta } \) occur at a very large loop order. One leading 
contribution is 12th order in the Yukawa coupling and second order in 
the \( U(1) \) gauge coupling. Another arises at  
14th order in Yukawa couplings \cite{Khriplovich:1993pf} due 
to Higgs exchange instead of vector exchange. The reason for the 
large order is that \( \overline{\theta } \) is sensitive to the rephasing of 
many fields. There is also a finite renormalization contribution of \( 
\delta \overline{\theta }=10^{-19} \) \cite{Ellis:1978hq,Khriplovich:1985jr},  
which is insignificant.

With the introduction of supersymmetry and the soft 
supersymmetry-breaking terms, gluino chiral rotations
can also contribute to the transformation of the \( \theta \) term, since
gluinos are additional fermions charged under \( SU(3)_{c} \).
Therefore, the analog of the SM formula Eq.~(\ref{eq:sminv}) for softly 
broken supersymmetry is
\begin{eqnarray} \overline{\theta } & = & \theta
-\textrm{Arg}[\textrm{Det}[Y_{u}Y_{d}]]- 
3\textrm{Arg}[m_{\widetilde{g}}]-3\textrm{Arg}[b].
\label{eq:susythetabar}
\end{eqnarray} 
In the above expression, the \( \textrm{Arg}[b] \) term is required by
rephasing invariance under the (anomalous) global \( U(1)_{PQ} \) 
described in Section~\ref{paramsect}.  This 
additional rephasing invariance owes its origin to the requirement of two 
Higgs doublets in the MSSM. Eq.~(\ref{eq:susythetabar}) is also invariant 
under the supersymmetry-native 
rephasing freedom \( U(1)_{R} \).

An advantage of supersymmetry for the strong CP problem is that \(
\overline{\theta } \) can be protected from UV sensitive divergent 
contributions by nonrenormalization theorems
\cite{Ellis:1982tk,Graesser:1997df} as long
as supersymmetry is spontaneously broken \cite{Akhoury:1983fn}. On the
flipside, however, there are more finite radiative contributions to \(
\overline{\theta } \). For example, there is a soft term-dependent 
contribution at one-loop order, whose magnitude is given by
\begin{equation}
\delta \theta _{soft}=\sum _{q}O \left (\frac{\alpha _{S}}{\pi }\textrm{or 
}\frac{\alpha }{\pi } \right )\textrm{Im}[UV^{\dagger }]_{qq}
[\Delta m_{sq}^{2}/(m_{sq}^{2}\textrm{ or }
m_{\widetilde{g}}^{2})]\frac{m_{\widetilde{g}}}{m_{q}},
\label{eq:susythetcor1}
\end{equation}
where \( U \) and \( V \) are the gaugino couplings to left- and
right-handed quark-squark combinations and the alternative denominators
apply when \( m_{sq}\gg m_{\widetilde{g}} \) or vice versa. 
Eq.~(\ref{eq:susythetcor1}) requires the phases to be smaller than about 
\( 10^{-8} \).\footnote{Hence, when discussing the possibility of nonzero 
\lsoft phases, one must presuppose that the strong CP problem is solved by 
one of the mechanisms discussed below.}  Even if all the phases are zero 
in the soft terms, because of the complex Yukawas presumably entering 
through the mass insertions, these one-loop diagrams still generate a \( 
\theta  \) term. The complex Yukawa contribution goes as 
 \begin{equation}
\textrm{Im}({\rm Tr}[Y^{\dagger }\widetilde{A}]),
\end{equation}
which vanishes if \( \widetilde{A}=0 \) or \( \widetilde{A}\propto Y \).  It 
should be noted that {\it e.g.} gauge mediated supersymmetry breaking 
gives the universality needed for this to vanish.

There are currently three widely known classes of proposed solutions to
the strong CP problem: (i) the axionic solution
\cite{Peccei:1977hh,Peccei:1977ur,Weinberg:1977ma,Wilczek:pj,Kim:1979if,Zhitnitsky:tq,Shifman:if,Dine:1981rt},
(ii) the Nelson-Barr solution \cite{Nelson:1983zb,Barr:qx}, and (iii) the 
\( m_{u}=0
\) solution \cite{'tHooft:1976up}.

The axionic solution states that the value of \( \overline{\theta } \)
is small because it is a dynamical variable which has the minimum
of its potential at \( \overline{\theta }=0 \). To make it a dynamical
variable, one associates it with the Goldstone boson of a broken \( U(1) \)
symmetry called a Peccei-Quinn (PQ) symmetry (\( U(1)_{PQ} \)). For
example, in the SM, one can minimally extend the Higgs sector to replace
the Higgs of the up-type quark Yukawa coupling with a second Higgs
\( H_{2} \) which transforms like \( i\tau ^{2}H_{1}^{*} \), where
\( H_{1} \) and \( H_{2} \) are now two independent \( SU(2) \)
doublet complex scalars. This simplest extension has \( U(1)_{PQ} \)
charges \( Q_{H_{1}}=1 \), \( Q_{H_{2}}=1 \), \( Q_{u}=-1 \), \( Q_{d}=-1 
\), and \( Q_{Q_{L}}=-1 \), where \( u \) and \( d \) are the right
handed \( SU(2) \) singlets and \( Q_{L} \) is the left-handed doublet.
In this setting, due to electroweak symmetry breaking, \( U(1)_{PQ} \)
is automatically broken, and the resulting Goldstone is the axion.
The axion is not massless, however, due to \( SU(3) \) instantons
which in the dilute gas approximation generate a periodic potential
schematically of the form\begin{equation}
V\sim \frac{2Z}{(1+Z)^{2}}m_{\pi }^{2}f_{\pi }^{2}\left (1-\cos
\left [\frac{a}{f_{PQ}} \right ] \right ),
\label{eq:axionpotential}
\end{equation}
in which \( f_{\pi } \) is the pion decay constant, \( m_{\pi } \) is
the pion mass, \( Z\equiv m_{u}/m_{d} \), and \( f_{PQ} \) is the scale
of PQ symmetry breaking ({\it e.g.} for the electroweak scale models of
\cite{Weinberg:1977ma,Wilczek:pj}, \( f_{PQ}\sim 246 \) GeV).
A more general argument for this potential can be found in 
\cite{Vafa:1984xg}. Given that \( a \) as written in
\eqr{eq:axionpotential} is the rephasing invariant strong CP phase, when 
\( a/f_{PQ} \) is in its ground state minimum of
$a/f_{PQ}=0$, the strong CP problem is solved. This model and similar
low $f_{PQ}$ scale models are ruled out because of laboratory
constraints
\cite{Donnelly:1978ty,Barshay:ky,Barroso:1981ta,Krauss:1986wx,Bardeen:1986yb}, but
there are viable extended models where \( f_{PQ}\gg 246
\) GeV (the cosmologically favored value of \(
f_{PQ} \) is around \( 10^{11} \) GeV). Because these viable axions
have suppressed couplings to quarks \( \propto
1/f_{PQ} \) (see Section~\ref{subsec:axino}), they are called
invisible axions.

The biggest challenge in axion model building is to protect the PQ
symmetry sufficiently. In other words, for this mechanism to work,
the dominant contribution to the potential has to be from the
QCD instantons in Eq.~(\ref{eq:thetaterm}). Since the PQ symmetry is a
global symmetry, it is expected to be broken by gravitational
interactions \cite{Giddings:cx,Coleman:1988tj,Gilbert:nq}. Any
explicit breaking of $U(1)_{PQ}$ is expected to shift the minimum
of \( a/f_{PQ}\) away from zero, which is dangerous for the solution to 
the strong CP problem. Even though gravitational interactions are weak 
because their effective interactions are Planck-suppressed nonrenormalizable 
operators, the required tolerance for \( a/f_{PQ}\) away from zero is so 
small that $U(1)_{PQ}$-violating nonrenormalizable
operators with coefficients less suppressed than \( 1/M_{Pl}^{6} \)
are disallowed \cite{Holman:1992us,Barr:1992qq}. If this 
must occur as an accidental result of the gauge symmetry and the
representation of the fields, it is a difficult challenge.  Another 
challenge is to set up the phenomenologically favored large hierarchy 
between \( M_{Pl} \) and \( f_{PQ} \); as stated above and argued below, 
the favored value of \( f_{PQ} \) is \( 10^{-8}M_{Pl} \).  For other 
issues, see {\it e.g.} \cite{Dobrescu:1996jp}.

Another generic prediction of axion models in the context of supersymmetry
is the existence of the axino, the fermionic partner of the
pseudoscalar axion, and a saxion, the scalar completing the
multiplet. These particles have mainly cosmological implications. For 
couplings and phenomenological implications, see 
Section~\ref{subsec:axino}.

The Nelson-Barr mechanism \cite{Nelson:1983zb,Barr:qx} assumes that CP
is a fundamental symmetry of the high energy theory and is broken
spontaneously by a complex VEV which is coupled to the quarks. The
spontaneous breaking induces complex mixings with heavy vectorlike
fermions assumed to exist. By an appropriate choice of quark masses and 
Yukawa couplings, a large CKM phase and \( \overline{\theta }=0 \) can be 
arranged.
Unfortunately, the biggest problem is to protect this solution from
loop corrections, particularly from squarks and gauginos
\cite{Dine:1993qm}.  Since squark mass degeneracy and tight
proportionality between the quark and squark mass matrices suppress
the loop effects, models which solve the supersymmetric flavor problem 
such as gauge mediation may help provided the needed suppression 
\cite{Hiller:2001qg,Hiller:2002um}.

The \( m_{u}=0 \) solution is not favored by chiral perturbation
theory \cite{Leutwyler:1990pn}. Lattice simulations may eventually settle 
this issue \cite{Cohen:1999kk}.
\setcounter{footnote}{0}
\section{Dark matter}
\label{darkmattersect}
The most favored cosmological model today inferred from WMAP and other
cosmological data\footnote{One must be careful in interpreting the error
bars offered by these experiments, since there are priors and 
model-dependent assumptions in the fits.} maintains a cosmological expansion 
driven by an energy density comprised of the following 
approximate fractions \cite{Bennett:2003bz,Spergel:2003cb} (see also {\it e.g.} 
\cite{Ellis:2003cw,Chattopadhyay:2003xi}):
\begin{itemize}
\item \( 0.73\pm 0.04 \) negative pressure dark energy

\item \( 0.22\pm 0.02 \) cold dark matter 

\item \( 0.05 \) of other components, of which baryons
contribute around \( 0.044\pm 0.004 \), massive neutrinos make up
around \( 0.006 \), photons contribute around \( 5\times 10^{-5} \),
and the relativistic neutrinos make up around \( 10^{-5}
\).
\end{itemize}
Let us consider each of these components in turn.

Negative pressure dark energy \cite{Huterer:1998qv} is defined to
be an energy density component whose pressure \( p \) to energy density
\( \rho  \) ratio ({\it i.e.}, its equation of state) is \( p/\rho <-1/3 
\). 
A cosmological constant can qualify as such an energy component, because  
its equation of state is \( -1 \). The most sensitive probe of this energy 
is the combination of CMB and supernovae data \cite{Frieman:2002wi}.  
Scalar fields whose potential energy dominates the kinetic energy can also 
be responsible for this energy component. If such fields are time
varying as well as weakly spatially varying, it is fashionable to
refer to these fields as quintessence \cite{Wang:1999fa}: for a
sense of the evolution of this idea, see \cite{Ratra:1987rm,Wetterich:fm}
and the review \cite{Binetruy:2000mh}.
As the required energy scale is far removed from the electroweak
scale, the MSSM fields are not likely candidates for quintessence fields. 
The only connection of quintessence with \lsoft is that supersymmetry 
breaking terms will induce radiative masses for such fields which are 
large compared to the Hubble
expansion rate today and generically give a
cosmological constant contribution which is at least of order \( 
\widetilde{m}^{4} \), where $\widetilde{m}$ denotes a typical scale of the \lsoft 
parameters.  Generically one might also expect a cosmological 
constant contribution of order $M_S^4$, where $M_S$ is the scale of 
supersymmetry breaking in the hidden sector 
\cite{Kolda:1998wq,Brax:1999yv,Chung:2002xj}.\footnote{Indeed, because
of its sensitivity, quintessence is a good probe of the K\"{a}hler
potential.}

Cold dark matter (CDM) is defined as matter which is nonrelativistic
at the time of matter-radiation equality: when the relativistic energy
density, characterized by its positive nonvanishing pressure, is equal
to the nonrelativistic energy density, which has vanishing
pressure. Similarly, hot dark matter is defined as matter which is
relativistic at the time of matter-radiation equality. In between
lies warm dark matter, which is similar to hot dark matter except that it 
becomes nonrelativistic at a much earlier epoch, and hence has a much 
smaller free-streaming scale of about 1 Mpc (3 million light 
years). Dark matter is categorized in this manner because the time of 
matter-radiation equality marks the beginning of the matter-dominated 
universe, which is the beginning of the time during which the 
universe is expanding slowly enough for matter to gravitationally cluster 
appreciably.\footnote{The 
physics of this gravitational clustering can be understood via a 
modified Jeans instability analysis, which is described in any standard 
textbook on gravity.} Whether the dark 
matter is relativistic or nonrelativistic changes the clustering property
during this matter domination period. A comparison of cosmological
observations, such as CMB and galaxy observations, with various
theoretical calculations (including numerical simulations) favors the
nonnegative pressure component of the dark matter to be CDM. As we
will see in detail, there are natural candidates for CDM in the MSSM.

Baryonic dark matter consists of white dwarfs, brown dwarfs, neutron
stars, and black holes. We will not discuss baryonic dark matter further
because it has little direct relation to the \lsoft parameters. The main
progress with respect to baryonic dark matter which is relevant for \lsoft 
is indirect, mainly pointing to the necessity of nonbaryonic CDM.

Among the various dark matter candidates, \lsoft has its closest connection
with cold dark matter because if R-parity is conserved, the lightest
supersymmetric particle (LSP) --- which has a mass controlled by the 
\lsoft parameters --- naturally provides just the right abundance today 
to account for the CDM if the LSPs were once in chemical thermal 
equilibrium with the background radiation. The beauty of LSP cold dark 
matter is that it was motivated mostly independently of any cosmological 
considerations. In the MSSM, the R-parity which guarantees LSP 
stability is needed to eliminate rapid proton decay, while the electroweak 
scale interactions and mass scales that determine the relic abundance are 
motivated from naturalness considerations of the SM. As there are strong 
bounds on charged dark matter 
\cite{DeRujula:1989fe,Dimopoulos:1989hk,Chivukula:1989cc},
the viable MSSM parameter region is usually that within 
which the LSP is neutral. Among the neutral LSP candidates, neutralinos 
and sneutrinos each have electroweak scale interactions that can naturally 
lead to dark matter densities consistent with observations. However, 
the possibility of sneutrinos as significant CDM is ruled out for most 
models from LEP constraints and direct detection \cite{Falk:1994es}. In 
the mass range allowed by these constraints, the sneutrinos annihilate too 
rapidly via s-channel \( Z \) exchange, and hence not enough remain today 
to make up the dark matter. However, sneutrinos can of 
course be the LSPs without violating experimental bounds if LSPs are not 
required to compose 
the CDM.

One particular LSP does not have electroweak scale interactions, but 
only gravitational interactions. This is the gravitino, which usually is 
the LSP in gauge mediation, as discussed 
in Section~\ref{lsoftmodelsect}. Even when the gravitino
is not the LSP and can decay, as in most gravity-mediated scenarios, its
lifetime can be very long due to its weak gravitational interactions,
leading to nontrivial consequences for late time cosmology. As we will
explain below, the typical upper limit on the temperature of the universe 
due to the gravitino decay constraint is about \( 10^{9} \) GeV.

Another well-motivated dark matter candidate, although not strictly
related to supersymmetry and the \lsoft parameters, is the axion.  
Remarkably, axions can 
still naturally live long enough to be the CDM even though they decay to
photons.  In many instances the axino, the supersymmetric partner of the
axion, can also serve as the LSP dark matter.  We discuss these candidates
below because (i) axions are arguably the most appealing solution
to the strong CP problem, and (ii) the interpretation of MSSM cosmology
can be misleading without taking axions and axinos into
consideration.

There are rare instances when the NLSP (the next-to-lightest 
supersymmetric particle) can be an absolutely stable dark matter 
candidate.  This can occur if the LSP is strongly interacting, such that 
its bound state to other strongly interacting fields has a mass large 
enough that kinematics allow a decay to the weakly interacting NLSP 
\cite{Farrar:1995pz,Chung:1997rq}.  We will not discuss this and other
rare situations in this review.  We will also not discuss the dark
energy connections with supersymmetry, primarily because they are of  
negligible relevance for the soft Lagrangian.

\subsection{\label{sec:lspdensity} Computing the LSP density}
The primary assumption in computing the LSP density in the standard 
cosmological scenario is to assume that the LSP initial abundance
is determined by the chemical thermal equilibrium condition. If two-body 
interactions comprise the dominant channel, the sufficient condition
for chemical equilibrium is \begin{equation}
\label{eq:equilib}
\sum _{i}\langle \sigma _{i}v\rangle n^{eq}_{LSP}\gg H,
\end{equation}
in which \( n^{eq}_{LSP} \) is the equilibrium LSP density, \( H 
\) is the Hubble expansion rate, $\sigma$ denotes the inelastic cross 
section of LSPs going into final states that are in equilibrium with the 
photon,  \( \langle
\sigma v\rangle \)  denotes the thermal averaging of $\sigma$ multiplied 
by the Moeller speed \( v \), and the summation is over all relevant cross
sections.  For nonrelativistic or mildly relativistic 
neutralinos, typically the higher the temperature, the easier it will be 
to satisfy this bound.  If the temperature of the background photons is 
not high enough, then one can of course still compute the LSP abundance 
today, but it will be sensitive to the mechanism through which the LSP is 
generated. In such situations, arguably the LSP dark matter candidates are 
not any more attractive, and perhaps are even less attractive, than other 
types of nonthermal dark matter.

Next, the Boltzmann equations are truncated to leading
hierarchical order. Although all chain reactions should in general be 
incorporated, for the purposes of an estimate is is sufficient to write
\begin{equation}
\label{eq:boltzmann}
\frac{df_{LSP}}{dx}=\sqrt{\frac{45}{4\pi ^{3}g_{*}}}\langle \sigma 
v\rangle m_{LSP}M_{Pl}f_{LSP}\left (f_{LSP}-f_{0}\frac{f_{0}}{f_{LSP}} 
\right ),
\end{equation}
in which \( f_{LSP}=n_{LSP}/T^{3} \) is the LSP volume density scaled
by the cube of the temperature \( T \) of the photons, \( \langle \sigma 
v\rangle  \) can be approximated as the summed cross section in 
Eq.~(\ref{eq:equilib}), \( x\equiv T/m_{LSP} \) is the 
temperature scaled by the LSP mass, \( f_{0}(x)\equiv 
x^{-3/2}e^{-1/x}/\sqrt{2\pi ^{2}} \) is the nonrelativistic approximation 
of the thermal equilibrium density (the LSP's are generally at most mildly 
relativistic), and \( g_{*} \) is a dimensionless number counting the 
field degrees of freedom. Eq.(\ref{eq:boltzmann})
demonstrates that as long as the annihilation reaction rates are large,
the LSP density \( f_{LSP} \) will follow the equilibrium density \( f_{0} 
\).\footnote{The equation is evolved backwards in $x$ since the 
temperature is getting cooler.}
Once the annihilation reaction becomes weak, the density will stop
following the equilibrium density and generically becomes much bigger
than the equilibrium density. This phenomenon is usually called 
``freeze-out.'' The LSP density today can thus be estimated as a  
fraction of the critical density \( \rho _{c}\) as follows:
\begin{equation}
\Omega \approx \frac{T^{3}}{M_{Pl}\rho 
_{c}}\sqrt{\frac{4\pi 
^{3}g_{*}(x_{F})}{45}}\left( \int _{x_{0}}^{x_{F}}\langle \sigma v\rangle 
dx\right) ^{-1}, 
\end{equation} 
in which 
\begin{equation}
\label{eq:interation}
x_{F}\approx \frac{1}{\ln [\xi m_{LSP}M_{Pl}\langle \sigma v(x_{F})\rangle 
]+\frac{1}{2}\ln x_{F}},
\end{equation}
with 
\begin{equation}
\xi \equiv \frac{1}{(2\pi 
)^{3}}\sqrt{\frac{45}{2g_{*}(x_{F})}}.
\end{equation}
In the expression above, the critical density \( \rho _{c}\approx 4\times 
10^{-47}\,\textrm{GeV}^{4} \),
the number of field degrees of freedom \( g_{*}\approx 100 \), and
the temperature today \( T\approx 2\times 10^{-13} \) GeV. 
The thermally averaged cross section can be estimated to be 
\begin{equation}
\label{eq:thermalaver}
\langle \sigma v\rangle \sim \frac{1}{64\pi 
}\frac{xm_{LSP}^{2}}{m_{\widetilde{f}}^{4}},
\end{equation}
in which \( m_{\widetilde{f}} \) is the mass of the intermediate sfermion 
through which the annihilation occurs. The appearance of 
\( x \) in the numerator in Eq.~(\ref{eq:thermalaver})
is due to the \( p \)-wave annihilation characteristic of light Majorana
particles. Although the \( p \)-wave does not always dominate over the
\( s \)-wave, we will consider this limit to keep the estimate simple.
Typically \( x_{F}\approx 1/20 \), as can be obtained by iteratively
solving Eq.(\ref{eq:interation}). Taking \( m_{LSP}=m_{\widetilde{f}}=100 \)
GeV, one finds \( x_{F}\approx 1/24 \) and \( \Omega \approx 0.4 \),
which is the right order of magnitude for the desired LSP density
(\( \Omega \approx 0.2 \) ). 

Technically the most difficult aspect of the calculation in practice is
the thermal averaging of the cross section 
\cite{Gondolo:1990dk,Griest:1990kh}. In most regions of parameter space,
the averaging is simple since \( \sigma v \) can be expanded in \( v^{2}
\) nonrelativistically. However, the thermal averaging can require some 
care because \( \sigma v \) cannot be expanded in \( v^{2} \) near 
nonanalytic points such as thresholds and poles of resonances. For more 
details about thermal averaging and the Boltzmann equations, see {\it 
e.g.} \cite{Gondolo:1990dk,Griest:1990kh,Chung:1997rq}.

There has been a great deal of activity in computing the relic density
for various regions of MSSM parameter space 
\cite{Ellis:1983wd,Ellis:1983ew,Srednicki:1988ce,Barbieri:zs,Griest:1989zh,Griest:1990kh,Gondolo:1990dk,Ellis:1991ni,Bottino:1992wj,Drees:1992am,Nath:1992ty,Bottino:1993zx,Berezinsky:1995cj,Roberts:1993tx,Kane:1993td,Baer:1995nc,Baer:1997ai,Edsjo:1997bg,Barger:1997kb,Arnowitt:1998uz,Ellis:1998kh,Pukhov:1999gg,Ellis:1999mm,Gomez:1999dk,Boehm:1999bj,Baer:2000jj,Ellis:2001ms,Gondolo:2000ee,Feng:2000gh,Feng:2000zu,Feng:2001ut,Arnowitt:2001yh,Gomez:2000sj,Roszkowski:2001sb,Djouadi:2001yk,Ellis:2001nx,Belanger:2001fz,Nihei:2002ij}.
The state of the art numerical programs take into account nearly \( 8000 
\) Feynman diagrams. Typically, the parameter exploration is done in
the context of mSUGRA/CMSSM models, in which the independent parameters at 
\( M_{GUT} \) are the universal scalar mass \( m_{0} \), gaugino mass \( 
m_{1/2} \),  trilinear scalar coupling \( A_{0} \), \( \tan \beta  \), and 
\( \textrm{sign}(\mu ) \).\footnote{Electroweak symmetry breaking 
constraints have allowed \( \tan \beta  \) and $m_Z$ to 
replace the $\mu$ and $b$ parameters, up to the sign of $\mu$; see the 
discussion of the mSUGRA scenario in Section~\ref{lsoftmodelsect}.}   
These parameters are then run from $M_{GUT}$ to low energies using the 
MSSM RGEs.   In CP-violating extensions of mSUGRA models, there are \lsoft 
phases present in the neutralino and sfermion mass matrices, which 
consequently affect the annihilation rate (see {\it e.g.} 
\cite{Falk:1995fk}).   

In practice, the network of relic abundance equations for the \( N \)
species with the same R-parity as the LSP is approximately replaced 
by a single evolution equation as in Eq.~(\ref{eq:boltzmann})
by defining an appropriate effective thermally averaged cross section 
\cite{Baer:2002fv}:
\begin{equation}
\langle \sigma _{\textrm{eff }}v\rangle =\frac{\int _{2}^{\infty
}K_{1}(a/x)\sum _{i,j=1}^{N }\lambda
(a^{2},b_{i}^{2},b_{j}^{2})g_{i}g_{j}\sigma _{ij}(a)da}{4x\left( \sum
_{i=1}^{N} K_{2}(b_{i}/x) b_{i}^{2} g_{i} \right)^2},
\end{equation}
in which \( g_{i} \) is the number of field degrees of freedom, \( 
\sigma _{ij} \)
is the annihilation cross section for \( ij\rightarrow X \), \( \lambda
(a^{2},b_{i}^{2},b_{j}^{2})=a^{4}+b_{i}^{4}+b_{j}^{4}-
2(a^{2}b_{i}^{2}+a^{2}b^{2}_{j}+b_{i}^{2}b_{j}^{2}) \) is a kinematic
function with \( b_{i}=m_{i}/m_{LSP} \), and \( a=\sqrt{s}/m_{LSP} \) is
the energy variable relevant for thermal averaging. 
In the expression above, \( K_{\nu } \) is the modified Bessel function of 
the second 
kind: its appearance is due to the more accurate expression for the 
thermal spectrum \(f_{0}=\sum _{i}^{N}g_{i}m_{i}^{2}K_{2}(m_{i}/T)/(2\pi 
^{2}T^{2}) \).  This evolution
equation governs \( f\equiv \sum _{i}^{N}f_{i} \), where the sum is over
the \( N \) sparticles.

\subsection{\label{sec:lspparam} Neutralino parameter dependence}
At the electroweak scale, the neutralino mass matrix depends
on \( M_{1} \), \( M_{2} \), \( \tan \beta  \), and \( \mu  \).
The masses and mixings have been analyzed in many papers; see {\it e.g.} 
\cite{Ellis:1983ew,Gunion:1984yn,Drees:1987jg,Griest:1988ma,Bartl:1989ms,Drees:1992am}.
In the limit in which \( |M_{1}|+|\mu |\gg M_{Z} \) and \( |M_{2}|>|M_{1}| \),
the LSP is either a pure bino (if \( |M_{1}|\ll \mu  \)), a higgsino
(if \( |M_{1}|\gg \mu  \) ), or a mixture (if $|M_{1}|\sim |\mu |$). 
When \( M_{Z} \) is comparable to the larger of \( |M_{i}| \)
or \( |\mu | \), \( \tan \beta  \) controls the mixing. The higgsino 
masses are somewhat sensitive to \( \tan \beta  \) in this limit. 

The renormalizable couplings of the neutralino are of the form 
neutralino-fermion-sfermion,
neutralino-neutralino-gauge boson, neutralino-chargino-gauge boson,
or neutralino-neutralino-Higgs. For annihilation reactions of 
neutralinos significant for the final dark matter abundance, one must
have either neutralino+neutralino, neutralino+sfermion, or neutralino+chargino
in the initial state. The annihilation reactions are broadly classified
into two categories:
\begin{itemize}
\item The LSPs are {\it 
self-annihilating}: {\it i.e.}, LSP+LSP in the initial state. 

\item The LSPs are {\it coannihilating}: {\it i.e.}, LSP 
+ other superpartner in the initial state. 
\end{itemize}
Due to the strong thermal suppression 
for initial states heavier than the LSP, the self-annihilation channels 
usually dominate in the determination of the relic abundance. However, if 
there are other superpartners with masses close to \( 
m_{LSP} \) (within an \({\cal O}(m_{LSP}/20)\) fraction of \( m_{LSP} \)), 
then the coannihilation channels become significant.

In typical nonresonant situations, the t-channel slepton exchange
self-annihilation diagrams dominate. However, many s-channel
contributions exist, and if the neutralino masses are light enough such 
that they sum approximately to the mass of one of the s-channel 
intermediate particles such as the Higgs or the \( Z \), the resonance 
contribution dominates the annihilation process.  When the resonance 
dominates, unless the resonance is wide as is possible {\it e.g.} for the 
Higgs, some fine tuning is required to obtain a nonnegligible final 
abundance of 
LSPs because the final relic density is inversely correlated with the 
strength of the annihilations. The relative strengths of the nonresonant
reactions are determined mostly by the mass of the
intermediate particle ({\it e.g.} suppressed if it is heavy) and the
kinematic phase space available for the final states ({\it i.e.}, their
masses relative to \( m_{LSP} \)).

Thus far, we have been discussing the effects of the low energy
parameters.  As mentioned previously, most of the parameter space
exploration in the literature is done within the 5-parameter mSUGRA model 
because of its relative simplicity compared to the general MSSM-124. Of 
course, in this context all of the above discussion applies: 
the low energy parameters are just functions of the 5 
mSUGRA parameters determined by using the RGEs. The
differences in the RGEs within the available computer codes in the 
literature appears to be the greatest source of discrepancy for the 
calculated dark matter abundance within the mSUGRA framework.

\begin{figure}
\centerline{
   \epsfxsize 3.3 truein \epsfbox {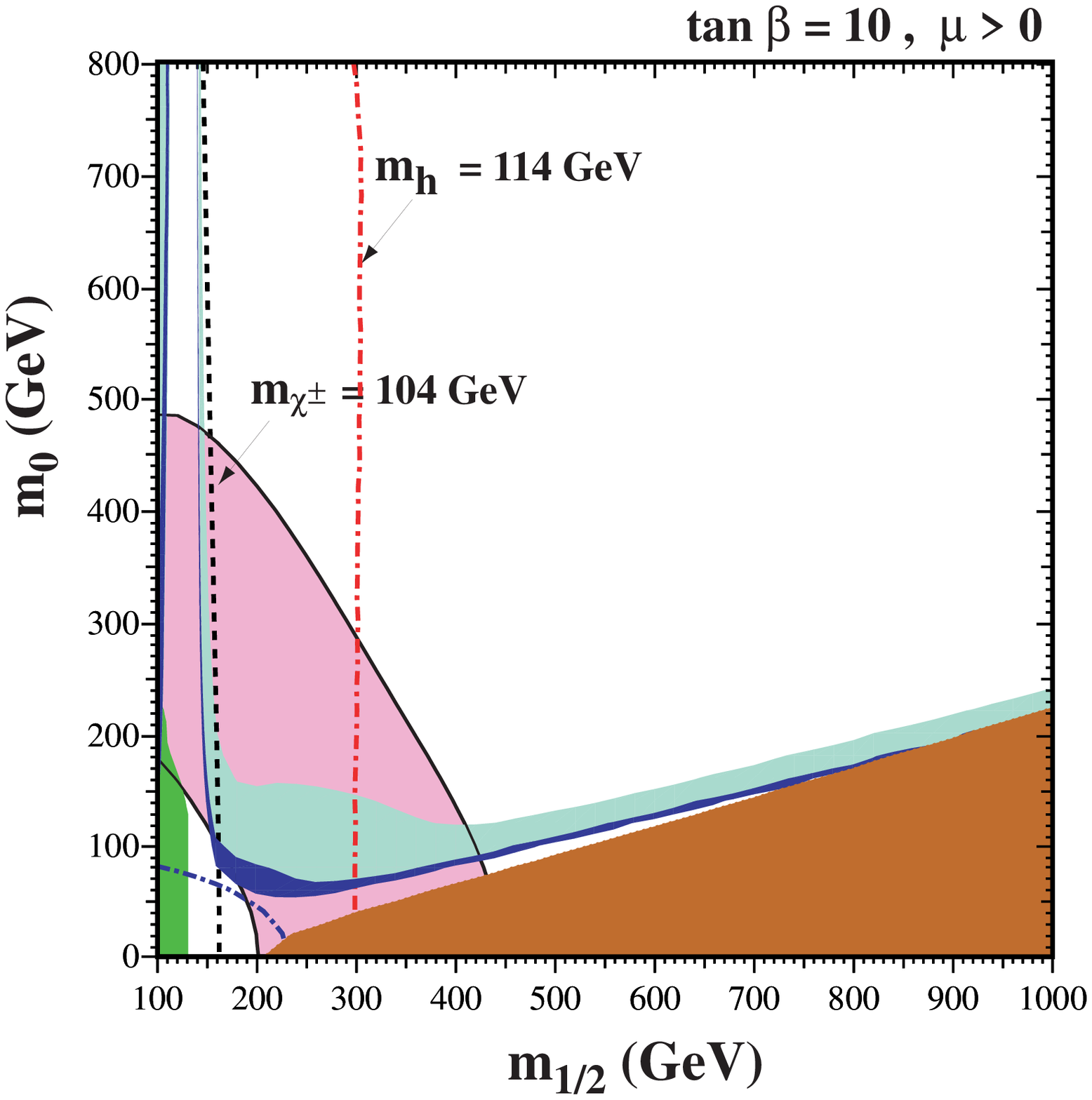}
    }

\centerline{
   \epsfxsize 3.3 truein \epsfbox {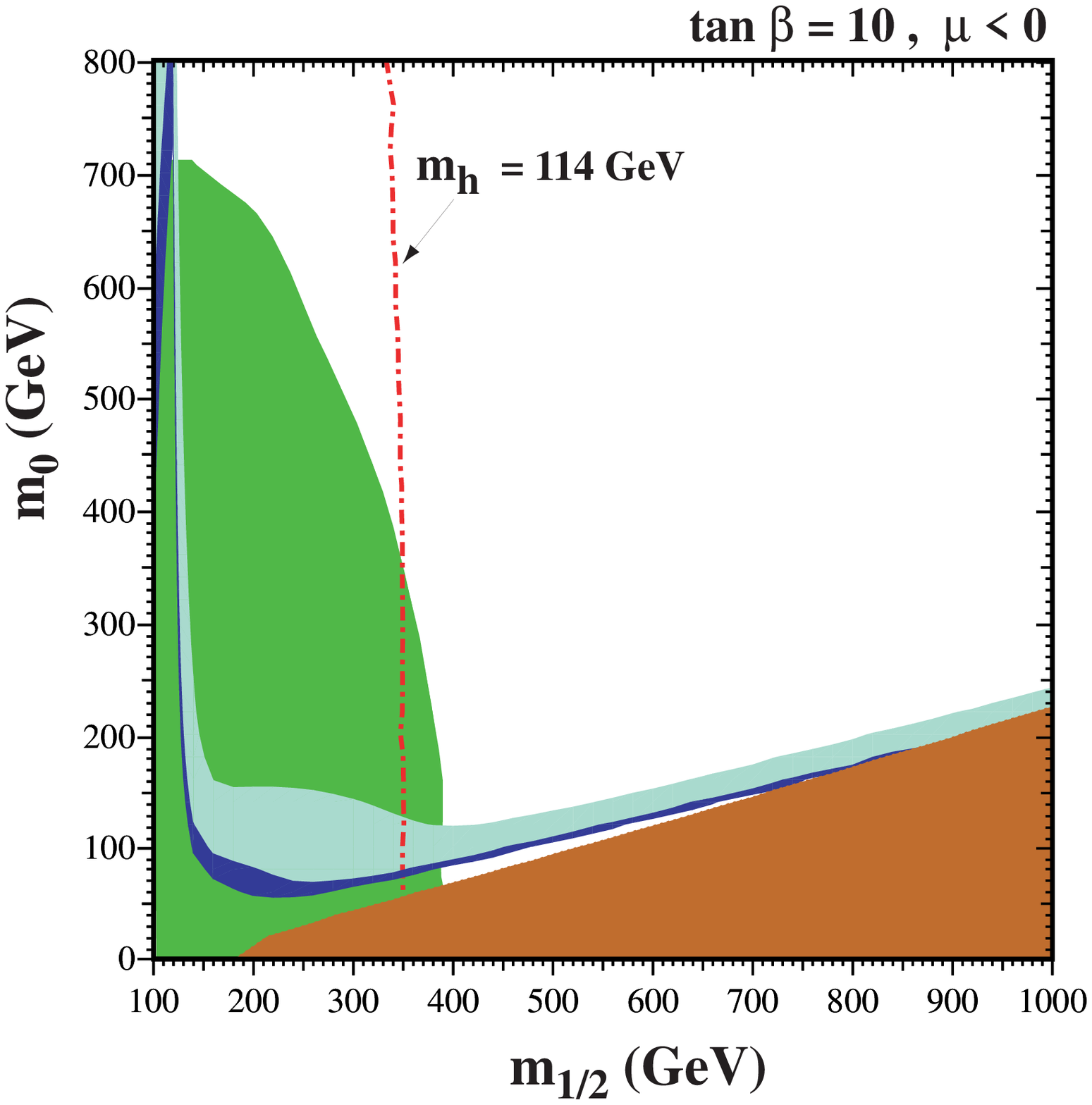}
    }



\caption{\label{fig:msugradmplots} mSUGRA/CMSSM parameter space
exclusion plots taken from \cite{Ellis:2003cw}, in which \protect\( 
A_{0}=0\protect \) and other parameters are as shown.  The
darkest ``V'' shaped thin strip corresponds to the region with 
\protect\( 0.094\leq \Omega h^{2}\leq 0.129\protect \), while a bigger
strip with a similar shape corresponds to the region with 
\protect\( 0.1\leq \Omega h^{2}\leq 0.3\protect \). (There are other
dark strips as well when examined carefully.) The triangular region in
the lower right hand corner is excluded by \protect\(
m_{\widetilde{\tau }_{1}}<m_{\widetilde{\chi }^{0}}\protect \), since DM cannot be
charged and hence is a neutralino \protect\( \widetilde{\chi }^{0}\protect
\)).  Other shadings and lines correspond to accelerator
constraints. In the lower figure ( \protect\( \mu
<0\protect \)), most of the DM favored region below \protect\(
m_{1/2}<400\protect \) GeV is ruled out by the \protect\( b\rightarrow
s\gamma \protect \) constraint.  In the upper figure, the medium
shaded band encompassing the bulge region shows that the region
favored by dark matter constraints is in concordance with the region 
favored by \protect\( g_{\mu }-2\protect \) measurements.  The Higgs and
chargino mass bounds are also as indicated: the parameter space
left of these bounds is ruled out. Unless excluded by 
accelerator constraints, the region below the darkest {}``V'' region is 
not excluded, but is not cosmologically interesting due to 
the small relic abundance.}
\end{figure}

Typical plots can be seen in Figure~\ref{fig:msugradmplots}.
Because of the tight constraints on \( \Omega h^{2} \) from the
recent WMAP fits \cite{Bennett:2003bz,Spergel:2003cb}, a fairly large
annihilation cross section is required for compatibility with
cosmology.  The cosmologically favorable vertical dark strips at \(
m_{1/2}<105 \) GeV are due to s-channel resonance annihilation through
the light Higgs and Z poles, and the horizontal strip between \( m_{0}=50 
\) and \( m_{0}=200 \) GeV is due to coannhilation channels as \(
m_{\widetilde{\tau }_{1}} \) approaches \( m_{LSP} \). 

As the masses of the pseudoscalar and the heavy scalar Higgs bosons 
decrease as \( \tan \beta  \) increases,  s-channel annihilation through 
very broad Higgs resonance dominates for high \( \tan \beta  \), 
giving an acceptable relic abundance. The allowed parameter space through
this resonance scattering is sometimes referred to as the funnel region.

There is another often discussed region of parameter space called the
{\it focus point} region \cite{Feng:2000gh}, which corresponds to very 
high values of \( m_{0} \), in the multi-TeV range.  In this parameter 
region the LSP becomes more and more higgsino-like due to the falling 
values of \( \mu \) consistent with radiative EWSB.  For moderate values 
of \( \tan \beta \), the growing higgsino component opens up new channels 
for annihilation that can bring down the final dark matter density.

Due to the lower bound on the Higgs mass, most of the mSUGRA parameter 
space is ruled out.  However, the smallness of 
the allowed regions in the mSUGRA scenario should not be too alarming for
considerations of neutralino dark matter.  If the universality assumptions 
of mSUGRA are relaxed, a much larger parameter region becomes viable 
\cite{Drees:1996pk,Nath:1997qm,Ellis:1998jk,Ellis:2000we}.  
Moreover, the smallness of the allowed parameter space also is partly a
reflection of the accuracy to which we know the phenomenologically
required CDM density.  In addition, if there are extra fields such as
the axino to which the neutralino can decay, a larger parameter space can
become viable, as discussed in Section~\ref{subsec:axino}.  Finally, 
there can be nonthermal production mechanisms for the LSP.

\subsection{Neutralino direct dectection}
A great deal of work has been done on both theoretical and
experimental aspects of direct detection (see {\it e.g.} the reviews
\cite{Primack:1988zm,Smith:1988kw,Lewin:1995rx,Bergstrom:2000pn}).
Direct detection of WIMPs can be accomplished through elastic scattering
off a nucleus in a crystal \cite{Drukier:gj,Goodman:1984dc,Wasserman:hh,Drukier:tm}.
The recoil energy is then measured by a variety of means: scintillation 
detection, cryogenic detection of phonons (usually relying 
on superconductor transitions), ionization detection, or some combination 
thereof. Inelastic nuclear scattering methods have also been considered 
\cite{Jungman:1995df}, but most of the proposed experiments use the 
elastic scattering method due to event rate considerations.

The typical elastic scattering cross section is of the order \( 10^{-10} \)
to \( 10^{-6} \) pb, and hence the expected event rate is about 1
kg/day or less. The recoil energy of the nucleus is also expected
to be very small, of order \( 10-100 \) keV. The background consists of 
neutrons, \( \gamma  \)-rays, and other cosmic rays. Neutrons are 
particularly
troublesome as the recoil induced by their scattering is difficult
to distinguish from the WIMP-induced recoil. Indeed, the background
reduction rather than larger exposure remains an important challenge for 
direct detection experiments. 

There are many experiments that have been or will be dedicated to
direct detection. DAMA, located in the Gran Sasso underground laboratory,
uses 58 kg of NaI \cite{Bernabei:vj,Bernabei:ft,Bernabei:1998td}; 
it has already claimed positive detection of dark matter \cite{Bernabei:2000qi}
(more below). The CDMS experiment \cite{Abusaidi:2000wg,Abrams:2002nb},
located at the Soundan mine in Minnesota, uses 100 g of Silicon and 495
g of germanium at 20 mK. The EDELWEISS experiment \cite{Benoit:2002hf}, 
located under the French-Italian Alps, uses three 320 g Ge detectors 
operating at 17 mK. The ZEPLIN I experiment uses liquid Xenon (a high mass 
nucleus) corresponding to 4 kg fiducial mass \cite{Luscher:yn} located in 
Boulby Mine (England). UKDMC NaI \cite{Kudryavtsev:2001hp} is also located
in Boulby Mine with a target of around 20 kg. The CRESST experiment
utilizes 262 g of sapphire cryogenic calorimeter operating at 15 mK
located in the Gran Sasso underground laboratory \cite{Probst:qb}.
Among the future experiments, GENIUS \cite{Klapdor-Kleingrothaus:2002me}
is a particularly prominent experiment progressing in the Gran Sasso
underground laboratory which will be able to directly test the DAMA
experimental results using 100 kg of natural Ge.

To determine the neutralino 
direct detection rates, the neutralino-quark elastic scattering amplitudes
as well as the one loop neutralino-gluon scattering amplitudes must be
computed. The parton level amplitudes are convoluted with quark and gluon
distribution functions in nucleons and some model of detector nucleus must
be used to account for detector-specific structure effects. This is
clearly a large source of uncertainty. Generically, there are both
spin-independent and spin-dependent contributions to the elastic cross
section.

The spin-independent or scalar part receives contributions from
neutralino-quark interactions via squark and Higgs exchange and from
neutralino-gluon interactions involving loop quarks, squarks, and Higgses.
This can be described in terms of an effective neutralino-nucleon
Lagrangian 
\begin{equation}
{\cal L}_{scalar}=f_{p}\overline{\chi }\chi \overline{\Psi }_{p}\Psi 
_{p}+f_{n}\overline{\chi }\chi \overline{\Psi }_{n}\Psi _{n},
\end{equation}
in which the nucleons are denoted by $\Psi_{n,p}$, and 
the neutralinos are collectively denoted by $\chi$.  In 
the above, the effective couplings \( f_{p,n} \) contain all the short
distance physics and nucleonic partonic structure information. The
differential cross section for scattering on a nucleus of charge \( Z \)
and atomic number \( A \) can then be written as \begin{equation}
\frac{d\sigma _{scalar}}{d|\vec{q}|^{2}}=\frac{m_{A}^{2}m_{\chi }^{2}}{\pi 
(m_{A}+m_{\chi })^{2}|\vec{q}|^{2}}\left[ Zf_{p}+(A-Z)f_{n}\right] 
^{2}F^{2}(Q_{r}), 
\end{equation}
where \( \vec{q}\equiv m_{A}m_{\chi }/(m_{A}+m_{\chi })\vec{v} \) is
the momentum transfer, \( Q_{r}=|\vec{q}|^{2}/(2m_{A}) \) is the
recoil energy, and \( F^{2}(Q_{r}) \) is the scalar nuclear form
factor. Note that the cross section grows with \( Z^{2} \) or \(
(A-Z)^{2} \).  There is significant uncertainty in \(
\{F^{2}(Q_{r}),f_{p},f_{n}\} \) because of the nuclear model
sensitivity, and hence the uncertainty should be at least a factor of
2. For intuitive purposes, one may estimate the dimensionless form factor 
as \begin{equation} F^{2}(Q)\sim \exp (-Qm_{N}R_{N}^{2}/3),
\end{equation}
where \( R_{N}\sim 5[0.3+0.91(m_{N}/\textrm{GeV})^{1/3}] \)GeV\( ^{-1}
\) is the nuclear radius. Similarly, the dimensionful
effective nucleon coupling parameters can be estimated as
\begin{equation} f_{p,n}\sim \left 
(\frac{m_{p}}{m_{W}} \right )\frac{10^{-1}\alpha
_{W}}{m_{H}^{2}}\sim 10^{-8}\textrm{ GeV}^{-2},
\end{equation}
in which we have assumed that the CP-even Higgs parton level exchange 
dominates and \(m_{H}\sim 100 \) GeV is the Higgs mass scale. $\tan \beta$ 
determines to a large extent which Higgs contribution dominates. 
In practice, the mass and mixing parameter dependence of these factors
are complicated and model dependent; {\it i.e.}, they are sensitive to the 
neutralino couplings to Higgs, squarks, and quarks.  For further details,  
see {\it e.g.} \cite{Drees:1993bu,Jungman:1995df,Lewin:1995rx,Baer:1997ai}.

The spin-dependent part receives contributions from squark and \( Z \)
exchange. The effective neutralino-nucleon Lagrangian is
\begin{equation}
{\cal L}_{spin}=2\sqrt{2}(a_{p}\overline{\chi }\gamma ^{\mu }\gamma _{5}\chi 
\overline{\Psi }_{p}s_{\mu }\Psi _{p}+a_{n}\overline{\chi }\gamma ^{\mu }\gamma _{5}\chi \overline{\Psi }_{n}s_{\mu }\Psi _{n}),
\end{equation}
where \( s_{\mu } \) is the nucleon spin vector and \( a_{n,p} \)
are the effective theory coefficients. Typically, \( a_{n,p}\sim \alpha 
_{W}/m_{\widetilde{q}}^{2} \) or \( \alpha _{W}/m_{W}^{2} \)
\cite{Drees:1993bu}. The spin interaction differential cross section
off of a nucleus with total angular momentum \( J \) is
\begin{equation}
\frac{d\sigma _{spin}}{d|\vec{q}|^{2}}=\frac{8m_{A}^{2}m_{\chi }^{2}}{\pi 
(m_{A}+m_{\chi })^{2}|\vec{q}|^{2}}[a_{p}\frac{\langle S_{p}\rangle 
}{J}+a_{n}\frac{\langle S_{n}\rangle }{J}]^{2}J(J+1)S_{1}(|\vec{q}|), 
\end{equation}
where \( S_{1}(|\vec{q}|) \) is the nuclear spin form factor normalized to 
\( 1 \) at \( |\vec{q}|=0 \) for pointlike particles and \( \langle S_{p}\rangle  \)
and \( \langle S_{n}\rangle  \) represent the expectation values of the 
proton and neutron spin content in the nucleus. Similar to the
spin-independent situation, \( \{a_{p},a_{n},S_{1}\} \)
have significant model dependence, but these quantities are generally 
believed to have uncertainties of about a factor of \( 2 \). However, 
in this case the cross section does not grow with \( Z^{2} \)
or \( (A-Z)^{2} \). Hence, unless the spin content of the nucleus
is large, the scalar interactions usually dominate (typically for \( 
A>30 \)). 
However, in certain regions of 
the parameter space, the 
spin-dependent part can play a significant role even when \( A>30 \). For 
example, for \( ^{73}\textrm{Ge} \), which has a nonzero nuclear spin 
of \( J=9/2 \), the spin-dependent contribution can give a significant 
contribution for \( \mu <0 \) and moderate values of \( \tan \beta  \).
Although not well-determined, one can approximate \( \langle S_{p}\rangle 
\approx 0.03 \) and \( \langle S_{n}\rangle \approx 0.378 \) 
\cite{Dimitrov:1994gc}.

The differential detection rate is given by
\begin{eqnarray}
\frac{dR}{dQ_{r}} & = & \frac{4}{\sqrt{\pi ^{3}}}\frac{\rho _{\chi
}}{m_{\chi }v_{0}}T(Q_{r})\left\{ [Zf_{p}+
(A-Z)f_{n}]^{2}F^{2}(Q_{r})\right. \nonumber \\
 &  & \left. + 8[a_{p}\frac{\langle S_{p}\rangle }{J}+a_{n}\frac{\langle
S_{n}\rangle }{J}]^{2}J(J+1)S_{1}(|\vec{q}|)\right\} ,
\end{eqnarray} 
in which 
\( v_{0}\approx 220 \) km/s is the speed of our sun 
relative
to the center of the galaxy, \( \rho _{\chi } \) is the local LSP
density, and \begin{equation}
T(Q_{r})=\frac{\sqrt{\pi }v_{0}}{2}\int _{v_{min}}^{\infty }dv\frac{f_{\chi }(v)}{v}\end{equation}
integrated over the neutralino velocity distribution \( f_{\chi } \).
The recoil energy \( Q_{r} \) is typically no more than \( 100 \)
keV. The greatest uncertainty in the differential detection rate is
from the uncertainty in the local LSP density \( \rho _{\chi } \) 
\cite{Brhlik:1999tt,Belli:1999nz,Green:2000jg,Vergados:sf,Vergados:1999rh,Vergados:2001rr,Green:2000ga,Vergados:2002hc,Qiao:2003et,Gelmini:2000dm,Belli:2002yt,Copi:2002hm,Green:2003yh};  
the answer is uncertain by a factor of a few. When
folded in with the nuclear physics uncertainties, the final theoretical
detection rate is uncertain by about a factor of 10 or more.

\begin{figure}
{\centering \includegraphics{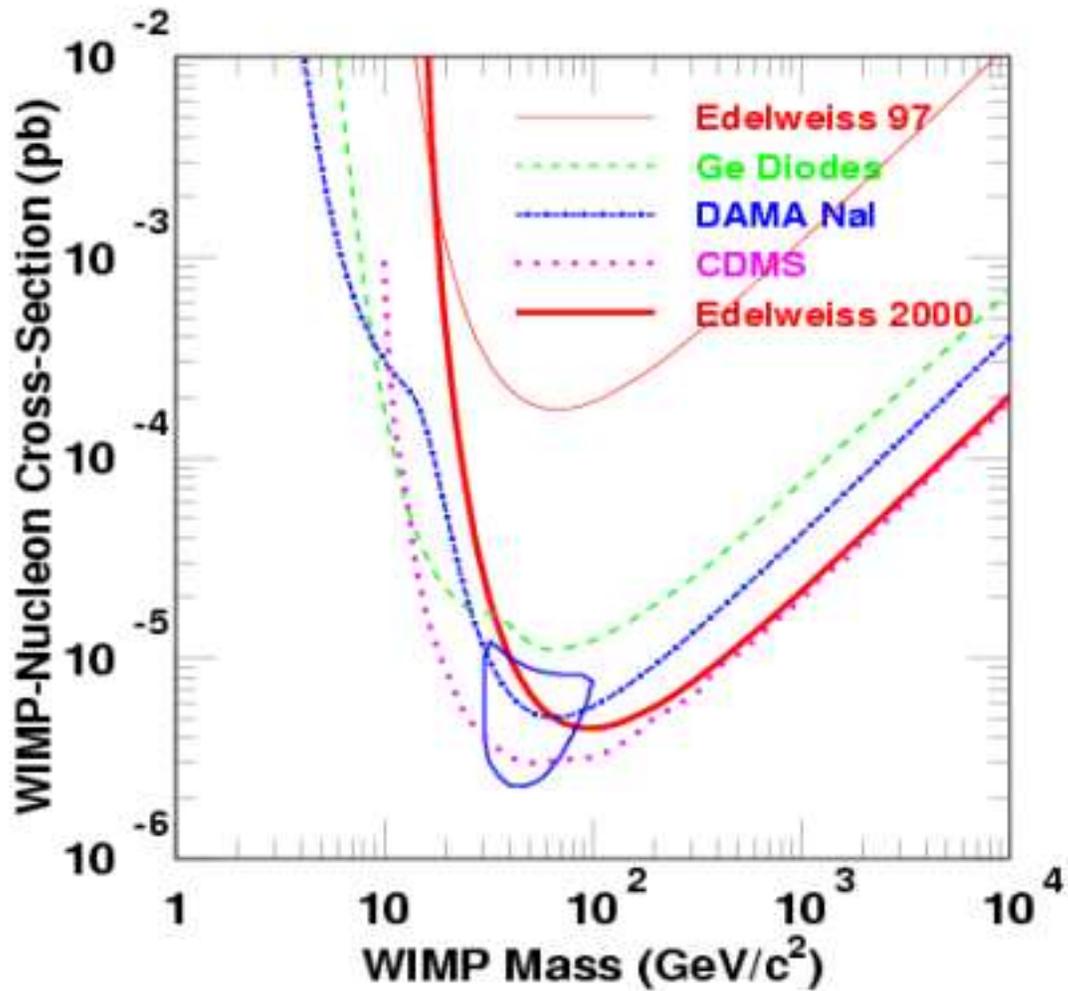} \par}

\caption{\label{Benoit}Typical exclusion plot taken from \cite{Benoit:2001zu}.
The region above the curves are excluded. The closed curve represents
the \protect\( 3\sigma \protect \) positive detection region of DAMA
experiment.}
\end{figure}

One way to enhance the detection signal above the background
is to look for modulations in the signal rate due to the detector's
time varying velocity relative to the dark matter halo. For example,
due to the earth's motion around the sun, the time of the maximum
velocity of the dark matter halo with respect to the terresterial
detector is six month separated from that of the minimum velocity of the 
dark matter halo with respect to the terresterial detector 
\cite{Drukier:tm,Freese:1987wu}.
This method has been the focus of the DAMA experiment 
\cite{Bernabei:vj,Bernabei:ft,Bernabei:1998td}, which 
has announced positive detection of the annual modulation 
\cite{Bernabei:2000qi}.
However, the discovery has been disputed by many experimental groups
and has still neither been undisputedly excluded nor confirmed 
\cite{Ramachers:2002kv},
despite many questionable claims to the contrary in the literature.

Another way to enhance the signal above the background is to resolve
the nuclear recoil direction as the dark matter elastically scatters
\cite{Spergel:1987kx}. Because of the strong angular dependence, 
generically the number of recoil events in the forward direction will 
significantly exceed the number of events in the backward direction for 
any energy threshold of the detector. Due to the daily rotation of the 
earth, the detector should then see a modulation between
the nighttime and the daytime (diurnal modulation). The proposed
experiment DRIFT \cite{Snowden-Ifft:1999hz} is thus far the only experiment
that has enough directional sensitivity to take advantage of diurnal
modulation. On the other hand, because this experiment relies on measuring
ionization tracks in a low pressure gas, one drawback 
is the low target mass required by the low pressure gas.

It has been argued that prospects for the discovery of supersymmetry
through the direct detection of LSP CDM are as good as or better than
through detection at the LHC (see {\it e.g.} \cite{Baer:2003jb}) in some
regions of parameter space, such as the focus point region. A typical
exclusion plot for data that has already been taken can be seen in
Figure~\ref{Benoit}. Of course, because different detectors have
different energy thresholds and detection techniques, one must be careful
to consider the details of the experiments before drawing conclusions from
these kinds of plots. Furthermore, recall from our previous discussion
that there is about a factor of 10 uncertainty in the final detection
rate. Given that this is an active area of experimental research, we
expect to see substantial improvements in the near future.

\subsection{Neutralino indirect detection}
The indirect detection processes are classified according to which
particles are actually interacting with the laboratory detector.  The 
detected particles are generally cosmic ray particles resulting from 
the annihilation of LSP neutralinos.  We will first discuss neutrino
telescopes, which arguably have the least number of astrophysical
uncertainties, and then mention the detection of other cosmic ray
particles.

\paragraph{Neutrino Telescopes}
LSP dark matter can accumulate in astrophysical bodies such as the sun or 
the earth by elastic scattering if the final state WIMPs have velocities 
less than the escape velocity 
\cite{Krauss:1985ks,Silk:1985ax,Gould:1987ir}. The
accumulated LSPs can annihilate, giving rise to observable final
products. Among the SM decay products of the primary
annihilation products, the muon neutrino can escape without being
absorbed by the core of the sun or the earth and can reach
terresterial detectors. Since \( \chi \chi \rightarrow \nu \nu \) is
suppressed by the small neutrino masses, the neutrinos primarily arise due 
to decays of primary products of annihilation with a mean 
energy of \(\sim m_{\chi }/2 \). In the water/terrestrial material 
immersing the detectors, the muon neutrinos induce muon production, which 
can easily be measured by its Cerenkov radiation.

The derivation of the capture rate (number per unit time) starts by 
writing the differential
scattering events per unit time as\begin{eqnarray}
d\dot{N} & = & (\#\textrm{ of nuclei })\times (\textrm{velocity 
differential flux})\times \nonumber \\
 &  & (\textrm{angular diff}\textrm{erential elastic  cross-section onto 
one nucleus}).
\end{eqnarray}
One then does the angular integration, restricting the 
final angle such that the final state particle is below the escape 
velocity, and performs the summation over the appropriate nuclei 
distributions. Thus computed, the capture
rate of neutralinos in an astrophysical body
of mass \( M \) (recall the mass of the sun is \( M=1.1\times 10^{57} \) 
GeV and the mass of the earth is \( M=3.4\times 10^{51} \) GeV) can be 
written as 
\cite{Gould:1987ir,Gould1991,Gould1992,Bottino:1994xp}
\begin{equation}
\label{eq:capturerate}
C\sim \frac{\rho _{\chi }}{v_{\chi }}M\sum _{i}f_{i}\frac{\sigma 
_{i}}{m_{\chi }m_{i}}\langle v_{\textrm{esc }}^{2}\rangle _{i}S(v_{\chi 
},v_{\textrm{esc }},m_{\chi },m_{i}), 
\end{equation}
where \( \rho _{\chi } \) and \( v_{\chi } \) denote the local neutralino
density and speed, \( f_{i} \) is the fraction of nucleus \( i \)
in the astrophysical body, \( v_{\textrm{esc}} \) is the escape speed, \( 
\langle ...\rangle  \) denotes averaging over the distribution of the element \( i 
\), \( \sigma _{i} \) is the nucleus-LSP elastic scattering cross section, 
and \( S(...) \) is a suppression factor which accounts for the additional 
kinematics of the neutralino-nucleus interaction.\footnote{The escape 
velocity is a local quantity,  given by\[
v_{esc}^{2}(r)\equiv \frac{2}{M_{Pl}^{2}}\int \frac{d^{3}x'\rho 
(\vec{x}')}{|\vec{x}-\vec{x}'|}, \]
where the integral is over the body with total mass \( M \). The
earth-sun distance is around \( 1.5\times 10^{13} \) cm, while the
earth radius is \( 6.4\times 10^{8} \) cm.} Typically, \( \rho _{\chi 
}\sim 3\times 10^{-42} \) GeV\( ^{4} \), 
\( v_{\chi }\sim 10^{-3} \), and \( v^{2}_{esc}\sim (4\times 10^{-5})^{2} \)
for the earth while it is \( v_{esc}^{2}\sim 10^{5}(4\times
10^{-5})^{2} \) for the sun. 

Because the local speed after the elastic scattering is 
\cite{Gould:1987ir,Damour:1998vg}
\begin{equation}
v^{2'}=v^{2}\left(1-\frac{2m_{\chi }m_{i}}{(m_{i}+m_{\chi})^{2}}[1- \cos 
\theta _{\textrm{cm }}] \right),
\end{equation}
where \( \theta _{\textrm{cm }} \) is the center of mass scattering
angle, there is a greater loss of energy after scattering when \(
m_{\chi }\approx m_{i} \) (and hence a ``resonant'' enhancement
\cite{Gould:1987ir} in the capture rate). Because the earth has heavy
elements, there is a resonant enhancement of capture for the mass
range\begin{equation} 10\textrm{ GeV}\leq m_{\chi }\leq 75\textrm{
GeV}, \end{equation}
with the peak near the iron mass of \( m_{Fe}\approx 56 \) GeV.

Although the sun does not have such heavy elements to cause resonant
scattering, the large quantity of the sun's hydrogen carries spin, 
allowing axial interactions to become important. Such interactions are 
particularly important if there is significant
\( Z \) coupling, which in turn depends on the higgsino fraction
of the neutralino. Due to the large solar mass and this spin-dependent
neutralino-quark cross section (\( \sigma _{\chi 
p}^{\textrm{scalar}}<\sigma _{\chi p}^{\textrm{spin }} \)),
the solar capture of the neutralinos is usually much more efficient
than neutralino capture in the earth. 

Given the capture rate of Eq. (\ref{eq:capturerate}), the annihilation
rate into neutrinos and the resulting neutrino flux near the detector
must be calculated. Following \cite{Griest:1986yu}, the annihilation
rate can be deduced from the simplified Boltzmann equation (neglecting 
evaporation): 
\begin{equation}
\dot{N}_{\chi }=C-C_{A}N_{\chi }^{2},
\end{equation}
where \( N_{\chi } \) is the number of neutralinos, and 
\begin{equation}
C_{A}\approx \frac{\langle \sigma _{A}v\rangle }{V_{0}}\left( \frac{m_{\chi }}{20\textrm{ GeV}}\right) ^{3/2}
\end{equation}
is the annihilation rate per effective volume of the body, with \(
V_{0}\sim 2.3\times 10^{25}\textrm{cm}^{3}(\approx 3\times 10^{66}\textrm{ GeV}^{-3}) \)
for the earth and \( V_{0}\sim 6.6\times 10^{28}\textrm{
cm}^{3}(\approx 8.6\times 10^{69}\textrm{ GeV}^{-3}) \)
for the sun. Assuming
that \( C \) and \( C_{A} \) remain constant, the total annihilation rate 
is \begin{equation}
\Gamma _{A}=\frac{1}{2}C_{A}N_{\chi }^{2}=\frac{C}{2}\tanh 
^{2}[t\sqrt{CC_{A}}],
\end{equation}
where \( t\approx 4.5 \) Gyr (\( \approx 2.2\times 10^{41} \)GeV\( ^{-1} \))
is the age of the macroscopic body. When accretion is efficient
such that \( \tanh ^{2}\approx 1 \), the annihilation rate \( \Gamma _{A} 
\) is independent of the annihilation cross section, but dependent
on the capture rate \( C \). For the sun, the neutralinos are nearly
in {}``equilibrium'' due to the large capture rate implying \( \Gamma _{A}\approx C/2 \).
However, when the higgsino component is small, for example as in the
low \( m_{0} \)-high \( m_{1/2} \) region of mSUGRA parameter space, \( 
\Gamma_{A} \) has a \( C_{A} \) dependence. Also, \( C_{A} \) is smaller 
when \( \tan \beta  \) is low, enhancing the \( C_{A} \) sensitivity of
\( \Gamma _{A} \). For the earth, neutrinos are not in equilibrium due to 
the generally smaller capture rate, leading to \( \Gamma _{A} \)
of the form \begin{equation}
\Gamma _{A}\sim \frac{C^{2}C_{A}t^{2}}{2}.
\end{equation}
This leads to enhancements in parameter regions where the annihilations
are large, as discussed in Section~\ref{sec:lspparam}.

\begin{figure}
{\centering
\resizebox*{1.2\columnwidth}{!}{\includegraphics{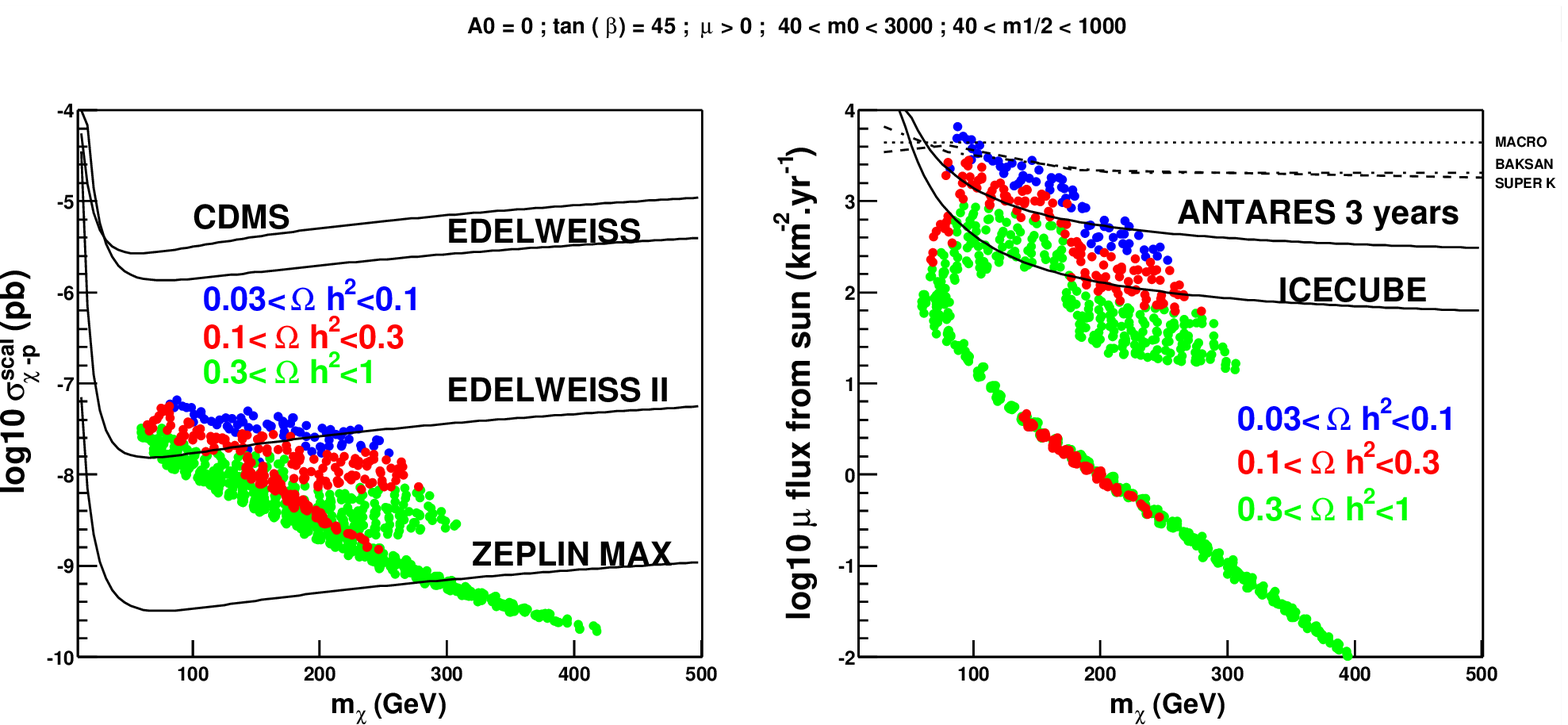}}
\par}

\caption{\label{fig:indirectvsdirectdetection}Taken from \cite{Bertin:2002ky},
the left figure shows the direct detection scalar elastic scattering
cross section for various neutralino masses, and the right figure
shows the indirect detection experiments' muon flux for various neutralino
masses. The scatter points represent ``typical'' class of models.
Specifically, the model parameters are \protect\( A_{0}=0,\tan \beta =45,\mu >0,m_{0}\in [40,3000],m_{1/2}\in [40,1000].\protect \)
The dotted curve, dot dashed curved, and the dashed curve on the right
figure represents the upper bound on the muon flux coming from Macro,
Baksan, and Super-Kamiokande experiments, respectively. This plot should 
be taken as an optimistic picture, because the threshold for detection was 
set 
at 5 GeV, where the signal-to-noise ratio is very low in practice.} 
\end{figure}

Given \( \Gamma _{A} \), the neutrino differential flux is 
\begin{equation}
\frac{d\Phi }{dE}=\frac{\Gamma _{A}}{4\pi R^{2}}\sum _{j}b_{j}\left( 
\frac{dN}{dE}\right) _{j},
\end{equation}
where \( R \) is the detector-(neutralino source) distance, \( b_{j}
\) is the branching ratio of annihilation channel \( j \), and \(
dN/dE \) is the differential neutrino spectrum. As mentioned
previously, the smallness of the neutrino mass suppresses annihilation
channels directly to neutrinos, and electron neutrinos scatter too
efficiently to reach the detector from the source. Therefore, the
neutrino-producing reactions of interest are secondary particle
decays. The hard (energetic) muon neutrinos come from \( WW \), \( ZZ
\), and \( t\overline{t} \) decays (assuming the neutralino mass is above
these thresholds), while the soft muons neutrinos are sourced by \( 
b\overline{b} \) decays. Since muons are the actual particles being detected 
and neutrino-induced production 
of muons grows with the neutrino energy, high energy neutrinos are easier 
to detect. This means that the muon flux will be larger for larger 
neutralino masses, which roughly translates to larger \( m_{1/2} \) in 
mSUGRA.  
Also, since an enhanced higgsino
component increases the annihilation into \( WW \) and \( ZZ \) which
gives more energetic neutrinos, increasing the higgsino component of
the neutralino enhances the muon signal as well. Although
the ratio of the mass of the sun to the mass of the earth is around \(
3\times 10^{5} \) and the distance-squared ratio between the earth-sun
distance and the earth radius is around \( 5\times 10^{8} \), because
\( \langle v_{esc}^{2}\rangle \) is also proportional to \( M \) and
the spin-dependent cross section is larger than the scalar cross
section, the flux of neutrinos originating from the earth is typically
much smaller than the flux originating from the sun.

The uncertainties in the theoretical calculations should be similar
to the direct detection case, since the quantities that enter are
similar: {\it i.e.}, most of the uncertainties stem from local
astrophysics. For example, even a small deviation from the usually
assumed Maxwellian distribution of neutralinos (caused by scattering
with the sun and interacting with large planets) can have an \( O(100)
\) effect on the indirect detection rates due to annihilation in the earth
for \( m_{\chi }<150 \) GeV \cite{Bergstrom:1999tk}.\footnote{The
corresponding effect for the direct detection is smaller because this
is a low momentum population with low momentum transfer.}

There have been several experiments under the category of neutrino
telescopes which had put bounds through indirect detection, including 
Macro \cite{Ambrosio:1998qj}, Baksan, Super-Kamiokande, and
AMANDA \cite{Andres:1999hp}. Future experiments have potential to
indirectly detect the neutralinos. One is the Antares 0.1 km\( ^{2} \)
project which covers a volume of around 0.02 km\( ^{3} \) (which
may be upgraded upgraded to 1 km\( ^{3} \) in the future) in the
Mediterranean sea at a depth of 2.4 km down south of France. Another
project, ICECUBE, will cover 1 km\( ^{3} \) volume under about 2.4
km of ice \cite{Barger:2001ur,Bergstrom:1998xh}. The reaches of these
experiments are compared to the direct detection experiments in 
Figure~\ref{fig:indirectvsdirectdetection}. The typical energy thresholds
are between 5 to 10 GeV.

\paragraph{Other cosmic rays}
In addition to the neutrino telescopes, there may also be the possibility 
possibility of indirect neutralino detection through other cosmic ray
particles \cite{Silk:zy}. Examples include gamma rays
\cite{Silk:zy,Urban:1992ej,Berezinsky:1992mx,Flores:1994gz,Bergstrom:mg,Bergstrom:1997fj,Bloom:1997vm,Strausz:1996ne,Tyler:2002ux,Aloisio:2002yq,Hooper:2002ru,Hooper:2002fx,Fornengo:2002tf},
lower energy photons such as radiowaves
\cite{Berezinsky:1992mx,Blasi:2002ct}, and antimatter such as positrons
and antiprotons
\cite{Silk:zy,Tylka:xj,Turner:1989kg,Kamionkowski:1990ty,Moskalenko:1999sb,Bottino:1994xs,Aguilar:ad,Battiston:1999yb,Pearce:ef}.
The source of these cosmic rays will be concentrated towards the center of
our galaxy. In fact, the recent positron excess reported by the HEAT
balloon borne experiment \cite{DuVernois:bb,Coutu:1999ws,Barwick:1997ig}
may be attributable to WIMP annihilations if certain nonstandard
astrophysical phenomena are assumed to take place
\cite{Moskalenko:1999sb,Kane:2001fz,Baltz:2001ir,Baltz:2002ua,Kane:2002nm}.  
Specifically, the HEAT collaboration has reported an excess of
positrons that are consistent with arising from LSP annihilation if the
LSP is heavier than the W.  While further study is needed to argue 
that this excess does not arise from conventional sources, there has not 
been a convincing alternative scenario which leads to an excess with a 
peak at an energy of order 10 GeV.  The excess has been seen in several 
sets of data with different detectors.

As far as theoretical predictions are concerned, there is greater
uncertainty in the non-neutrino signals since they involve greater
model dependence of the galactic halo. For example, consider the photons.
The computation of the differential flux is usually done using the 
approximate formula
\begin{equation}
\label{eq:photonflux}
\frac{dF_{\gamma }}{d\Omega dE}=\sum _{i}\frac{dN_{\gamma }^{i}}{dE}\sigma 
_{i}v\frac{1}{4\pi }\int n_{\chi }^{2}dl,
\end{equation}
where \( dl \) is the line of sight integral, \( dN_{\gamma }^{i}/dE \)
is the photon spectrum injected per annihilation channel \( i \)
(this includes any secondary particle decay probability), \( \sigma _{i}v \)
is the usual annihilation cross section times the Moeller speed factor,
and \( n_{\chi } \) is the neutralino density in the halo. The strong
model dependence is in the \( n_{\chi }^{2} \) integral.
The fiducial value is usually taken to be
\begin{equation}
\int n_{\chi }^{2}dl\sim \left (\frac{0.3\textrm{GeV}}{m_{\chi 
}}\frac{1}{\textrm{cm}^{3}} \right )^{2}(8.5\textrm{kpc}),
\end{equation}
which corresponds to the critical density being made up by the dark 
matter, and \( 8.5 \) kpc is the distance of the sun from the Galactic
center. There is at least a factor of \( 10^{3} \) (perhaps even as large 
as \( 10^{5} \)) uncertainty in
this integral \cite{Bergstrom:1997fj}.
The line signal (neutralino annihilation directly into photons \cite{Bergstrom:1997fh,Bern:1997ng,Ullio:1997ke,Berezinsky:1991sn}) is
a loop-suppressed process and is generically smaller relative to the
continuous spectrum signal (dominated by \( \pi ^{0}\rightarrow \gamma \gamma  \))
in the parameter region of interest. On the other hand, because it
is difficult to mimic a line signal by astrophysical processes not
involving heavy WIMPs, the line signal is more robust in terms of being
able to claim discovery of a heavy relic. 

The positron flux predictions stem from a equation similar to
Eq.(\ref{eq:photonflux}), except with an additional convolution of a
nontrivial Green's function for the positron propagation. On the other
hand, because only the high energy positrons (with energies above the soft
positrons coming from the solar wind) are easily measurable above the
background and since the high energy positrons lose energy efficiently,
the source of measurable positron flux cannot be as far away as the
galactic center, and instead must be within a few kpc of the earth. 
This makes the calculation less sensitive to the uncertainties of the 
matter distribution at the galactic center compared to the photon case. 
The positron flux can then be written as
\begin{equation}
\frac{dF_{e^{+}}}{d\Omega dE}=n_{\chi }^{2}(x_{0})\sum \sigma _{i}v\int 
dE'\frac{dN_{e^{+}}^{i}(E')}{dE'}G(E,E'),
\end{equation}
where \( n_{\chi }^{2}(x_{0}) \) is the local neutralino density,
\( \frac{dN_{e^{+}}^{i}(E')}{dE'} \) is the positron injection spectrum
at the neutralino annihilation source, and \( G(E,E') \) embodies the
propagation of the postirons and any remaining uncertainties in the
halo profile models. An example of \( G(E,E') \) for a {}``leaky
box'' toy model \cite{Tylka:xj,Turner:1989kg,Kamionkowski:1990ty}
is \begin{equation}
G(E,E')\approx \frac{1}{4\pi \xi }\frac{1}{E^{2}}\theta 
(E'-E)e^{\frac{E^{'-1}-E^{-1}}{\tau _{0}\xi }}, 
\end{equation}
which at best can give a reasonable order of magnitude estimate with \( 
\xi =1.11\times 10^{-9}\textrm{yr}^{-1}\textrm{GeV}^{-1} \), \( \tau 
_{0}=10^{7}\textrm{ yr} \). For a better model and further discussions, 
see \cite{Moskalenko:1999sb}.

Regarding photon detection, among the various future experiments the outer 
space experiment GLAST will 
have the greatest sensitivity and will have a good chance of seeing a 
signal because of its 
wider angular acceptance and better energy resolution and reach 
\cite{Morselli:2002nw}. As previously stated, the most clean signal is the
line (narrow width) spectral signal, which corresponds to at least one of
the primary annihilation products of the neutralinos being a photon. Other
photon-sensitive experiments that have already run or are planning to
run include STACEE, CELESTE, ARGO-YBJ, MAGIC, HESS, VERITAS, AGILE,
CANGAROO, and AMS/\( \gamma \). The most promising experiments as far as 
the positron (and other antimatter) signal is concerned are the space 
borne experiments PAMELA \cite{Pearce:ef} and AMS-02 \cite{Coutu:1999ws},
both of which are sensitive to high positron energies, as large
as 200 and 1000 GeV. Unfortunately, the positron signal-to-background 
ratio is generically extremely low, typically less than \( 0.01 \) 
\cite{Feng:2000zu}. An antiproton signal also must fight a large 
background 
\cite{Bergstrom:1999jc,Orito:1999re,Fornengo:2002tf,Bieber:1999dn}.

\subsection{Complementarity}
Not surprisingly, direct detection, indirect detection, collider
detection, and constraints from SM precision data play complementary roles  
--- mutually checking as well as having different parameter reaches ---
in the search for supersymmetry. 
This can be understood by
examining the schematic dependence on physical quantities controlling the
magnitude of the direct and indirect signals, as shown in
Table~\ref{tbl:schematicdependences}.

\begin{table} 
\begin{tabular}{|c|c|c|c|c|} 
\hline 
&\, p-elastic \,& \, low \(
\sqrt{s} \) annhilation \,& \, abundance \,& \, error \,\\ \hline 
direct
detection& \( \sigma _{\chi p} \)& & \( n_{\chi }(\textrm{local}) \)& \(
10 \)\\ 
\hline 
neutrino telesc.& \( \sigma _{\chi p} \)& little for sun&
\( n_{\chi }(\textrm{local}) \)& \( 10 \)\\ 
\hline 
\( \gamma \) (line,
continuum)& & \( \sigma _{\chi \chi \rightarrow \gamma X},\sigma _{\chi
\chi \rightarrow \pi ^{0}\pi ^{0}} \)& \( n^{2}_{\chi }(\textrm{core}) \)&
\( 10^{3} \)\\ 
\hline \( e^{+}(E>10\textrm{ GeV}) \)& & \( \sigma _{\chi
\chi \rightarrow WW,ZZ} \)& \( n_{\chi }^{2}(\textrm{nearby}) \)& \( 100
\)\\ 
\hline 
collider
& \( \sigma _{\chi p} \)& \( \sigma _{\chi \chi \rightarrow XX} \)& small& 
small\\  
\hline
\end{tabular} 
\caption{\label{tbl:schematicdependences} A schematic
picture of the various search processes. The column labeled
{}``p-elastic'' gives the dependence on proton-neutralino elastic-cross
section; {}``low \protect\( \sqrt{s}\protect \) annihilation'' refers to
the dependence on various self-annihilation cross section at very low
momenta (characteristic of the dark matter temperature in the halo);
\protect\( n_{\chi }(\textrm{local})\protect \) refers to the density of
the neutralinos in our solar system; \protect\( n_{\chi
}(\textrm{core})\protect \) refers to the density at the center of the
galaxy; \protect\( n_{\chi }(\textrm{nearby })\protect \) refers to the
halo density within few kpc of the solar system (not at the core of the
galaxy). The {}``error'' refers to a minimal multiplicative
uncertainty in the theoretical predictions. The table is not precise for
all parts of the MSSM parameter space and is merely meant to provide 
an elementary picture of the typical situation.
Collider data obviously does not directly
involve the proton-neutralino elastic cross section nor the 
self-annihilation cross section at nonrelativistic energies.  However, 
collider sensitivity generically is enhanced with light superpartners, 
which also tend to enhance both the elastic and the self-annihilation 
cross sections.} 
\end{table}

Collider and electroweak precision searches prefer lighter superpartners.
In mSUGRA, this corresponds to smaller \( m_{0} \) and \( m_{1/2} \)
parameters. On the other hand, indirect searches are typically enhanced
for a larger higgsino component, which in mSUGRA corresponds to the large
\( m_{0} \) region. In fact, if the LSP has a large higgsino component and
is heavier than a few TeV, the detection of gamma rays through the \( 
\chi \chi \rightarrow \gamma \gamma \) and \( \chi \chi \rightarrow \gamma 
Z \) channels \cite{Bergstrom:2000pn} may be the only way to discover
supersymmetry in the foreseeable future because the accelerator, direct
detection, and indirect neutrino dection may not have the required
sensitivities. Of course, such heavy neutralinos may be disfavored from
fine-tuning arguments.  Even for such large mass neutralinos, the
annihilation can be strong enough to not overclose the universe if there
is a sufficient higgsino component. The direct detection searches, which 
are sensitive to \( \sigma _{\chi p} \), have an inverse correlation in 
some regions of the parameter space with the indirect detection 
searches through \(\sigma _{\chi \chi \rightarrow \gamma \gamma } \), as 
higgsino-like neutralino models with \( m_{\chi }>400 \) GeV 
which have a small \( \sigma _{\chi p} \) generally have large \( \sigma 
_{\chi \chi \rightarrow \gamma \gamma } \) \cite{Bergstrom:2000pn}.

The neutrino telescope searches tend to complement the direct detection 
searches by having some overlap in sensitivity, as both are very sensitive 
to \( \sigma _{\chi p} \) \cite{Kamionkowski:1994dp}.\footnote{Of course, 
there are parameter space regions, such as \( m_{0}<500 \) GeV and \( 
m_{1/2}>800 \)
GeV, where the neutrino telescopes will be also sensitive to the
self-annihilation \cite{Bertin:2002ky}.} In fact, there is a possibility
of measuring \( m_{\chi } \) by detecting the angular distribution of the
muons in the neutrino telscope \cite{Edsjo:1995zc,Bergstrom:1997tp}.

Generically, there is an inverse correlation of the elastic scattering
cross section with the cosmological relic abundance of the neutralinos. By
looking at Table~\ref{tbl:schematicdependences}, one would then naively
conclude that the direct detection process and indirect detection to some
extent can still detect neutralinos even if neutralino LSPs did not
dominate the CDM composition.  Indeed, direct dark matter searches have
sensitivity in both the light LSP and the heavy LSP region, as can be seen
in Figure~\ref{Benoit}.  Even for the indirect detection,
\cite{Duda:2002hf} demonstrates that an LSP halo fraction as small as 1\%
can be indirectly detected with the current generation of experiments.

However, collider measurements of LSP neutralinos and their couplings 
relevant for self-annihilation do not imply that the dark matter abundance 
can be computed, because R-parity violation, light axinos (see 
Section~\ref{subsec:axino}), or a low
reheating temperature may spoil the standard LSP dark matter scenario.  
In practice, even within the standard cosmological scenario, the situation
with collider measurements alone is even worse than what it naively 
would seem because the relevant 
parameters needed to calculate the relic density must be measured to an
accuracy of order 5\% to obtain a useful answer for the relic density 
\cite{Brhlik:2000dm}.  

Remarkably, even with LHC discovery of supersymmetry and LSP 
neutralinos and with the assumptions of a standard cosmological scenario 
and R-parity conservation, we still may not be able to know whether the 
bulk of the CDM is composed of LSP neutralinos.  Hence, direct and 
indirect detection of dark matter are important to
ascertain the identity and the fraction of CDM in LSP neutralinos. On the
flipside, having direct and indirect detection of the LSP neutralino dark
matter by themselves do not specify the fraction of CDM in LSP neutralinos
because the local astrophysical uncertainties are unlikely to be smaller
than a factor of 2 in the near future and because the relevant 
\lsoft parameters must be measured to interpret the detection 
meaningfully. Therefore, very accurate collider and other
measurements of the parameters that are essential for the relevant kind of 
dark matter which can allow computations of Section~\ref{sec:lspdensity} 
are essential to determine the LSP fraction of the CDM.  This will most 
likely require colliders beyond the LHC.

\subsection{\label{sec:gravitino} Gravitinos}
In scenarios such as gauge-mediated supersymmetry breaking, the gravitino
naturally is the LSP, and hence becomes a dark matter candidate
\cite{Pagels:ke,Weinberg:zq}. For example, if the supersymmetry-breaking 
scale is of order \( \sqrt{F}\sim 10^{6} \) GeV, the gravitino mass 
is of the order \( F/\overline{M}_{pl}\sim 10^{-6} \) GeV ($F$ is the F 
term VEV which characterizes supersymmetry breaking, as discussed in 
Section~\ref{lsoftmodelsect}). The helicity \( 1/2 \) (goldstino) 
component has gravitational interactions of dimension 4 and 5, with  
coefficients of the order \( (m_{\chi }^{2}-m_{\phi 
}^{2})/(m_{3/2}\overline{M}_{pl}) \) and \( m_{\lambda
}/(m_{3/2}\overline{M}_{pl}) \) \cite{Fayet:1986zc} (here 
$m_{\phi}$ and $m_{\lambda}$ denote scalar and gaugino masses, 
respectively). This allows it to interact much more strongly when \( 
m_{\lambda,\chi,\phi}\gg m_{3/2} \) than the helicity \( 3/2 \) component, 
for which the gravitational interactions are not similarly enhanced. 
Without this enhancement, as in {\it e.g.} gravity-mediated 
supersymmetry breaking scenarios in which \( m_{3/2}\sim 
O(\textrm{TeV})\), gravitinos never reach thermal equilibrium below 
Planck scale temperatures. The enhancement allows thermal equilibrium to 
be reached, such that the gravitinos can go through the usual freeze out 
process to act as warm dark matter candidates. The relic abundance can be 
calculated as
\begin{equation} \Omega _{\widetilde{g}(th)}h^{2}\approx
\left (\frac{m_{3/2}}{1\textrm{
keV}} \right ) \left (\frac{g_{*}(T_{f})}{100} \right )^{-1},
\end{equation} 
which requires \( m_{3/2} \) to be less than about \( 0.1 
\) keV if the Hubble parameter today is given by \( h\approx 0.7 \) and \( 
g_{*}(T_{f})=100 \).  This may cause problems in the context of gauge 
mediation \cite{Pagels:ke,Moroi:1993mb,deGouvea:1997tn}, 
because such low values for the \( F \) term are unattractive in some 
gauge-mediated models.  One may need to invoke methods to dilute the 
gravitino abundance \cite{Baltz:2001rq} or have a low reheating 
temperature \cite{Moroi:1993mb,deGouvea:1997tn}. In certain special 
arrangements of the sparticle mass spectrum, there can be a secondary 
population of nonthermal gravitinos from NLSP decay \cite{Borgani:1996ag}. 
Due to their nonthermal momentum distribution, this secondary population 
can mimic hot dark matter consisting of eV range neutrinos. There are 
other ways to generate a nonthermal distribution of gravitinos as well
\cite{Kallosh:1999jj,Giudice:1999yt,Bastero-Gil:2000je}. Even when the
reheating temperature is small enough that there is no 
overclosure of the universe with LSP gravitinos, there may be a
cosmological problem with the decay of NLSPs (which typically have long 
lifetimes) into gravitinos, because such decays are generally accompanied 
by decay products which can spoil big bang nucleosynthesis (BBN) 
\cite{Khlopov:pf,Ellis:1984eq,Kawasaki:1994af,Dimopoulos:1987fz,Dimopoulos:1988ue,Reno:1987qw,Bolz:1998ek,Holtmann:1998gd,Gherghetta:1998tq}.

\begin{figure}

\centerline{
   \epsfxsize 5.3 truein \epsfbox {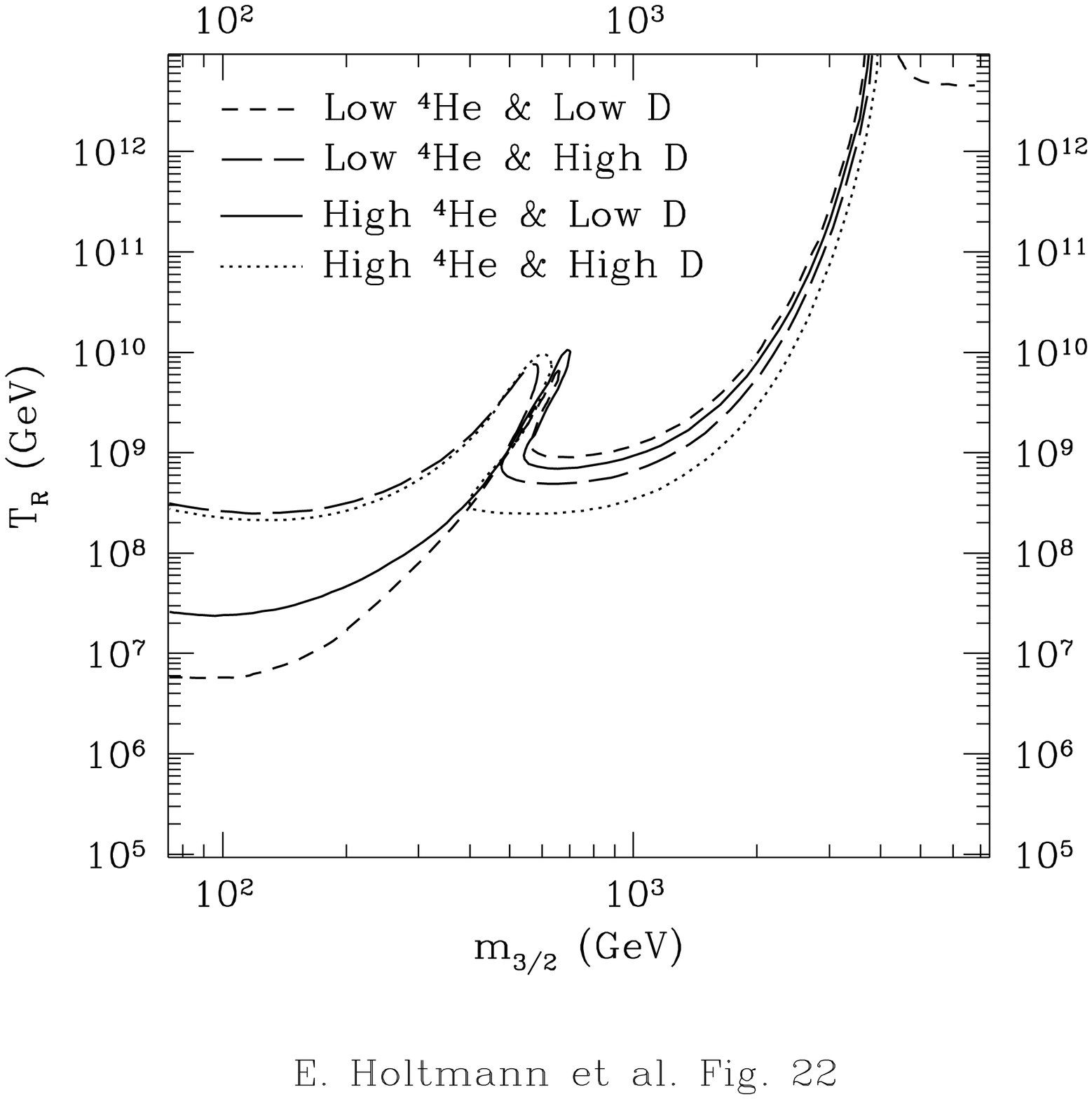}
    }


\caption{\label{fig:bbnbound} Reheating temperature upper bound
constraints from BBN as a function of the gravitino mass taken from
\cite{Holtmann:1998gd}. The various ``high'' and ``low'' values refer to
the usage of observationally deduced light nuclei abundances in deducing
the upper bound. Hence, the discrepancy can be seen as an indication of
the systematic error in the upper bound constraint from observational
input uncertainties.} \end{figure}

In gravity-mediated supersymmetry-breaking scenarios, the mass of the gravitino
is generically close to a TeV and it usually is not the LSP. In such 
scenarios,  there may be several cosmological problems caused by gravitino 
decay products 
which can dissociate nuclei during BBN, destroying its successful
predictions 
\cite{Kawasaki:1994af,Kawasaki:1994bs,Ellis:1984eq,Dimopoulos:1987fz,Reno:1987qw,Holtmann:1998gd,Dimopoulos:1988ue,Ellis:1990nb,Sarkar:1995dd}. 
In general,
successful BBN requires the photons to have a nearly thermal spectrum,
while the gravitino decay products may induce sufficient departures from
the thermal spectrum to ruin the successful ratios of element abundances.
The disruption of the photon spectrum can occur through
primary decay products as well through particles farther down the
cascade of reactions. Assuming that the gravitino decays to a photino
and a photon, its lifetime is given by \cite{Kawasaki:1994af}
\begin{equation}
\tau _{3/2}=3.9\times 10^{8}\left (\frac{m_{3/2}}{100\textrm{
GeV}} \right )^{-3}\textrm{ sec}.
\end{equation}
The decay has a long time scale because it originates from a dimension 
five (\( 1/M_{Pl} \) suppressed) operator. For reference, 
BBN occurs during \( \tau _{BBN}\approx 1-10^{2} \) sec (\( 
T\approx 1-0.1 \) MeV).

Assuming the gravitinos are produced thermally (although they never reach 
thermal equilibrium unless \( m_{3/2}\ll 100 \) GeV),
the gravitino abundance can be calculated as a function of 
the reheating temperature of the universe \( T_{RH} \) to be 
\cite{Kawasaki:1994af}
\begin{equation}
\frac{n_{3/2}}{n_{\gamma }}\approx 2.14\times 
10^{-11}\left (\frac{T_{RH}}{10^{10}\textrm{GeV}} \right )\end{equation}
for \( T\ll 1 \) MeV but for time \( t<\tau _{3/2} \). This is a 
significant number and energy density since the baryon-to-photon ratio is 
\( n_{B}/n_{\gamma }\sim 10^{-10} \) and \( m_{3/2} \gg  
m_{p} \). This large number of gravitinos will decay to photons, which 
will cause the dissociation of BBN nuclei through reactions such as \( 
D+\gamma \rightarrow n+p \)
or \( ^{4}He+\gamma \rightarrow n+^{3}He \) . An example of bounds
coming from successful BBN can be seen in Figure~\ref{fig:bbnbound}.

\setcounter{footnote}{0}
\subsection{\label{subsec:axino} Axion, axino, and saxion}
As discussed previously, the axion field \( a \) is a 
pseudo-Nambu-Goldstone
boson of the broken \( U(1)_{PQ} \) symmetry which solves the strong CP 
problem; its presence changes the usual \( \overline{\theta } \) 
to\begin{equation}
\overline{\theta }_{a}=\overline{\theta }-\frac{a(x)}{f_{PQ}/N},\end{equation}
where \( f_{PQ} \) is the PQ breaking scale and $N$ is defined below. Its 
properties depend most strongly on only one unknown parameter, the axion 
mass \( m_{a} \) or equivalently the PQ breaking scale \( f_{PQ} \): 
\begin{equation}
m_{a}\approx \frac{\sqrt{Z}}{1+Z}\frac{f_{\pi }m_{\pi }}{(f_{PQ}/N)},
\end{equation}
where the pion decay constant is \( f_{\pi }\approx 93 \) MeV, the
pion mass is \( m_{\pi }\approx 135 \) MeV, the dimensionless ratio
\( Z\equiv m_{u}/m_{d} \), and \( N =\textrm{Tr}[Q^{PQ}(Q^{\alpha
}_{\textrm{SU}(3)_{C}\textrm{ }})^{2}]\) is the 
color anomaly of the PQ
symmetry 
\cite{Weinberg:1977ma,Sikivie:yu,Bardeen:1977bd,Ellis:1978hq,Donnelly:1978ty,Srednicki:xd}.
Its interactions include its coupling to the gluon  \begin{equation}
\frac{\alpha _{3}g_{agg}}{8\pi }\frac{a}{(f_{PQ}/N)}\widetilde{G}^{a}_{\mu \nu 
}G^{\mu \nu }_{a},
\end{equation} 
the nucleon and electron
\begin{equation}
i\frac{1}{(f_{PQ}/N)}\partial _{\mu }a\left[ g_{ann}(\overline{n}\gamma ^{\mu} 
\gamma _{5}n)+g_{aee}(\overline{e}\gamma ^{\mu }\gamma _{5}e)\right], 
\end{equation} and the photon
\begin{equation}
\frac{\alpha _{EM}g_{a\gamma \gamma }}{2\pi }\frac{a}{(f_{PQ}/N)}\widetilde{F}^{\mu \nu }F_{\mu \nu }\end{equation}
where \( g_{aii} \) are model-dependent $O(1)$ coefficients. Two
standard models of axions are the KSVZ
\cite{Kim:1979if,Shifman:if} and DFSZ
\cite{Dine:1981rt,Zhitnitsky:tq} models.  Models such as KSVZ models with 
\( g_{aee}=0 \) at tree level are called ``hadronic'' 
because they lack direct couplings to leptons. As all the couplings are 
suppressed by ${\rm momentum}/(f_{PQ}/N)$, the axion can be
essentially ``invisible'' if \( f_{PQ}/N \) is large
enough. However, as will be explained below, \( f_{PQ}/N \) is severely
constrained by various measurements.

Since the interaction strength becomes larger as \( f_{PQ}/N \) is
lowered, the lower bound on \( f_{PQ}/N \) is determined both indirectly
and directly by observable particle reactions that can produce axions
\cite{Turner:1989vc,Raffelt:1990yz}. One example is Supernova 1987A
(SN1987A) which yielded a total of 19 detected neutrino events
spanning a time period of about 12 seconds which was in accord with
the expectations. For axions in the mass range \( f_{PQ}/N\approx
4\times 10^{9} \) GeV to \( f_{PQ}/N\approx 2\times 10^{6} \) GeV, the
cooling due to axion emission through bremsstrahlung from nucleons
would shorten the duration of the neutrino emission to unacceptable 
values much smaller than 12 seconds, according to the standard picture 
\cite{Burrows:1988ah}.  
The main reason why SN cannot rule out smaller values of \( f_{PQ}/N \) is 
because at these smaller values, the interactions become sufficiently 
strong such that the axions become trapped in the supernova core, causing 
the axion-mediated cooling to be inefficient.
For smaller \( f_{PQ}/N \), stellar processes provide 
constraints. Axion emission from the stellar core accelerates stellar 
evolution (more intense burning to compensate for the axion
emission energy loss), shortening the lifetime of red giants. For
hadronic axions, this gives a bound of \( 22\textrm{
GeV}<f_{PQ}/N<9\times 10^{6}\textrm{ GeV} \)
\cite{Dicus:fp,Dicus:1979ch,Raffelt:1987yu}. The lower bound is due
to the red giant core temperature scale of \( 10 \) keV being too
small to excite heavier axions. The upper bound is from the requirement
of the axion being sufficiently strongly coupled to be produced. 
Because the He core is supported by the electron degeneracy pressure for 
the DFSZ type of axions, the axion coupling to the electrons can cool the 
He core to such an extent that the He burning never takes
place \cite{Dearborn:gp}. This extends the upper bound from red giants
on \( f_{PQ}/N \) to \( 22\textrm{ GeV}<f_{PQ}/N<4\times 10^{8} \)GeV.
Finally, for even more strongly coupled, heavier axions, a variety
of lab experiments \cite{Caso:tx} put constraints of \( f_{PQ}/N>86 \)
GeV. Therefore, the combined experimental results exclude a broad range 
of scales, leading to a lower bound on the axion scale of \( 
f_{PQ}/N>4\times 10^{9} \) GeV.

The upper bound on \( f_{PQ}/N \) is given by cosmology from dark
matter constraints. Since axions have a long lifetime
\begin{equation}
\tau _{a}\sim 10^{17}\textrm{ yrs }\left( \frac{m_{a}}{1\textrm{ 
eV}}\right) ^{-5},
\end{equation}
axions can be good dark matter candidates.  The long
lifetime compared to that of the pion is due to the enhancement \(
(m_{\pi }/m_{a})^{5} \).  The cosmology of axions 
depends on the
inflationary history of the universe: we will assume throughout this 
review that inflation took place. If inflation reheats to a temperature 
larger than the lower bound of \( f_{PQ}/N \) of \( 4\times 10^{9} \) GeV,
gravitinos tend to disrupt the successes of standard cosmology (see
Section~\ref{sec:gravitino}). Furthermore, if the reheating
temperature is above \( f_{PQ} \), there may be a problem with domain
wall formation; this leads to at best a complicated, more 
model-dependent cosmology \cite{Sikivie:qv}. To keep the model dependence 
down and the physics simple for this review, we will focus on situations 
where the reheating temperature is lower than the PQ transition.  Even 
then, there are inflationary model dependent constraints due to the 
quantum fluctuations of the axion field during inflation
\cite{Lyth:1992tx,Lyth:1992tw}, which we will not discuss here.

Because the interaction rate is extremely small ({\it e.g.}, for quark 
mass \( m_q \), \( \langle \sigma v\rangle \sim \alpha 
(m_q/(f_{PQ}/N))^{2}/T^{2} \)
for \( T>m_q \), which is again strongly suppressed by \( 
f_{PQ}/N \)), the axions typically cannot be in thermal equilibrium for \( 
f_{PQ}/N>4\times 10^{9} \)
GeV \cite{Turner:1986tb}. Furthermore, one can estimate that the relic
density of thermally produced axions will be a negligible component
of the CDM, typically close to the energy density contribution of the 
cosmic microwave background (CMB) radiation.
However, axions can be a large source of CDM from the condensate oscillation
contribution, {\it i.e.} essentially, homogeneous classical axion field 
oscillations in time. The reason why the axion field will generically 
have such oscillations is that before the QCD phase transition, the 
axion has a relatively flat potential, such that its value (call it
\( a_{i} \)) can be anywhere of \( O(f_{PQ}) \). After the QCD phase
transition, instanton effects will generate a potential for \( a \).
Since the minimum of the potential \( a_{m} \) will be different
from \( a_{i} \), the axion will undergo a damped oscillating motion
about the minimum of the potential with the maximum initial amplitude
of \( a_{i}-a_{m} \). This oscillation will contribute an energy
density \cite{Preskill:1982cy,Abbott:af,Dine:1982ah,Ipser:mw,Asztalos:2001tf}
\begin{equation}
\Omega _{a}\approx \frac{1}{6}\left( \frac{a_{i}-a_{m}}{f_{PQ}/N}\right) 
^{2}\left( \frac{f_{PQ}/N}{10^{12}\textrm{GeV}}\right) ^{7/6}\left( \frac{0.7}{h}\right) ^{2},
\end{equation}
which would generically give a large contribution if \( f_{PQ} \) is
large with the oscillation amplitude \( (a_{i}-a_{m})/(f_{PQ}/N) \)
fixed (which naively is naturally expected to be of $O(1)$). In the
absence of fine tuning \( a_{i} \), the \( U(1)_{PQ} \) breaking scale
is then bounded to be \( f_{PQ}/N<10^{12}\textrm{GeV}. \) Therefore,
remarkably, the scale of new physics is known to be within a small
window \begin{equation} 10^{9}\textrm{GeV}<f_{PQ}/N<10^{12}\textrm{ 
GeV}.\end{equation} 
However, there is some room for adjustment (particularly at the upper 
end), if  there is a method to relax \( a_{i} \) to \( a_{m} \) during 
inflation 
or if there is a way to introduce extra entropy after the oscillations
begin. If the axion condensate oscillations make
up the CDM, there will be spatially dependent fluctuations that must
necessarily participate in structure formation \cite{Preskill:1982cy}.

Upon supersymmeterization, the pseudoscalar axion field, which is one real 
degree of freedom, attains a fermionic superpartner, the axino \( 
\widetilde{a} \), and a real scalar, the  saxion \( s \), to match the 
axino degrees of freedom.
Since the axion supermultiplet clearly involves physics beyond the MSSM,
it is difficult to justify the inclusion of this topic in a review
of the \lsoft parameters. Nonetheless, since the strong CP problem exists 
in the MSSM, one cannot justify a phenomenological/cosmological discussion 
of the MSSM without at least briefly considering what the effects of a 
strong CP problem solution may be.\footnote{For other more general reviews
on theory and astrophysics of axions, see {\it e.g.} 
\cite{Cheng:1987gp,Kim:ax,Turner:1989vc,Raffelt:1990yz}.}

The saxion-axino interactions include (see {\it e.g.} 
\cite{Covi:2001nw,Chun:1995hc})\begin{equation}
\frac{\alpha _{3}}{8\pi f_{PQ}/N}\left[ sF^{(a)}_{\mu \nu }F^{(a)\mu \nu 
}+\frac{1}{2}\overline{\widetilde{a}}\gamma _{5}[\gamma _{\mu },\gamma _{\nu 
}]\widetilde{g}^{(a)}F^{(a)\mu \nu }\right] 
\end{equation}
for the strong gauge group and related couplings for other gauge groups.
The first term allows the saxion to decay to gluons (pions) while
the second term allows the axino to scatter with gluinos into quarks
via s-channel gluons. There will also be couplings to the matter sector.
The interaction strengths should be similar to those of the axion.  On the 
other hand, the masses are very different. The saxion can have a soft 
breaking mass term, in analogy with the usual \lsoft terms, and thus is 
naturally  expected to have a mass at least the order of \( m_{3/2} \). 
The axino also might naively be expected to have a mass of order \( 
m_{3/2} \). However, explicit models (see {\it e.g.} 
\cite{Moxhay:1984am,Goto:1991gq,Chun:1992zk,Rajagopal:1990yx})
demonstrate that the axino mass can be smaller, depending on the model
(not surprisingly): the axino can even be lighter than the
lightest neutralino. Hence, with R-parity conservation, the axino
can be the dark matter. 

The axino has difficulty reaching thermal equilibrium because of its
weak interactions ({\it e.g.} see \cite{Rajagopal:1990yx,Covi:2001nw}).
Indeed, the axino fails to reach equilibrium unless the reheating
temperature \( T_{RH} \) of the universe is \begin{equation}
T_{RH}>10^{10}\textrm{GeV}\left( \frac{f_{PQ}/N}{10^{11}\textrm{GeV}}\right) \left( \frac{\alpha _{s}}{0.1}\right) ^{-3}\equiv T_{D}.
\end{equation}
This is typically in conflict with the gravitino bound. If this condition
is satisfied, then the relic abundance of axinos can be written as
\begin{equation}
\Omega _{a}h^{2}\sim \left( \frac{m_{\widetilde{a}}}{12.8\textrm{ eV}}\right) 
\left( \frac{g_{eff}}{g_{*}(T_{D})}\right), 
\end{equation}
where the effective number of degrees of freedom \( g_{eff}=1.5 \)
for axinos and \( g_{*}(T_{D}) \) is the number of relativistic degrees
of freedom when \( T=T_{D} \) ($\sim 230$ in the MSSM).

If the axinos never reach chemical thermal equilibrium, the details
of their production mechanisms become relevant in determining their 
final density. One class of production mechanisms that has
been explored is when the production occurs through interactions of
particles that were \emph{once} in thermal equilibrium \cite{Covi:2001nw}.
In such scenarios, the actual axino production can occur through the
decay and scattering of particles that continue to be in equilibrium
or have fallen out of equilibrium. When the reheating temperature
\( T_{RH} \) is above the squark and the gluino masses such that
they are in equilibrium, the thermal scattering processes involving
the axino-gluino-gluon vertex will result in
\begin{equation}
\Omega _{\widetilde{a}}h^{2}\sim 0.05\left[ \frac{\alpha 
_{s}(T_{RH})}{0.3}\right] ^{3}\left[ 
\frac{10^{12}\textrm{GeV}}{(f_{PQ}/N)}\right] ^{2}\left[ 
\frac{T_{RH}}{1\textrm{ TeV}}\right] \left[ 
\frac{m_{\widetilde{a}}}{\textrm{GeV}}\right],  
\end{equation} 
where the strong coupling is evaluated at \( T_{RH} \) \cite{Covi:2001nw}.
When the reheating temperature is in the range \( m_{\chi }\leq T_{RH}\leq 
m_{\widetilde{q},\widetilde{g}} \) with gluinos in thermal equilibrium, the axino 
abundance can be written as \cite{Covi:2001nw}
\begin{eqnarray*}
\Omega _{\widetilde{a}}h^{2} & \sim  & 0.3\left[ \frac{\alpha _{s}(T_{RH})}{0.3}\right] ^{2}\left[ \frac{10^{12}\textrm{GeV}}{(f_{PQ}/N)}\right] ^{2}\left[ \frac{m_{\widetilde{g}}}{1\textrm{ TeV}}\right] ^{3}\left[ \frac{1\textrm{ TeV}}{T_{RH}}\right] ^{2}\left[ \frac{m_{\widetilde{a}}}{\textrm{GeV}}\right] \times \\
 &  & \left[ 1-\frac{m_{\widetilde{a}}^{2}}{m_{\widetilde{g}}^{2}}\right] ^{3}\exp 
\left[ -\frac{m_{\widetilde{g}}}{T_{RH}}\right] . 
\end{eqnarray*}
Finally, if the decays of {}``frozen-out'' neutralinos \( \chi  \)
dominate the axino abundance, the axino abundance is \begin{equation}
\Omega _{\widetilde{a}}h^{2}=\frac{m_{\widetilde{a}}}{m_{\chi }}\Omega _{\chi }h^{2}\end{equation}
where \( \Omega _{\chi }h^{2} \) can be taken from neutralino CDM
calculation of Section~\ref{sec:lspdensity}.

Axinos must have several other properties in order to be cosmologically 
consistent dark matter candidates.
For example, for the axino to be cold dark matter instead of hot or warm
dark matter, its mass must be sufficiently large. Since BBN 
strongly constrains the number of relativistic species in excess of those 
in the SM at temperatures of order \( T=10 \) MeV, the axino mass must 
also be heavy enough to be nonrelativistic by that time. These 
considerations lead to a lower bound on the axino mass of around \( 300 \) 
keV \cite{Covi:2001nw}. Because axinos are weakly coupled, light negative 
R-parity particles such as the lightest neutralinos that decay to them can 
be very long lived. This poses a danger to BBN through the decay products 
destroying delicate light elements, leading to a model-dependent bound 
of order \( m_{\widetilde{a}}\geq 360 \) MeV for light binos (see {\it e.g.} 
\cite{Covi:2001nw,Asaka:2000ew}).

In contrast to the axion and the dark matter axino, the saxion (of 
mass $m_s$) decays relatively quickly 
\begin{equation}
\tau _{a}=3\times 10^{-6}\textrm{sec}\left( 
\frac{f_{PQ}/N}{10^{11}\textrm{GeV}}\frac{0.1}{\alpha _{s}}\right)^{2} 
\left( \frac{m_{s}}{1\textrm{ TeV}}\right) ^{-3} 
\end{equation}
because of its typical \( m_{3/2} \) scale mass. If the saxion energy
dominated during its decay, the decay could introduce significant
entropy into the universe, possibly diluting unwanted gravitational
moduli and/or relaxing the cosmological bound on \( f_{PQ}/N \).

In axion-axino cosmology, both the gravitino bound and the LSP 
overclosure bound can be relaxed to a certain extent. The gravitino 
problem of dissociating the BBN elements through energetic
decay photons can also be evaded in the context of the axino model
\cite{Asaka:2000ew}, since the gravitinos would then decay primarily
through \( \psi _{3/2}\rightarrow \widetilde{a}+a \) without creating
a strong cascade in the SM channel. Finally, the most direct influence
on \lsoft is that the usual \( \Omega _{CDM} \) bounds
constraining the MSSM parameter space can be relaxed by large factors
(100 or more) once the neutralinos can decay into axinos.

In collider phenomenology, the effects of the axino are typically
negligibly small since it is very weakly coupled. One must only keep in
mind that because the neutralinos can be long lived even without being the
LSP, neutralinos at colliders can be mistaken for a stable particle
even if they are are not stable and axinos are the stable LSP
\cite{Rajagopal:1990yx}. Since axinos with R-parity conservation cannot be 
detected by the usual direct/indirect detection experiments due to the \( 
1/(f_{PQ}/N) \) suppressed coupling, a positive detection of 
neutralinos by such experiments can rule out axino CDM as a significant 
dark matter component.  Of course, axino decays 
may be detectable if R-parity is violated.

\section{Baryogenesis}
\label{bgensect} 
Phenomenologically, there are many reasons to believe that we live
in a baryon asymmetric universe. One strong piece of evidence is from the 
acoustic peaks --- early universe baryon-photon plasma oscillations ---  
inferred from CMB measurements (see {\it e.g.} \cite{Bennett:2003bz}), 
which give the baryon-to-photon ratio: \begin{equation} \eta \equiv 
\frac{n_{B}}{s}\equiv 
\frac{n_{b}-n_{\overline{b}}}{s}=6.1\times 10^{-10}\, _{-0.2\times
10^{-10}}^{+0.3\times 10^{-10}}, \end{equation}
in which \( s \) is the entropy density (roughly the photon density),
and \( n_{b} \) and \( n_{\overline{b}} \) are the number densities of
baryons and antibaryons, respectively. This data agrees well with
big bang nucleosynthesis (BBN), which requires the baryon-to entropy 
density ratio to be (see e.g. \cite{Olive:1999ij,Fields:cn}
)\begin{equation}
2.6\times 10^{-10}\leq \eta \leq 6.2\times 10^{-10}.
\end{equation}
The problem of baryogenesis \cite{Sakharov:dj} is to explain the
origin of this small number starting from the natural initial
condition of \( \eta =0 \), which in most cases is attained at high
enough temperatures.\footnote{People also often state that the
sign of $\eta$ must be explained.  From an empirical 
point of view, this sign is of course an arbitrary convention.  On the 
other hand, the problem of baryogenesis may be restrictively redefined 
to include the goal of relating the observed signs and magnitudes of the 
short distance CP-violating phases with the sign of the baryon asymmetry.}

Assuming CPT is preserved, there are three necessary conditions for
baryogenesis, usually referred to as the Sakharov requirements 
\cite{Sakharov:dj}: 
\begin{enumerate}
\item Baryon number violation 
\item Departure from thermal equilibrium 
\item  C and CP violation 
\end{enumerate}
The first requirement is obvious, since the production of a 
nonzero baryon number requires baryon number violation by definition. The 
second requirement follows from considering the thermal equilibrium 
average of the baryon number- violating operator\begin{equation}
\langle B\rangle ={\rm Tr}[e^{-\beta H}B]={\rm
Tr}[(\mbox{CPT})(\mbox{CPT})^{-1}e^{-\beta   
H}B],\end{equation}
using the cyclic property of the trace and that \( B \) is odd 
under CPT.  The third requirement arises because for every \( B \) 
increasing reaction there is an exactly equivalent \( B \) decreasing 
reaction if  C and CP are exact symmetries, as these reactions are related 
by C and CP transformations. 

Several mechanisms have been proposed for baryogenesis (for reviews, see
{\it e.g.} \cite{Riotto:1999yt,Trodden:1998ym,Dine:2003ax}).  Among the
available possibilities, electroweak baryogenesis is by far the most
relevant mechanism with respect to the parameters of \lsoft (as
measureable today). We review electroweak baryogenesis in the MSSM in the
next subsection.  We will also review two other popular baryogenesis
mechanisms, the leptogenesis and Affleck-Dine scenarios, although neither
provide many direct constraints for the \lsoft parameters.

\subsection{Electroweak baryogenesis}
The mechanism of electroweak baryogenesis is simple to understand
heuristically.  At high temperatures, {\it i.e.}, early in the universe, 
the electroweak symmetry is typically restored. As the universe cools to
\( T_{c}\sim 100 \) GeV, there is a first order phase transition
breaking the electroweak symmetry, resulting in the formation of bubbles 
of the broken phase. During this time, particles interact CP
asymmetrically with the bubble walls, causing a buildup of a nonzero
quark-antiquark asymmetry: a left-handed quark-antiquark density and 
an equal and opposite right-handed quark asymmetry. At this point, the
baryon asymmetry vanishes, but there is a nonzero chiral asymmetry.  The 
left-handed quark-antiquark asymmetry \( n_{q_{L}}\), which we will 
loosely refer to as the chiral asymmetry for reasons explained below, 
then flows and diffuses into the unbroken phase --- {\it i.e.}, in 
``front'' of the bubble walls. Nonperturbative baryon number processes 
called {\it sphaleron} processes  then convert the chiral asymmetries 
into baryon number asymmetries in the unbroken phase.  Finally, the generated
baryon asymmetry is transported back to the broken phase (through the
bubble wall sweeping over the baryon asymmetry generated region and
diffusion) where the sphaleron rate is suppressed, thereby protecting
the baryon number. 

Parametrically, the baryon asymmetry can be estimated as follows: 
\begin{equation}
\eta \sim \frac{(k\alpha _{w})\alpha _{w}^{4}\delta _{CP}}{g_{*}}f,
\end{equation}
in which \( k\alpha _{w}\sim 1 \), \( g_{*} \) is the number of
relativistic field
degrees of freedom at the critical temperature, \( \delta _{CP} \)
denotes the relevant rephasing-invariant CP-violating phase of the theory, 
and \( f \) is a factor that characterizes the variation of the Higgs 
expectation value in a moving bubble wall. Let's see how this
parametric estimate arises. The sphaleron transition rate, which is 
proportional to $k\alpha _{W}^{5}$, yields the requisite baryon number 
violation. The  factor \( f \) accounts for the out-of-equilibrium 
condition, since $f$ determines the protection of the baryon number in the 
broken phase (of course \( f \) depends on the bubble wall velocity \( v_{w} \), but 
not monotonically). The CP-violating quantity \( \delta _{CP} \) satisfies
Sakharov's third requirement. Finally, since the entropy \( s \)
counts the relativistic degrees of freedom through \( g_{*} \), the
ratio \( n_{B}/s \) should be proportional to \( 1/g_{*} \).  Since \(
\alpha _{w}^{4}\sim 10^{-6} \) and \( g_{*}\sim 10^{2} \), there is
not much room for \( \delta _{CP}f \) to be small. 
Most of the labor and complexity in the computation of \( \eta \) is
involved in determining \( f \), which is associated with nonequilibrium 
physics. We summarize these issues in the next subsection.

Electroweak baryogenesis in the SM is (most likely) impossible because
of two reasons.  Firstly, 
the CP violating phase 
\begin{eqnarray}
\delta _{CP}&=&\left( \frac{g_{W}}{2m_{W}} \right )^{12} 
(m_{t}^{2}-m_{u}^{2})(m_{t}^{2}-m_{c}^{2})(m_{c}^{2}-m_{u}^{2})(m_{b}^{2}-m_{d}^{2})(m_{b}^{2}-m_{s}^{2})(m_{s}^{2}-m_{d}^{2})j\nonumber 
\\
&\sim& 10^{-22},\label{eq:smdelCP}
\end{eqnarray}
characterized by the Jarlskog invariant \cite{Jarlskog:1985cw}
\begin{equation}
j={\rm Im}[V_{cs}V_{us}^{*}V_{ud}V_{cd}^{*}]\sim 10^{-4},
\end{equation}
is too small.  Secondly, the phase transition is too weak, resulting
in a washout of baryon asymmetry.  The weak phase transition, which is closer 
to second order than first order, essentially means 
that there is a smooth transition from the broken to the unbroken phase 
without  a bubble wall to protect the baryon asymmetry, which should 
result in \( f\ll 10^{-2} \). 

Before passing off on the SM baryogenesis, couple of remarks are in
order regarding the smallness of the CP violation argument. Firstly,
another way to see why the SM \( \delta _{CP} \) is too small is
simply that the rephasing invariance requires many Yukawa couplings to
be multiplied together and the Yukawa couplings are small.  Secondly,
although one must be careful to interpret the dimensionless phase
parameter to be that of Eq.~(\ref{eq:smdelCP}), because the dominant
quantum coherent energy scale is the critical temperature $T_c \sim
m_W$, perturbation in the mass parameter as in Eq.~(\ref{eq:smdelCP})
gives a good estimate wher $m_W$ represents the the critical
temperature scale.  Possible low energy coherent effect which evades
the naive estimate of Eq.~(\ref{eq:smdelCP}) is given in
\cite{Farrar:sp,Farrar:hn} which has been refuted for example by
\cite{Gavela:dt}.

The MSSM has two main advantages over the SM for electroweak baryogenesis:
\begin{enumerate}
\item Supersymmetry has additional sources of CP violation, and hence \( 
\delta _{CP} \) is no longer suppressed as in the SM.
\item The Higgs sector of MSSM allows a first order phase transition, such 
that $f$ is relatively unsuppressed.
\end{enumerate}
To explain these advantages of the MSSM, let us look at the three 
conditions necessary for baryogenesis in more detail. Readers interested 
in the electroweak baryogenesis constraints on the MSSM parameter space 
only can skip the next subsection.

\subsubsection{Basics of electroweak baryogenesis}
\paragraph{Baryon number violating operator}
In both the SM and MSSM, there is a nonperturbative baryon number
violating operator arising from the topological term
\begin{equation}
\label{eq:topologicalsu2}
\int d^{4}x\widetilde{F}_{\mu \nu }F^{\mu \nu },
\end{equation}
in which \( F_{\mu \nu } \) is the field strength for the \( 
SU(2)_{L} \) gauge fields and \( \widetilde{F}_{\mu \nu } \) is its dual.   
Among the SM gauge groups, only \( SU(2)_{L} \) contributes to the 
baryon number violating operator, because it is the only non-Abelian gauge 
group with chiral couplings.   To clarify this point, consider the baryon 
number \( U(1)_{B} \) rotation 
\begin{equation} q(x)\rightarrow e^{i\frac{1}{3}\theta 
(x)}q(x),
\end{equation}
corresponding to the baryon current\begin{equation}
J_{B}^{\mu }=\sum _{q}\frac{1}{3}\overline{q}\gamma ^{\mu }q.
\end{equation}
Due to the transformation of the path integral measure, there is
an induced anomaly term \begin{equation}
\label{eq:anomalycontribution}
\delta S_{1}=i\int d^{4}x\frac{1}{3}\theta (x)\left[ \frac{1}{8\pi 
^{2}} {\rm Tr} F^{(L)}\widetilde{F}^{(L)}-\frac{1}{8\pi 
^{2}}{\rm Tr}F^{(R)}\widetilde{F}^{(R)}\right],  
\end{equation}
in which \( F^{(L,R)} \) denote gauge field strengths coupled to
the left- and the right-handed quarks. Under \( SU(2)_{L} \),
the second term in Eq. (\ref{eq:anomalycontribution}) is absent, and 
hence there is a nonvanishing anomaly term.  Although this term is a
total derivative, the nontrivial topological property (winding) of the
\( SU(2)_{L}\) vacuum renders the term physical.

On the other hand, since \( SU(3)_{c} \) couples to both the left-
and the right-handed fermions equally, 
Eq.~(\ref{eq:anomalycontribution}) vanishes, and thus 
there is no baryon number violating operator coming from \( SU(3)_{c} \). 
However, as we have seen in our discussion of the strong CP problem, 
transitions from one \( SU(3)_{c} \) vacuum to another induce changes in 
the chiral density because \( SU(3)_{c} \) has a chiral anomaly. \( 
U(1)_{Y} \) does not couple to the left and the right equally, but there  
still is no nonperturbative baryon number violating operator contribution
for the same reason that there are no $ U(1)_{Y} $ instantons.

At zero temperature, the topological term Eq.(\ref{eq:topologicalsu2}) can
only induce baryon number violation through \( SU(2)_{L} \) instantons,
which have exponentially suppressed amplitudes. However, above a critical
temperature of around \( 100 \) GeV, the \( SU(2)_{L} \) vacuum transition
rate can occur without any tunneling through thermally excited modes
callled ``sphalerons''
\cite{Manton:1983nd,Klinkhamer:1984di,Kuzmin:1985mm}. Roughly speaking,
these modes are thermally energetic enough to go over the potential
barrier separating the \( SU(2)_{L} \) vacua.

The actual magnitude of the baryon number change per sphaleron
transition is given by the current equation
\begin{equation}
\partial_\mu j^\mu_B = \frac{3}{8 \pi^2} {\rm Tr}[F {\tilde F}].
\end{equation}
This leads to an effective operator \cite{'tHooft:1976up,'tHooft:fv}
\begin{equation}
\sim c O \Pi_i q_{L_i}q_{L_i}q_{L_i}l_{L_i},
\end{equation}
where the product index runs over the number of generations, the
operator $O$ corresponds to non-baryonic/leptonic fermions charged
under $SU(2)_L$, and the coefficient $c$ can be an exponentially
suppressed coefficient. Note that in MSSM, $O$ consists of winos and
higgsinos.  When folded in with the transition rate, the chemical
potential of the left handed particles participating in the sphaleron
transitions gives the baryon number changing rate as
\cite{Joyce:kd,Riotto:1999yt}
\begin{equation}
\dot{B} = - N_F \frac{\Gamma}{2T} \sum_i \mu_i,
\end{equation}
in which $N_F$ is the number of families, $\mu_x$ denotes the chemical
potentials for left-handed $SU(2)$ charged fermions, and $\Gamma$ is
the sphaleron transition rate.

The sphaleron-induced baryon number violating transition rate at finite 
temperature with the electroweak symmetry broken \( (T<T_{c}\approx 100 \) 
GeV) is \cite{Carson:1990jm} 
\begin{equation}
\label{eq:sphaleronrate}
\Gamma \approx 2.8\times 10^{5}T^{4}\left( \frac{\alpha _{W}}{4\pi
}\right) ^{4}\kappa \frac{\zeta ^{7}}{{\cal B}^{7}}e^{-\zeta }
\end{equation}
in which \( \zeta =E_{sph}(T)/T \), \( 10^{-4}\leq \kappa \leq 10^{-1} \),
\( {\cal B} \) is a radiative correction factor, and \( E_{sph}(T) \) is
the energy of the sphaleron solution. When the electroweak symmetry
is unbroken (\( T>T_{c} \)), the sphaleron-induced baryon number
violation rate is\begin{equation}
\label{eq:unbrokenrate}
\Gamma \approx k\alpha _{W}^{5}T^{4},
\end{equation}
where \( k\alpha _{W}\sim O(1) \)
\cite{Yaffe:2001xu,Bodeker:1999gx,Moore:1998zk}. In front of the bubble 
wall (unbroken phase), the sphaleron converts the chiral 
asymmetry (or more precisely \( n_{q_{L}} - n_{\bar{q}_L} \)) into baryon number.  This calculation will be 
described in more detail below. 

Regarding the baryon number violation rate, the MSSM differs from the SM 
primarily in \( E_{sph}(T) \) and \( {\cal B}
\) in \eqr{eq:sphaleronrate}, possibly enhancing the final baryon
asymmetry. Hence, the MSSM primarily affects the sphaleron
transition rate in the broken phase (Eq.~(\ref{eq:sphaleronrate})) and
not in the unbroken phase (Eq.~(\ref{eq:unbrokenrate})).  The
suppression of the broken phase transition rate is mostly an issue of
the out-of-equilibrium condition.

The weak sphaleron only participates in violating baryon number with the
left-handed quarks through reactions such as $t_L t_L b_L \tau_L
\leftrightarrow 0$ and $t_L b_L b_L \nu_L \leftrightarrow 0$.  Hence, if 
a left-handed baryon number can be built {\em without} violating total 
baryon number, {\it i.e.} if the right-handed and left-handed 
baryon numbers cancel, sphalerons can act on the left-handed baryon
number to produce a net right-handed baryon number (see
\eqr{eq:sphaleronprocess}, which includes additional terms associated with
the washout as well as diffusion).  This is the key to the electroweak
baryogenesis mechanism.

This can be seen symbolically as follows.  Let there be a nonvanishing
$n_L - n_{\overline{L}} = x \neq 0$.  Baryon number conservation would
imply $x = n_{\overline{R}} - n_R$, which in turn implies 
\begin{equation}
\label{eq:chiralasymmetry}
n_L - n_R = 2 x + n_{\overline{L}} - n_{\overline{R}}.
\end{equation}  
A chiral asymmetry must be set up starting from a nonzero left-handed 
baryon number despite the total baryon number conservation.  
This left-handed baryon number $x$ is what is  processed by the sphaleron.  
Following \eqr{eq:chiralasymmetry}, we will loosely refer to the process 
of building a nonvanishing $x$ as accumulating chiral asymmetry.\\

\paragraph{The out-of-equilibrium condition}
If the temperature of the plasma exceeds the critical temperature \(
T_{c} \sim  100 \) GeV, there is an electroweak phase
transition, with the Higgs field VEV as the order parameter, due to the
interaction of the SM plasma with the Higgs field. 
As the out-of-equilibrium necessary for sufficient
baryogenesis requires a first order phase transition (explained
below), the strength of the out-of-equilibrium can be characterized by
two physical observables: (i) the velocity of the bubble wall, and (ii) 
the suppression of the baryon number violation in the broken phase
(Eq. (\ref{eq:sphaleronrate})).  The bubble wall velocity \( v_{w} \)
has a large uncertainty. Its value is typically somewhere between
\( 0.01 \) and \( 0.1 \) and has only a mild dependence on the Higgs
mass \cite{Moore:2000wx,John:2000zq}.

The suppression of baryon number violation in the broken phase, on the
other hand, is more sensitive to the MSSM Higgs mass. The factor
controlling the protection of the baryon number, {\it i.e.}, the 
suppression of baryon number violation in the broken phase, is given in
Eq.~(\ref{eq:sphaleronrate}).  To have sufficient protection, the
sphaleron energy needs to be large enough:
\begin{equation}
\label{eq:protbound}
\frac{E_{sph}(T)}{T_{c}}\geq 45.
\end{equation}
The sphaleron energy has been computed at finite 
temperature:
\begin{equation}
\label{eq:sphaleronenergy}
E_{sph}(T)=\frac{H(T)g_{W}}{\alpha _{W}}B(m_{h}/m_{W}),
\end{equation}
in which  \( H(T) \) is the VEV of the lightest Higgs field, \( B(x) \)
has been computed in the SM to be a function of order
1 ( \( B(x)\approx 1.58+0.32x-0.05x^{2} \)), and \( g_{W} \) is
the weak coupling. Eqs.~(\ref{eq:protbound}) and 
(\ref{eq:sphaleronenergy})
therefore translate into the bound\begin{equation}
\label{eq:higgstempbound}
\frac{H(T_{c})}{T_{c}}\geq 1.
\end{equation}
More intuitively, this condition ensures that the first
order phase transition described by a potential of the form\begin{equation}
V(H,T)=D(T^{2}-T_{0}^{2})H^{2}-ETH^{3}+\frac{\lambda 
(T)}{4}H^{4},\end{equation}
with \( E\neq 0 \) controlling the height of the bubble wall potential,
is strong enough to protect the newly-created baryon number, since 
\begin{equation}
\label{eq:eoverlambda}
\frac{H(T_{c})}{T_{c}}\sim \frac{E}{\lambda (T_{C})}.
\end{equation}

To compute \( H(T_{c})/T_{c} \), the finite temperature effective
action must be computed near \( T=T_{c} \). This computation is technically
difficult because infrared resummations as well as two-loop order
calculations must be performed in the parameter ranges of interest.
Because the validity of the perturbation series was not obvious, 
lattice computations have been employed as well as a check. Except
for special points in the parameter space, the lattice seems to be
in agreement with the two loop computation.

For right-handed stop masses below or of order the top quark mass, and
for large values of the CP-odd Higgs mass \( m_{A}\gg M_{Z} \), the
one-loop improved Higgs effective potential can be expanded in \( 1/T
\) (keeping only the top contribution) \cite{Carena:1997ki,Carena:1996wj}:
\begin{equation}
\label{eq:1looppotential}
V_{0}+V_{1}=-\frac{m^{2}(T)}{2}H^{2}-T[E_{SM}H^{3}+2N_{c}\frac{(m_{\widetilde{t}}^{2}+\Pi _{\widetilde{t}_{R}}(T))^{3/2}}{12\pi }]+\frac{\lambda (T)}{8}H^{4}+...
\end{equation}
\begin{equation}
\Pi _{\widetilde{t}_{R}}=\frac{4}{9}g_{s}^{2}T^{2}+\frac{1}{6}h_{t}^{2}\left[ 
1+\sin ^{2}\beta (1-X_{t}^{2}/m_{Q}^{2})\right] T^{2}+\left( 
\frac{1}{3}-\frac{1}{18}|\cos \beta |\right) {g'}^{2}T^{2},\end{equation}
in which \( N_{c}=3 \) is the number of colors, \( X_{t}\equiv A_{t}-\mu 
/\tan \beta  \)
is the effective stop mixing parameter,  \( E_{SM}\approx
\frac{1}{4\pi v^{3}}(2m_{W}^{3}+m_{Z}^{3}) \) 
is the small cubic term coefficient in the SM
case, and $\Pi_{\widetilde{t}_R}$ is the thermal contribution to the stop
mass. Since \begin{equation}
\label{eq:stopmass}
m_{\widetilde{t}}^{2}\approx m_{U}^{2}+0.15M_{Z}^{2}\cos 2\beta + 
m_{t}^{2}(1-\widetilde{A}_{t}^{2}/m_{Q}^{2}),
\end{equation}
a cubic term can arise (thereby enhancing the first order phase 
transition) if there is a cancellation between \( m_{U}^{2} 
\) and \( \Pi _{\widetilde{t}_{R}}(T) \), since both \( m_{t}
\) and \( m_{Z} \) are proportional to \( H \). 
Only the bosonic thermal contributions give rise to this cubic term. 
However, the one-loop induced cubic term
alone is insufficient since this cancellation effect is restricted 
because too negative values of \( m_{U}^{2} \) can induce color
breaking minima. Fortunately, there are regions of parameter space
where two-loop contributions (of the double sunset type) with the
gluon or Higgs line becomes important \cite{Espinosa:1996qw}.  Its
contribution to the effective potential is of the form \(
H^{2}T^{2}\ln H \), which enhances the first order phase
transition:\begin{equation}
\label{eq:2loop}
V_{2}(H,T)\approx \frac{H^{2}T^{2}}{32\pi ^{2}}\left[ 
\frac{51}{16}g^{2}-3h_{t}^{4}x^{2}\sin ^{4}\beta + 
8g_{s}^{2}h_{t}^{2}x\sin ^{2}\beta \right] \ln \left[ \frac{\Lambda 
_{H}}{H}\right],
\end{equation}
in which \( x\equiv 1-\widetilde{A}_{t}^{2}/m_{Q}^{2}\) 
\cite{Espinosa:1996qw,Carena:1997ki}.
The first term of Eq.~(\ref{eq:2loop}) is present in the SM,  while the  
others are due to the superpartners. The validity of the two-loop 
effective potential approach to studying the MSSM electroweak phase 
transition has been supported by a lattice study \cite{Laine:2000rm}. 

In summary, the light right-handed stop loops enhance the strength of
the first order phase transition, and hence give the electroweak 
baryogenesis scenario a sufficient out-of-equilibrium condition in the 
MSSM. The first order phase transition is also enhanced with a smaller 
Higgs mass at zero temperature (\( m_{H}^{2}(T=0) \) ) because of 
Eq.~(\ref{eq:eoverlambda}) and the relation  \begin{equation}
\label{eq:higgsrunmass}
m_{H}^{2}(T=0)\sim \lambda v^{2}
\end{equation}
where \( v=246 \) GeV is the zero temperature Higgs VEV.\\

\paragraph{CP violation}
CP violation enters the electroweak baryogenesis calculation in
building up the chiral asymmetry in the bubble wall region (more
discussion of this point will follow when we discuss the baryon asymmetry
calculation).  Although spontaneous (also often called 
``transient'') CP violation without any explicit CP violation could in
principle occur during the out-of-equilibrium period of the 
electroweak phase transition, the requirement of a strong enough first 
order phase transition essentially prevents the utility of 
this scenario for electroweak baryogenesis (see {\it 
e.g.} \cite{Huber:2001xf}).
Due to the large top Yukawa coupling, which aids in efficiently
transferring the CP-violating charges from the superpartners to the 
quarks, the most important superpartner currents involve stops and
higgsinos.  The right-handed stop and the higgsino CP-violating
currents source through the top Yukawa interaction a chiral asymmetry
for the left-handed quarks ({\it i.e.}, a nonzero left-handed baryon 
number although the total baryon number is zero). This chiral asymmetry in 
turn gets converted into a total nonzero baryon number by the sphalerons, 
which only act on the left-handed particles.

\begin{figure}
\centerline{
   \epsfxsize 3.3 truein \epsfbox {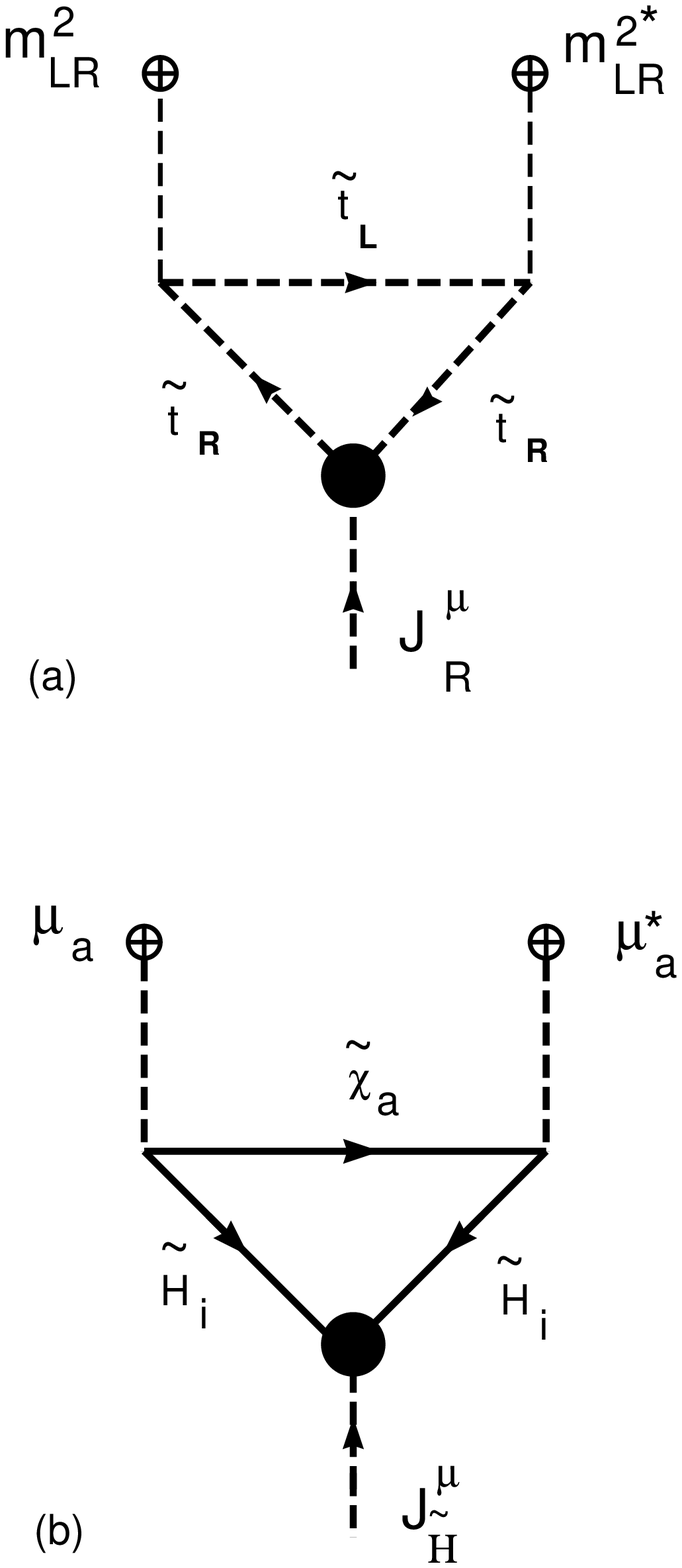}}
\caption{\label{fig:currenttriangle}  The leading
diagrams contributing to the CP-violating currents that eventually
sources the quark chiral asymmetry.  The diagram a) corresponds to the
right-handed squark current $J_R^\mu$ and the diagram b) corresponds
to the higgsino current $J_{{\tilde H}}^\mu$.  The effective mass
terms correspond to $m_{LR}^2 =Y_t (A_t H_u - \mu^* H_d)$ and $\mu_a =
g_a (H_d P_L + \frac{\mu}{|\mu|}H_u P_R)$ where $P_{L,R}$ are chiral 
projectors and
$g_a=g_2$ for $a=1,2,3$ and $g_a=g_1$ for $a=4$. }
\end{figure}

In the parameter regime of interest, the chiral asymmetry sourcing
current of stops tends to be subdominant to the higgsino current
\cite{Carena:1997gx}.  This can be seen from the mass insertion
diagrams of Fig.~\ref{fig:currenttriangle}, which are taken from
\cite{Carena:1997gx}.  Note that the left-handed squark mass $m_Q$
enters the propagator of one of the legs for the right-handed squark
current.  For large $m_Q$, 
the CP-violating piece of the squark current, which is proportional to 
${\rm Im}(A_t\mu)$, is suppressed relative to the higgsino current, which 
is proportional to ${\rm Im}(\mu (M_2g_2^2))$ or ${\rm Im}(\mu( M_1 
g_1^2))$.  The dominant phase then is naturally \( \phi _{\mu }+\phi 
_{M_{2}}\), which is strongly constrained by laboratory EDM bounds, as 
discussed in Section~\ref{EDMsect}.  In the WKB approach, the squark 
current is absent to leading derivative order, while the higgsino 
current is present.\\

\paragraph{Baryon number calculation}
As previously mentioned, the process of baryon number production
involves the accumulation of a chiral asymmetry in front of 
the bubble wall, sphaleron transitions converting the chiral asymmetry 
into baryon number, and then the bubble wall moving past the converted
baryons to protect it. All of these processes can be approximately
computed using the Boltzmann equation. One of the first uses of the
diffusion equation for electroweak baryogenesis can be found in
\cite{Cohen:1994ss}.  Another nice recent summary of the computations
(using the WKB approach) can also be found in
\cite{Cline:2000kb}. Here we will follow the semiclassical
presentation of \cite{Carena:2002ss}, which agrees with
\cite{Cline:2000kb,Cline:2001rk} except
in certain details that we will specify below. The discrepancy is
rooted in arguments about the consistency of various approximations, which 
should be sorted out in the near future.

Starting from the usual classical Boltzmann equation,
\begin{equation}
\frac{p^{\mu }}{E}\partial _{\mu }f_{i}+F_{i}^{\mu }\nabla _{\mu
}f_{i}=C_{i}[f], \end{equation} 
where \( p^{\mu }/E\equiv dx^{\mu }/dt \) is the 4-velocity, \( F^{\mu }\equiv dp^{\mu }/dt \)
is the force generated by the spatially dependent background Higgs VEV,
and \( C_{i} \) are collision integrals, 
the diffusion equation can be derived \cite{Carena:2002ss} 
\begin{equation}
\label{eq:diffusiongeneq}
-v_{w}\partial _{z}n_{i}+D_{i}\partial _{z}^{2}n_{i}+\Gamma _{ij}\frac{n_{j}}{k_{j}}=S_{i}[n^{(B)}],
\end{equation}
after making several assumptions about interactions. Note that
diffusion greatly enhances the efficiency of the chiral asymmetry to
move out of the wall and into the unbroken phase. In the expression above,
\begin{equation} n_{i}\equiv \int \frac{d^{3}p}{(2\pi )^{3}}f_{i},
\end{equation} \(\Gamma _{ij} \) is the averaged interaction rate 
for the inelastic reaction channel \( i\rightarrow j \), \( k_{j}=2 \) for 
bosons while \( k_{j}=1 \) for fermions, \( D_{i} \) are diffusion 
coefficients defined as\begin{equation} D_{i}=\frac{1}{\Gamma 
_{i}^{T}}\frac{\int
\frac{d^{3}p}{(2\pi )^{3}}\frac{p_{z}^{2}}{E^{2}}\frac{\partial
f_{0}}{\partial E}}{\int \frac{d^{3}p}{(2\pi )^{3}}\frac{\partial
f_{0}}{\partial E}},\end{equation}
\( \hat{z} \) is the direction perpendicular to the plane of the
bubble wall, \( f_{0} \) is the equilibrium distribution, and \(
\Gamma _{i}^{T} \) is the total interaction rate. \( \Gamma _{ij} \)
includes the strong sphaleron transitions
\cite{Giudice:1993bb,Mohapatra:1991bz}, which participate in relaxing
the chiral asymmetry, although they preserve baryon number.
Specifically, the strong sphaleron  induces the condition
\begin{equation}
\sum_{i=1}^3 (n_{u_L}^i - n_{u_R}^i + n_{d_L}^i - n_{d_R}^i) =0
\end{equation}
when in equilibrium. The source terms \( S_{i}[n^{(B)}] \) are given 
by
\begin{equation}
\label{eq:sourceterm}
S_{i}[n^{(B)}]\equiv D_{i}\partial _{z}^{2}n_{i}^{(B)}-v_{w}\partial 
_{z}n_{i}^{(B)}, 
\end{equation}
in which \( n_{i}^{(B)} \) is the density in the absence of interactions
other than the background Higgs VEV. The source term, which contains
all the CP violation information, can be roughly interpreted as the
integrated current flowing from the wall due to the \( z \)-varying
Higgs VEV, or simply as the force exerted by the \( z \) varying
background Higgs VEV. As discussed earlier, the strongest source for 
baryogenesis is from the higgsino current and is proportional to 
\(\phi_{\mu}+\phi_{M_2}\). The reason for its importance is because the 
higgsinos have a strong coupling to the top quark, and it is the quark 
chiral charge which is converted into baryons ({\it i.e.}, CP violation 
must be fed into the quarks from the chargino sector). As argued 
previously, the squark source current is suppressed in the parameter range 
of interest.  The background Higgs field variation ({\it i.e.} the 
bubble) is approximated 
as \cite{Moreno:1998bq,Carena:2000id}\begin{equation}
H(z)=\frac{1}{2}v(T)\left( 1-\tanh \left[ \alpha (1-2z/L_{w}) \right] 
\right) \end{equation}
\begin{equation}
\beta (z)=\beta -\frac{1}{2}\Delta \beta \left( 1+\tanh \left[ \alpha 
(1-2z/L_{w}) \right] \right) \end{equation}
where \( \alpha \approx 3/2 \), \( L_{W}\approx 20/T, \) \( \tan \beta  \)
is the usual ratio of Higgs VEVs, and \( \Delta \beta \sim O(10^{-2}) \)
is the \( \beta  \) difference between the broken phase and the unbroken
phase. 

The background density for the species \( i \) in the presence of the
background fields is computed
\cite{Landsman:uw,Henning:sm,Carena:1997gx,Riotto:1997gu,Riotto:1998zb,Riotto:1997vy} by evaluating 
\( \langle J_{(i)}^{\mu }\rangle  \) in perturbation theory, in which
the background Higgs field variation is Taylor
expanded to linear order \cite{Carena:2000id,Carena:1997gx} (the free
part of the Lagrangian corresponding to the kinetic term with a constant
mass, while the interacting piece is the first derivative piece of
the mass with a linear spatial variation). The background density
then is \begin{equation}
\label{eq:background}
n_{i}^{(B)}=\langle J_{(i)}^{0}\rangle .
\end{equation}
In computing \( \langle J_{(i)}^{\mu }\rangle  \), 
\cite{Cline:2000kb,Cline:2001rk} uses the WKB approximation instead of 
doing a linear expansion of the background.

Using the set of diffusion equations Eq.~(\ref{eq:diffusiongeneq}) and 
neglecting the slow sphaleron rate, \cite{Carena:2002ss} solves for the 
chemical potential of the quarks. This is summarized in the quantity \( 
\mu ^{\textrm{diff}}_{L} \), which is the sum of chemical potentials over 
the three generations of the left-handed up and down quarks. The final 
equation describing
the conversion of \( \mu _{L}^{\textrm{diff}} \) into baryon number
can be written as \begin{equation}
D\partial _{z}^{2}n_{B}(z)-v_{w}\partial _{z}n_{B}(z)=\theta (-z)\left( 
\frac{3T^{2}\mu _{L}^{\textrm{diff }}(z)}{4}+\frac{24}{7}n_{B}(z)\right) 
\Gamma _{ws}, 
\label{eq:sphaleronprocess}
\end{equation}
in which \( \Gamma _{ws}=6k\alpha _{w}^{5}T \) is the weak sphaleron
rate in the unbroken phase (derived from Eq.~(\ref{eq:unbrokenrate})) and
\( D\sim 6/T \). This is then integrated to obtain the baryon asymmetry.

As alluded to previously, the specific form of the CP-violating sources
(the details of evaluating Eq.~(\ref{eq:background}) and 
Eq.~(\ref{eq:sourceterm}))
is still controversial \cite{Carena:2002ss}. The question is regarding
the existence of the source term \begin{equation}
\label{eq:antisymmetricsource}
\epsilon _{ij}H_{i}\partial ^{\mu }H_{j},
\end{equation}
in which \( H_{i} \) here denotes the neutral components of the two Higgs 
doublets. If such a source term is absent and the dominant source term is 
instead proportional to \begin{equation}
\label{eq:symmetricsource}
H_{1}\partial ^{\mu }H_{2}+H_{2}\partial ^{\mu }H_{1},
\end{equation}
then sufficient baryogenesis is essentially unattainable \cite{Carena:2002ss}
within most if not all of the allowed parameter region of the MSSM.

\subsubsection{Valid MSSM parameter space}
The analysis of \cite{Cline:2000kb} reported that sufficient
baryogenesis requires $\phi_{\mu }+\phi_{M_2}$ to be larger than \(
0.15 \) even for the extreme (and probably now excluded by LEP) case
of very light charginos (\( \mu \sim m_{2}\sim 50 \) GeV).  As
discussed in Section~\ref{EDMsect}, experimental EDM bounds constrain
this phase, which implies that MSSM electroweak baryogenesis is
tightly bounded and ruled out in a large region of the parameter
space. The EDM constraints on this phase vary in the literature
depending on how the uncertainties inherent in the atomic and hadronic
EDMs are implemented, as discussed in Section~\ref{EDMsect},
resulting in various boundaries of the MSSM parameter space with
sufficient electroweak baryogenesis.  For example, using the MSSM EDM
analysis of \cite{Abel:2001vy} (which yields the strongest bound on
$\phi_{\mu }+\phi_{M_a} \leq 10^{-2}$ at the GUT scale for sparticle
masses consistent with naturalness) leads to the conclusion
\cite{Cline:2000kb,Cline:2001rk} that the $O(10^{-1})$ phase required for
baryogenesis is only possible in models with most superpartner masses
above the TeV range. However, the EDM bounds on this phase presented
in \cite{Barger:2001nu} are about an order of magnitude less
stringent, which may alleviate the restrictions on the MSSM parameter
space somewhat in the case of light superpartner masses.  Note there is no
analysis of the MSSM parameter space yet in the literature in which
the collider, EDM, and electroweak baryogenesis constraints are all
rigorously implemented simultaneously. The conclusion of 
\cite{Cline:2000kb} is based on the nonexistence of the controversial
source term proportional to Eq.~(\ref{eq:antisymmetricsource}) (recall
\cite{Carena:2002ss} and \cite{Cline:2000kb,Cline:2001rk} disagree
about whether this term exists); however, for parameter ranges away
from \( |M_{2}|\approx |\mu | \) this conclusion is more robust
because in this parameter regime, the feasibility of EW baryogenesis
does not significantly depend on the existence of the controversial
source term. 

When the controversial source proportional to
Eq.~(\ref{eq:antisymmetricsource}) is included, the baryon asymmetry
has an order of magnitude resonant enhancement at \( |M_{2}|=|\mu | \)
when \( m_{A}\leq 300 \) GeV. Hence, sufficient baryogenesis seems
possible without resorting to large scalar masses, but
\cite{Pilaftsis:2002fe,Chang:2002ex} have recently argued that the
requisite phase of \(\phi _{\mu }+\phi_{M_2}>0.1 \) may still be too
large to satisfy the EDM bounds. On the other hand, even if the
antisymmetric source proportional to
Eq.~(\ref{eq:antisymmetricsource}) is neglected there is a corner of
parameter space in which sufficient baryogenesis is possible. This
corresponds to the regime in which large first and second generation
masses suppress the one-loop EDMs while a large pseudoscalar mass \(
m_{A} \) suppresses the two-loop contributions which become enhanced
at larger \( \tan \beta \). The results of \cite{Carena:2002ss}
demonstrate that sufficient baryogenesis is possible with \( \phi
_{\mu }\approx O(1) \), \( m_{A}=100 \)0 GeV, \(\tan \beta =10 \), and
a large range of \( \mu \).  One should of course keep in mind, however, 
that given the uncertainties inherent in the electroweak 
baryogenesis calculation, an additional factor of ten uncertainty 
should be assigned to the phase constraints, which would significantly 
increase the allowed parameter space.

Aside from phases, another parametric requirement for electroweak
baryogenesis is that one stop be mainly right-handed 
and its mass be small to make the
phase transition sufficiently first order
\cite{Espinosa:1993yi,Carena:1996wj,Carena:1997gx,Delepine:1996vn}:
\( 120\textrm{ GeV }\leq m_{\widetilde{t}_{R}}\leq m_{t} \). The upper
bound on the stop mass is reasonable in light of Eqs. (\ref{eq:1looppotential})
and (\ref{eq:stopmass}) and recalling that the \( H^{3} \) term enhancement
requires a partial cancellation between \( m_{U}^{2} \) and \( \Pi 
_{\widetilde{t}_{R}} \).
The lower bound on the stop mass is constrained by the requirement of 
no color breaking minima and also possibly \( b\rightarrow s\gamma  \) 
\cite{Carena:2000id}.

A final crucial ingredient for successful baryogenesis is that the
Higgs must be light because of the out of equilibrium condition explained
in Eq.~(\ref{eq:higgsrunmass}). Unfortunately, the LEP bounds push up the
acceptable Higgs mass to be above around \( 113 \) GeV, which pushes
the allowed parameter region to a corner. To achieve such a scenario
with {}``large'' Higgs mass, several conditions are required: \( \tan 
\beta >5 \),
\( m_{Q}\geq 1 \) TeV, and \( A_{t}\geq 0.2\textrm{ }m_{Q} \) GeV
\cite{Carena:2000id}. Also, to preserve sufficiently large 
Eq.~(\ref{eq:higgstempbound}),
\( A_{t}\leq 0.4m_{Q} \). There is an upper bound on \( \tan \beta  \)
as well since both the antisymmetric source 
Eq.~(\ref{eq:antisymmetricsource})
and the symmetric source Eq.~(\ref{eq:symmetricsource}) vanish as \( \beta 
\rightarrow \pi /2 \) \cite{Carena:2002ss,Murayama:2002xk,Cline:2000nw}.

Hence, if electroweak baryogenesis is correct, experimental
{}``predictions'' would include observations of a light stop and a light
Higgs. To give more support to the electroweak baryogenesis
scenario, it is also crucial to find evidence for phases in the
chargino sector. A linear collider would be of great assistance
in this direction \cite{Murayama:2002xk}.

\subsection{Leptogenesis}
The basic idea of leptogenesis is to generate a nonvanishing baryon number 
by first creating a nonzero \( B-L \) density and converting the \( B-L \)
into \( B \) using weak sphalerons (which preserve \( B-L \)
while violating \( B+L \)). Given a \( B-L \), the equilibrium sphalerons
converts it into a baryon asymmetry \cite{Khlebnikov:sr,Plumacher:1997ru}: 
\begin{equation}
\label{eq:leptontobaryon}
B=\left( \frac{8N_{f}+4N_{H}}{22N_{f}+13N_{H}}\right) (B-L),
\end{equation}
in which \( N_{f} \) is the number of fermion families and \( N_{H} \)
is the number of Higgs doublets coupled to \( SU(2)_{L} \).

There are a couple of reasons why it is advantageous to create \( B-L 
\) first, instead of \( B \) directly as in electroweak baryogenesis.  
First, typically there is enough time to convert lepton 
number to baryon number in equilibrium. The baryon number generation does 
not suffer from the sphaleron rate suppression of \( O(1)\alpha 
_{w}^{4}\approx 10^{-6} \) as in Eq.~(\ref{eq:unbrokenrate}).
A second advantage is that there is a natural \( B-L \) violating
operator which arises in a very natural 
solution to the problem of the origin of the light neutrino masses.
This operator is \( M\overline{\nu _{R}^{c}}\nu _{R} \),
which leads to the seesaw mechanism \cite{seesaw1,seesaw2} when combined
with a Dirac mass term \( m\overline{L}_{L}\nu _{R} \). For \( m\sim 1 \)
GeV and \( M\sim 10^{10} \) GeV, the seesaw mechanism gives a light 
neutrino mass of the order
\begin{equation}
m_{\nu }\sim \frac{m^{2}}{M}\sim 10^{-1}\textrm{ eV},\end{equation}
which seems to be the neutrino mass scale that experiments are finding 
(see {\it e.g.} \cite{Altarelli:2003vk} for a review of neutrino 
phenomenology). 
The beauty of this operator is that it
also gives the needed large mass for the right-handed neutrinos to
go out of equilibrium at very high temperatures, long before the onset
of the electroweak phase transition. This will allow the equilibrium 
sphalerons to convert the lepton number to baryon number without any 
suppression (Eq.~(\ref{eq:leptontobaryon})). Using this operator 
for leptogenesis was first suggested by \cite{Fukugita:1986hr}. We will 
focus on such seesaw scenarios for this review since that seems to be the 
best experimentally motivated scenario, and hence has been receiving 
increased attention lately.

The general physics of leptogenesis is very much similar to the GUT
baryogenesis scenario, for which the general physics has been
carefully studied and beautifully presented in \cite{Kolb:qa}. The
Boltzmann dynamics here are very similar to that of neutralino LSP
abundance computation (see Section~\ref{sec:lspdensity}). First, one
assumes that the temperatures are high enough such that the
right-handed neutrinos are in thermal equilibrium.\footnote{For a
recent paper carefully addressing the leptogenesis dependence on the
reheating temperature, see \cite{Giudice:2003jh}.}  Without this high
temperature starting point, there is a loss of predictivity since the
neutrino production history must be taken into account. The lepton
number conserving processes with reaction rate \( \langle \sigma
v\rangle n_{\nu _{R}} \) usually keep the right-handed neutrinos in
equilibrium (the lepton number violating processes are typically
suppressed relative to the conserving processes). When the temperature
falls to the extent that \begin{equation}
\label{eq:outofeq}
\langle \sigma v\rangle n_{\nu _{R}}<\frac{T^{2}}{M_{Pl}},
\end{equation}
the right-handed neutrinos go out of equilibrium. During this time,
lepton number is created through CP and lepton number violating reactions
of the right-handed neutrinos. Then typically, when the heavy right
handed neutrino abundance falls below the \( B-L \) density (due {\it 
e.g.} to its decay), the baryon asymmetry approximately freezes
out. If the right-handed neutrino goes back in equilibrium
before its density falls below the \( B-L \) density, a noticeable
part of the \( B-L \) is erased.

Typically, there will be more than one right-handed neutrino that
will undergo leptogenesis out of equilibrium. In that case, the last
right-handed neutrino to decay (usually the lightest one) will determine
the bulk of the baryon asymmetry, since the \( B-L \) violating reactions
of the lightest right-handed neutrino will erase the previously existing
\( B-L \) density \cite{Plumacher:1996kc}.

There is a large literature on lepton asymmetry computations (see {\it
e.g.} \cite{Buchmuller:2000as,Pilaftsis:1998pd,Hamaguchi:2002vc} and references
therein). The parametric dependence estimate can be written
as\begin{equation}
\label{eq:handwaiving}
\eta \sim \frac{\delta _{CP}m_{\nu }M}{g_{*}v^{2}}\sqrt{\frac{M}{T}}
e^{-M/T_{c}},
\end{equation}
in which \( m_{\nu } \) is the neutrino mass scale, \( M \) is the
right-handed Majorana neutrino mass scale (seesaw scale), \( v\approx 246 
\) GeV is the Higgs VEV, \( T_{c} \) is the temperature at which 
\eqr{eq:outofeq} is first satisfied (decoupling temperature), \( g_{*} \) 
is the number of degrees of freedom at \( T=T_{c} \), and \( 
\sqrt{\frac{M}{T}} e^{-M/T_{c}} \) is the Boltzmann suppression factor 
associated with the number density divided by the entropy.  One can 
substitute \(
\delta _{CP}\sim 10^{-1}
\), \( m_{\nu }\sim 10^{-1}\textrm{eV} \),
\( M\sim 10^{9}\textrm{GeV} \), \( g_{*}\sim 100 \), \( m_{W}\sim 100 \)
GeV, and \(\sqrt{\frac{M}{T}} e^{-M/T_{c}}\sim 10^{-1} \) to obtain \(
\eta \sim 10^{-10} \).  Note that the lepton number violating
reaction, which goes like \( m_{\nu }M/v^{2} \), is not strongly
suppressed (only quadratic in the Yukawa coupling).

The CP-violating phase \( \delta _{CP} \) is unfortunately not strictly
measurable from low energy data. This is obvious because 
the matrix \( M \) at the seesaw scale breaks part of the rephasing
invariance that existed in the absence of this matrix. 
Defining the orthogonal complex matrix \( R \) by 
\begin{equation}
m_{\nu }=\left( U_{MNS}\sqrt{(m_{\nu })_{diag}}R\right) \left( 
R^{T}\sqrt{(m_{\nu })_{diag}}U^{\dagger }_{MNS}\right), 
\end{equation}
the phases of $R$ are what enters \( \delta _{CP} \). Therefore, low 
energy data of the neutrinos alone (which specify \( U_{MNS} \), the 
matrix which diagonalizes the light neutrino mass matrix) cannot specify 
\( \delta _{CP} \) and hence \(\eta  \)
(see for example a good discussion in \cite{Branco:2002xf}).

By assuming a minimal seesaw model, hierarchical neutrino mass pattern,
and dominance of the lightest neutrino for generating the correct
baryon asymmetry, upper bounds can be set on all light neutrino masses of 
about \( 0.23 \) eV \cite{Buchmuller:2003gz}. 
There have also been attempts to connect leptogenesis with lepton-flavor 
violation experiments \cite{Davidson:2003cq} and CP violation experiments 
\cite{Branco:2002xf}. However, as one can guess from \eqr{eq:handwaiving}, 
there does not seem to be a large difference whether or not a 
supersymmetric embedding of leptogenesis is implemented. 

One of the strongest cosmological constraint on the leptogenesis
scenario comes from the reheating temperature.  As mentioned
previously, the standard scenario assumes thermal equilibrium initial
conditions for the right handed neutrinos.  However, because the right
handed neutrinos must be heavy for successful see-saw and for
sufficient $B-L$ asymmetry generation (for a recent paper on lower
bound on the right handed neutrino mass, see
{\it e.g.} \cite{Davidson:2002qv}), $T_{RH}$ typically must be large as
$10^{10}$ GeV.  As we discuss in subsection
\ref{sec:gravitino}, such large reheating temperatures may be
difficult to reconcile with a successful cosmological scenario.

\subsection{Affleck-Dine}
Affleck-Dine baryogenesis refers to the scenario in which a scalar
condensate charged under baryon number, initially
displaced away from its potential minimum, attains field motion equivalent
to a nonzero baryonic current, and then decays to produce ordinary
baryons \cite{Affleck:1984fy}. Thus, the heart of the physics of 
Affleck-Dine baryogenesis resides in the initial conditions and the 
variety of ways the scalar condensate can decay. We will refer to the 
baryon number carrying condensate as the Affleck-Dine condensate (ADC) 
and use the variable \( \widetilde{C} \) to denote it. (The baryonic charge 
density carried by \( \widetilde{C}\equiv \rho  e^{i\theta } \)
is approximately \( \rho ^{2}\dot{\theta } \) where \( \{\rho ,\theta \} \)
are real.) It should also be kept in mind that the baryon number can
be replaced by lepton number and leptogenesis then carried out using a 
similar setup.

In terms of Sakharov's conditions, the out-of-equilibrium condition
is that the ADC is initially displaced away from the true
minimum. The CP violation comes from the combinations of 
parameters of the potential (such as \( A \)-term phases) and any 
spontaneous CP violation induced by VEVs. CP violation biases the \( 
\widetilde{C} \) motion to have nonvanishing baryonic current. The baryon 
number violation is contained in the baryon number carrying condensate and 
its interactions. 

The physical mechanism that displaces the ADC is generically attributed
to the physics that gives rise to a large Hubble expansion rate in
the early universe. Any scalar field with a mass much smaller than
\( H \) will have quantum fluctutations of order \( H \).  Due to the
expansion of the universe, this quantum fluctuation converts to
classical displacement (fluctuation) of order \( H \). Somewhat more
concrete scenarios \cite{Dine:1995uk,Allahverdi:2001is} have the
supersymmetry breaking during inflation generate a negative curvature
of the potential at what will eventually be the stable minimum
(with positive curvature) after the end of inflation. This
will then determine the inital displacement of the ADC (if one assumes
that the ADC is at the minimum of its potential in the early
universe). The field \( \widetilde{C} \) will adiabatically track \( H
\) (say during and after inflation) due to the friction term provided
by \( H \) until \( H \) falls below its mass of order
\( m_{3/2} \), at which time \( \widetilde{C} \) will attain motion
and induce the baryonic current.

In supersymmetric models, there are many baryon number or lepton number
carrying renormalizable flat directions --- {\it i.e.}, field directions
in which the potential vanishes --- which are lifted by nonrenormalizable
operators and supersymmetry-breaking terms. (We will generically refer to
these as just flat directions, although this can refer to field directions
whose flatness is broken only by supersymmetry-breaking operators.) The
flat directions in the MSSM have been classified in
\cite{Gherghetta:1995dv}.  Since the final baryon asymmetry is
proportional to the initial ADC field displacement, flat directions are
useful for obtaining a large baryon asymmetry.  The initial displacement
will then be determined by the cosmological dynamics and the 
nonrenormalizable operators, both in the superpotential and the 
supersymmetry-breaking sector.

The decay/evolution channels of the ADC can be quite complicated. Because
\( \widetilde{C} \) is typically large, the particles that are coupled
to the ADC will obtain large masses and thereby prolong its lifetime. In 
the case that the decay is suppressed, the primary conversion
of \( \widetilde{C} \) into ordinary baryons (or leptons for a leptogenesis
scenario) will then transpire through scattering of the condensate
with thermal particles. The scattering effects which induce plasma
mass can also suppress the baryon number by causing \( \widetilde{C} \)
to oscillate early \cite{Allahverdi:2000zd}. Unlike in other baryogenesis
scenarios, the final baryon asymmetry can be typically very large
\cite{Dine:2003ax}:\begin{equation}
\eta \sim 10^{-10}\left( \frac{T_{R}}{10^{9}\textrm{GeV}}\right) 
\left( \frac{M_{p}}{m_{3/2}}\right) ^{\frac{n-1}{n+1}}\sin \delta_{CP}.
\end{equation} 
In the above expression, it has been assumed that the initial conditions 
were fixed 
by the minimum of\begin{equation}
V\sim 
-H^{2}|\widetilde{C}|^{2}+\frac{1}{M^{2n}}|\widetilde{C}|^{2n+4},\end{equation}
with \( H\sim m_{3/2} \), and the CP-violating phase is \( \delta _{CP} \)
is assumed to be from a supersymmetry breaking sector coupling to \( 
\widetilde{C} \).
An unacceptably large baryon asymmetry may be brought to tolerable
levels by additional cosmological events such as gravitational
moduli decay, which can dump extra entropy and hence dilute the baryons. 

In addition to the usual particle decay/evaporation channel, because \(
\widetilde{C} \) can develop inhomogeneities which can become unstable, it can
fragment into smaller condensates if the baryon number carried by the
condensate is too big \cite{Kusenko:1997si,Enqvist:2000cq,Kasuya:2000wx}.
The fragmentation can lead to formation of Q-balls, which are
nontopological solitons whose stability against decay into scalar
particles is guaranteed by there being a global minimum of \(
V(\widetilde{C})/|\widetilde{C}|^{2} \) \cite{Coleman:1985ki}. If the mass per 
baryon number is less than the proton mass in the
Q-ball, it is stable, even against decay to fermions \cite{Cohen:1986ct}.
For gauge mediated supersymmetry breaking models, this leads to a bound on the
large number of charges necessary for the stability of Q-balls:\begin{equation} Q_{B}\gg
\left( \frac{M_{S}}{1\textrm{ GeV}}\right) ^{4}\geq 10^{16},\end{equation}
 which is quite large \cite{Kusenko:1997si}. Such stable Q-balls can
compose dark matter. The gravity mediated supersymmetry breaking models do
not possess such absolutely stable Q-balls \cite{Enqvist:1997si}.
Unstable Q-balls can decay to LSPs and still provide a source of dark
matter. Meanwhile, some of the baryon number can evaporate to
contribute to the baryon asymmetry. This possible connection between
the dark matter abundance and the baryonic abundance has intrigued
many researchers \cite{Banerjee:2000mb,Enqvist:1997si,Laine:1998rg,Fujii:2002aj}.

As the Affleck-Dine baryogenesis scenario depends in a crucial way on the 
introduction of the inflaton and its consequent inflationary and reheating 
history, it does not by itself provide direct constraints on \lsoft. 
Nonetheless, Q-balls carrying baryon number do make good dark matter 
candidates.
Although the flux is low, their detection \cite{Arafune:2000yv} at large
detectors like ANTARES and Ice Cube would give spectacular support for
the Affleck-Dine baryogenesis scenario since the creation of stable 
Q-balls is otherwise quite difficult 
\cite{Griest:1989bq,Postma:2001ea}. We
refer the interested reader to the comprehensive reviews
\cite{Dine:2003ax,Enqvist:2003gh} for more details.

\setcounter{footnote}{0}
\section{Inflation}
\label{inflationsect}
The benefits of inflationary cosmology in alleviating the cosmological 
initial data problems are by now standard textbook knowledge 
(see {\it e.g.} \cite{Kolb:vq}). Standard inflationary cosmology is 
defined by the condition that there was some period of time in the early 
universe when energy density with a negative equation of state, typically
associated with a scalar field called the {\it inflaton},
dominated the universe, inducing an approximately de Sitter-like
metric long enough to solve the cosmological problems. As the cosmological
initial condition problems are associated with the SM-motivated 
restrictions to particular types of stress tensor, by extending the SM 
one can arrange for the stress tensor to have the negative 
pressure dominated phase behavior required for inflation to take place.  

A remarkable prediction of inflationary cosmology (rather than a 
postdiction of solving the initial data problems) is the generation of 
scale-invariant energy density perturbations on superhorizon scales which 
may eventually become seeds for structure formation (for reviews, see {\it 
e.g.} \cite{Mukhanov:1990me,Lidsey:1995np,Liddle:1993fq,Gong:2002cx}).
These density perturbations are also manifest as temperature
fluctuations on the cosmic microwave background (CMB) radiation.
Various experiments, such as COBE DMR, DASI, MAXIMA, BOOMERANG, CBI,
and WMAP, have measured these CMB temperature fluctuations. The  
qualitative features are in agreement with what one expects from 
most inflationary scenarios. 

Hence, there is a strong motivation to take inflation seriously. In 
the context of supersymmetric extensions of the SM such as the MSSM, one 
might imagine that inflation may yield insights into the soft 
supersymmetry-breaking Lagrangian.  However, the connections are somewhat 
tenuous, as we will explain. One basic 
difficulty in connecting inflation directly with \lsoft is related to the 
observationally and theoretically constrained scales for inflation. For 
most models, a SM singlet sector needs to be introduced; in many cases, this 
sector is tied with the intermediate scale of supersymmetry breaking. 
Indeed, inflationary models require physics beyond the 
MSSM by definition.  Currently, there are no compelling models of 
inflation connected to high energy physics, although some models are more 
plausible than others. We thus see a great opportunity for significant 
progress in the future, since it is quite unlikely that particle physics 
does not have anything to do with the observationally favored paradigm of 
inflationary cosmology.

\subsection{\label{subsec:req} Requirements of inflation}
\label{inflationreq}
To discuss the requirements of inflation, for simplicity we start
with the simplest semirealistic parameterization that captures the
essential physics during the inflationary epoch. Consider a homogeneous
and isotropic metric \begin{equation}
ds^{2}=g_{\mu \nu }dx^{\mu }dx^{\nu }=dt^{2}-a^{2}(t)d\vec{x}^{2},
\end{equation}
in which \( a(t) \) is the scale factor of the universe.  The Hubble
expansion is \( H=\dot{a}/a \). The equation of motion for
\( a \) is governed by one of the Einstein equations\begin{equation}
\left (\frac{\dot{a}}{a} \right )^{2}=\frac{8\pi }{3M_{Pl}^{2}}\rho, 
\end{equation}
in which \( \rho  \) is the energy density dominated by the inflaton
field(s) \( \phi_i  \). The final equation in the set is the equation
of motion for the fields composing \( \rho  \). Both the inflaton field(s)
\( \phi_i  \) and corresponding energy density \( \rho  \) are assumed to 
be homogeneous to leading approximation: {\it i.e.} \( \phi_i 
(t,\vec{x})\approx \phi_i (t) \) and \( \rho (t,\vec{x})\approx \rho (t) 
\).

Inflation requires the following {\it qualitative} elements:
\begin{enumerate}
\item Negative pressure must dominate, such that \( \ddot{a}>0 \) for 
about \( N>60 \) e-folds. By \( N \) e-folds, we mean that $a(t)$ must 
be smaller at the time of the beginning of inflation $t_i$ 
than it is at the time of the end of inflation $t_f$ by an exponentially 
large factor: \( a(t_i)/a(t_f)=e^{-N} \). 
\item Inflation must end. 
\item Writing the inflaton fields as \( \phi_i(t,\vec{x})=\phi_i(t) + 
\delta \phi_i(t,\vec{x})\), the inhomogeneous fluctuations \( \delta 
\phi_i (t,\vec{x}) \) which perturb the background inflaton field(s) \( 
\phi_i (t) \) must generate sufficiently small perturbations \( \delta 
\rho (t,\vec{x}) \) of the energy density \( \rho  \) on largest observable 
scales with a scale-invariant spectrum.\footnote{Standard structure 
formation scenarios prefer that \( \delta \rho (t,\vec{x}) \) has a 
certain value. However, alternative structure formations have been 
proposed which do not lead to such restrictions on \( \delta \rho 
(t,\vec{x}) \).}

\item After the end of inflation, the universe must release entropy and
heat to a temperature of at least \( 10 \) MeV for successful
nucleosynthesis of the heavy elements
\cite{Sarkar:1995dd,Olive:1999ij}. The photon energy density must also
dominate by the temperature of \( 10 \) MeV, and a successful
baryogenesis mechanism must be possible. When the energy density
becomes radiation-dominated, the temperature at that time is referred
to as the ``reheating'' temperature
\( T_{RH} \).
\item After the end of inflation, thermodynamics and particle interactions
must not generate unobserved heavy particles, solitons, or other ``relics.'' 
\end{enumerate}
In the crudest attempts at model building, requirements 4 and 5 are 
neglected because they depend on necessarily small couplings of \( \phi_i  
\), and require a more detailed field content. Requirements 1,
2, and 3 generically require the presence of small parameters and tuned 
initial conditions, which are the main challenge for model building.

As an example, consider the action for a single scalar field \( \phi  \)
(the inflaton):
\begin{equation}
S_{\phi }=\int d^{4}x\sqrt{g}\left [\frac{1}{2}\dot{\phi }^{2}-V(\phi ) 
\right ],
\end{equation}
in which \( \phi  \) to leading approximation only depends on time,
consistent with the symmetry of the metric. In this toy model, \( \rho 
=\frac{1}{2}\dot{\phi }^{2}+V(\phi ) \).
The qualitative requirements 1, 2, and 3 can be translated into 
approximate quantitative requirements in terms of the 
``slow-roll'' parameters as follows:
\begin{enumerate}
\item Negative pressure amounts to \begin{equation}
\label{eq:defeps}
\epsilon \equiv \frac{\overline{M}_{p}^{2}}{2}\left (\frac{V'}{V} 
\right )^{2}\ll 1
\end{equation}
 and \begin{equation}
\label{eq:defeta}
\eta \equiv \overline{M}^{2}_{p}\left (\frac{V''}{V} \right )\ll 1,
\end{equation}
where \( \overline{M}_{p}\equiv M_{Pl}/\sqrt{8\pi } \). The 60 e-foldings
amount to \begin{equation}
\label{eq:req60}
N(\phi _{i})\equiv \left |\int 
_{\phi(t_i)}^{\phi(t_f)}\frac{1}{\sqrt{2\epsilon 
}}\frac{d\phi }{\overline{M}_{p}} \right |>60, 
\end{equation}
where \( \phi(t_{i,f}) \) is the value of the inflaton field at the 
beginning and end of inflation, respectively.
and \( \phi(t_f) \) is at the end of inflation. 
\item The end of inflation is reached when \( \phi =\phi(t_f) \) 
satisfying 
\begin{equation}
\epsilon (\phi(t_f))\approx 1.\end{equation}
In some cases, the end of inflation can be signaled by \( \eta 
(\phi(t_f))\approx 1 \) as well. In addition, 
\( V(\phi _{min})\approx 0 \) at the minimum of the potential. 
\item The density perturbation amplitude is given by \begin{equation}
\label{eq:densityperturb}
\sqrt{P_{k}^{\zeta }}\approx \sqrt{\frac{16}{3\pi ^{2}\epsilon (\phi 
_{60})}\frac{V}{\overline{M}_{p}^{4}}}\sim 10^{-5},  
\end{equation}
in which\( \phi _{60} \) is the value of the field 60 e-folds before
the end of inflation (somewhere between \( \phi(t_i) \) and \( \phi(t_f)\))
and is defined by \( N(\phi _{60})\approx 60 \), with \( N(\phi ) \)
defined in 1 above. The scale invariance is characterized by \begin{equation}
|2\eta (\phi _{60})-6\epsilon (\phi _{60})|<0.3. \end{equation}

\end{enumerate}
Note that requirement 1 forces the potential to be flat amd inflaton to 
have a small mass: \( (V''\sim m_{\phi }^{2})\ll 
(V/\overline{M}_{p}^{2}\sim 
(\dot{a}/a)^2=H^2) \). Satisfying this small mass constraint will be 
aided significantly by supersymmetry, although supergravity corrections 
also generically cause difficulties.
The number 60 in requirement 1 depends on postinflationary cosmology, 
but is typically between 30 and 60. Since \( N(\phi _{i})>60 \) is
a history-dependent requirement ({\it i.e.} an integration over \( \phi 
\)), it requires a fine tuning of the initial conditions
for \( \phi  \). Conditions 2 and 3 sets a limit on the absolute
magnitude of the potential, and thus are primarily responsible for
requiring a small dimensionless parameter. Furthermore, the latter
part of the requirement 2 contains the cosmological constant problem,
which remains one of the greatest unsolved problems of high energy 
physics. However, the 
challenge of building a compelling model of inflation is surprisingly 
difficult even if one is freely allowed to throw out the cosmological 
constant.

The slow-roll formulae (see {\it e.g.} \cite{Liddle:1992wi,Lidsey:1995np})
presented above represent a leading approximation and can break down
in many instances such as nonanalytic points in the potential or points
where the slow-roll parameters vanish 
\cite{Inoue:2001zt,Wang:1997cw,Dodelson:2001sh}.
The state of the art in slow-roll formulae can be found in 
\cite{Gong:2002cx}.

Although there are some new features in the more realistic multifield
inflationary scenario, most of the local physics remains the same
as in the single field model except for density perturbations which
can have contributions from fluctuations in all the light directions.
A more general formula for density perturbations in the case of multifield
inflation can be found in \cite{Sasaki:1995aw}. One elementary but
important consequence of a multifield inflationary scenario is its
ability to lower the required field values to be much smaller than
\( M_{Pl} \). The reason why this is important is because in an effective
field theory with \( M_{Pl} \) as the cutoff scale, the nonrenormalizable
operators whose coefficients we cannot generally obtain from low energy
data become important if \( \phi \geq M_{Pl} \). For related discussions,
see for example \cite{Lyth:1997ai}.

Another unsettled and dubious issue within the inflationary paradigm is
the necessary conditions for starting inflation. Although some potentials
are more likely to have the inflaton field sitting far away from the
minimum, if there is a nonzero probability of inflation taking place (even
if it is small), inflation can take place within a finite time. For any
set of fixed assumptions about the probability space of the potentials,
there may be a well-defined probability for inflation taking place, but
such assumptions are difficult to justify rigorously.

\subsection{Scales}
Although the scales are model dependent, one can make some general 
statements. By considering a single inflaton potential \( 
V(\phi )\sim \lambda \overline{M}_{p}^{4}(\frac{\phi }{\overline{M}_{p}})^{n} \)
in Eqs.~(\ref{eq:defeps}),(\ref{eq:densityperturb}), and (\ref{eq:req60}),
it can be shown that the energy scales are \begin{equation}
\phi \gg \overline{M}_{p}
\end{equation}
 and \begin{equation}
\lambda \ll 10^{-10}
\end{equation}
where \( \overline{M}_{p}\equiv M_{Pl}/\sqrt{8\pi }\sim 10^{18} \) GeV.
Hence, the potential energy scale is close to the GUT scale and the
dynamical scale \( H \) is around \( 10^{13} \) GeV. 

Another prototypical model is called the hybrid inflationary model
\cite{Linde:1993cn}, in which one field \( \sigma  \) being away
from the minimum gives the vacuum energy density while the fluctuations
of \( \phi  \) slowly rolling gives the density perturbations. For
example, consider\begin{equation}
V(\phi ,\sigma )=\frac{1}{4\lambda }(m_{\sigma }^{2}-\lambda \sigma ^{2})^{2}+\frac{1}{2}m_{\phi }^{2}\phi ^{2}+\frac{1}{2}g^{2}\phi ^{2}\sigma ^{2}\end{equation}
 where because initially \( \phi >\phi _{c}=m_{\sigma }/g \), the
field \( \sigma  \) sits at 0, and the potential looks like \( V(\phi 
,\sigma 
)\sim \frac{m_{\sigma }^{4}}{4\lambda }+\frac{1}{2}m_{\phi }^{2}\phi ^{2} \)
initially. This means that when \( m_{\phi }^{2}\ll g^{2}m_{\sigma }^{2}/\lambda  \)
(and moderate values of \( \phi >\phi _{c} \)), the vacuum energy
will be dominated by a constant term \( m_{\sigma }^{4}/(4\lambda ) \).
Inflation ends when \( \phi <\phi _{c} \), since at that time \( \sigma  
\)
acquires a negative mass squared and rolls down to its minimum at
\( m_{\sigma }/\sqrt{\lambda } \). Here, \eqr{eq:densityperturb}
gives\begin{equation}
\label{eq:densitypertrel}
\frac{g}{\lambda ^{3/2}}\frac{m_{\sigma }^{2}}{m_{\phi }^{2}}m_{\sigma 
}^{3}\sim 10^{-3}\overline{M}_{p}^{3}
\end{equation}
which implies that \( m_{\sigma } \) can be at a much lower scale
than \( \overline{M}_{p} \) if \( \overline{M}_{p}\gg m_{\sigma }\gg 
m_{\phi }^{2} \). For example, if we choose the electroweak scale for \( 
m_{\phi }=100 \) GeV, then \eqr{eq:densitypertrel} implies \( m_{\sigma 
}\sim 10^{11} \) GeV, which is the intermediate scale that may be 
associated with gravity-mediated supersymmetry breaking. Hence, in this 
case there need not be small couplings or transPlanckian field values. 
The potential energy can
be naturally as low as the intermediate scale, with \( V\sim 
(10^{11}\textrm{ 
GeV})^{4} \), and the dynamical scale naturally as low as \( H\sim 1 \) 
TeV. Thus, from a simple consideration of scales, hybrid inflation is 
a much more ``natural'' model than a single field model.

As far as the reheating temperature is concerned, if one assumes a
perturbative decay of the inflaton with decay width \( \Gamma _{\phi } 
\) over several oscillations after the inflaton reaches its minimum,
the temperature is given by\begin{equation}
T_{RH}\approx 0.2\left (\frac{200}{g_{*}}\right )^{1/4}\sqrt{M_{Pl} 
\Gamma_{\phi }},\end{equation}
where \( g_{*} \) is the number of relativistic degrees of freedom
(see {\it e.g.} \cite{Kolb:vq}). If the inflaton interacts fairly strongly
with the decay particles, the oscillating time-dependent mass of the
particles to which the inflaton couples can induce a parametric resonance-like
phenomenon which can significantly increase the efficiency of reheating
and raise the temperature of the ensuing radiation domination period
(for the seminal, original papers, see \cite{Kofman:1994rk,Shtanov:1994ce,Khlebnikov:1996wr,Kofman:1997yn,Felder:2000hj}).
Although a rich and fascinating subject in its own right, reheating 
dynamics will not be addressed in this review due to its marginal 
connection to \lsoft in the literature.

\subsection{Implications for supersymmetry}
From even the field content point of view, supersymmetry is attractive for
inflation, as it contains as many scalar degrees of freedom as fermionic 
degrees of freedom. Hence, in supersymmetric models there may be 
plenty of inflaton candidates without condensation of higher spin fields, 
unlike the meager choice of the Higgs boson in the SM.  Furthermore, there 
are a great number of field directions called flat directions in 
which the potential receives nonvanishing
VEV contributions only from nonrenormalizable operators and supersymmetry 
breaking: see {\it e.g.} \cite{Gherghetta:1995dv} for a catalog of flat 
directions in the MSSM. Since inflation
potential needs to be flat, these flat directions are very attractive
for building inflationary models.

As we have seen, one of the primary requirements of inflation is
keeping a flat potential (small slope and mass, see ``slow-roll''
requirement 1 in subsection \ref{subsec:req}) over a range of field
values during inflation. Even allowing for fine tuning at tree level,
the flatness of the potential is generically spoiled by radiative
corrections. Without supersymmetry, for each degree of freedom that
can generate loops coupled to the inflaton field \( \phi \), there is
a contribution to the effective potential of the form\begin{equation}
\pm \frac{1}{64\pi ^{2}}M^{4}(\phi )\ln \left [\frac{bM^{2}(\phi 
)}{Q^{2}}\right ],
\end{equation}
in which \( Q \) is the renormalization scale and \( M^{2}(\phi ) \)
is the coupling-generated effective mass. For example, in \( \lambda \phi 
^{4} \) theory, \( M^{2}(\phi )=12\lambda \phi ^{2} \), which generates
a \( \phi ^{4}\ln (\phi /Q) \) type correction. On the other hand,
with supersymmetry, there is a generic contribution \begin{equation}
\frac{1}{64\pi ^{2}}STr \left [{\mathcal{M}}^{4}\left (\ln 
\left [\frac{{\mathcal{M}}^{2}}{Q^{2}} \right]-\frac{3}{2}\right ) \right 
],
\end{equation}
where the fermionic contribution can cancel the bosonic contributions.
With only soft supersymemtry breaking, one typically has \( \phi ^{2}\ln 
(\phi /Q) \) and with spontaneous breaking in which \( 
STr{\mathcal{M}}^{2}=0 \),
the corrections go as \( \ln (\phi /Q) \), which is functionally
a much milder correction \cite{lythriotto}. This cancellation (the
heart of the nonrenormalization theorem) is one of the key advantages
of supersymmetric inflationary models.

A related advantage of supersymmetric models is the possibility
of motivating large field initial conditions, which generically help
in attaining a sufficient number of e-folds (requirement 1 in
subsection \ref{subsec:req}). Supersymmetric models generally have 
a plethora of scalars and the nonrenormalization theorems which protect 
the superpotential to all orders in perturbation theory in the limit of 
unbroken supersymmetry combine to give many directions in scalar field 
space which are flat (up to supersymmetry-breaking and nonrenormalizable 
terms), allowing the scalar fields to move far away from the minimum of 
the potential without costing much energy.

An important feature of supersymmetric inflation is the SUGRA structure.
The SUGRA structure becomes particularly important for cases
in which the inflaton field \( \phi  \) has a value close to or larger
than \( M_{Pl} \). As previously discussed, the most general 4D \( N=1 \) 
SUGRA scalar sector Lagrangian is specified by the 
K\"{a}hler potential, the 
superpotential, and the gauge kinetic function.  In principle, there 
also may be a nonvanishing FI term. Of course, looking at the bosonic 
sector alone, the structure is only slightly more rigid than the most 
general nonrenormalizable local effective field theory.  The main 
difference is that certain scalar couplings in the potential
are tied together because of the \( F \) term and \( D \) term contributions.
The SUGRA structure, however, is neither generically bad or generically
good for inflation. The verdict lies in the structure of the nonrenormalizable
terms generated by the K\"{a}hler potential and the gauge kinetic
function. In the fermionic sector, there is an important generic cosmological
implication from the SUGRA structure. Namely, the existence of the 
gravitino in the spectrum often plays an important
role in satisfying requirements 4 and 5 of subsection \ref{subsec:req}.
We discuss the gravitino problem in 
Section~\ref{sec:gravitino}.

In the context of SUGRA, people also often refer to the inflationary
\( \eta  \) problem \cite{Copeland:1994vg,Stewart:1994ts,Stewart:1994pt} 
(for related literature, see 
\cite{Coughlan:yk,Dine:1995uk,Dine:1983ys,Barbieri:1982eh}),
where \( \eta  \) is defined in \eqr{eq:defeta}. This arises because
if the inflaton potential energy density is dominated by the \( F \)
term, then the minimal K\"{a}hler potential \( K \) generically leads
to \( \eta \sim O(1) \) because of the \( \exp [K/\overline{M}_{p}^{2}] \)
in the potential\begin{equation}
\label{eq:gensugrapot}
V\sim e^{K/\overline{M}_{p}^{2}} \left[ 
(K^{-1})^{j}_{i}F_{i}F^{j}- \frac{3|W|^{2}}{\overline{M}_{p}^{2}}\right] 
+\frac{g^{2}}{2}\textrm{Re}f_{AB}^{-1}D^{A}D^{B},
\end{equation}
where \( f_{AB} \) is the gauge kinetic function and \( D^{i} \)
is the D term. However, this should be seen as a challenge
rather than a no-go since the K\"{a}hler potential (in conjunction
with the superpotential) may satisfy conditions such that \( \eta \ll 1 \)
can be achieved \cite{Stewart:1994ts}. Futhermore, the K\"{a}hler
potential can flatten the potential (see {\it e.g.} \cite{Chung:2002xj})
just as easily as ruining the flatness. Unfortunately, the K\"{a}hler
potential generically is not fully computable without a UV complete
theory.  Even in string models, it is difficult to compute in practice.

To evade the \( \eta  \) problem, 
it was pointed out in \cite{Stewart:1994ts,Binetruy:1996xj,Halyo:1996pp} 
(see also \cite{Casas:1988du,Casas:1988pa}) that if the vacuum
energy is dominated by a \( U(1) \) Fayet-Iliopoulos D term \( \xi 
^{A} \)
\begin{equation}
D^{A}=K^{i}(T^{A})^{j}_{\, \, i}\phi _{j}+\xi ^{A},
\end{equation}
inflation can occur even with the offending \( \exp 
(K/\overline{M}_{p}^{2}) \) term equal to zero. This scenario,
called the D term inflationary scenario, has an inflaton (and hence
an end to inflation) due to the one-loop generated dependence of the
potential on a \( U(1) \) neutral field \cite{Binetruy:1996xj,Halyo:1996pp}.
In models with an anomalous \( U(1) \) symmetry, the vacuum energy 
determining the \( \xi  \) magnitude is fixed by the Green-Schwarz 
mechanism, but generically the magnitude of this term \begin{equation}
\xi _{GS}=\frac{g^{2}M_{Pl}^{2}TrQ}{192\pi ^{2}}\end{equation}
is too large. There has been much model-building activity in this
direction \cite{Jeannerot:1997is,Dvali:1997mh,Casas:1997uk,Lyth:1997pf,Kolda:1998kc,Espinosa:1998ks},
but these generally have very little connection with the MSSM and 
the \lsoft parameters. As pointed out by \cite{Lyth:1997ai}, D term 
inflation also is sensitive to nonrenormalizable operators through the 
gauge kinetic function.

\subsection{Models related to the soft parameters}
Since there is a large literature of supersymmetric inflationary models 
(some of the literature that we will not discuss below includes
\cite{Berezhiani:2001xx,Dimopoulos:1997fv,Dvali:1997wz,Dvali:ms,Ross:1995dq,Adams:1996yd,Adams:1997de,German:1999gi,German:2001tz,Randall:1995dj,Linde:1997sj,Stewart:2000pa,Gaillard:1995az,Kallosh:2003ux,Greene:2002ku,King:1998uv,Dvali:1997uq,Covi:1997my,Lazarides:1996dv,Lazarides:1995vr}),
and since most of them do not have a direct link with the MSSM and
\lsoft, we review a few representative models to illustrate
some of the attempts to connect the MSSM and inflation.

\subsubsection{\protect\( \phi \protect \) NMSSM}
The next-to-minimal supersymmetric standard model (NMSSM) is
a model which has a superpotential of the form (in addition to the usual 
quark/lepton Yukawa terms): 
\begin{equation} W=\lambda
\hat{N}\hat{H}_{u}\hat{H}_{d}-k\hat{N}^{3}, \end{equation}
where \( H_{u,d} \) are the usual Higgs fields and
\( N \) is a SM gauge singlet field. The NMSSM is described in more 
detail in Section~\ref{nmssm}.  The main motivation of the model is to 
generate the \( \mu  \) term in the MSSM by giving a VEV to the scalar 
component of \( N \). However, the \( k N^{3} \) term has a discrete \( 
Z_{3} \) symmetry which can generate 
cosmologically unattractive domain walls if the symmetry is broken 
spontaneously after inflation. Therefore,   
this superpotential can be modified 
\cite{Bastero-Gil:1997vn,Bastero-Gil:1998te}
to be  
\begin{equation}
W=\lambda \hat{N}\hat{H}_{u}\hat{H}_{d}-k\phi \hat{N}^{2},
\end{equation}
where \( \phi  \) is a SM gauge singlet inflaton (for a related
model, see \cite{Dine:1997kf}). Now the term with coefficient \( k \)
has a global \( U(1) \) PQ symmetry instead of the discrete \( Z_{3} \)
symmetry.\footnote{Even if strings formed after inflation ended by 
the spontaneous breaking
of the \( U(1) \), they would not cause much harm to cosmology.} Just as 
in the MSSM, soft supersymmetry-breaking terms are added containing
the new fields \( N \) and \( \phi  \), requiring dimensionful parameters
\( m_{i} \) and \( A_{k} \). One can of course assume that these terms 
come from gravity mediation. This gives generic 
values\begin{equation}
m_{i}\sim A_{k}\sim 1\textrm{ TeV},
\end{equation}
but peculiarly not for the mass \( m_{\phi } \) of the inflaton
field, which is fixed by the density perturbation amplitude.

As the \( U(1) \) PQ symmetry is spontaneously broken in the true
vacuum by the VEVs of \( \phi  \) and \( N \), there is an axion
in the low energy spectrum. Since at the minimum of the potential
the axion VEV scale is \( A_{k}/k \) and is preferred (for dominant
axion dark matter) to be around \( 10^{13} \) GeV, the dimensionless
coupling \( k \) is forced to take on a tiny \( \sim 10^{-10} \)
value.  \( \lambda  \) is then constrained as well to obtain a 
reasonable value for the effective \( \mu  \) parameter. These small 
values may be explained by discrete symmetries.
Since the inflaton VEV scale is tied with the axion VEV scale, the
inflaton VEV is also \( 10^{13} \) GeV. Finally, a constant term \( V_{0} 
\) must be added to enforce that the potential is zero at the minimum. The 
value of \( V^{1/4}_{0}\sim A_{k}/\sqrt{k}\sim  10^{8} 
\) GeV. The potential generated by the superpotential for \( N \) and
\( \phi  \) naturally gives rise to hybrid inflation \cite{Linde:1993cn}
with \( \langle N\rangle  \) acting as the switch field for \( \phi  \),
if a constant potential \( V_{0} \) is added to the system. During
inflation, when the VEV of \( \phi  \) is beyond some critical value,
the VEV of \( N \) sits at the origin (the Higgs VEVs are assumed
to be at the usual electroweak symmetry breaking values, and hence
are negligible). This gives the potential \begin{equation}
V_{0}+\frac{1}{2}m_{\phi }^{2}\phi ^{2}.
\end{equation}
Inflation ends when \( \phi  \) reaches a critical value, effectively
governed by requirement 2 discussed in Section~\ref{inflationreq}.

The required amplitude of density perturbations force \( m_{\phi } \) to 
be very light: \( m_{\phi }\sim 1 \) eV. (Even if just the slow-roll 
conditions were imposed, the mass \( m_{\phi } \) would be only \( 100 \) 
keV.) Because \( k \) is very small, \emph{if} \( m_{\phi } \) is forced 
to vanish at some high renormalization scale, the running will only 
generate a tiny mass of the order \( k\times 1\textrm{TeV} \) which is 
close to the requisite \( m_{\phi }\sim 1 \) eV. It is then supposed that
the inflaton is massless at the high energy scale and the mass is 
generated radiatively. This vanishing mass can be justified in a situation 
in which the potential only receives contributions from 
vanishing modular weight terms \cite{Bastero-Gil:1998te}. However, 
this is not generic \cite{lythriotto}.

However, if \( m_{\phi }\sim 1 \) eV and thus is much smaller
than the spacetime curvature scale \( H\sim 1 \) MeV during inflation,
graviton loops (which were not discussed in the original papers since
these corrections are separate from those related to the usual \( \eta  \)
problem, as they are too small to cause the \( \eta  \) problem)
may give significant contributions to the inflaton mass.  These graviton 
loop contributions can even possibly destabilize the inflaton mass.  Such 
graviton loop corrections are suppressed by a loop factor, and hence are 
not a problem when \( |m_{\phi }/H|>0.01 \). However, they can pose 
a problem here because \(m_{\phi }/H\sim 10^{-3} \) in this model.
Discussions related to this one-loop effect can be found
in \cite{Gaillard:1995az}.

In summary, the only connection of inflation with the soft parameters 
in this scenario is the scale of \( 1 \) TeV, and the flatness
of the inflaton potential is not due to cancellation properties of
supersymmetry, but rather special discrete symmetries that protect
the tuning of a small coupling \( k \). The weakest points are the
justification of a small inflaton mass and the smallness of the coupling
constant.\footnote{A lack of explanation of the origin of \( V_{0} \)
is also a problem in the context of SUGRA. Furthermore, because \( \lambda  \) is
forced to be tiny, the \( \mu  \) magnitude is not controlled by
the VEV of \( N \). Hence, the \( \mu  \) problem really is not
solved unless a dynamical mechanism is given for the smallness of
\( \lambda  \). } The strong features are that \( \phi  \) does not
take transPlanckian values typical of hybrid inflation, and that the 
model connects inflationary physics with possibly observable axion 
physics. This is to be considered
a very low scale inflationary model since \( H\approx 1 \) MeV. Some
modifications can be made to make some of the extraordinarily small
dimensionless and dimensionful parameters more natural. For example,
extra dimensions much larger than the inverse GUT scale can be 
invoked to suppress couplings by the large volume factor 
\cite{Bastero-Gil:2002xs}.
To raise the inflaton mass from \( O(1) \) eV to \( O(100) \) keV, the 
idea of isocurvature perturbations converting
into curvature perturbations on superhorizon scales due to nonadiabatic
physics \cite{Lyth:2001nq,Mollerach:hu,Hamazaki:1996ir,Kodama:bj,Moroi:2001ct,Enqvist:2001zp}
also has been implemented \cite{Bastero-Gil:2002xr} by requiring
the Higgs to be almost massless (with a mass of order of the \( 100 \) keV 
inflaton mass) during inflation and tuning the Higgs field initial 
conditions appropriately to make it the source of large isocurvature 
perturbations.

\subsubsection{Chaotic inflation with right-handed sneutrino}
Here the main idea is to try to connect the seesaw scale of \( 10^{13} 
\) GeV with the chaotic inflationary scale \( H \) 
\cite{Murayama:xu,Murayama:1992ua}.
The starting point is a PQ invariant extension of the MSSM 
including right-handed neutrinos \cite{Murayama:1992dj}.
The superpotential of the theory includes the usual Yukawa couplings for 
the quarks, leptons, and the neutrinos (note that a bare $\mu$ term is 
disallowed), and has an additional set of PQ-breaking terms.  Denoting 
these terms collectively as $W_2$, they are given by 
\begin{equation}
W_{2}=\frac{1}{2}h_{M}^{i}N_{i}^{c}N_{i}^{c}P+\frac{f}{M_{Pl}}P^{3}P'+\frac{g}{M_{Pl}}PP'H_{u}H_{d}, 
\end{equation}
such that the PQ symmetry breaking is at an intermediate scale, near
\( 10^{12} \) GeV.

Considering the flatness of the potential, the upper bound of the 
potential of \( M_{Pl}^{4} \), the large field value required for the 
chaotic inflationary scenario (large means \( >O(M_{Pl}) \)), and the 
relative lightness of the sneutrino, \cite{Gherghetta:1995jx} concludes 
that chaotic inflation occurs with a quartic potential associated with
the right-handed electron sneutrino whose VEV is transPlanckian \( 
\widetilde{N}_{1}\gg M_{Pl} \).
The effective potential essentially becomes\begin{equation}
V(\phi )=\frac{1}{4}h_{1}^{2}|\widetilde{N}_{1}^{c}|^{4}
\end{equation}
where \( h_{1}=10^{-7} \) is required to generate the observationally
required density perturbations. Since \( h_{1} \) is akin to the
electron Yukawa coupling, the as of yet unknown
reason for the smallness of the electron Yukawa may be responsible for
the smallness of \( h_{1} \). Here the radiative corrections associated 
with soft supersymmetry breaking can induce an intermediate 
scale breaking \( \langle P\rangle \approx 10^{12} \) GeV, giving
an electron Majorana neutrino mass scale of \( M_{N_{1}}\approx 
h_{1}\langle P\rangle \approx 10^{5}\textrm{ GeV}. \)

In summary, the only connection of inflation to
the soft supersymmetry-breaking Lagrangian in this scenario is the 
radiative breaking of $U(1)_{PQ}$, leading
to \( \langle \widetilde{P}\rangle \approx 10^{12} \) GeV. 
One of the most observationally promising implications
of this model is through flavor phenomenology. The general difficulty with
inflationary models in which the inflaton has a VEV much larger than \( 
M_{Pl} \) is that the nonrenormalizable operators
that have been neglected are important, making such simple
scenarios unlikely. Since \( H\sim 10^{13} \) GeV, this scenario
is a prototypical {}``high'' scale inflationary scenario.

\subsection{Outlook}
Inflation is a paradigm that has been attaining increasing observational
support \cite{Bennett:2003bz}. Although there are many analyses of
supersymmetric inflationary models that we did not touch upon
\cite{Berezhiani:2001xx,Dimopoulos:1997fv,Dvali:1997wz,Dvali:ms,Ross:1995dq,Adams:1996yd,Adams:1997de,German:1999gi,German:2001tz,Randall:1995dj,Linde:1997sj,Stewart:2000pa,Gaillard:1995az,Kallosh:2003ux,Greene:2002ku,King:1998uv,Dvali:1997uq,Covi:1997my,
Lazarides:1996dv,Lazarides:1995vr}, there is little direct connection with
the MSSM and \lsoft in most cases.

The reason can be stated schematically as follows. Single field
inflationary models generically require fine tuning of the couplings as
well as transPlanckian field values. The only source of sufficient fine
tuning within the MSSM is the Yukawa couplings. (We have given an example
of such a scenario above.) However, here the transPlanckian values
require a determination of the nonrenormalizable operators, which is
impossible without a UV complete framework. As we have seen, the hybrid
inflationary scenario can phenomenologically accommodate the electroweak
scale and the intermediate scale.  However, if the flat directions involve
only MSSM fields, the VEVs that are tuned to be the inflaton tend to be
unacceptably large at the end of inflation and/or break unwanted gauge
groups \cite{Randall:1995dj}.

\setcounter{footnote}{0}
\section{How do the soft parameters show up in collider experiments?}
\label{expsect}
We now turn to the direct production of superpartners at 
colliders, and
how one can learn about the low energy values of the 
${\mathcal{L}}_{soft}$ parameters from the data. As explained  in
Section~\ref{paramsect},  at most one parameter of
${\mathcal{L}}_{soft}$ is directly measurable, the gluino 
mass (which could have up to $25 \%$ radiative correction
\cite{Martin:1993zk}). Before considering how to extract the Lagrangian
parameters from data after a discovery, let us first examine the 
current
experimental and theoretical limits on superpartner masses (as of 2003).

\subsection{Current limits on superpartner masses}
The general limits from direct experiments that could produce
superpartners are not very strong.  They are also all model dependent, 
sometimes a little and sometimes very much.  Limits from
LEP on charged superpartners are near the kinematic limits except for
certain models, unless there is close degeneracy of the charged
sparticle and the LSP, in which case the decay products are very soft
and hard to observe, giving weaker limits.  In most scenarios 
charginos and charged sleptons have limits of about 100 GeV.  Gluinos
and squarks have typical limits of about 250 GeV, except that if one
or two squarks are lighter the limits on them are much weaker.  For
stops and sbottoms the limits are about 85 GeV separately.

There are no general limits on neutralinos, though sometimes such limits
are quoted.  For example, suppose
the LSP was pure photino.  Then it could not be produced at LEP through a
Z which does not couple to photinos.  If selectrons are very
heavy, photino production via selectron exchange is very small in pair or
associated production.  Then no cross section at LEP is large enough to
set limits.  There are no general relations between neutralino masses and
chargino or gluino masses, so limits on the latter do not imply limits on
neutralinos.  In typical models the limits are $m_{LSP}\gtrsim 40$ GeV,
$m_{\widetilde{N}_{2}}\gtrsim 85$ GeV.  



Superpartners get mass from both the Higgs mechanism and from
supersymmetry breaking, so one would expect 
them to typically be heavier than SM particles.  All SM particles would be
massless without the Higgs mechanism, but superpartners would not.  Many
of the quark and lepton masses are small presumably because they do not
get mass from Yukawa couplings of order unity in the superpotential, so
one would expect naively that the normal mass scale for the Higgs
mechanism was of order the Z or top masses.  In many models, the chargino 
and neutralino masses are often of order Z and top masses, while the 
gluino mass is a few times the Z mass.

There are no firm indirect limits on superpartner masses.  If 
supersymmetry explains the origin of electroweak symmetry breaking, there 
are rather light upper
limits on certain superpartner masses, but they are not easily made
precise, as discussed in Section~\ref{finetunesect}. Radiative 
electroweak symmetry breaking produces the Z mass in terms of soft 
supersymmetry-breaking masses, so if the 
soft supersymmetry-breaking masses are too large such an explanation does not make 
sense.  The soft parameters most sensitive to this issue are $M_{3}$ 
(the gluino mass parameter) and $\mu$ (which enters the chargino 
and neutralino mass matrices).   Qualitatively, one then expects rather 
light gluino, chargino, and neutralino masses.  Taking this argument 
seriously, one is led to expect $m_{\widetilde{g} }\lesssim 500$ GeV, 
$m_{\widetilde{N}_{2}}$, $m_{\widetilde{C}}\lesssim 250$ GeV, and 
$m_{\widetilde{N}_{1}}\lesssim 100$ GeV.  These are upper limits, seldom 
saturated in typical models of the soft parameters.  There are
no associated limits on sfermions.  They suggest that these gaugino states
should be produced in significant quantities at the Tevatron. Recently, 
these arguments for light superpartners have been
examined to study whether  cancellations among different soft
parameters such as $\mu$ and $M_3$, or scalars, could weaken the
constraints. Based on typical models, particularly string-motivated 
models, cancellations are arguably very unlikely 
because $\mu$ and the different soft masses on which electroweak symmetry 
breaking depends typically arise from rather different physics 
\cite{Kane:2002ap}. 

There are other clues that some superpartners may be light.  If
the baryon number is generated at the electroweak phase transition then 
the 
lighter stop and charginos should be lighter than the top.  If the
LSP is indeed the cold dark matter, then at least one scalar fermion   
is probably light enough to allow enough annihilation of relic LSPs, 
but there are loopholes to this argument.

\subsection{After the discovery: deducing \lsoft}
\label{expdisc}
Suppose superpartners and Higgs bosons are found.  First, there will be a
great celebration. Next, it will be time to study the signals in order to 
learn the values of $\tan\beta$ and the Lagrangian parameters,
and to study how the patterns point to the underlying theory.  In a
sense the main result from study of the Standard Model at LEP is that
the data point toward a perturbative, weakly coupled origin of electroweak
symmetry breaking. Similarly, ${\mathcal{L}}_{\mathtt{soft}}$ will
point toward some underlying theories and away from others. Consider the 
particles that will eventually be seen.  There are 4
neutralino masses, associated with the soft terms from $W^0$, $B^0$,
$H_u^0$, $H_d^0$ (or, in the electroweak mass eigenstate basis, 
$\gamma$, $Z$, $H_u^0$, $H_d^0$). The neutralino superpartners mix, 
with the physical
neutralino mass eigenstates denoted as $ \widetilde{N}_{1,2,3,4}$. Similarly,
there are two chargino mass eigenstates from the chargino mass matrix 
$\widetilde{C}_{1,2}$.  
There are four Higgs boson masses, for $ h^{0}$, $H^{0}$, 
$A^{0}$, $H^{\pm}$.  There
is one gluino mass and one gravitino mass.  The squark mass matrix for
up-type squarks has six independent eigenvalues, the superpartners of the
left- and right-handed quarks $u$, $c$, $t$: $\tilde {u}_{L}$,
$\widetilde{c}_{L}$, $\widetilde{t}_{L}$, $\widetilde{u}_{R}$, $\widetilde{c}_{R}$,
$\widetilde{t}_{R}$. Similarly, there are 6 down-type mass eigenstates and 6
charged lepton mass eigenstates.  In the MSSM there are only the three
left-handed neutrinos and their sneutrinos.  Including the gravitino 
mass, these add up to 33 physical
masses that can be measured if all the states are found in experiments. If
the gravitino is not the LSP then it may not be possible to measure its
mass since it couples too weakly to be produced directly at colliders and
affects only certain aspects of early-universe cosmology, perhaps rather
indirectly.

Another important parameter is $\tan\beta$, the ratio of the 
VEVs of the two Higgs fields: $\tan \beta \equiv
\langle H_{u}\rangle /\langle H_{d}\rangle$. $\tan \beta $ is
intrinsically a low energy parameter, since the Higgs fields do not have
VEVs until the RG running induces them somewhat above the electroweak 
scale.  As
will be explained below, in general measuring $\tan\beta $ is difficult
and cannot be done accurately, {\it i.e.}, without model-dependent 
assumptions, without a lepton collider with a polarized beam that is above 
the threshold for several superpartners.  When trying to deduce the
unification scale Lagrangian, $\tan \beta $ can be traded for a high scale
parameter in the Higgs sector. Perhaps with luck $\tan\beta$ has a 
value that leads to effects that do allow its determination.  For 
example, large values of $\tan\beta$ have distinctive phenomenological 
implications (see Section~\ref{ltbpheno}).

The form for ${\mathcal{L}}_{soft}$ is rather general and allows for other 
effects, such as D terms (from the breaking of extra
U(1) symmetries) that give contributions to squark and slepton masses
(Section~\ref{EWSBsect}), or Planck scale operators that lead to
contributions to masses when some fields get VEVs.  Extra U(1)'s or extra
scalars can lead to a larger neutralino mass matrix than the $4
\times 4$ one
expected here.  Terms of the form $\phi ^{\ast }\phi ^{2}$ (rather than
$\phi ^{3}$) are generally allowed
\cite{Jones:cu,Hall:1990ac,Jack:1999ud,Diaz-Cruz:1999fx} in gauge theories
where the scalars are charged under some broken gauge group, but no models are
yet known where such terms give significant effects.  They can be added if
necessary. It is extremely important to allow for the possibility that
effects such as these are present, by not overconstraining the form of
${\mathcal{L}}_{soft}$ too stringently with assumptions.

Let us turn in the following sections to how to connect the soft
parameters with observables.  The essential point is that 
at colliders experimenters only measure kinematic masses, and cross sections
times branching ratios, {\it etc.}, 
which must be expressed in terms of soft parameters to extract the values
of the soft parameters from data.  The gravitino mass can probably only be
measured if it is the LSP and then only very approximately.  The soft
parameter $M_3$ can be deduced from the gluino mass to about 20\% accuracy
from theoretical uncertainties \cite{Martin:1997ns} due to large loop
corrections depending on squark masses (not counting experimental
uncertainties).

43 of the parameters in ${\mathcal{L}}_{soft}$ are phases.  As explained
previously, a certain subset of the phases affect essentially all
observables.  Phenomenologically, life would be much simpler if the phases
were zero, or small.  It would be much easier to determine the soft
parameters from data, to measure $\tan \beta $, {\it etc}. There are 
arguments
that the MSSM phases are small, but it is certain that sources of CP
violation beyond the CKM phase are necessary for baryogenesis ({\it i.e.},
it is known that the Standard Model cannot explain baryogenesis).  
If the baryon asymmetry is generated at the electroweak phase 
transition ({\it i.e.} in the standard picture of electroweak 
baryogenesis, there must be phases of ${\mathcal{L}}_{soft}$ associated 
with the stop and chargino sector.  
Until the values of the phases are measured, or understood theoretically, 
in principle one must allow for their effects in relating data and 
theory. For our purposes it is only necessary to allow for the 
possibility that the phases are not small (recall that this is not ruled 
out, although such  points do appear to represent exceptional points of 
the MSSM parameter space), and consider the question of how the presence 
of the phases complicates the extraction of the Lagrangian parameters 
from low energy data.

There has been a significant amount of research effort studying the issue
of reconstructing the soft Lagrangian from data; see {\it e.g.}
\cite{Zerwas:2002as,Barger:1998bj,Barger:1999tn,Gunion:2002ip,Kalinowski:2002gc,Choi:2002mc,Barger:2000fi,Choi:2000ta,Mrenna:1999ai} 
and references therein for further details.  In this
section, we will illustrate the general issues and complications, such as
nontrivial phases and large $\tan\beta$, involved in this reconstruction
process.\\

\paragraph {Charginos}
The simplest example is the chargino sector. This is
treated in many places in the literature; more details are given in {\it e.g.}
\cite{Haber:1984rc,Martin:1997ns} as well as in Appendix~\ref{gauginoapp}.  
The superpartners of $W^{\pm }$ and of the charged Higgs bosons $H^{\pm}$
are both spin-1/2 fermions and they mix once the electroweak symmetry is
broken, {\it i.e.} once the neutral Higgs field get VEVs.  There is a
$\widetilde{W} \widetilde{W}$ mass term $M_{2}e^{i\phi _{2}},$ a
higgsino mass term 
$\mu e^{i\phi _{\mu }}$, and a mixing term, so the chargino mass matrix is
\begin{eqnarray} 
M_{\widetilde{C}}=\left( \begin{array}{cc} M_{2}e^{i\phi
_{2}} & \sqrt{2}m_W\sin \beta \\ \sqrt{2}m_W\cos \beta & \mu e^{i\phi
_{\mu }} \end{array} \right). 
\end{eqnarray} 
The eigenvalues of this matrix (since it is not symmetric one usually
diagonalizes ${M_{\tilde c}}^\dagger{M_{\tilde c}})$ are the physical mass
eigenstates, $M_{ \widetilde{C}_{1}}$ and $M_{\widetilde{C}_{2}}.$ \ The formulas
are a little simpler after rewriting in terms of the trace (sum of
eigenvalues) and determinant (product of eigenvalues),
\begin{eqnarray}
{\rm 
Tr}M_{\widetilde{C}}^{\dagger}M_{\widetilde{C}}&=&M_{\widetilde{C}_{1}}^{2}+
M_{\widetilde{C}_{2}}^{2}=M_{2}^{2}+\mu ^{2}+2m_W^{2}\\
{\rm Det}M_{\widetilde{C}}^{\dagger }M_{\widetilde{C}}&=&M_{\widetilde{C}
_{1}}^{2}M_{\widetilde{C}_{2}}^{2}=M_{2}^{2}\mu ^{2}+2m_W^{4}\sin
^{2}2\beta\nonumber\\ &-&2m_W^{2}M_{2}\mu \sin 2\beta \cos (\phi
_{2}+\phi _{\mu }). 
\end{eqnarray}
The physical masses $M_{\widetilde{C}_{1}}$ and $M_{\widetilde{C}_{2}}$ will be
what is measured, but what must be known to determine the Lagrangian are
$M_{2},\mu ,$ the phases, as well as $\tan \beta$. The phases enter in the
reparameterization invariant (and hence observable) combination $\phi
_{2}+\phi _{\mu }$.  While generally the presence of nonzero phases are
linked to CP-violating phenomena, they also have an impact on 
CP-conserving quantities (here the masses also depend strongly on the
phases).

After diagonalizing this matrix, the gauge eigenstates can be expressed in
terms of the mass eigenstates, which will be linear combinations of gauge
eigenstates whose coefficients are the elements of the eigenvectors of the
diagonalizing matrix.  These coefficients, which also depend on $\tan
\beta$ and the phases, enter the Feynman rules for producing the mass
eigenstates.  Thus the cross sections and decay branching ratios (BR) also
depend on the phases and $\tan \beta$. To measure any of the parameters
it is necessary to invert the equations and measure all of them.  Since
there are four parameters here one has to have at least four observables.  
In practice more observables will be necessary since there will be
quadratic and trigonometric ambiguities, and experimental errors will lead
to overlapping solutions.  Thus from the masses alone it is not 
possible to measure $\tan \beta $ in a model-independent way 
\cite{Brhlik:1998gu}.
We elaborate on this point because the results 
of many phenomenological analyses have made the erroneous claim that 
tan$\beta$ can be measured in various sectors.  Whenever this claim has 
been made (except at a lepton collider  
with polarized beams or by combining a variety of Higgs sector data ---
see below), the analysis has actually assumed various soft terms are zero
or equal to reduce the number of parameters.  While such assumptions
may (or may not) be good guesses, once there is data it is important to
measure such parameters without assumptions.

The next thing to try is to add the (presumed) cross section data.  The
dominant processes are s-channel Z and $\gamma $, and squark exchanges for
hadron colliders. The couplings to Z and $\gamma $ are determined by the
diagonalized mass matrix, but now the squark masses and couplings enter,
giving new parameters. If chargino decays are not considered, there are
three cross sections,
$\widetilde{C}_{1}\widetilde{C}_{1},\widetilde{C}_{2}\widetilde{C}_{2},
\widetilde{C}_{1}\widetilde{C}_{2}.$ In principle, one can imagine measuring
differential cross sections, obtaining several angular bins.  In practice,
with limited statistics and backgrounds, usually at best one only measures
total cross sections and forward-backward asymmetries $A_{FB}$.  At a
hadron collider it would be very hard to measure even the asymmetries
(because of difficulties in reconstructing the superpartners from their
decay products, because of large backgrounds, and because more than one
superpartner channel may contribute to a given signal) and before they
were included in the counting a careful simulation would have to be done.  
Thus, if the produced charginos can be reconstructed, it may be possible
to measure $\tan \beta $ at an electron collider (see {\it e.g.}
\cite{Choi:2000ta,Barger:2001nu}), but probably still not at a
hadron collider.  However, it needs to be shown that the produced
charginos can be reconstructed even at a lepton collider.

\begin{figure}
\centerline{
   \epsfxsize 3.3 truein \epsfbox {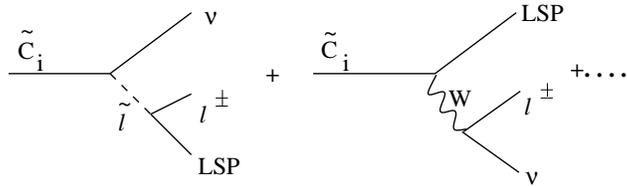}
    }
\caption{Possible mechanisms for chargino decay.}
\label{f2}
\end{figure}

Further, the charginos of course decay. There are a number of possible
channels, a few of which are shown in Figure~\ref{f2}.
These introduce new parameters, slepton and squark masses and
couplings, and the LSP mass and couplings, even assuming the prompt decay
to the LSP dominates over decay cascades through other neutralinos.  
Unless one decay dominates, too many parameters may enter to measure $\tan
\beta $ from these channels even at a higher energy lepton collider. If
the decay via an intermediate $W$ dominates, some final polarization can
be obtained, but if sleptons and squarks are light and contribute to the
decays then no polarization information is transmitted to the final state
because they are spinless.  Their chirality can still enter since the wino
component of charginos couples to left-handed sfermions, while the 
higgsino component couples to right-handed sfermions.

In general then it is not possible to measure $\tan \beta $ or the soft
phases or other soft parameters from chargino channels alone, though if
squarks and sleptons are heavy or if charginos can be reconstructed
experimentally it may be possible (see {\it e.g.} 
\cite{Choi:2000ta,Barger:2001nu} and
references therein).  If one assumes values for phases or assumes
relations for parameters the results for tan$\beta$ and other parameters
are not true measurements and may not correspond to the actual values.
However, it is still worthwhile to make certain assumptions and learn as 
much as possible within that context.  For example, one standard 
set of assumptions includes assuming that the three sneutrinos are 
approximately degenerate, that $\widetilde{e}_{L},\widetilde{\mu}_{L,}\widetilde{\tau}
_{L}$ are approximately degenerate and similarly
$\widetilde{e}_{R},\widetilde{\mu} _{R},\widetilde{\tau}_{R}$ are approximately  
degenerate, with similar assumptions for the first two squark families.
Also, for collider physics the first two families can be taken to have
small LR mixing, since LR mixing is expected to be proportional to the
mass of the associated fermions. 
Under these assumptions it will be possible to measure $\tan \beta $ and 
the soft phases at lepton colliders that can produce at least a subset of 
the superpartners, when the extra observables from beam polarization and a 
second energy are included, even if the collider does not have enough 
energy to produce many superpartners (see Section~\ref{tevatronsect}). 
With such assumptions it may even be possible to measure tan$\beta$ and 
certain phases at hadron colliders.  Several of the assumptions can be 
checked independently.

Here only the chargino channels have been looked at so as to have a simple
example, but of course all the accessible superpartners will be produced
at any collider, leading to more parameters and more observables. Only   
with good simulations (or of course real data) can one be confident about
counting observables.  Conservatively, with hadron colliders true
measurements of $\tan \beta $ and soft phases and other soft parameters  
are not possible, but they may be possible for reasonable approximate 
models depending on the actual values of the parameters, or by combining a
number of measurements.  For lepton colliders with a polarized beam, above
the threshold for some superpartners, the parameters of
${\mathcal{L}}_{soft}$ can be measured, as discussed below.

\paragraph{Neutralinos}
Of course, if charginos are produced, neutralinos will also be produced,
leading to more observables (masses, cross sections, asymmetries).  There
are more parameters in the neutralino sector, but not as many new
parameters as new observables.  The neutralino mass matrix is 
(see Appendix~\ref{gauginoapp}):
\begin{eqnarray}
M_{\widetilde{N}}=\left(
\begin{array}{cccc}
M_{1}e^{i\phi _{1}} & 0 & -m_Z\sin \theta _{W}\cos \beta & m_Z\sin
\theta _{W}\sin \beta \\
& M_{2}e^{i\phi _{2}} & m_Z\cos \theta _{W}\cos \beta & -m_Z\cos 
\theta
_{W}\sin \beta \\
&  & 0 & -\mu e^{i\phi _{\mu }} \\
&  &  & 0
\end{array}
\right) ,
\end{eqnarray}
in the basis ($\widetilde{B},\widetilde{W},\widetilde{H}_{U},\widetilde{H}_{D})$. Even 
when the elements are complex it can be diagonalized by a single  
unitary matrix.  For simplicity, here a phase in the
Higgs VEVs is being ignored that will in general be present.

The chargino sector depended on a single physical phase, the
reparameterization invariant combination $\phi _{2}+\phi _{\mu }.$
Similarly there are two physical phases that cannot be rotated away in the
neutralino mass matrix.  One can see this by simply calculating 
observables, or one can redefine the basis by multiplying by 
\begin{eqnarray}
\left(
\begin{array}{cccc}
e^{+i(\phi _{\mu }-\phi_2 )/2} &  &  &  \\
& e^{+i(\phi _{\mu }-\phi_2 )/2} &  &  \\
&  & e^{-i(\phi _{\mu }-\phi_2 )/2} &  \\
&  &  & e^{-i(\phi _{\mu }-\phi_2 )/2}
\end{array}
\right) ,
\end{eqnarray}
such that the resulting matrix depends explicitly only the physical
phases.  Thus there is one new soft mass $M_{1}$ and one new physical
phase $\phi_{1}+\phi _{\mu }.$ In principle the masses of the four mass
eigenstates can be measured, as well as the cross sections
$\widetilde{N}_{1}+\widetilde{N}_{1},\widetilde{N}_{1}+
\widetilde{N}_{2},\widetilde{N}_{2}+\widetilde{N}_{2},$ {\it etc.}, and associated
asymmetries.  The number of new observables is different at different
colliders.

If only two new masses are measured, there is no progress in inverting the
equations to solve for $\tan \beta$, {\it etc}. If cross sections are used 
there
are additional parameters, from squark or selectron exchange. The number
of parameters and observables arising from the Higgs sector will also be
counted below explicitly.  It is extremely important for detector groups
at various colliders to count the number of observables they can expect to
measure.  This has to be done using models, of course, but the models 
should be quite general, so that parameters are not defined away by 
arbitrary assumptions. The models should also be able to accommodate 
electroweak symmetry breaking without excessive fine tuning.  Of course, 
the models should also be consistent with LEP data.

\paragraph{Gluinos}
We now consider the effects of phases in the gluino sector,
which nicely illustrates the subtleties of including and measuring the
phases \cite{Mrenna:1999ai}.  In general, there can be a phase $\phi_3$ 
associated with the soft supersymmetry-breaking gluino mass parameter 
$M_3$.  However, this phase is not by itself an observable phase.  As 
shown in Appendix~\ref{gauginoapp}, it is convenient to redefine the 
gluino field to absorb the phase of $M_3$ as follows: 
\begin{eqnarray}
\lambda_{\widetilde{g}}=G\lambda'_{\widetilde{g}}, 
\overline{\lambda}_{\widetilde{g}} =G^{\ast }\overline{\lambda}'_{\widetilde{g}}   
\end{eqnarray}
where $G=e^{-i\phi _{3}/2}$.  Then for any flavor quark the Feynman rules
introduce factors of $G$ or $G^*$ at the vertices in addition to the color 
factors.
\begin{figure}
\centerline{
   \epsfxsize 3.3 truein \epsfbox {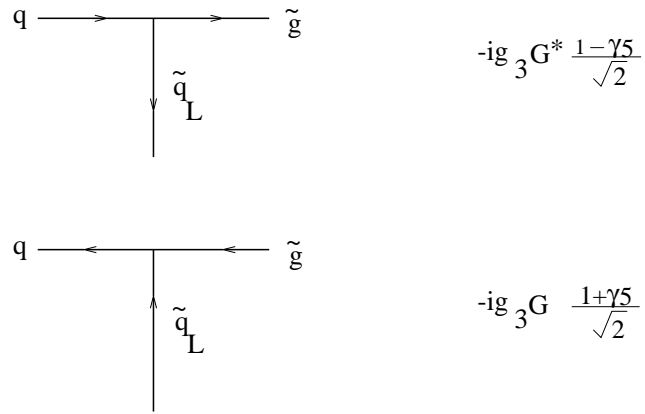}
    }
\caption{Feynman Rules after redefining the gluino filed so that
    gluino mass is real and the phase shows up at the vertices.}
\label{f5}
\end{figure}

Now consider a simple version of gluino production $q+\overline{q}\rightarrow
\widetilde{g}+\widetilde{g}$.
Factors of $G$ and $G^{\ast }$ enter so that there is no dependence on
the phase from these two diagrams.
\begin{figure}
\centerline{
   \epsfxsize 3.3 truein \epsfbox {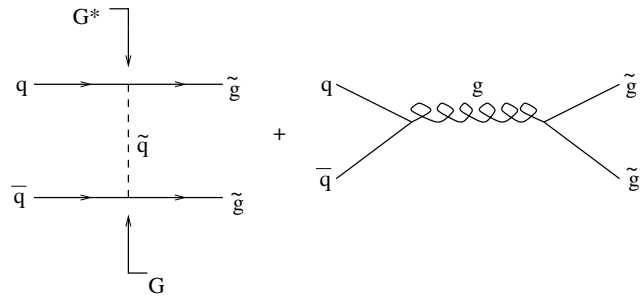}
    }
\caption{How phases enter from gluino production.}
\label{f6}
\end{figure}
Next consider $q+g\rightarrow \widetilde{q}+\widetilde{g}.$   Production of   
$\widetilde{q}_{L}$ leads to an overall factor of $G^{\ast },$ while
production of $\widetilde{q}_{R}$ gives an overall $G$.  This overall phase 
is combined with the phase of the LR mixing part of the squark masses; the  
relevant phases of the LR sector are the phase of $\widetilde{A}$ and 
$\mu$.  Effects of the reparameterization invariant phase combinations 
$\phi_3-\phi_{\widetilde{A}}$ and $\phi_3+\phi_{\mu}$ are then
observable in principle, but LR mixing is expected to be
very small for the first two families (which are the constituents of the 
beams used in experiments) because LR mixing is typically 
proportional to the associated fermion mass (see 
Eq.~(\ref{pok1})).\footnote{However, this is not necessarily true if the 
$\widetilde{A}$ parameters are not factorizable in a particular way with 
respect to the Yukawa matrices.}  Thus the effects of the phases will in 
general be suppressed in gluino production.

\begin{figure}
\centerline{
   \epsfxsize 3.3 truein \epsfbox {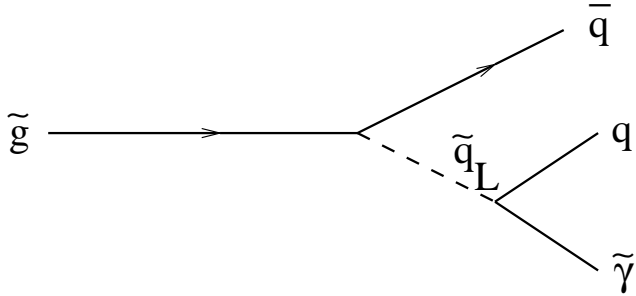}
    }
\caption{Gluino production and decay. Phase factors
    enter at the vertices, as described in the text.}
\label{f7}
\end{figure}
But gluinos have to decay, and then the phases enter.  To illustrate
what happens, imagine the gluino decay is via a squark to 
$q\overline{q}\widetilde{B}$, as shown for $\widetilde{q}_{L}$. Then a factor 
$e^{i\phi _{3}/2}$ enters at the
gluino vertex and a factor $e^{-i\phi_1 /2}$ at the bino vertex.
The resulting differential cross section is
\begin{eqnarray}
\frac{d\sigma}{dx}&\propto&(\frac{1}{m_{\widetilde{q}_{L}}^{4}}+
\frac{1}{m_{\widetilde{q}_{R}}^{4}}){m_{\widetilde{g}}^{4}x\sqrt{x^{2}-y^{2}}}
\nonumber \\
&\times& (x-\frac{4}{3}x^{2}-\frac{2}{3} 
y^{2}+xy^{2}+y\left(1-2x+y^{2}\right)
\cos(\phi_3 -\phi_1))
\end{eqnarray}
where $x=E_{\widetilde{B}}/m_{\widetilde{g}},y=m_{\widetilde{B}}/m_{\widetilde{g}}.$
The physical, reparameterization invariant
phase which enters is $\phi _3 - \phi_1 $. This is a simplified  
discussion assuming no CP-violating phases are present in the squark 
sector and the LSP is a  bino. More generally, additional 
reparameterization invariant combinations can enter. The 
ways in which various distributions depend on this phase (and on $\tan 
\beta $ and the soft masses) have been studied in \cite{Mrenna:1999ai} so 
measurements can be made at the Tevatron and the LHC. 

\paragraph{Higgs bosons} 
In a similar manner, let us consider the Higgs sector in 
further detail.  In Section~\ref{EWSBsect} the Higgs sector and 
electroweak symmetry breaking were
discussed. Here we include the quantum corrections and explain how in 
practice the Higgs sector depends on a minimum of seven
parameters. The dominant radiative corrections come from the
top quark loops (see e.g. \cite{Carena:2002es} for a review), and in 
general have large effects on 
the spectrum and couplings. It is beyond the scope of this review to 
provide a comprehensive and thorough presentation of the Higgs 
sector; a starting point to the relevant literature can be found in 
the recent report of the
Tevatron Higgs Working Group \cite{Carena:2000yx}, which summarizes these 
effects thoroughly (except for phases), including numerical studies. The 
recent comprehensive Higgs sector review \cite{Carena:2002es} 
includes CP-violating effects and is an excellent reference for those 
interested in studying the Higgs sector.  Here we simply wish to 
reiterate the point that it is crucial to include the radiative 
corrections (which are functions  of the \lsoft parameters) when embarking 
on phenomenological analyses of the MSSM Higgs sector. In addition, if 
$\tan \beta\gtrsim 4$ there can also be important effects from 
gluino loops  that affect $m_b$ and $h b \overline{b}$ couplings and other
quantities. These are also studied in \cite{Carena:2000yx}, 
and a more recent summary is given in \cite{Logan:2000cz}. The
phases of the soft supersymmetry-breaking parameters can significantly affect the 
physics of the Higgs sector
\cite{Pilaftsis:1998dd,Pilaftsis:1998pe,Brhlik:1998gu,Babu:1998bf,Demir:1999zb,Demir:1999hj,Choi:2000wz,Grzadkowski:1999wj,Carena:2000yi,Kane:2000aq}. 
At tree level it has long been understood that all the
quantities that affect the Higgs physics can be chosen to be real.  The
phase effects enter at one loop order, because the stop loops are a
large contribution \cite{Brhlik:1998gu,Pilaftsis:1998dd,Pilaftsis:1998pe}.
The stop loops involve phases because the $2\times 2$  stop mass
matrix is given by  
\begin{eqnarray}
m^2_{\widetilde{t}}=\left (\begin{array}{cc} (m^2_Q)_{33}+m^2_t +\Delta_{u}& 
v^*\widetilde{A}^*_t\sin\beta-v\mu Y_t \cos\beta\\ 
v\widetilde{A}_t\sin\beta-v^*\mu^* Y_t \cos\beta & 
(m^2_U)_{33}+m^2_t+\Delta_{\overline{u}} 
\end{array} \right), 
\end{eqnarray}  
where $\Delta_{u}=(\frac{1}{2}- \frac{2}{3}\sin^{2}\theta_{W})\cos2\beta 
m_{Z}^{2}$, $\Delta_{\overline{u}}=\frac{2}{3}\sin^{2}\theta_{W}%
\cos2\beta m_{Z}^{2}$. $Y_t = Y_{u_{33}}$ ({\it i.e.}, we assume nonzero 
Yukawas for only the third generation) and 
$\widetilde{A}_t= (\widetilde{A}_u)_{33}$, which should be a good
approximation in this context.\footnote{This can be obtained from 
Eq.~(\ref{pok1}) dropping all but third generation quantities.} 
Writing the Higgs fields in the standard way as 
\begin{eqnarray}
\hspace{.45in}H_{d}=\frac{1}{\sqrt{2}}\left(
\begin{array}{c}
{v}_{d}+h_{1}+ia_{1} \\
h_{1}^{-}
\end{array}
\right),\nonumber\\
H_{u}=\frac{e^{i\theta }}{\sqrt{2}}\left(
\begin{array}{c}
h_{2}^{+} \\
{v}_{u}+h_{2}+ia_{2} 
\end{array}
\right),  
\end{eqnarray}
(with the VEVs taken to be real and $\tan \beta
\equiv v_{u}/v_{d}$), the phase $\theta$ is zero at tree level but 
generally nonzero if radiative corrections are included.  $\tan\beta$ can 
be chosen to be a real quantity, but both 
$\tan\beta$ and $\theta$ are necessary to specify the vacuum.

As the stop mass matrix has off-diagonal LR mixing entries, the phases of 
the trilinear coupling $\widetilde{A}_{t}$ and of $\mu $ and the
relative phase $\theta $ enter the stop mass eigenvalues 
$m_{\widetilde{t}_{i}}$.  The effective potential at one loop includes terms 
with stop loops as follows:   
\begin{equation}
V_{1-{\rm loop}}\sim \sum m_{\widetilde{t}_{i}}^{4}\ln m_{\widetilde{t}_{i}}^{2},
\end{equation}
such that $V=V_{{\rm tree}}+V_{1-{\rm loop}}$. Two of the four
minimization conditions ($\partial V/ \partial h_{1}$, $\partial V/
\partial h_{2}$, $\partial V/ \partial a_{1}$, $\partial V/ \partial
a_{2}=0$) are redundant, so three conditions remain.

The Higgs sector thus has 12 parameters, $v_{u},$ $v_{d}$, $\phi
_{\widetilde{A}_{t}}$,$\phi_{\mu }$, $\theta$, $\widetilde{A}_{t}$, 
$\mu$,
$m_{\widetilde{Q}}^{2}$, $m_{\widetilde{u}}^{2}$, $b$, $m_{H_{u}}^{2}$,
$m_{H_{d}}^{2}$ from ${\mathcal{L}}_{soft}$, and the renormalization scale
$Q$ since the parameters run.  Three can be eliminated by the three
equations from minimizing $V$.  The scale $Q$ is chosen to minimize higher
order corrections.  The conditions that guarantee radiative
electroweak symmetry breaking occurs allow $v_{u},$ $v_{d}$ to be replaced
by $m_{Z}$ and $\tan \beta $ as usual.  Thus there are 7 physical
parameters left,
including $\tan \beta $ and one physical phase $\theta$ which is
determined as a function of the (reparameterization-invariant) phase
$\phi_{\widetilde{A}_{t}} +\phi_{\mu }$ and other parameters.  This number 
cannot be
reduced without new theoretical or experimental information. Any
description of the Higgs sector based on fewer than 7 parameters has made
arbitrary guesses for some of these parameters and may be wrong.  If $\tan
\beta $ is large, then sbottom loops can also enter $V$ and additional
parameters are present. Chargino and neutralino loops also enter and may
not be negligible \cite{Ibrahim:2000qj}. This counting is done assuming a
phenomenological approach. In a top-down theory $\tan \beta$ and other
parameters will be predicted.

If the phase is nonzero it is not possible to separate the
pseudoscalar $A=\sin \beta a_{1}+\cos \beta a_{2}$ from $h,H$ so it is
necessary to diagonalize a $3\times 3$ mass matrix.  For this section, we 
name the 
three mass eigenstates $H^{i}$; in the limit of no CP-violating phase
$H^{1}\rightarrow h, H^{2}\rightarrow A,H^{3}\rightarrow H.$  Generally, 
all three mass eigenstates can decay into any given final state or be 
produced in any channel, so there could be three mass peaks
present in a channel such as $Z+$Higgs (wouldn't that be nice).  All   
production rates and branching ratios depend on the phase and can change
significantly as the couplings of Higgs bosons to the SM gauge
bosons and chiral fermions depend sensitively on the CP-violating
phases (see {\it e.g.} \cite{Carena:2000yi,Kane:2000aq}). 

The phases also have a significant impact on how to extract the parameters 
from experimental results of Higgs searches (discovery or exclusion) 
\cite{Kane:2000aq}.  For example, If no Higgs boson is found, there is an 
experimental limit on $\sigma (H^{1}) \times BR(H^{1} \rightarrow 
b\overline{b})$. The resulting lower limits on $m_{H^{1}}$ and $\tan \beta$ 
in the full seven parameter theory change significantly compared with the 
CP-conserving MSSM. For example, if the model is CP-conserving the 
lower limit on the lightest Higgs mass is 
about 10\% below the SM limit, but if the Higgs sector is CP-violating the 
lower limit can be an additional 10\% lower (see also 
\cite{unknown:2001xw,Barate:2000ts,Acciarri:2000ke}).  
If a Higgs boson is found, then $m_{H^{1}}$ and its $\sigma \times
BR$ have been measured.  The allowed region of the full seven 
parameter space is quite different for the CP-violating and
CP-conserving models. Thus once there is a discovery it could be
misleading to not include this phase in the analysis.

If the heavier Higgs bosons are heavy and decouple, the effects
for both questions decrease for the lower limit on the mass of the
lightest eigenstate (and the effects of CP violation on the other
properties of the lightest eigenstate also decouple in this limit). There 
is still CP-violating mixing between the two heavy 
eigenstates. However, this can only be carefully studied after
the production of those states. 

With full parameters space for the Higgs potential,  we would need at
least seven or more observables in order to 
determine $\tan \beta $ or any of the ${\mathcal{L}}_{soft}$ parameters
from the Higgs sector alone. For example, consider the following
collection of possible observables: three neutral scalar mass eigenstates, 
the charged
Higgs mass, the 
three $\sigma \times BR$ for $Z+$Higgs and three $\sigma \times BR$ for  
channels $H^{i}+H^{j}$, and the two stop mass eigenstate masses.
Probably in addition one can add the ratio $r=\sigma (gg\rightarrow
H^{2}\rightarrow b\overline{b})/\sigma (gg\rightarrow H^{1}\rightarrow
b\overline{b})$. Which observables can be measured depends on the masses, $\tan
\beta$, {\it etc.} If the Tevatron and its detectors function well,
several observables can be measured. The WWh and ZZh couplings, which
are the most important Higgs couplings, since they
confirm the Higgs mechanism (because they are not gauge invariant), can be
detected. Once $m_h$ is known the inclusive production can be used. As
many as 50,000 Higgs bosons could eventually be produced and studied at
the Tevatron (if sufficient integrated luminosity is gathered), and it 
should be possible to confirm h couples proportional
to mass. Ratios of $\sigma \times BR$ for several channels may provide
independent observables. The states $A$, $H^0$, $H^{\pm}$ could be
observed there. Combining LHC and Tevatron data may lead to enough
observables to invert the equations and measure $\tan \beta$,
$\phi_{\widetilde{A}_t}+\phi_{\mu}$, and other ${\mathcal{L}}_{soft}$
parameters.

There are two recent pieces of information about Higgs physics that
both independently suggest it will not be too long before a confirmed
discovery (of course the discovery of the Higgs is such an important 
question that solid data is needed).  

First, there is an upper limit on $m_{h}$ from the global 
analysis of
precision LEP (or LEP + SLC +Tevatron) data \cite{lepewwg}.  There are a 
number of independent measurements of SM observables, and every parameter 
needed to calculate at the observed level of precision is measured except 
$m_{h}.$ Hence, one can do a global fit to the
data and determine the range of values of $m_{h}$ for which the fit is
acceptable.  The result is that at 95\% C.L. $m_{h}$ should be below about
200 GeV.  The precise value does not matter for us, and because the data
really determines $\ln m_{h}$ the sensitivity is exponential so it moves
around with small changes in input.  What is important is that there is an
upper limit.  The best fit is for a central value of order 100 GeV, but
the minimum is fairly broad.  The analysis is done for a SM Higgs but is
very similar for a supersymmetric Higgs over most of the parameter space.

An upper limit of course does not always imply there is something below
the upper limit.  Here the true limit is on a contribution to the
amplitude, and maybe it can be faked by other kinds of contributions that
mimic it.  However, such contributions behave differently in other
settings, so they can be separated.  If one analyzes the possibilities
\cite{Kane:2000di,Peskin:2001rw} one finds that there is a real upper
limit of order 450 GeV on the Higgs mass, if (and only if) additional new
physics is present in the TeV region. That new physics or its effects
could be detected at LHC and/or a 500 GeV linear electron collider, and/or
a higher intensity Z factory (``giga-Z'') that accompanies a linear
collider.  So the upper limit gives us powerful new information.  If
no other new physics (besides supersymmetry) occurs and conspires in just
the required way with the heavier Higgs state, the upper limit really is
about 200 GeV.

Second, there was also a possible signal from LEP \cite{unknown:2001xw} in
its closing weeks for a Higgs boson with $m_{h}$=115 GeV.  
It was not possible to run LEP to get enough data to confirm
this signal.  Fortunately, its properties are nearly optimal for early
confirmation at the Tevatron, since its mass is predicted and its cross
section and branching ratio to $b\overline{b}$ are large.  Less is required to
confirm a signal in a predicted mass bin than to find a signal of unknown
mass, so less than 10 $fb^{-1}$ of integrated luminosity will be required
if the LEP signal is correct.  If funding and the collider, detectors,
{\it etc.}, all work as planned, confirming evidence for $h$ could occur 
in 2004.

If the LEP $h$ is indeed real, what have we learned 
\cite{Kane:2000kc,Ellis:2000ig}?  Of course, first we have learned that a
fundamental Higgs boson exists.  The Higgs boson is point-like because its
production cross section is not suppressed by structure effects.  It is a
new kind of matter, different from the known matter particles and
gauge bosons.  It completes the SM and points to how to extend the theory. 
It confirms the Higgs mechanism, since it is produced with the
non-gauge-invariant $ZZh$ vertex, which must originate in the
gauge-invariant $ZZhh$ vertex with one Higgs having a VEV.

The mass of $115$ GeV can potentially tell us important information. 
First, one can obtain information about the nature of the Higgs sector by 
the requirement that the potential energy not be unbounded from below.   
To derive bounds on the Higgs mass, different types of criteria for  
stability may be used.  Requiring absolute stability naturally leads to
the strongest bounds; however, as this assumption is not experimentally
required, somewhat weaker bounds can be obtained by requiring stability
with respect to either thermal or quantum fluctuations.  The bounds most
often discussed in the literature are those derived by requiring that the
potential remain stable with respect to thermal fluctuations in the early
universe, where it can be shown that a $115$ GeV Higgs boson is 
not a purely SM one, since the potential energy would be unbounded from
below at that mass.  The argument is
\cite{Weinberg:pe,Sher:1988mj,Casas:1994qy,Casas:1994us,Altarelli:1994rb}
that the corrections to the potential from fermion loops dominate because
of the heavy top and can be negative if $m_{h} $ is too small.  The SM
potential is
\begin{eqnarray}
V(h)=-\mu^{2}h^{2}+\left\{  \lambda+\frac{3m_Z^{4}+6m_W^{4}+m_{h}
^{4}-12m_{t}^{4}}{64\pi^{2}\rm{v}^{4}}\ln(\textit{ \ })\right\}
h^{4},
\end{eqnarray}
where the argument of the logarithm is a function (of the masses) larger
than one.  
In the usual way $\lambda=m_{h}^{2}/2 v^{2}.$ The second term
in the brackets is negative, so $\lambda$ (and $m_{h}$) has to be large
enough. A careful calculation yields that $m_{h}$ must be larger than
about 125 GeV if $h$ can be a purely SM Higgs boson, and hence an
experimentally confirmed Higgs boson mass less than this value would be a
signal of new physics.\footnote{However, this conclusion may not hold if
certain assumptions are relaxed.  For example, see \cite{Isidori:2001bm}
for weaker lower bounds on the Higgs mass derived by requiring that the
Higgs potential remain stable with respect to quantum fluctuations at zero
temperature.}

Second, 115 GeV is a possible value of $m_{h}$ within the MSSM, but 
only if $\tan\beta$ is constrained to be larger than about 4.  That is 
because as described above, the tree level contribution is proportional to 
$\left| \cos2\beta\right| $ and to get a 
result as large as 115 it is necessary that $\left| \cos2\beta\right| $  
be essentially unity, giving a lower limit on $\tan\beta$ of about 4 .   
Even then the tree level piece can only contribute a maximum of $m_Z$ to   
$m_{h}$, and the rest comes from radiative corrections (mainly the 
top loop).  Numerically one gets
\begin{eqnarray}
m_{h}^{2}\approx(91)^{2}+(40)^{2}\left\{\ln\frac{m_{\widetilde{t}}^{2}}
{m_{t}^{2}}+...\right\},
\end{eqnarray}
where $m_{\widetilde{t}}^{2}$ is an appropriate average of the
two stop mass eigenstates.  The second term must supply about $(70\,{\rm 
GeV})^2$,
which is 
possible but constraining, and somewhat fine tuned.  When the MSSM
Higgs sector is extended, there  
are additional contributions to $m_h$ at tree level and $\tan\beta$ can be 
closer to unity.

\subsection{The large $\tan\beta$ regime}
\label{ltbpheno}
Phenomenologically there are a number of effects if $\tan \beta $ is 
large.  If any of these effects are seen they will greatly help determine 
the numerical value of $\tan \beta$. \ First, there are large 
(nondecoupling) radiative corrections
to the down-type quark masses (in particular the $b$ quark mass) and
couplings which then affect a number of observables
\cite{Hall:1993gn,Babu:1999hn,Logan:2000cz}.  The radiative
corrections are large because the $\tan\beta$ enhancement can
compensate the suppression from loop factors. Both $m_{b}$ and $b$
couplings can change significantly, with the signs of the change not
determined. In particular, Higgs couplings to $b\overline{b}$ can
change, which in turn changes Higgs branching ratios to photons and
other channels \cite{Akhoury:2001mz}. In the large $\tan\beta$ limit Higgs 
couplings are no longer simply proportional to mass \cite{Babu:1999hn}; 
for example, because certain enhanced corrections involve gluinos they 
contribute more to $h\longrightarrow b\overline{b}$ than to $h\longrightarrow 
\tau \overline{\tau}$ so the ratio of these branching ratios is no longer in 
the ratio of the masses squared. In many processes in addition $\tan 
\beta$ enters explicitly.  The large $\tan\beta$ corrections also have
considerable effects on FCNC, as will be discussed in
Section~\ref{CPflsect1}. To summarize briefly, the branching ratios
for rare decays such as {\it e.g.} the branching ratio for
$B_{s}\longrightarrow \mu ^{+}\mu ^{-} $ or $B_{d}\longrightarrow \tau
^{+}\tau ^{-}$ can be greatly enhanced
\cite{Babu:1999hn,Dedes:2001fv}, but there is little effect on
$B-\overline{B}$ mixing \cite{Babu:1999hn}. Studies of the important flavor
changing decay $b\longrightarrow s\gamma $ must be done carefully and
include resumed contributions if $\tan \beta $ is large.  Other
questions such as relic density calculations for neutralino cold dark
matter can be significantly affected by large $\tan \beta $.

There can be a variety of effects on collider signatures in the large 
$\tan\beta$ regime. The reason is that large $\tan \beta $ leads to both 
$\widetilde{\tau}$ and $\widetilde{b}$ \ having lighter masses 
than the other sleptons and squarks from two effects --- larger
off-diagonal terms in their mass matrices proportional to $m_{\tau }$ or
$m_{b}$ give a lighter eigenvalue, and RG running from 
a common mass at a high scale pushes the $\widetilde{\tau}$ and 
$\widetilde{b}$ masses lower. \
Effects have been studied in detail in \cite{Baer:1998bj} (see also 
\cite{Ellis:2001ms}). They lead to   
$\tau $-rich and $b$-rich events because branching ratios such as 
$\widetilde{C} \longrightarrow \widetilde{\tau}(\longrightarrow \tau 
\widetilde{N_{1}})\nu 
_{\tau }$ and $\widetilde{N}_{2}\longrightarrow
\widetilde{\tau}(\longrightarrow \tau
\widetilde{N_{1}})\overline{\tau},\widetilde{N}_{2}\longrightarrow \widetilde{b}
(\longrightarrow
b\widetilde{N_{1}})\overline{b},\widetilde{N}_{2}\longrightarrow
h\widetilde{N_{1}}$ are enhanced. That also reduces the particularly   
good trilepton signature since there are fewer $e\mu \mu $ and $ee\mu $ 
trileptons, but if the tau detection is good enough the signal can still
be seen in the $l\tau \tau ,l\overline{l}\tau ,b\overline{b}\tau $ etc channels  
($l=e,\mu )$. The production cross section for the Higgs state A grows  
with $\tan \beta $ so A may be visible at the Tevatron. The dominant
decay of stops may be $\widetilde{t_{1}}\longrightarrow \widetilde{\tau}\nu 
_{\tau }b.$

\subsection{From Tevatron and LHC data to ${\mathcal{L}}_{soft}$}
\label{tevatronsect} 
At present, all evidence for low energy supersymmetry is indirect.  
Although the evidence is strong, it could in principle be a series of
coincidences. Additional indirect evidence could come soon from FCNC rare
decays at the b-factories, proton decay, better understanding of the
$g_{\mu}-2$ SM theory (hadronic vacuum polarization and light-by-light
scattering), or CDM detection. However, finally it will be necessary to
directly observe superpartners and to show they are indeed superpartners.
This could first happen at the Tevatron collider at the Fermi National
Accelerator Laboratory, and is later expected to happen at the Large 
Hadron Collider (LHC) at CERN. Indeed, if supersymmetry is really the 
explanation for electroweak symmetry breaking then the soft masses should 
be $O(m_Z)$, as discussed in Section~\ref{finetunesect}.  
Furthermore, 
if the cross sections for superpartner production are typical electroweak
ones (or larger for gluinos), superpartners should be produced in
significant quantities at the Tevatron and the LHC. This subsection is 
dedicated to an examination of how superpartners might appear at the 
Tevatron and the LHC. We emphasize the lighter states here; of course, 
the possibility remains that superpartners are heavier than one might 
expect from fine-tuning, but below their natural upper limits of a few 
TeV, in which these states would be detectable first at the 
LHC.\footnote{However, one of us would like to emphasize that taking the 
fine-tuning arguments one step further and assuming the luminosity and the 
detectors are good enough to separate signals from backgrounds, it is 
possible to make the argument that if direct evidence
for superpartners does not emerge at the Tevatron (assuming it 
achieved design luminosity) then either nature does
not have low energy supersymmetry or there is something missing from our
understanding of low energy supersymmetry.  If superpartners do not appear
at the Tevatron, many will wait until the LHC has taken data to be
convinced nature is not supersymmetric, but one could argue (and one of 
us would like to stress this point) that it is 
unlikely that superpartners could be produced at the LHC if at least a 
few of them are not 
first produced at the Tevatron.}

The very nature of supersymmetry (accepting R-parity conservation) implies
that (with one possible exception) there can be no elegant, clear
signal that can convince an uninformed 
observer that a dramatic discovery has occurred, because superpartners are
being produced in pairs. Each decays into an LSP that escapes the  
detector, so there are two escaping particles carrying away mass and
energy.  No distribution can show a sharp peak, but rather several event
topologies will show excesses over the expected number of events from the 
SM.  Nevertheless, if the backgrounds are accurately known, as expected
since the backgrounds arise from (in principle) calculable SM 
processes, it will be possible to discover compelling evidence for signals 
beyond the SM. (The possible exception is that prompt photons could be
present for some signatures and is briefly described below.) After the 
excitement of that discovery the challenge of learning the underlying 
physics will begin.

Accepting that supersymmetry explains the origin of electroweak symmetry
breaking, the gluinos, neutralinos, and charginos are expected to be
rather light. Typically the lighter stop may be light as well due to
strong LR mixing in the top-squark sector. Sleptons may also be light,
though there is somewhat less motivation for that. One can list a number
of possible channels and look at the signatures for each. Almost all cases
require a very good understanding of the SM events that resemble the
possible signals, both in magnitude (given the detector efficiencies) and
the distributions. Missing transverse energy will be denoted by $\notE$.  
Until the ordering of the superpartner masses is known, it is necessary to
consider a number of alternative decays of $\widetilde{N}_{2},$
$\widetilde{C}_{1},$ $\widetilde{t} _{1},$ $\widetilde{g}$, {\it etc.}

An immediate complication is that certain excesses will come mainly from
one channel but others will have significant contributions from several. \
There will be too few events to make sharp cuts that might isolate one
channel \cite{Kane:1997yr}. \ Consequently it will be necessary to study
``inclusive signatures'' \cite{Kane:1997fa}.  Possible channels include
$l\notE$, $l\overline{l}\notE$, $\gamma l\notE$,$ j\notE$,$jj\notE$, {\it 
etc.}, 
where $l$
represents a charged lepton, $j$ an isolated, energetic jet, $\gamma$ an 
isolated, energetic photon, and $\notE$ missing transverse
energy.  They can arise from a variety of superpartner channels, such as
production of $\widetilde{C}_{1}^{+}+\widetilde{C}_{1}^{-}$,
$\widetilde{C}_{1}^{\pm}+\widetilde{N}_{1,2}$, 
$\widetilde{N}_{1}+\widetilde{N}_{2,3}$, $\widetilde{g}+\widetilde{g}$, 
$\widetilde{g}+\widetilde{C}_{1}^{\pm }$, $\widetilde{l}^{+}+\widetilde{l}^{-}$, 
{\it etc.}  If 
the excess arises
mainly from one channel it may be possible by kinematic methods such as
endpoints of spectra to deduce the masses of a certain 
subset of the superpartners. The following survey is meant to
illustrate the types of signals that could arise, not to be a full
catalog of possible signals for all theories. 

\paragraph{Neutralinos, charginos, and sleptons}
Let us consider several channels in detail, assuming $\widetilde{N}_1$ is 
the LSP.
\begin{itemize}
\item $\widetilde{N}_{1}+\widetilde{N}_{1}$: This channel is very hard to 
tag at a hadron collider, since both LSPs escape.

\item $\widetilde{N}_{1}+\widetilde{N}_{2,3}$: These channels can be
produced through an s-channel Z or a t-channel squark exchange. The
signatures depend considerably on the character of $\widetilde{N}_{2},$
$\widetilde{N}_{3}$. $\widetilde{N}_{1}$ escapes. If $\widetilde{N}_{2}$
has a large coupling to $\widetilde{N}_{1}+Z$ (for real or virtual Z)
then the $\widetilde{N}_{1}$ will escape and the Z will decay to $e$ or
$\mu$ pairs each 3$\%$ of the time, so the event will have missing energy
and a prompt (``prompt'' means energetic, appearing to originate in the
main event vertex and not a delayed one, and for leptons or photons,
isolated, {\it i.e.}, not in a jet of hadrons) lepton pair. There will 
also be
tau pairs but those are somewhat harder to identify. Or, perhaps
$\widetilde{N}_{2}$ is mainly photino and $\widetilde{N}_{1}$ mainly
higgsino, for which there is a large BR for
$\widetilde{N}_{2}\rightarrow\widetilde{N}_{1}+\gamma$ 
(see \cite{Baer:2002kv} and references therein
to the history of the calculation) and the signature
of $\widetilde{N}_{2}$ is one prompt $\gamma$ and missing energy. The
production cross section can depend significantly on the wave functions of
$\widetilde{N}_{1},\widetilde{N}_{2}.$  If the cross section is small for
$\widetilde{N}_{1}+\widetilde{N}_{2}$ it is likely \ to be larger for
$\widetilde{N}_{1}+\widetilde{N}_{3}.$  Most cross sections for lighter
channels will be larger than about 50 ${\rm fb}$, which corresponds to 200
events for an integrated luminosity of 2 ${\rm fb}^{-1}$ per detector.

\item $\widetilde{N}_{1}+\widetilde{C}_{1}$:  These states are produced
through s-channel $W^{\pm}$ or t-channel squarks. The $\widetilde{N}_{1}$
escapes, so the signature comes from the $\widetilde{C}_{1}$ decay, which
depends on the relative sizes of masses, but is most often
$\widetilde{C}_{1}\rightarrow l^{\pm}+\notE$. This is the signature if
sleptons are lighter than charginos ($\widetilde{C}
_{1}\rightarrow\widetilde{l}^{\pm}+\nu,$ followed by
$\widetilde{l}^{\pm}\rightarrow l^{\pm}+\widetilde{N}_{1})$, or if sneutrinos
are lighter than charginos by a similar chain, or by a three-body decay
($\widetilde{C}_{1}\rightarrow \widetilde{N}_{1}+{\rm virtual}
\hspace{0.2cm} W$,  
$W\rightarrow l^{\pm}+\nu)$. However, it is not guaranteed --- for example
if stops are lighter than charginos the dominant decay could be
$\widetilde{C}_{1}\rightarrow\widetilde{t}+b$. If the lepton
dominates, the event signature is then $l^{\pm}+\notE$, so it is necessary
to find an excess in this channel. Compared to the SM sources of such
events the supersymmetry ones will have no prompt hadronic jets.  The
supersymmetry events also have different distributions for the lepton
energy and for the missing transverse energy.

\item $\widetilde{N}_{2}+\widetilde{C}_{1}$: If $\widetilde{N}_{2}$ decays 
via a Z to $\widetilde{N}_{1}+l^{+}+l^{-}$ and $\widetilde{C}_{1}$ decays 
to $\widetilde{N}_{1}+l^{\pm},$   this channel gives the 
well-known {\it tri-lepton} 
signature: three charged leptons, $\notE$ , and no prompt jets, which may 
be relatively easy to separate
from SM backgrounds (see 
\cite{Barger:1998wn,Barger:1998hp,Matchev:1999yn,Baer:1999bq} 
for recent 
discussions of
the signature and backgrounds for the trileptons). 
But it may be that
$\widetilde{N}_{2}\rightarrow\widetilde{N}_{1}+\gamma,$ so the signature
may be $l^{\pm}+\gamma+\notE$.

\item $\widetilde{l}^{+}+\widetilde{l}^{-}$: Sleptons may be light enough to be
produced in pairs. Depending on masses and whether lepton-L or slepton-R 
is produced, they could decay via
$\widetilde{l}^{\pm}\rightarrow l^{\pm}+\widetilde {N}_{1},$
$\widetilde{C}_{1}+\nu$, $W+\widetilde{\nu}$. If $\widetilde{N}_{1}$ is 
mainly higgsino decays to it are suppressed by lepton mass factors, so
$\widetilde{l}^{\pm}\rightarrow l^{\pm}+\widetilde{N}_{2}$ may dominate,
followed by $\widetilde{N}_{2}\rightarrow\widetilde{N}_{1}+\gamma$.
\end{itemize}
For a complete treatment one should list all the related channels and
combine those that can lead to similar signatures.  The total sample may
be dominated by one channel but have significant contributions from
others, {\it etc.}  It should also be emphasized that these 
``backgrounds'' are 
not junk backgrounds that cannot be calculated, but
from SM events whose rates and distributions can be understood
if the appropriate work is done.  Determining these background rates is
essential to identify a signal and to identify new physics. This requires
powerful tools in the form of simulation programs, which in turn require
considerable expertise to use correctly.  The total production cross
section for all neutralino and chargino channels at the Tevatron collider
is expected to be between 0.1 and 10 ${\rm pb}$, depending on how light 
the superpartners are, so even in the worst case there should be several
hundred events in the two detectors (at design luminosity), and of course 
many more at the LHC.  
If the cross sections are on the 
low side it will require combining inclusive signatures to demonstrate 
new physics has been observed.

\paragraph{Gluinos} 
Gluinos can be produced via several 
channels, 
$\widetilde{g}+\widetilde{g}$, 
$\widetilde{g}+\widetilde{C}_{1}$, $\widetilde{g}+\widetilde{N}_{1}$, {\it etc.} 
As 
previously stated, if 
supersymmetry indeed explains electroweak symmetry breaking it would
be surprising if the gluino were heavier than about 500 GeV.  For such 
light gluinos the total production cross section should be large enough 
to observe gluinos at the Tevatron. The LHC will be sensitive to 
considerably
larger gluino masses, over 2 TeV. If all of its decays 
are three-body decays, {\it e.g.}
$\widetilde{g}\rightarrow \widetilde{q}+\overline{q}$ followed by $\widetilde{q}\rightarrow
q+\widetilde{C}_{1},$ etc, then the signature has energetic jets, $\notE$,
and sometimes charged leptons. There are two channels that are
particularly interesting and not unlikely to occur --- if $t+\widetilde{t}$ or
$b+\widetilde{b}$ are lighter than $\widetilde{g}$ then they will dominate because
they are two-body. \ The signatures can then be quite different, with
mostly $b$ and $c$ jets, and different multiplicity.

Gluinos and neutralinos are Majorana particles, and thus can decay either 
as particle or antiparticle.  If, for example, a decay
path $\tilde {g}\rightarrow\overline{t}(\rightarrow W^{-}\overline{b})+\widetilde{t}$
occurs, with $W^{-}\rightarrow e^{-}\nu,$ there is an equal probability
for $\tilde {g}\rightarrow e^{+}+\ldots$. This indicates that a pair of 
gluinos can give same-sign or opposite sign 
dileptons with equal probability! This result holds for any way of tagging 
the electric charge --- here leptons have been focused on since their 
charges are easiest to identify. The same result  
holds for neutralinos.  The SM allows no way to get prompt same-sign
leptons, so any observation of such events is a signal of physics beyond 
the SM and is very likely to be a strong indication of supersymmetry. 

\paragraph{Squarks}
Stops can be rather light, so they should be looked
for very seriously. They can be pair produced via gluons, with a cross
section that is about $1/8$ of the top pair cross section; the cross
section is smaller because of a p-wave threshold suppression for scalars
and a factor of four suppression for the number of spin states.  Stops 
could also be produced in top decays if they are lighter than
$m_{t}-M_{\widetilde{N}_{1}},$ and in gluino decays if they are lighter than
$m_{\widetilde{g}}-m_{t}$ (which is not unlikely).  Their most obvious decay
channel is $\widetilde{t}\rightarrow\widetilde{C}+b$, which will indeed
dominate if $m_{\widetilde{t}}>m_{\widetilde{C}}$.  If this relation does not
hold, it may still dominate as a virtual decay, followed by
$\widetilde{C}$ real or virtual decay (say to $W+\widetilde{N}_{1}),$ 
such that the final state is 4-body after $W$ decays and suppressed by
4-body phase space. That may allow the one-loop decay
$\widetilde{t}\rightarrow c+\widetilde{N}_{1}$ to dominate stop decay. As an
example of how various signatures may arise, if the mass ordering is
$t>\widetilde{C}_{1}>\widetilde{t}>\widetilde{N}_{1}$ and
$t>\widetilde{t}+\widetilde{N}_{1},$ then a produced $t\overline{t}$ pair will
sometimes (depending on the relative branching ratio, which depends on the
mass values) have one top decay to $W+b $ and the other to $c+\widetilde
{N}_{1},$ giving a $W+2$ jets signature, with the jets detectable by $b$
or charm tagging, and therefore excess $Wjj$ events.\\

An event was reported by the CDF collaboration \cite{Acosta:2001ib} from
Tevatron Run 1, $p\overline{p}\rightarrow ee\gamma\gamma\notE$, that is
interesting both as a possible signal and to illustrate a few pedagogical
issues.  The possibility that such an event might be an early signal of
supersymmetry was suggested in 1986 \cite{Haber:ht}.  Such an event can 
arise \cite{Ambrosanio:1996jn} if a selectron pair is produced and if the 
LSP is higgsino-like, for which the decay of the selectron to
$e+\widetilde{N}_{1}$ is suppressed by a factor of $m_{e}$. Then 
$\widetilde{e}\rightarrow e+\widetilde{N}_{2}$ dominates, followed by
$\widetilde{N} _{2}\rightarrow\widetilde{N}_{1}+\gamma$.  The only way to
get such an event in the SM is production of $WW\gamma\gamma$ with both
$W\rightarrow e+\nu,$ with an overall probability of order 10$^{-6}$ for
such an event in Run 1.  Other checks on kinematics, cross section for
selectrons, {\it etc.}, allow for an interpretation in the context of 
supersymmetry, and the resulting
values of masses do not imply any that must have been found at LEP or as
other observable channels at CDF.  There are many consistency conditions
that must be checked if such an interpretation is allowed and a number of
them could have failed but did not.  
If this event were a signal additional ones would
soon occur in Run 2.  Because of the needed branching ratios there would
be no trilepton signal at the Tevatron, since $\widetilde{N}_{2}$ decays
mainly into a photon instead of $l^{+}l^{-},$ and the decay of
$\widetilde{N}_{3}$ would be dominated by $\widetilde{\nu}\nu$. Even with 
limited luminosity at the Tevatron it
will be cleaner there if such an event is real well before the LHC takes
data.

Once the signals are found, experimenters will be able to make some
determinations of some superpartner masses and cross sections (times BR).
Our real goal is to learn the Lagrangian parameters which will be
difficult from limited data. In spite of the difficulty in measuring the
needed parameters, a number of aspects of the data will allow one to make
progress toward learning how supersymmetry is broken and how the breaking
is transmitted.  Different mechanisms imply various qualitative features
that can point toward the correct approach.  For example, one clue is
whether the events have prompt photons, {\it i.e.} isolated energetic
photons emerging from the superpartner decays and therefore the primary
event vertex.  Gravity-mediated supersymmetry breaking with large $\mu $
gives a bino-like LSP, so decays of heavier produced superpartners to the
LSP do not give photons.  If $\mu $ is small the LSP is higgsino-like so
decays to light quarks and leptons are suppressed and decays of heavier
neutralinos give photons.  In gauge-mediated models the gravitino is light
so any neutralinos lighter than the Z, as the LSP is likely to be, decay
to photon plus gravitino so every event has two photons unless the NLSP
happens to be very long lived and does not decay in the detector.  While
an explicit measurement of $\mu $ is difficult because of the inability to
invert the equations relating observables and parameters, the combination
of information from knowing the dominant inclusive signatures and
approximate superpartner masses may allow an approximate determination of
the value of $\mu$.  A brief summary is presented in \cite{Kane:1997fa}
and in Table 5. 


\smallskip
\hspace{-.4in}
\begin{tabular}{ccccccc}
Inclusive & ${\rm SUGRA},$ & ${\rm SUGRA},$ & ${\rm GMSB},$ &
Unstable &  $\underline{\left\langle D\right\rangle}$ \\ 
\underline{Signatures} & \underline{large $\mu$} & \underline{small
$\mu$} & \underline{low scale} & \underline{LSP} & \\ 
Large E$_{T}$ & yes & yes & yes & no &  &  \\
Prompt $\gamma^{\prime}s$ & no & sometimes & yes (but...) & no &  &  \\
Trilepton events & yes & no & no & no &  &  \\
Same-sign dileptons &  &  &  &  &  &  \\
Long-lived LSP &  &  &  &  &  &  \\
$\tau-$ rich &  &  &  &  &  &  \\
$b-$ rich &  &  &  &  &  &  
\end{tabular}

$\bigskip$

One can add both rows and columns --- this is work in progress.  This
approach also shows how to combine top-down and bottom up approaches ---
one uses top-down analysis to identify the columns and fill in the
missing entries in the table.  By simply identifying qualitative
features of the channels with excesses one can focus on a few or even
one type of theory.  Then detailed study can let one zoom in on the
detailed structure of the underlying theory and its high energy
features.  With such an approach one can partly bypass the problem of
not being able to fully isolate the Lagrangian explicitly.  One will not
be able to prove that specific superpartners are being observed with
this ``inclusive'' analysis, but we can gamble and leave the proof for
later.  
In this table 
${\rm SUGRA}$ stands for gravity-mediated supersymmetry breaking, 
${\rm GMSB}$ for gauge-mediated supersymmetry breaking,
$\left\langle D\right\rangle
$ for supersymmetry breaking by an D term VEV, {\it etc.}
Each inclusive observation allows one to carve away part of the
parameter space, and the remaining parts point toward the underlying
high scale theory.  One does not need to measure every soft parameter to
make progress, because the patterns, the mass orderings, {\it etc.}, imply
much about the underlying theory --- if one understands the theory.

What we want to emphasize is that since supersymmetry is a well-defined 
theory it 
is possible to calculate its predictions for many processes and use them
all to constrain parameters.  Because of this even at hadron colliders
the situation may not be so bad.  By combining information from several
channels each with almost-significant excesses we can learn a lot about
the parameters and perhaps about the basic theory itself.   In practice
we may be lucky, and find that some parameters put us in a 
region of parameter space where measurements are possible.  For example,
if $\tan\beta$ is very large it may be possible to observe 
$B_{s}\rightarrow \mu^+ \mu^-$ at the Tevatron (see {\it e.g.}
\cite{Babu:1999hn})  
and  therefore get a measurement of $\tan\beta$.  Data from the Higgs 
sector, the way the electroweak symmetry is broken, how the hierarchy 
problem is solved, gauge coupling unification, the absence of LEP signals, 
rare decays, cold dark matter detectors, $g_{\mu}-2,$ proton decay, the 
neutrino sector, and other non-collider physics will be very important to 
combine with collider data to make progress.

Although it might look easy to interpret any
nonstandard signal or excess as supersymmetry, a little thought
shows not because supersymmetry is very constrained.  As illustrated in 
the above examples, a
given signature implies an ordering of superpartner masses, which implies
a number of cross sections and decay branching ratios. All must be right.
All of the couplings in the Lagrangian are determined, so there is little
freedom once the masses are fixed by the kinematics of the candidate
events. Once masses are known, contributions to rare decays, CDM
interactions, $g_{\mu}-2$, {\it etc.}, are strong constraints.  To prove a
possible signal is indeed consistent with supersymmetry one has also to
check that certain relations among couplings are indeed satisfied.  Such
checks will be easy at lepton colliders, but difficult at hadron
colliders; however, hadron collider results are likely to be available at
least a decade before lepton collider results. There can of course be
alternative interpretations of any new physics. However, it
should be possible to show whether the supersymmetry interpretation is
preferred --- a challenge which would be enthusiastically welcomed.
 
In 2008 or soon after we will have data about superpartner and Higgs
boson production at LHC. Assuming weak-scale supersymmetry is indeed
present, the LHC will be a superpartner factory. \ There has been a great
deal of study of how to measure certain superpartner masses (and mass
differences) at LHC, and some study of how to measure superpartner cross
sections. \ The literature can be traced from the summary given in
\cite{Branson:2001pj}.

But almost none of this work by the detector groups and theorists has
studied the questions on which this review is focused, namely how to learn
the parameters of the soft Lagrangian. \ The issues raised particularly in
Section~\ref{expdisc} about inverting the equations relating data to soft 
parameters have hardly been addressed yet and there is a great deal of 
work to do here.  The first goal is to find
direct signals of supersymmetry at colliders --- that is paramount, and
deserves the emphasis it has. \ Ideally, next one would measure masses and
cross sections, with methods based on extensive study
\cite{Branson:2001pj}.  But first,
only 32 of the 105 soft parameters are masses, and second, at hadron 
colliders there are in principle not enough observables to invert the 
equations to go from masses and cross sections
(assuming those can be measured) to $\tan \beta $ and soft parameters. \
Very little study has been devoted to this inversion problem, and to
relating the data to the physics of the underlying theory. \ Some activity
can be traced from \cite{Allanach:2001qe}.

Linear collider data will be essential for more complete measurements of 
the soft parameters. \ Several groups
have addressed inverting the equations to obtain the soft parameters using
future linear collider data \cite{Kneur:1999ab,Kneur:1999nx,Kneur:1998gy}.
Most of this work relies on measurements at lepton colliders, in practice
future linear $e^{+}e^{-}$ colliders. \ The extra observables arising from
polarized beams, the small errors that can be achieved there, and the
ability to measure cross sections combine to give sufficient data in some
cases to carry out the inversions. \ Additional information will come from
running the linear collider at more than one energy, which gives additional
independent observables since the coefficients depend on energy; this
additional information does not seem to have been used so far in the
studies. \ Learning the soft parameters from linear collider data,
particularly the phases, has also been studied 
in \cite{Barger:2001nu,Choi:2001ww,Choi:2000ta}.

A somewhat different and useful approach has been begun by Zerwas and
collaborators, who specify the soft parameters at a high scale, run them
down to the electroweak scale, assume they are somehow measured with 
assumed
errors at LHC plus a linear collider, and run back up to see how well the
parameters can be recovered at the high scale. \ They have studied some
obstacles to doing this, such as infrared fixed point behavior, though
they have not studied most of the obstacles which are described 
briefly in the conclusions of this review and more extensively in {\it 
e.g.} in \cite{Kane:2002ss},
nor have they studied how to actually measure the soft parameters
at the electroweak scale from LHC. \ A basic result of these analyses is 
that measurement accuracy will be very valuable in making progress.

Recently there has been some discussion \cite{Kane:2002ss} of the more
general problem of going from limited data on superpartners, plus data on
rare decays, magnetic moments, electric dipole moments, cold dark matter
data, and more to the soft Lagrangian and perhaps to learning aspects of
the underlying theory without complete measurements of \lsoft. We will 
briefly return to such issues at the conclusion of the review.

\subsection{Benchmark models}
\qquad Benchmark models can be of great value. \ They force one to
understand the theory well enough to produce concrete models, and help
theorists gain insight into which features of the theory imply certain
phenomena and vice versa. \ They help plan and execute experimental
analyses, allow quantitative studies of triggers and detector design, and
can affect setting priorities for experimental groups. \ They suggest what
signatures can be fruitful search channels for new physics, and provide
essential guidance about what backgrounds are crucial to understand, and
what systematic errors need to be controlled. \ To be precise, here we
define a benchmark model as one in the framework of softly-broken
supersymmetry and based on a theoretically motivated high scale approach. \
At the present time such models cannot be specified in sufficient detail to
determine a meaningful spectrum of superpartners and their interactions
without assumptions and approximations, and those should be ones
that make sense in the context of the theory rather than arbitrary ones. \
As theory improves it should be increasingly possible to derive the main
features of the models. \ Eventually it would be good to have $\mu $ and $%
\tan \beta $ determined by the theory instead of being fixed by 
electroweak symmetry breaking conditions.

In this section we give a brief survey of some of the benchmark models
proposed in recent years (see \cite{Allanach:2002nj} for a synthesis of 
many of the proposed benchmarks).  
The proposed benchmark models generally fall in two classes: (i)  
supergravity models, and (ii) models based on alternative supersymmetry
mediation scenarios. The supergravity benchmarks (see {\it e.g.} 
\cite{Battaglia:2001zp,points_dAix,Allanach:2002nj}) typically 
encode the minimal choice of supergravity couplings.  This class of models 
is known as minimal supergravity (mSUGRA), or in a slightly broader sense, 
the constrained MSSM (CMSSM).
With a number of universality assumptions (see the discussion in 
Section~\ref{validmssmsect} ), these models contain the
following four parameters: 
\begin{equation}
m_{1/2}, \hspace{0.3cm}, m_0 \hspace{0.3cm}, \tan \beta,
\hspace{0.3cm} {\rm sign}(\mu ).
\end{equation}
There are also benchmarks based on other popular alternative
supersymmetry-breaking scenarios, such as gauge mediation and anomaly
mediation, with generically different patterns of soft mass 
parameters, as discussed in Section~\ref{lsoftmodelsect}.

\begin{table}[t]
\begin{center}
\begin{tabular}{cccccccc}
\hline\hline
SPS & \multicolumn{6}{c}{Point \hspace{3em}} & Slope\\ 
\hline\hline
mSUGRA: & $m_0$ & $m_{1/2}$ & $A_0$ & $\tan\beta$ & & & \\
\hline
1a & 100  & 250 &   -100 & 10 & & & $m_0 = -A_0 = 0.4 \, m_{1/2}$, 
                                    $\; m_{1/2}$ varies\\
1b & 200  & 400 &      0 & 30 & & & \\
2  & 1450 & 300 &      0 & 10 & & & $m_0 = 2  m_{1/2} + 850 $ GeV ,
                                    $\; m_{1/2}$ varies\\
3  &   90 & 400 &      0 & 10 & & & $m_0 = 0.25  m_{1/2} - 10$  GeV,
                                    $\; m_{1/2}$ varies \\
4  &  400 & 300 &      0 & 50 & & & \\
5  &  150 & 300 &  -1000 &  5 & & & \\
\hline\hline
mSUGRA-like: & $m_0$ & $m_{1/2}$ & $A_0$ & $\tan\beta$ & 
               $M_1$ & $M_2 = M_3$ & \\
\hline      
6  &  150 & 300 &      0 & 10 & 480 & 300 & $M_1 = 1.6 M_2$, 
                              $m_0 = 0.5 M_2$, $\; M_2$ varies \\ 
\hline\hline         
GMSB: & $\Lambda/10^3$ & $M_{\rm mes}/10^3$ & $N_{\rm mes}$ & $\tan\beta$ & & & \\ 
\hline     
7  &   40 &  80 &      3 & 15 & & & $M_{\rm mes}/ \Lambda = 2$, 
                                    $\; \Lambda$ varies \\    
8  &  100 & 200 &      1 & 15 & & & $M_{\rm mes}/ \Lambda = 2$, 
                                        $\; \Lambda$ varies \\   
\hline\hline 
AMSB: & $m_0$ & $m_{\rm aux}/10^3$ & & $\tan\beta$ & & & \\ 
\hline         
9  &  450 &  60 &        & 10 & & & $m_0 = 0.0075 \, m_{\rm aux}$,
                                    $\; m_{\rm aux}$ varies \\
\hline\hline   
\end{tabular}
\end{center}
\caption{The parameters (which refer to 
{\sl ISAJET} version 7.58) for the Snowmass Points and Slopes (SPS). 
The masses and scales are given in GeV.
All SPS are defined with $\mu > 0$. The parameters $M_1$, $M_2$, $M_3$
in SPS 6 are understood to be taken at the GUT scale. The value of the
top-quark mass for all SPS is $m_t = 175$~GeV.
\label{params}}
\end{table}
A typical collection of those benchmark models, the Snowmass
Points and Slopes,  are collected in Table~\ref{params}, taken from 
\cite{Allanach:2002nj}. The low energy spectra which result from 
these points can be found in \cite{Ghodbane:2002kg}.  The
bounds which have been used in the selection of model 
points include: (i) The relic abundance, (ii) LEP exclusion limits for the 
Higgs mass, (iii) the $ b \rightarrow s \gamma$ constraint, and (iv) 
the muon $g_{\mu}-2$ constraint. 
The phenomenological analyses of such models has evolved into a 
sophisticated industry. Several well-developed codes exist to handle 
different parts of the calculation with high accuracy. The resulting 
benchmark models pass all the existing known experimental bounds. Such 
models can clearly serve as a very useful guide for present, future, and 
forthcoming experimental searches. 

We now comment on several features of these benchmark models,
focusing on their fine-tuning properties. In the mSUGRA models, larger
gaugino masses, in particular the gluino mass, are quite typical.  This
feature is due to the imposed degeneracy between the input values of the
gluino and other gaugino masses and the experimental limits on the
chargino mass.  Another underlying factor here is the rather stringent
Higgs mass bound from LEP. Within the MSSM, the current Higgs lower bound
from direct searches points to heavier squark masses, particularly
for the stops. This will in turn require heavier gluino masses, because
the gluino mass has a dominant role in the RG running of the squark soft
masses. However, it is known that a larger gluino mass will imply a larger
fine-tuning for electroweak symmetry breaking, which represents a 
potential problem. The higher fine-tuning would appear to 
require certain nontrivial relations to exist between the soft mass 
terms.\footnote{However, the ``focus-point'' region, point 2 in the 
SPS table, is a possible solution to this problem. In this region, 
the low energy value of the Higgs soft parameter $m^2_{H_u}$ is relatively 
insensitive to the input value in the focus point region 
\cite{Feng:1999zg}. Thus, within this region when
the focus point conditions are satisfied, the electroweak symmetry
breaking is not fine-tuned.}  
In the gauge mediation and anomaly
mediation models, the patterns of the gaugino masses are quite different 
than in the mSUGRA models.  Unfortunately, in both of those scenarios, the 
gluino is typically even heavier and thus the fine-tuning problem is not 
in general mitigated. However, gauge 
mediation models generically have a much lower supersymmetry breaking 
scale than the mSUGRA models, which can change the
analysis of fine-tuning significantly \cite{Kane:2002ap}. On the other
hand, electroweak symmetry breaking naively may be harder to achieve  
because $m_{H_u}^2$ will run less negative.

Arguably, all of the above benchmark models are intrinsically
``bottom-up'' models, with their main motivation arising solely from low
energy phenomenology. One can then consider the question of whether such 
scenarios are also motivated from the ``top-down"  perspective, {\it e.g.} 
within a more fundamental theory such as string theory.  Given 
what is currently known about the moduli space of the string theory vacua, 
one can ask the question of whether models resembling some of the 
above benchmark points are generic. mSUGRA models do represent a 
particular corner of that
(very big) moduli space. However, it is fair to say there are other points
at least as natural as the mSUGRA point from a model building point of
view. The same question must be addressed for gauge mediation and
anomaly mediation as well.

Another recently-proposed set of benchmark models which attempts to 
address these issues was presented in \cite{Kane:2002qp}.  This analysis 
uses full one-loop expressions for soft parameters and incorporates three 
classes of string-based models. 
The assumptions are different from the more familiar
constrained MSSM scenarios. 
\ One class of models assumes the dilaton is
stabilized by nonperturbative contributions to the K\"{a}hler potential. 
In this class  model the vacuum energy is set to zero 
and the models are determined by only three parameters: $\tan \beta ,$ 
$m_{3/2},$ and a parameter called $a_{np}$ related to nonperturbative
corrections. \ A further class of models is based on string approaches 
where supersymmetry breaking is due to VEVs of moduli fields. The
``racetrack'' method for dilaton stabilization is used in this class of models
\ They are
parameterized by $\tan \beta ,$ $m_{3/2},$ a moduli VEV, and a Green-Schwarz
coefficient $\delta _{GS}.$ \ The final class is based on partial
gauge-mediated models where the mediating particles are high scale ones that
actually arise in the spectrum of the models. \ They are parametrized
again by $\tan\beta ,$ $m_{3/2},$ and by three parameters that
determine the quantum numbers of the high-scale fields. \  
\begin{table}[th]
{\begin{footnotesize} {\begin{center}
\begin{tabular}{|l|c|c|c|c|c|c|c|} \cline{1-8}
Point & A & B & C & D & E & F & G \\  \cline{1-8} $\tan\beta$ & 10
& 5 & 5 & 45 & 30 & 10 & 20
\\ $\Lambda_{UV}$ & $2 \times 10^{16}$ & $2 \times 10^{16}$ & $2 \times
10^{16}$ & $2 \times 10^{16}$ & $2 \times 10^{16}$ & $8 \times
10^{16}$ & $8 \times 10^{16}$
\\ \cline{1-8} $M_{1}$ & 198.7 & 220.1 & 215.3 & 606.5 & 710.8
& 278.9 & 302.2
\\ $M_{2}$ & 172.1 & 162.3 & 137.3 & 195.2 & 244.6 & 213.4 & 231.2
\\ $M_{3}$ & 154.6 & 122.3 & 82.4 & -99.2 & -89.0 & 525.4 & 482.9
\\ \cline{1-8} $A_t$ & 193.0 & 204.8 & 195.4 & 286.0 & 352.5 & 210.7 & 228.2
\\ $A_b$ & 205.3 & 235.3 & 236.3 & 390.6 & 501.5 & 211.6 & 229.2
\\ $A_{\tau}$ & 188.4 & 200.0 & 188.9 & 158.1 & 501.5 & 210.3 & 227.8
\\ \cline{1-8} $m_{Q_3}^{2}$ & $(1507)^2$ & $(3216)^2$ &
$(4323)^2$ & $(2035)^2$ & $(2144)^2$ & $(286)^2$ & $(276)^2$
\\ $m_{U_3}^{2}$ & $(1504)^2$ & $(3209)^2$ & $(4312)^2$ &
$(1487)^2$ & $(1601)^2$ & $(290)^2$ & $(281)^2$
\\ $m_{D_3}^{2}$ &  $(1505)^2$ & $(3213)^2$ & $(4319)^2$ & $(1713)^2$ &
$(1870)^2$ & $(287)^2$ & $(277)^2$
\\ $m_{L_3}^{2}$ & $(1503)^2$ & $(3208)^2$ & $(4312)^2$ & $(1361)^2$ &
$(1489)^2$ & $(125)^2$ & $(135)^2$ \\ $m_{E_3}^{2}$& $(1502)^2$ &
$(3206)^2$ & $(4308)^2$ & $(756)^2$ & $(1139)^2$ & $(140)^2$ &
$(152)^2$
\\ \cline{1-8} $m_{Q_{1,2}}^{2}$ & $(1508)^2$ & $(3220)^2$ & $(4328)^2$ &
$(2347)^2$& $(2347)^2$ & $(286)^2$ & $(276)^2$
\\ $m_{U_{1,2}}^{2}$ & $(1506)^2$ & $(3215)^2$ & $(4321)^2$ & $(2050)^2$
& $(2050)^2$ & $(290)^2$ & $(281)^2$ \\ $m_{D_{1,2}}^{2}$ &
$(1505)^2$ & $(3213)^2$ & $(4319)^2$ & $(1919)^2$ & $(1919)^2$ &
$(287)^2$ & $(277)^2$
\\ $m_{L_{1,2}}^{2}$ & $(1503)^2$ & $(3208)^2$ &
$(4312)^2$ & $(1533)^2$ & $(1533)^2$ & $(125)^2$ & $(135)^2$
\\ $m_{E_{1,2}}^{2}$ & $(1502)^2$ & $(3206)^2$ & $(4308)^2$ & $(1252)^2$
& $(1252)^2$ & $(140)^2$ & $(152)^2$   \\ \cline{1-8}
$m_{H_u}^{2}$ & $(1500)^2$ & $(3199)^2$ & $(4298)^2$ & $-(797)^2$
& $-(331)^2$ & $(125)^2$ & $(135)^2$
\\ $m_{H_d}^{2}$ & $(1503)^2$ & $(3208)^2$ & $(4312)^2$ & $(858)^2$
& $(1392)^2$ & $(125)^2$ & $(135)^2$ \\ \cline{1-8}
\end{tabular}
\end{center}}
\end{footnotesize}}
{\caption{\label{tbl:inputs}\footnotesize{\bf Soft Term Inputs}.
Initial values of supersymmetry-breaking soft terms in GeV,
including the full one-loop contributions, at the initial scale
given by $\Lambda_{UV}$. All points are taken to have $\mu > 0$.}}
\end{table}

\begin{table}[p]
{\begin{footnotesize} {\begin{center}
\begin{tabular}{|l|c|c|c|c|c|c|c|} \cline{1-8}
Point & A & B & C & D & E & F & G \\ \cline{1-8} $\tan\beta$ & 10
& 5 & 5 & 45 & 30 & 10 & 20
\\ $\Lambda_{UV}$ & $2 \times 10^{16}$ & $2 \times 10^{16}$ & $2 \times
10^{16}$ & $2 \times 10^{16}$ & $2 \times 10^{16}$ & $8 \times
10^{16}$ & $8 \times 10^{16}$
\\ $m_{3/2}$ & 1500 & 3200 & 4300 & 20000 & 20000 & 120 & 130 \\
\cline{1-8} $M_1$ & 84.0 & 95.6 & 94.7 & 264.7 & 309.9 & 106.2 &
115.7
\\ $M_2$ & 133.7 & 127.9 & 108.9 & 159.0 & 198.5 & 154.6 & 169.6
\\ $M_3$ & 346.5 & 264.0 & 175.6 & -227.5 & -203.9 & 1201 & 1109 \\
\cline{1-8} $m_{\widetilde{N}_1}$ & 77.9 & 93.1 & 90.6 & 171.6 & 213.0 &
103.5 & 113.1
\\ $m_{\widetilde{N}_{2}}$ & 122.3 & 132.2 & 110.0 & 264.8 & 309.7 &
157.6 & 173.1
\\ $m_{\widetilde{C}_{1}^{\pm}}$ & 119.8 & 131.9 & 109.8 & 171.6 & 213.0 & 157.5 & 173.0
\\ $m_{\widetilde{g}}$ & 471 & 427 & 329 & 351 & 326 & 1252 & 1158
\\ $\widetilde{B} \; \% |_{\rm LSP}$ & 89.8 \% & 98.7 \% &  93.4 \% &
 0 \% & 0 \% & 99.4 \% & 99.4 \%
\\ $\widetilde{W}_{3} \% |_{\rm LSP}$ & 2.5 \% &  0.6 \% & 4.6 \% & 99.7 \% & 99.7 \% &
0.1 \% & 0.06 \%
\\ \cline{1-8} $m_{h}$ & 114.3 & 114.5 & 116.4 & 114.7 & 114.9 &
115.2 & 115.5
\\ $m_{A}$ & 1507 & 3318 & 4400 & 887 & 1792 & 721
& 640
\\ $m_{H}$ & 1510 & 3329 & 4417 & 916 & 1821 & 722 & 644 \\
$\mu$ & 245 & 631 & 481 & 1565 & 1542 & 703 & 643
\\ \cline{1-8} $m_{\widetilde{t}_{1}}$ & 947 & 1909 & 2570 & 1066
& 1105 & 954 & 886 \\ $m_{\widetilde{t}_{2}}$ & 1281 & 2639 & 3530 &
1678 & 1897 & 1123 & 991
  \\ $m_{\widetilde{c}_{1}}$, $m_{\widetilde{u}_{1}}$ & 1553 & 3254
& 4364 & 2085 & 2086 & 1127 & 1047
\\ $m_{\widetilde{c}_{2}}$, $m_{\widetilde{u}_{2}}$ & 1557 & 3260 & 4371 & 2382 &
2382 & 1132 & 1054
\\ \cline{1-8} $m_{\widetilde{b}_{1}}$ & 1282 & 2681 & 3614 & 1213 &
1714 & 1053 & 971 \\ $m_{\widetilde{b}_{2}}$ & 1540 & 3245 & 4353 &
1719 & 1921 & 1123 & 1037
\\ $m_{\widetilde{s}_{1}}$, $m_{\widetilde{d}_{1}}$ & 1552 & 3252 & 4362
& 1950 & 1948 & 1126 & 1045
\\ $m_{\widetilde{s}_{2}}$, $m_{\widetilde{d}_{2}}$ & 1560 & 3261 & 4372 & 2383 &
2384 & 1135 & 1057
\\ \cline{1-8} $m_{\widetilde{\tau}_{1}}$ & 1491 & 3199 & 4298 & 559 & 1038
& 153 & 135
\\ $m_{\widetilde{\tau}_{2}}$ & 1502 & 3207 & 4308 & 1321 & 1457 &
221 & 252 \\ $m_{\widetilde{\mu}_{1}}$, $m_{\widetilde{e}_{1}}$ & 1505 &
3207 & 4309 & 1274 & 1282 & 182 & 196
\\ $m_{\widetilde{\mu}_{2}}$, $m_{\widetilde{e}_{2}}$ & 1509 & 3211 & 4313 & 1544 &
1548 & 200 & 217 \\ $m_{\widetilde{\nu_3}}$ & 1500 & 3206 & 4307 &
1314 & 1453 & 183 & 198  \\ \cline{1-8}
\end{tabular}
\end{center}}
\end{footnotesize}}
{\caption{\label{tbl:spectraSUM}\footnotesize {\bf Sample
Spectra}. All masses are in GeV.  For the purposes of calibrating
these results with those of other software packages we also
provide the running gaugino masses at the scale $M_Z$, which
include NLO corrections.}}
\end{table}

The phenomenology of benchmark models is most strongly determined by
whether they have gaugino mass degeneracy or not. 
\ In the set of benchmark models mentioned above, tree-level
contributions to gaugino masses are suppressed, so one-loop 
contributions are significant and remove degeneracy. \ One might worry that
gaugino mass degeneracy is implied by gauge coupling unification. \ That is
not so because the tree-level suppression of gaugino masses happens but not
the tree-level suppression of gauge couplings. 
\ More theoretically, gaugino
masses arise from one VEV of the F components of the moduli fields 
(including the dilaton), while the gauge 
couplings from the VEV of the scalar component of the dilaton
supermultiplet. \  The RG invariance of $M_{a}/g_{a}^{2}$
only holds at tree level as well. \ Further, gaugino mass degeneracy plus
constraints from data on $M_{1}$ and $M_{2}$ necessarily lead to fine-tuning
with respect to electroweak symmetry breaking, so phenomenologically there 
is good reason to be concerned about
imposing gaugino mass degeneracy and about taking its implications too
seriously. \ While the models of \cite{Kane:2002qp} do not require large
cancellations to get the value of $m_Z,$ several still have a large 
$m_{3/2}.$ \ At the present time there are no benchmark models in 
the literature that have all soft parameters and superpartner masses of 
order at most a few times $m_Z.$ \ 

For concreteness, we reproduce here the soft parameters in
Table~\ref{tbl:inputs} and the resulting low energy MSSM parameters in
Table~\ref{tbl:spectraSUM} of the seven benchmark models 
of \cite{Kane:2002qp}. \ These allow the reader to get a feeling for the
parameter values that such models give.\footnote{Although both the soft
term inputs and resulting mass spectra look rather complicated, recall 
that these models are specified in terms of only a few fundamental 
parameters (similar to the more familiar minimal SUGRA models), with the 
soft term inputs given by specific functions of these parameters.} \ These 
models are consistent with all collider constraints
and indirect constraints such as cold dark matter, loop-induced rare
decays, $ g_{\mu }-2,$ {\it etc.} \ They all have some superpartners light
enough to give signals observable at the Tevatron collider with a few
$fb^{-1}$ of integrated luminosity, with signatures that can be studied. 
One possible signature of gluinos studied in \cite{Kane:2002qp}, four jets 
plus large missing transverse energy plus two soft isolated prompt charged 
pions, was suggested by the string-based partial gauge-mediation models 
and had not previously been thought of phenomenologically. \ It is 
encouraging that such stringy approaches can lead to new phenomenology. \ 
Further phenomenology is studied in
\cite{Kane:2002qp}. \ They also begin study of a possibly useful approach
to relating limited data to the underlying theory --- if one makes scatter
plots of which theories give various inclusive signatures (such as the
number of trilepton events versus the number of events with opposite sign
dileptons plus jets) one finds that different string-based approaches lie
in different parts of the plots. \ If such plots can be made for several
inclusive signatures, and for rare decays or quantities such as 
$g_{\mu}-2$ that are sensitive to supersymmetry, the results may help 
point to the type of string-based models which might be relevant, and help 
focus attention toward fruitful directions.

\setcounter{footnote}{0}
\section{Extensions of the MSSM}
\label{extensionsect}
Throughout most of this review, we have assumed that MSSM is the
correct and complete parameterization of the low energy effective
Lagrangian with softly broken supersymmetry. Although this is quite a 
well-motivated assumption, extensions of this model may prove to be 
inevitable theoretically or experimentally. In this
section, we discuss several simple extensions of the MSSM
(though we admittedly do not provide an exhaustive or comprehensive
survey), with an emphasis on how the phenomenology can change with respect
to the MSSM.

\subsection{The minimal supersymmetric seesaw model}
\label{neutrinosect}
This review has mainly focused on the MSSM, in which
there are no right-handed neutrinos below the GUT scale but well above 
the electroweak scale.  If the slepton
mass matrices at the GUT scale are diagonal in flavor space, three
separate lepton numbers $L_e, L_{\mu}, L_{\tau}$ would be conserved also
at low energies since the RGEs would preserve these symmetries just as in
the SM. The convincing recent evidence for atmospheric
\cite{Fukuda:1998mi} and solar neutrino \cite{Ahmad:2001an} oscillations
seems to imply the existence of neutrino masses.  An attractive
interpretation of the smallness of neutrino masses is in terms of a
seesaw mechanism \cite{seesaw1,seesaw2,Mohapatra:1979ia}, which, together
with the atmospheric neutrino data, implies that there is at least one
right-handed neutrino with a lepton number violating Majorana mass below
the GUT scale.\footnote{Such an extension of the MSSM is 
also well-motivated in particular from a supersymmetric grand unification 
model (SUSY-GUTs) point of view, as many GUT models (such as 
$SO(10)$) naturally contains heavy right-handed neutrinos. There are many 
studies along this direction in the literature 
\cite{Babu:1998wi,Mohapatra:1999em,Albright:1999ux,Babu:2002en,Mohapatra:1999zr,Fritzsch:1999ee}.} 
In the framework of seesaw model, the requirement of a high energy scale 
at which lepton number is violated lends support to the notion of at least 
one physical high energy scale in nature which is hierarchically much 
larger than the electroweak scale, in addition to the scale where the 
gauge couplings unify and the Planck scale.  However it does mean that the 
discussion in this review must be extended to include the presence of 
right-handed neutrinos below the GUT scale. The purpose of this section is 
to discuss the new phenomenological features that this implies.

Consider for definiteness the addition of three right-handed
neutrinos to the MSSM, and work in the diagonal basis 
of right-handed Majorana masses where the three right-handed neutrinos 
have large Majorana
mass eigenvalues $M_{R_1},M_{R_2},M_{R_3}$. Such a framework has been 
called 
the
{\it minimal supersymmetric seesaw model}. The three right-handed 
neutrinos  
couple to the lepton doublets via a new Yukawa matrix $Y_{\nu}$ and the   
soft supersymmetry-breaking Lagrangian will involve a new soft trilinear
mass matrix $\widetilde{A}_{\nu}$ and a new soft mass matrix for the
right-handed sneutrinos $m_N^2$. The new terms which must be added to the
superpotential and the soft supersymmetry-breaking Lagrangian are
\begin{equation}
\Delta W = -\epsilon_{ab}\hat{H}_u^a\hat{L}^b_iY_{{\nu}_{ij}}\hat{N}^c_j 
+\frac{1}{2}\hat{N}^c_i M_{R_i}\hat{N}^c_i
\label{deltaW}
\end{equation}
\begin{equation}
\Delta V_{soft} =
[-\epsilon_{ab}{H}_u^a\widetilde{L}^b_i\widetilde{A}_{{\nu}_{ij}}\widetilde{N}^c_j  
+\frac{1}{2}\widetilde{N}^c_i b^\nu_i\widetilde{N}^c_i
+{\rm h.c.}]
+\widetilde{N}^{c\ast }_im_{Nij}^2\widetilde{N}^{c}_j.
\end{equation}
It is also often convenient to work in the basis
where the charged lepton Yukawa matrix $Y_e$ is real and diagonal.  
In this case, the remaining phase freedom can be used to remove three 
phases from the neutrino Yukawa matrix $Y_{\nu}$, so that the number of 
free parameters in the neutrino Yukawa sector of the superpotential 
consists of
6 complex plus 3 real Yukawa couplings, together with the 3 real diagonal
heavy right-handed Majorana masses.\footnote{One can of course also do
the counting without specifying a particular basis ({\it i.e.} the
Majorana mass term is $\frac{1}{2}\hat{N}^c_iM_{R_{ij}}\hat{N}^c_j$)  
\cite{Ellis:2001xt}. After utilizing all possible field redefinitions,
there are 21 parameters: 3 charged lepton masses, 3 light neutrino masses,
3 heavy Majorana neutrino masses, 3 light neutrino mixing angles, 3
light neutrino mixing phases, and 3 mixing angles and 3 phases associated 
with the heavy neutrino sector.} Eq~\ref{deltaW} also shows that the 
theory contains right-handed neutrino and sneutrino masses even when 
supersymmetry is not broken.

In such an extension of the MSSM with right-handed neutrinos (which is 
often labeled as the $\nu$MSSM), there are modifications 
of the MSSM RGEs which have signifiant phenomenological implications. 
These terms have already been included in the RGEs stated in 
Appendix~\ref{rgeapp}. One immediate implication of these additional 
terms is that even if the soft 
slepton masses are diagonal at the GUT scale, the three
separate lepton numbers $L_e, L_{\mu}, L_{\tau}$ are not 
generically not conserved at low energies if 
there are right-handed neutrinos below the GUT scale. Below the
mass scale of the right-handed neutrinos we must decouple the heavy
right-handed neutrinos from the RGEs and then the RGEs return to those of
the MSSM. Thus the lepton number violating additional terms are only
effective in the region between the GUT scale and the mass scale of the   
lightest right-handed neutrino and all of the effects of lepton number
violation are generated by RG effects over this range. The effect of RG
running over this range will lead to off-diagonal slepton masses at high
energy, which result in off-diagonal slepton masses at low energy, and
hence observable lepton flavor violation in experiments.

For example, the RGE for the soft slepton doublet mass contains
the additional terms 
\begin{eqnarray}
& & \frac{d m_L^2}{dt}  =  \left(\frac{d m_L^2}{dt}\right)_{Y_{\nu}=0}   
\nonumber \\
& - & \frac{1}{32\pi^2}\left[
Y_{\nu}Y_{\nu}^\dagger m_L^2+m_L^2Y_{\nu}Y_{\nu}^\dagger
+2Y_{\nu}m_N^2Y_{\nu}^\dagger+2(m_{H_u}^2)Y_{\nu}Y_{\nu}^\dagger
+2\widetilde{A}_{\nu}\widetilde{A}_{\nu}^\dagger
\right].
\end{eqnarray}
The first term on the right-hand side represents terms which do not depend
on the neutrino Yukawa coupling. If we assume for illustrative purposes
universal soft parameters at $M_{GUT}$, $m_L^2(0)=m_N^2(0)=m_0^2I$, where
$I$ is the unit matrix, and $\widetilde{A}_{\nu}(0)=A Y_{\nu}$, then 
\begin{equation}
\frac{d m_L^2}{dt}  =  \left(\frac{d m_L^2}{dt}\right)_{Y_{\nu}=0}
 -  \frac{(3m_0^2+A^2)}{16\pi^2}\left[ Y_{\nu}Y_{\nu}^\dagger \right] .
\label{simple}
\end{equation}
The first term on the right-hand side of Eq.~(\ref{simple})
represents terms which do not depend on the neutrino Yukawa coupling;
in the basis in which the charged lepton Yukawa couplings are
diagonal, these terms are also diagonal.
In running
the RGEs between $M_{GUT}$ and a right-handed neutrino mass $M_{R_i}$, 
the neutrino Yukawa couplings generate off-diagonal contributions to the
slepton mass squared matrices,
\begin{equation}
m_{{L}_{ij}}^2 \approx -\frac{1}{16\pi^2}
\ln \left(\frac{M_{GUT}^2}{M_{i}^2}\right)
(3m_0^2+A^2)
\left[ Y_{\nu}Y_{\nu}^\dagger \right]_{ij},\;\;i\neq j,
\label{offdiagslepton}
\end{equation}
to leading log approximation. In the simplest case, the right-handed 
neutrino couplings may represent the only source of LFV in the model.  
There has been a great deal of work examining the phenomenological 
implications of this case since, in this way, LFV can be communicated very 
efficiently from the neutrino sector to the charged lepton sector.  This 
is in strong contrast to the SM, where the known LFV in the neutrino 
sector has essentially no observable impact on the charged lepton sector.  
Thus, supersymmetry may provide a window into the Yukawa matrices that 
would not be available in the SM alone 
\cite{King:1998nv,Casas:2001sr,Blazek:2001zm,Blazek:2002,
Lavignac:2001vp,Davidson:2001zk,Ellis:2001xt}.

\subsection{R-parity violation}
\label{Rparityviolation}
\label{RPsect}
In the SM, gauge invariance implies that all operators of dimension 
less than 4 automatically (but accidentally) preserve both baryon number 
and lepton number.  However, supersymmetric extensions of the SM have the 
additional complication that in general there are additional 
renormalizable terms that one could write in the superpotential that are 
analytic, gauge invariant, and Lorentz invariant, but violate B and/or L.  
These terms are
\begin{equation}
W_{R}=\lambda_{ijk}\hat{L}_{i}\hat{L}_{j}\hat{E}^c_{k}+\lambda_{ijk}^{\prime} 
\hat{L}_{i}\hat{Q}_{j}\hat{D}^c_{k}+ 
\lambda_{ijk}^{\prime\prime}\hat{U}^c_i\hat{D}^c_{j}\hat{D}^c_{k}.
\end{equation}
The couplings $\lambda,\lambda^{\prime},\lambda^{\prime\prime}$ are
matrices in family space.  If both the second and third terms are present
in $W_R$, there is a new tree-level mechanism for proton decay which 
predicts microscopically short proton lifetimes.  To
avoid this phenomenologically disastrous result, it is necessary that one
or both of these couplings vanish.  Therefore, the usual expectation is 
that a symmetry of underlying fundamental theory forbids {\it all} of the 
terms in $W_{R}$, although this is not phenomenologically required (see 
below).  

There are two approaches to dealing with $W_{R}$. As previously mentioned, 
a symmetry, called R-parity or a variation called matter parity, can be 
added to the effective low energy theory. Presumably 
this symmetry arises from new physics at higher energy scales, 
such as an extended gauge group or discrete symmetries from string theory.  
R-parity is defined as follows:
\begin{equation}
R=(-1)^{3(B-L)+2S},
\end{equation}
where $S$ is the spin. This is a discrete Z$_{2}$ symmetry (a parity) in 
which the SM particles and Higgs fields are even and
the superpartners are odd. [Recall that such symmetries that treat
superpartners differently from SM particles and therefore do not commute
with supersymmetry are generically called R symmetries.] Equivalently, one
can use {\it matter parity},
\begin{equation}
P_{m}=(-1)^{3(B-L)}.
\end{equation}
A term in $W$ is only allowed if $P_{m}=+1.$  Gauge fields and Higgs
are assigned $P_{m}=+1,$ and quark and lepton supermultiplets $P_{m}=-1.$   
$P_{m}$ commutes with supersymmetry and forbids $W_{R}.$\footnote{Matter
parity and R-parity are equivalent because $(-)^{2S}=1$ for any vertex of
any theory which conserves angular momentum.} Matter parity could be an
exact symmetry of nature and such symmetries do arise in string theory. If
R-parity or matter parity holds there are major phenomenological
consequences:
\begin{itemize}

\item  At colliders (or in loops) superpartners are produced in pairs.

\item Each superpartner decays into one other superpartner (or an odd
number of superpartners).

\item The lightest superpartner (LSP) is stable.  This feature determines
supersymmetry collider signatures and makes the LSP a good candidate for
the cold dark matter of the universe.

\end{itemize}

The second approach to dealing with $W_R$ is very different and does not
have any of the above phenomenological consequences. In this approach,
$\lambda^{\prime}$ and/or $\lambda^{\prime\prime}$ are
arbitrarily set to zero\footnote{Recall that the nonrenormalization
theorem ensures that these terms are not regenerated through radiative
corrections.} so there is no observable violation of baryon number or
lepton number.  The other terms in $W_R$ are then allowed and one sets
limits on their coupling strengths when their effects are not observed,
term by term.  If we only have MSSM particle content R-parity must be
broken explicitly if it is broken at all.  If it were broken
spontaneously, {\it e.g.} by a nonzero VEV for the sneutrino, there would be a
Goldstone boson associated with the spontaneous breaking of lepton number
(the Majoron)  and certain excluded Z decays would have been observed.

Although this approach has been pursued extensively in the literature 
(see {\it e.g.} \cite{Dreiner:1997uz} for a review, and the references therein),   
R-parity violation is often considered to be less theoretically appealing 
because of the loss of the LSP as a cold dark matter candidate. 
Many people feel
that the often {\it ad hoc} nature of the second approach, where one of 
the $\lambda'$ or $\lambda''$ is set to zero without theoretical motivation, 
means R-parity violation is unlikely to be a part of a basic theory. 
Arguments are further made that large classes of theories
do conserve R-parity or matter parity. For example, often theories
have a gauged U(1)$_{B-L}$ symmetry that is broken by scalar VEVs and
leaves $P_{m}$ automatically conserved. In string models, examples exist 
which conserve R-parity, as do examples with R-parity violation (which 
still have proton stability). Within 
this framework the compelling question is how R-parity might 
arise within string theory.  For example, issues include how the 
string construction distinguishes between lepton and down-type Higgs 
doublets, or whether the discrete symmetries often present in 4D string 
models can include R-parity or matter parity. 
 In general, when supersymmetry is
viewed as embedded in a more fundamental theory, R-parity conservation 
is often easily justified, but is not guaranteed. \ Ultimately, of course, 
experiment will decide between among the options.

\subsection{The NMSSM}
\label{nmssm}
Probably the simplest direction in which the MSSM can be extended, and   
the most studied, is the addition of a gauge singlet chiral superfield to
the MSSM matter content \cite{Fayet:1974pd,Nilles:1982dy,Frere:ag,Derendinger:bz,Durand:1988rg}, 
\cite{Drees:1988fc,Ellis:1988er}. Such an addition is particularly 
well-motivated by solutions to the $\mu$ problem which replace the 
explicit $\mu$ term 
with a field $N$. If $N$ receives a VEV during electroweak symmetry 
breaking, the size of the $\mu$ term is automatically tied to the 
electroweak scale, as desired \cite{Hall:iz,Kim:1983dt,Inoue:1985cw,Anselm:1986um,Giudice:1988yz}.
Such a model is known as the 
next-to-minimal supersymmetric standard model MSSM (NMSSM). We will
discuss in this section a few of the phenomenological issues which arise
in the NMSSM.\footnote{Before there was experimental
evidence for a heavy top quark, the NMSSM was also invoked as the minimal
supersymmetric model which naturally broke the electroweak symmetry. The 
heavy top quark, coupled with radiative electroweak symmetry breaking, has 
eliminated this particular argument for the NMSSM.}

The superpotential for the NMSSM replaces
the $\mu$ term of the MSSM superpotential as follows:
\begin{equation}
\label{muterm}
-\epsilon_{ab}\mu\hat{H}_d^a\hat{H}_u^b
\rightarrow
\epsilon_{ab}\lambda \hat{N}\hat{H}_d^a\hat{H}_u^b
- \frac{1}{3} k \hat{N}^3
\end{equation}
where $\lambda$ and $k$ are dimensionless couplings\footnote{In
principle, we could consider more general scalar potential
$V(\hat{N})$. We could even include more complicated scalar potential
involving other fields. We use cubic coupling here as an illustrative
example. Therefore, any statement depending specifically on the form
of cubic coupling, such as discrete symmetry, should be considered to
be model dependent. }.
The soft supersymmetry-breaking Lagrangian term associated with the Higgs sector
of the NMSSM is given by 
\begin{eqnarray}
-{\mathcal{L}}_{soft}^{NMSSM} &=& 
-\epsilon_{ab}[  \lambda {A}_{\lambda} N H^a_dH^b_u 
+ \frac{1}{3} k A_k N^3  +{\rm h.c.}]\nonumber\\
&+&m_{H_{d}}^{2}|H_{d}|^2+m_{H_{u}}^{2}|H_{u}|^2+ m_N^2 |N|^2. 
\label{NMSSMsoft}
\end{eqnarray}
The low energy spectrum of the NMSSM contains three CP-even Higgs
scalars, two CP-odd Higgs scalars, and two charged Higgs scalars.
The phenomenology of the Higgs mass spectrum in the NMSSM,
including the dominant radiative corrections, 
was first studied in \cite{Elliott:ex,Elliott:uc,Elliott:bs,Ellwanger:1993xa}.
The constrained version of the NMSSM, analogous to the constrained MSSM, 
was first studied in 
\cite{Elliott:1994ht,King:1995vk,Ellwanger:1995ru,Ellwanger:1996gw}. 

The $N^3$ term in the NMSSM superpotential is necessary in order to avoid
a $U(1)$ Peccei-Quinn symmetry which, when the fields acquire their VEVs,
would result in a phenomenologically unacceptable axion. However, a $Z_3$
symmetry still remains under which all the matter and Higgs fields $\Phi$
transform as $\Phi \rightarrow \alpha \Phi$, $\alpha^3=1$.  This $Z_3$
symmetry may be invoked to banish such unwanted terms in the
superpotential as $\hat{H}_d \hat{H}_u$, $\hat{N}^2$ and $\hat{N}$, all of
which would have associated mass parameters.

Despite the obvious usefulness of the NMSSM, it is not without its 
own unique set of problems. For example, models of physics at high
energies generically contain hard supersymmetry breaking terms which are   
suppressed by powers of the Planck scale. Usually such terms are harmless.
But in the presence of a gauge singlet field they become dangerous because
together they can form tadpoles \cite{Ellwanger:mg,Bagger:1993ji,Bagger:1995ay}
which violate the $Z_3$ symmetry and drag the singlet VEV up to the Planck 
scale, destabilizing the gauge hierarchy 
\cite{Ferrara:1982ke,Polchinski:1982an,Nilles:1982mp,Lahanas:1982bk,Alvarez-Gaume:1983gj}. 
A second problem is that spontaneous breaking of the $Z_3$ after 
electroweak symmetry breaking can
generate domain walls in the universe, with disatrous consequences for
cosmology \cite{Vilenkin:ib}. We will return to this below.

Unlike the MSSM, where it is possible to derive simple
constraints which test whether electroweak symmetry breaking
will occur (at least at tree level in the Higgs sector), the possible 
vacuum structure of the NMSSM is very complicated. One must always check 
that a particular selection of parameters in the low energy Higgs 
potential will not result in the VEVs breaking electromagnetism.  The 
condition that electromagnetism is not broken simply reduces to requiring 
that the physical charged Higgs mass squared is nonnegative 
\cite{Ellis:1988er}. 
It can be shown, at tree level, that spontaneous ${CP}$
violation does not occur in a wide range of supersymmetric models including the
NMSSM \cite{Romao:jy}. Given that these conditions are satisfied, we
are left with a choice of VEVs for $H_u$, $H_d$ and $N$.
One defines $\tan\beta$ as usual, and introduces the ratio of VEVs 
$r\equiv x/\nu$, with $<N>=x$.  

As in the MSSM, there is always the possibility of squark and/or slepton 
VEVs breaking  electromagnetism or color (or both). 
The authors of \cite{Derendinger:bz} have
formulated simple conditions which determine in which regions of
parameter space such VEVs do not occur. The condition that we have no 
slepton VEVs is
\begin{equation}
A_e^2 < 3 (m_e^2 + m_L^2 + m_{H_d}^2),
\label{slepvev}
\end{equation}
This constraint is derived from the tree-level
potential under certain approximations, and should be tested at a scale
of order $A_e/h_e$, a typical slepton VEV.  A similar
condition on squark parameters will ensure the absence of
color-breaking squark VEVs:
\begin{equation}
A_t^2 < 3 (m_t^2 + m_Q^2 + m_{H_u}^2).
\end{equation}
The reliability of these results has been discussed in the literature 
\cite{Ellis:1988er}. 

There is a well-defined limit of the NMSSM in
which the components of the singlet decouple from the rest of the
spectrum which therefore resembles that of the MSSM (assuming no
degeneracies of the singlet with the other particles of similar spin
and CP quantum numbers which may lead to mixing effects which will
enable the NMSSM to be distinguished from the MSSM even in this limit). 
This limit is simply \cite{Ellis:1988er}: $k\rightarrow 0, \lambda
\rightarrow 0, x\rightarrow \infty$ with $kx$ and $\lambda x$ fixed.  

In general, however, the neutral Higgs bosons will be mixtures of the
singlet and the neutral components of the usual MSSM Higgs doublets. One
might worry then that the LHC would not be capable of discovering the
NMSSM Higgs. This question has recently been addressed in
\cite{Ellwanger:2001iw}, where a number of difficult points were studied.
It was concluded that LHC will discover at least one NMSSM Higgs boson
unless there are large branching ratios for particular superpartner decays
\cite{Ellwanger:2001iw}.

It has also been pointed out that the failure to discover the Higgs
boson at LEP2 increases the motivation for the NMSSM \cite{Bastero-Gil:2000bw}.
The argument is twofold. Firstly fine-tuning is significantly smaller
in the NMSSM than the MSSM for a given Higgs boson mass, 
essentially because the tree-level Higgs
boson mass is larger in the NMSSM than the MSSM.
The tree-level Higgs boson mass bound in the NMSSM is given by
\begin{equation} 
m_h^2 \leq M_Z^2
\left( \cos^2 2\beta + \frac{2\lambda^2}{g^2+{g'}^2} \sin^2 2\beta
\right)  
\label{two} 
\end{equation} 
which contains an additional term proportional to 
$\lambda^2$. The extra tree-level term  means that for a given
Higgs boson mass, less of a contribution is required from
radiative corrections in the NMSSM than the MSSM, and thus the stop mass 
parameters in the NMSSM may be smaller than in the MSSM, leading to 
reduced fine-tuning. The second argument in favor
of the NMSSM is that electroweak baryogenesis is much easier to 
achieve in the NMSSM than in the MSSM. The failure to discover 
Higgs or stops at LEP2 severely constrains the MSSM parameter space
consistent with electroweak baryogenesis. However, 
the tree-level cubic coupling of the Higgs bosons to singlets in the NMSSM 
enhances the first order nature of the electroweak phase transition
without providing any constraints on the stop parameter space.

A phenomenological comparison of the MSSM to the NMSSM, including Higgs
mass bounds, can be found in \cite{Carena:1996bj}. Typically the Higgs
mass bound in the NMSSM is about 10 GeV higher than in the MSSM
\cite{Elliott:bs}. The increase in the Higgs mass in extensions with gauge
singlets was first observed in \cite{Kane:1992kq,Espinosa:1992hp}.  
Assuming only perturbative unification, the Higgs mass could be as heavy
as 205 GeV in more general frameworks than the MSSM or NMSSM ({\it i.e.}
with additional nonsinglet Higgs representations)  
\cite{Espinosa:1998re,Quiros:1998bz}.  
Given the constraints placed on the MSSM parameter space from the current
LEP Higgs mass bounds, there is certainly a strong motivation to consider
models such as the NMSSM which have extended Higgs sectors.

Finally, let us return to the problem of the domain walls 
created in the early universe due to the discrete $Z_3$ symmetry which is 
broken at the electroweak scale in the NMSSM.  This 
cosmological catastrophe can of course be avoided by allowing explicit 
$Z_3$ breaking by terms suppressed by powers of the Planck mass which will 
ultimately dominate the wall evolution 
\cite{Gelmini:1988sf,Rai:1992xw,Abel:1995uc,Abel:1995wk}
without affecting the phenomenology of the model. One can also construct 
variations of the NMSSM which solve this domain wall problem. There are 
several classes of solutions:
\begin{itemize}

\item Break the $Z_3$ symmetry explicitly by retaining
the $\mu$ term, together with additional $\mu$-like terms of the form
$\mu'N^2$, $\mu''N$ \cite{King:1995ys}. Such a model clearly does not 
solve the $\mu$ problem, but remains a possible alternative to the MSSM.

\item Remove the $\hat{N}^3$ term and gauge the PQ $U(1)$ symmetry 
\cite{Cvetic:1997ky}. This introduces a $Z'$ gauge boson with interesting
electroweak scale phenomenology \cite{Cvetic:1997ky}.

\item Remove the $\hat{N}^3$ term and break the PQ $U(1)$ symmetry 
with a discrete $R$ symmetry \cite{Panagiotakopoulos:1999ah}. 
This allows loop-suppressed tadpole terms which have acceptable
electroweak phenomenology \cite{Dedes:2000jp}.

\item Replace the $\hat{N}^3$ term by a $\hat{\phi}\hat{N}^2$ term
where $\phi$ is a second singlet which is identified as
an inflaton field in a hybrid inflation scenario \cite{Bastero-Gil:1997vn}.
With a second singlet the PQ symmetry remains, and the VEVs of the
$N,\phi$ scalars are assumed to be at a high energy scale associated
with the PQ solution to the strong CP problem. Inflation also occurs
at that scale which inflates away any unwanted relics. In this
version of the model, the $\mu$ term requires a very small value of 
$\lambda \sim 10^{-10}$, which must be explained ({\it e.g.} as 
originating from effective nonrenormalizable operators 
\cite{Bastero-Gil:1997vn}). \end{itemize}

\setcounter{footnote}{0}

\section{Conclusions and outlook: from data to the fundamental theory}
In addition to the very strong indirect phenomenological evidence for low 
energy supersymmetry and its considerable theoretical attractiveness, 
supersymmetry is probably the only meaningful
approach that will allow us to connect data at the energies where
experiment is possible with a fundamental short distance theory that
includes gravity.  Traditionally data plus theory provoked ideas that led
to tests and to progress in understanding, but always at the same scale.
Today we are in a new kind of situation where the fundamental theory is
expected to be at short distances but the data is not.  If there is
indeed low energy supersymmetry in nature we have the exciting opportunity
to scientifically connect these two realms and to effectively be doing
physics at or near the Planck scale.

Traditionally one approach was the gradual bottom-up one where data was
gathered and studied and analyzed, leading to clues about the underlying
theory.  Alternatively, studies of the theory with little regard for the
data (top-down) led to major progress too, teaching us about such things
as the Higgs mechanism, Yang-Mills theories, and more.  Of course, both of
these approaches have inherent limitations.  The main limitation of the
purely top-down approach is obvious. One must guess the form of the
underlying theory, and hence progress may require compelling theoretical
guidelines (and ideally new fundamental principles) which render this
process less arbitrary.  Since our main emphasis in this review has been
along the lines of the bottom-up approach, we now pause to elaborate on
the limitations inherent within the purely bottom-up framework, and
discuss why a closer connection of the two approaches will be necessary
for progress now and in the future.

Suppose we succeed in measuring the low energy soft supersymmetry-breaking
Lagrangian parameters.  What obstacles exist to deducing a more
fundamental, high scale theory?  In a purely bottom-up approach, the
measured parameters must be extrapolated to higher scales using the
renormalization group equations.  In this lies the basic limitation: the
running of the RGEs must be stopped and modified when new light degrees of
freedom enter the theory, but low energy data alone can not tell us at
what scale such states appear or the details of the new particle content.
More explicitly, without any knowledge of the high energy theory, we have
the freedom to stop running the RGEs at any scale and declare that should
be where the fundamental or embedding theory is defined.  

Initial studies along this direction
\cite{Blair:2000gy,Blair:2002pg,Kalinowski:2002uk,Kane:2002ss,Martin:2001zw,Allanach:2001qe}
typically assume there is a desert between the TeV scale and the GUT
scale, where the RG running is stopped. Even then, there are limitations
associated with the experimental uncertainties in the low energy data.  
For example, the low energy parameters can be close to an infrared
(quasi-)fixed point which would make them insensitive to their high scale
values (this is certainly true for the top Yukawa coupling). In this case,
a small uncertainty due to experimental error will translate into large
uncertainties in the extrapolated values of the high scale parameters.

Setting aside the issue of how to guess the ``fundamental'' scale, it is
well known that the presence of new light degrees of freedom at
intermediate scales in general has a significant impact on the RG running
of the parameters from low to high scales.  For example, if 
arbitrary gauged degrees of freedom with intermediate scale masses exist 
between the electroweak scale and the GUT scale,
the successful MSSM gauge unification is generically spoiled.  
Intermediate states can also destroy the perturbativity of the gauge
interactions at a lower scale, {\it i.e.}, the RG evolution of the gauge
couplings can encounter a Landau pole.  Of course, not all choices of
intermediate states destroy gauge unification and/or perturbativity, and
in fact such states may even be phenomenologically desirable in top-down 
constructions.

In this context, there is a related issue which does not appear to have
been addressed much in the literature.  In particular top down
supersymmetry breaking scenarios, supersymmetry is broken spontaneously
(for example through gaugino condensation) at an energy scale $\Lambda$
far below the GUT scale.  This naively implies that when the RGE is
evolved above the scale $\Lambda$, there are no longer any soft breaking
terms in the effective theory.  In such cases, it is not clear exactly
what one can learn by evolving the soft parameters above the scale
$\Lambda$.

Due to the above ambiguities, a purely bottom-up approach cannot
provide sufficient information about the embedding theory. Insisting on
using this approach only with oversimplified assumptions can lead to
misleading results.  Not surprisingly, it is most prudent to adopt an
approach which combines the top-down and the bottom-up methods, which has
led to progress throughout the history of physics.

There is a great deal of work to be done along this direction.  One should
construct top-down models which include information such as the
supersymmetry breaking scale, possible additional particle content and
intermediate scales, {\it etc.}, enough to resolve the ambiguities in the
running-up process. This information can then be combined with the usually
incomplete, low energy experimental results to obtain further information
about the embedding theory which is not fully specified in the original
model. However, the new information may not be consistent with the 
original model: {\it e.g.}, certain patterns of couplings may
not exist in a particular model setting. In such situations, one should
improve the model and repeat the process.  Gradually, with the
accumulation of experience with models and experimental inputs, one can
hope to close in on a more fundamental theory.
 
Ideally we would have been able to present plans and algorithms that could
be applied to point towards the underlying theory as data from colliders
and virtual superpartner effects become available.  But we cannot say so
much about how to do that because these are not yet solved problems.  
Much important work needs to be done here by experimenters and
phenomenological theorists and formal theorists.  We urge that the
powerful opportunities provided by supersymmetry be studied much more
thoroughly than they have been, even before the data requires such
studies.  In the review we have often pointed out aspects of the
data-theory connection that needed better understanding.

In this review, our goal has been to bring together much of what is
currently known about the supersymmetry soft-breaking Lagrangian, and to
describe the opportunities that may emerge as particle physics enters a
new data-rich era.  We also believe that we will soon enter an era 
where basic connections of the superpotential and \lsoft to an underlying 
embedding theory such as string theory can be deduced.  If the description 
of nature indeed includes low energy supersymmetry, apart possibly from a 
few cosmological observations almost all phenomena (collider data, rare
decays, dark matter detection, neutrino physics, magnetic and electric
dipole moments, and more) measurable by experiment beyond the standard
models of particle physics and cosmology can be interpreted as
measurements of the superpotential and \lsoft parameters.  Our goal has
been to stimulate and facilitate those interpretations.

In the present era it is possible for the first time that all of the 
basic questions about the laws of nature and the universe can be the 
subject of scientific research. String theory is exciting
because it is a framework which can address how to explain the Standard 
Model forces and particles and relate them to gravity in a
quantum theory.  The Standard Model is exciting because it provided a
description that summarized four centuries of physics and told us how the
world we see works.  Supersymmetry is exciting perhaps most because it,  
and probably only it\footnote{Other approaches are sometimes stated to be
competitive. However, when the full set of questions are 
included, {\it e.g.} dark matter, inflation, baryogenesis, the origin 
of flavor and CP violation, collider opportunities, and electroweak 
symmetry breaking, {\it etc.}, then no other approach is as
successful as low energy supersymmetry.}, provides the opportunity to 
combine these 
approaches and extend the Standard Model by giving us a window on the 
Planck scale.

%

\vspace{0.5in}
\leftline{\large\bf Acknowledgments}
\medskip

We thank J. Cline, D. Demir, T. Han, C. Kolda, B. Nelson,
A. Pilaftsis, M.  Pl\"{u}macher, P. Ramond, R. Rattazzi, S. Rigolin,
A. Riotto, S. Su, C. Wagner, T.  Wang, and J. Wells for helpful
conversations and suggestions about the work presented herein.  We
especially thank C. Kolda and P.  Ramond for detailed critiques of
sections of this review and D. Demir for many helpful
correspondences. S. King is supported in part by a Senior PPARC
Fellowship.  We also thank the Aspen Center for Physics for support.

\newpage
\appendix
\setcounter{footnote}{0}

\setcounter{footnote}{0}
\section{Global supersymmetry basics}
\label{backgroundsect}

This section of the review aims to provide the reader with a basic
overview of the properties of $N=1$ supersymmetric quantum field theories
and soft supersymmetry breaking, with a few relevant details.  For more
comprehensive and pedagogical approaches, there are many textbooks
\cite{Bagger:1990qh,West:tg,Weinberg:cr,Gates:nr} and reviews, including
two classic reviews of the early 1980s \cite{Nilles:1983ge,Haber:1984rc}
as well as more recent theoretical and phenomenological reviews
\cite{Ramond:1994fm,Lykken:1996xt,Drees:1996ca,Martin:1997ns}.

Supersymmetry avoids the restrictions of the Coleman-Mandula theorem
\cite{Coleman:ad} by extending the structure of Lie algebra to include
anticommutators and successfully embeds Poincare group into its larger 
group structure without modifying the usual notions of local quantum field
theory. Although not invented for this purpose, supersymmetry has
unique high energy properties in comparison with generic
(nonsupersymmetric) quantum field theories: in particular, 
supersymmetry has the ability to stabilize large hierarchies of scales
even in the presence of fundamental scalar fields.  In this way,
supersymmetric theories provide a resolution to the hierarchy problem
which plagues ordinary (nonsupersymmetric) QFTs.

Given its importance, let us consider the hierarchy problem 
in greater detail. Suppose an effective quantum field theory is defined at 
a cutoff scale $\Lambda$, beyond which new ultraviolet physics sets in 
such that the effective low energy description is no longer valid. At 
the scale $\Lambda$, the theory is given by ${\mathcal{L}}_\Lambda (m_\Lambda, 
\lambda_\Lambda)$, where $m_\Lambda$ and $\lambda_\Lambda$
collectively denote the masses,  
coupling constants, and other parameters at that scale. Consider an 
example in which the high energy theory is a scalar $\phi^4$
model:
\begin{equation}
{\mathcal{L}}_\Lambda=\frac{1}{2} \partial^{\mu}\phi_\Lambda 
\partial_{\mu}\phi_\Lambda +  \frac{1}{2} m^2_\Lambda \phi^2_\Lambda + 
\frac{\lambda_\Lambda}{4!} \phi^4_\Lambda,
\end{equation} 
Because of quantum fluctuations and self interactions, the low energy
observed mass is $m_\Lambda^2+ \lambda_\Lambda \Lambda^2$, where we
have absorbed possible loop factors into a redefinition of $\lambda_\Lambda$.  However, the 
physical mass $m$ must be small if the low energy effective theory is to 
describe a light degree of freedom relevant for low energy experimental
processes. This requires that $m_\Lambda^2\sim O(\Lambda^2)$ must be
fine-tuned such that $m_\Lambda^2$ and $\Lambda^2$ cancel to a precision of
$m^2$.  This is the statement of the hierarchy problem: the physical
scale $m$ is unstable with respect to quantum corrections
if the ratio $\Lambda/m$ is large. This problem exists in the SM
because the electroweak scale fixed by the Higgs mass $m_H \sim 10^2$ GeV 
is much smaller than the cutoff scale suggested by the grand unification 
scale of $10^{16}$ GeV or the quantum gravity scale of $10^{19}$ GeV.
This fine tuning problem applies to any
term in the Lagrangian with a dimensionful parameter which is measured to
be much less than the cutoff scale of the effective theory.  The hierarchy
problem is a generic feature of nonsupersymmetric quantum field theories
with fundamental scalar fields and cutoff scale much larger than the
electroweak scale.

One way to alleviate the hierarchy between the scales $\Lambda$ and $m$ is 
to eliminate the unwanted quantum fluctuations that generate the large 
``corrections'' above the scale $m$ using a fundamental symmetry
of the Lagrangian.  Since the supersymmetry algebra contains both commuting and
anticommuting generators, there is a natural pairing between the bosonic
and fermionic degrees of freedom whose quantum fluctuations come with
opposite signs but with equal magnitudes such that the quantum
fluctuations that generate corrections to dimensionful parameters sum up
to zero. Supersymmetry thus provides the necessary cancellations to 
stabilize the
low energy scale $m$. Due to the paucity of alternative mechanisms for
such natural cancellations, it seems highly probable that supersymmetry will play a
role in extensions of the SM if the cutoff scale is really much larger
than the electroweak scale.

\subsection{Renormalizable models}
Supersymmetry is a symmetry under which bosons can transform into fermions 
and vice versa. Therefore, the irreducible representations of 
supersymmetry, the supermultiplets, contain both fermions and bosons.  
We will illustrate the basic ideas of constructing a supersymmetric 
interacting quantum field theory by presenting a review of the Wess-Zumino 
model \cite{Wess:tw}.  The building blocks of this model are the 
fields $\{\phi,\psi,F\}$, where $\phi$ and $F$ are complex scalars and 
$\psi$ is a spinor. For simplicity, assume for now that these fields have 
no gauge interactions. Under supersymmetry, these fields 
transform as $\phi \rightarrow \phi + \delta \phi$, 
$\psi \rightarrow \psi + \delta \psi$, $F \rightarrow F + \delta F $, with
\begin{eqnarray}
\delta \phi &=& \epsilon \psi \nonumber \\
\delta \psi&=&i(\sigma^{\mu}\epsilon^{\dagger}) \partial_{\mu} \phi + 
\epsilon F \nonumber \\
\delta F &=& i \epsilon^{\dagger} \overline{\sigma}^{\mu} \partial_{\mu} \psi 
\label{susytransformation}
\end{eqnarray}
plus the conjugates of the equations above (see Appendix~\ref{spinorapp} 
for a discussion of spinor conventions). In the expression above, 
$\epsilon$ is a 
two-component spinor which is the supersymmetry transformation parameter. 
Bosons 
and fermions are mixed 
in specific ways under supersymmetry transformations. The 
renormalizable Lagrangian left 
invariant (up to total derivatives) with respect to these transformations is 
\begin{eqnarray}
\label{susyL1}
{\mathcal{L}} &=& -(\partial^{\mu} \phi^* \partial_{\mu} \phi + 
i\psi^{\dagger} \overline{\sigma}^{\mu} \partial_{\mu} \psi )\nonumber \\
&-&(\frac{1}{2} m \psi \psi + 
\frac{1}{2}m^*\psi^{\dagger}\psi^{\dagger} )
\nonumber \\
&-& FF^*- F (m\phi+ \frac{y}{2} \phi^2) - F^* (m\phi^* + 
\frac{y^*}{2} \phi^{* 2})\nonumber \\
&-&\frac{1}{2}y \phi \psi \psi -\frac{1}{2} y^* \phi^* 
\psi^{\dagger} \psi^{\dagger}.
\end{eqnarray}
Eq.~(\ref{susyL1}) includes kinetic terms for $\phi$ and 
$\psi$, fermionic and bosonic mass terms, and interaction terms. However, 
since $F$ does not have a kinetic term, it does not represent a physical 
degree of freedom (it is an auxiliary field). $F$ can thus be integrated 
out of the theory, effectively replaced by the solution of its classical 
(Euler-Lagrange) equation of motion $F=-m\phi^*-\frac{y}{2} 
\phi^{* 2}$. Upon this replacement of $F$ by its equation of motion, 
the third line of Eq.~(\ref{susyL1}) becomes
\begin{equation} 
V(\phi, \phi^*)=FF^*=|m|^2 |\phi|^2+ \frac{1}{2}my^*\phi 
\phi^{* 2}+ \frac{1}{2}m^* y \phi^* \phi^{2} + 
\frac{yy^*}{4} 
\phi^2 \phi^{* 2}. 
\end{equation}
These terms are usually called the F term contributions to the scalar 
potential. 

Supersymmetry constrains the parameters of the Lagrangian  since different 
terms transform into each other under supersymmetric transformations. A 
Lagrangian with similar couplings could have 7 parameters, one for 
the strength of each term after the kinetic terms, while in 
Eq.~(\ref{susyL1}) these couplings are determined in terms of 3 real 
parameters ($m$ and complex 
$y$). This feature is not an artifact of the Wess-Zumino model, but is 
also true for a more general supersymmetric model. The 
interactions in an $N=1$ supersymmetric Lagrangian involving only 
gauge-neutral chiral supermultiplets (assuming canonical kinetic terms)
can be summarized efficiently through the 
introduction of a function 
called the {\it superpotential}. 
In the Wess-Zumino model, the superpotential is 
\begin{equation}
W= \frac{y}{6} \Phi^3 + \frac{m}{2} \Phi^2,
\label{sptoy}
\end{equation} 
in which $y$ is a dimensionless coupling and $m$ has dimensions of mass. 
Note that the superpotential has dimensions of [mass]$^3$, assuming $\Phi$ 
has canonical mass dimension 1.
The superpotential contains all of the couplings
necessary to describe all renormalizable interactions except gauge 
interactions. In this respect, the
superpotential can be viewed as a concise way of summarizing the
interactions of a renormalizable supersymmetric theory. 
The Lagrangian can be obtained from the superpotential using a set of 
rules, discussed later in this section. The $\Phi$'s are called
chiral superfields; chiral superfields contain all of the fields in a 
chiral supermultiplet ($\phi$, $\psi$, and $F$) as its 
components.\footnote{An elegant way to derive and present supersymmetric 
interactions uses superfields and an extended version of ordinary 
spacetime called superspace \cite{Salam:1974yz,Ferrara:1974ac}. See {\it 
e.g.} \cite{Bagger:1990qh} for a detailed and pedagogical presentation of 
this formalism.}
Superfield techniques will not be discussed in this review. Rather, the 
superfields  will only serve a symbol and a reminder that within 
this model (and all supersymmetric theories involving only gauge-neutral 
chiral supermultiplets with canonical kinetic terms), the superpotential 
contains the information about all the interactions between 
all the fields, both bosonic and fermionic.  

The rules for obtaining the Lagrangian
from the superpotential are as follows.
Define the quantities:
\begin{equation}
W_i=\frac{\partial W}{ \partial \phi^i},\hspace{1cm}
W_{ij}=\frac{\partial^2 W}{ \partial \phi^i \partial \phi^j},
\end{equation}
where the superscript $i$ labels the quantum numbers of $\phi_i$. Note 
that in computing these two quantities, 
the superfields $\Phi_i$ are replaced with their bosonic components 
$\phi_i$ and the derivatives are taken with respect to the bosonic 
components. The Lagrangian is then given by 
\begin{eqnarray}
{\mathcal{L}} &=& -\partial^{\mu} \phi^{*i} \partial_{\mu} \phi^i - i
\psi^{\dagger i} \overline{\sigma}^{\mu}\partial_{\mu} \psi^i \nonumber \\
&-& F^i F^{*i} -(\frac{W_{ij}}{2}\psi_i \psi_j + W_i F^i + {\rm h.c.}).
\label{WtoLrule}
\end{eqnarray}
The solutions of the equation of motion of the auxiliary fields are
$W_i=-F^{i*}$ (the $W_i$ are often labelled as F terms). 
The Lagrangian is obtained upon substitution of this 
solution into Eq.~(\ref{WtoLrule}). It is a good exercise to check
that the interactions of the Wess-Zumino model can be reproduced by 
applying this rule to the superpotential presented in Eq.~(\ref{sptoy}).   

One property of the superpotential warrants further 
comment. Suppose the superpotential is not given by Eq.~(\ref{sptoy}), but 
instead is  
\begin{equation}
W=\frac{y}{6} \Phi^3 + \frac{m}{2} \Phi \Phi^*. 
\end{equation}
This ``superpotential'' only differs from Eq.~(\ref{sptoy}) by 
the term $\Phi \Phi^*$ rather than $\Phi^2$. However, it can be verified 
using the supersymmetric transformations that the Lagrangian obtained 
by applying the rules of Eq.~(\ref{WtoLrule}) is NOT
supersymmetric. This is an example of the following general rule: The
superpotential must be {\it holomorphic} (analytic) in all superfields 
to yield a Lagrangian which respects supersymmetry. 

It is straightforward to include gauge symmetries, which commute with
supersymmetry.\footnote{An exception is the general coordinate
transformation, which is a gauge symmetry.  These 
transformations are generated automatically by gauging supersymmetry since 
general coordinate invariance is a subgroup of local supersymmetry.}  In 
$N=1$ supersymmetric
theories, the gauge boson $A^a_{\mu}$ is always accompanied by its
superpartner, a spin $\frac{1}{2}$ particle called the gaugino $\lambda^a$
(here $a$ labels the generators of the gauge group). Together they form 
the physical degrees of freedom of a superfield known as the {\it vector
multiplet}. Like the gauge boson, the gaugino transforms under the
adjoint representation of the gauge group.  Like the chiral multiplet, the
vector multiplet contains a complex scalar auxiliary field $D^a$, whose
purpose is to make supersymmetry manifest without using equation of motion.

To construct supersymmetric models with gauge interactions, a well-defined
procedure can be followed. Rather than going through the derivation
here, we will just present the results here as most of them are
straightforward to understand.   One first includes the
supersymmetric interactions for the vector multiplet:
\begin{equation}
{\mathcal{L}}_{{\rm gauge-kinetic}}=-\frac{1}{4} F^a_{\mu \nu} F^{a \mu 
\nu} -i
\lambda^{a \dagger} \overline{\sigma}^{\mu} D_{\mu} \lambda^a + \frac{1}{2}
D^a D^a,
\end{equation}
where covariant derivatives for the gauginos are 
\begin{equation}
D_{\mu} \lambda^a = \partial_{\mu} \lambda^a -g f^{abc} A^b_{\mu}\lambda^c.
\end{equation}
$f^{abc}$ are the structure constants of the gauge group since
gauginos are transformed under the adjoint representation of the gauge
group. 

The next step is to replace all the other ordinary derivatives for the
matter fields of Eq.~(\ref{susyL1}) by covariant derivatives, 
which yields the couplings of the gauge bosons to the chiral matter:
\begin{equation}
\partial_{\mu} \rightarrow \partial_{\mu} + ig A^a_{\mu} T^a,
\end{equation}
where $T^a$ is the generator of the gauge group written in the proper
representation of the matter field. However, 
supersymmetry requires similar couplings between the gauginos and 
the  chiral matter. These couplings are
\begin{equation}
{\mathcal{L}}_{\lambda \phi} = -\sqrt{2}[(\phi^* T^a \psi) \lambda^a +
\lambda^{a \dagger} (\psi^{\dagger} T^a \phi)].
\end{equation}
There is also an interaction between the chiral matter fields
and the auxiliary fields:
\begin{equation}
{\mathcal{L}}_{aux-\phi}=g(\phi^* T^a \phi) D^a.
\end{equation}
Both of the two couplings above can be obtained by supersymmetric
transformation of the kinetic terms containing the couplings between
the gauge bosons and matter fields. Therefore, they can be regarded as
supersymmetric generalizations of the usual gauge couplings. 

Combining ${\mathcal{L}}_{{\rm gauge-kinetic}}$ and  other terms involving
the auxiliary field, we obtain the equation of motion 
\begin{equation}
\label{dterm1}
D^a=-g (\phi^* T^a \phi).
\end{equation}

Another useful form for the supersymmetric interactions of the
vector multiplet is obtained by redefining the fields $A^a_{\mu} 
\rightarrow g
A^a_{\mu}$, $\lambda^a \rightarrow g \lambda^a$ and $D^a \rightarrow g
D^a$, 
\begin{eqnarray}
\label{susygaugeL}
{\mathcal{L}}_{{\rm gauge-kinetic}}&=&\frac{1}{g^2}\left[-\frac{1}{4} 
F^a_{\mu \nu}
F^{a \mu \nu} -i
\lambda^{a \dagger} \overline{\sigma}^{\mu} D_{\mu} \lambda^a + \frac{1}{2}
D^a D^a \right] \nonumber \\
&-& \frac{\theta_G}{32 \pi^2}F^a_{\mu \nu}\widetilde{F}^{a \mu \nu},  
\end{eqnarray}
where $\widetilde{F}^{a \mu \nu} = \frac{1}{2}\epsilon^{\mu \nu \rho
\sigma}F^a_{\rho \sigma} $. Included in Eq.~(\ref{susygaugeL})
is a term
corresponding to a nontrivial vacuum configuration of Yang-Mills
fields (for example, the $\theta$-vacuum of QCD). Obviously, this part
of the Lagrangian contains the usualy kinetic terms for the usual
gauge couplings and their supersymmetric generalizations.

\subsection{Nonrenormalizable models}
The most general renormalizable supersymmetric model of chiral and vector 
supermultiplets can be specified by the generic superpotential
\begin{equation}
W=\frac{Y_{ijk}}{6} \Phi^i \Phi^j \Phi^k + \frac{M_{ij}}{2} \Phi^i \Phi^j,
\label{spgeneral}
\end{equation}
where $i$, $j$, and $k$ label all quantum numbers 
of $\Phi$, and minimal coupling of gauge and matter fields.
The superpotential of the Minimal 
Supersymmetric Standard
Model is of this form. In the MSSM, two of the indices
of the Yukawa couplings $Y_{ijk}$ label family indices, 
while the third denotes the Higgs fields. The second (mass) term in the 
superpotential will vanish by gauge invariance for all of the MSSM 
fields except the Higgs doublets  $H_u$ and $H_d$.  Mixed 
lepton doublet--$H_u$ terms are also possible in theories with R-parity 
violation.

The superpotential presented in Eq.~(\ref{spgeneral}), together with
the gauge interactions, gives the most general supersymmetric
renormalizable couplings of chiral supermultiplets with standard
kinetic terms.  Since phenomenologically realistic theories require
that supersymmetry be softly broken, \lsoft must be added, leading to
an effective theory such as the MSSM-124 specified by its
renormalizable superpotential (Eq.~(\ref{superpot})) and soft 
supersymmetry-breaking Lagrangian (Eq.~(\ref{lsoftexpr})).

In practice, nonrenormalizable operators will also be present in the
superpotential because such
terms are generic in effective theories as a result of integrating out
heavy degrees of freedom.  The nonrenormalizable terms are
suppressed by powers of the scale at which the new physics becomes
relevant and thus involve assumptions as to the magnitude of this energy
scale.  For most phenomenological studies of supersymmetric theories, the
nonrenormalizable operators involving only the MSSM fields can be safely
neglected because the new physics energy scale is generically much larger
than the electroweak scale.  However, certain highly suppressed processes
are sensitive to higher dimensional operators.  The classic example of
this is proton decay, which probes superpotential terms of up to dimension
8 when the scale of the new physics is as low as is phenomenologically
allowed ($O({\rm TeV})$).  Nonrenormalizable superpotential terms 
involving
the MSSM fields and additional fields are also often used to generate
small effective renormalizable couplings when the additional heavy
fields are replaced by nonzero VEVs.  For example, this approach is used
to understand the origin of small Yukawa couplings in the SM and MSSM (see
{\it e.g.}  \cite{Froggatt:1978nt,Leurer:1992wg,Leurer:1992wg}).

Nonrenormalizable couplings do not have to appear only in the
superpotential.  They can also appear in the noncanonical kinetic
terms for the chiral and the vector superfields.  For the chiral
superfields, such operators can be encoded by a function called the
{\it K\"ahler potential} $K(\Phi,\Phi^{*})$, while for the vector
superfields such terms arise from the {\it gauge kinetic function}
$f_a(\Phi)$, where $a$ labels the gauge groups.

Let us first discuss the K\"ahler potential.  The K\"ahler potential
has dimensions of [mass]$^2$ and is a real valued function of the
superfields $\Phi_i$ and $\Phi^*_i$.  The simplest K\"ahler potential is
$K=\sum_i \Phi_i \Phi^*_i$, which leads to canonical kinetic terms.  A
more general K\"ahler potential leads to noncanonical kinetic terms
through the field-dependent prefactor (known as the K\"ahler metric)
$g_{ij^*}\equiv \partial^2 K/\partial \Phi_i \partial \Phi^*_j$ of the
kinetic terms.  

Besides giving noncanonical kinetic structure, the K\"ahler potential
can generate nonrenormalizable interactions as well.
If we denote the inverse K\"ahler metric by $g^{ij^*}$, we can write
\begin{eqnarray}
{\mathcal{L}}&=& -g_{ij^*} \partial_{\mu} \phi^i \partial^{\mu}
\phi^{*j}  - i g_{ij^*} \psi^{j \dagger} \overline{\sigma}^{\mu} D_{\mu}
\psi^i \nonumber \\
&+&\frac{1}{4} R_{ij^* kl^*} \psi^i \psi^k \psi^{\dagger j}
\psi^{\dagger l} \nonumber \\
&-&\frac{1}{2} D_{ij} W  \psi^i \psi^j + {\rm h.c.} \nonumber \\
&-& g^{ij^*} W_i W_j^*, 
\end{eqnarray}
where
\begin{eqnarray}
g_{ij^*, k} &=& \frac{\partial}{\partial \phi^k} g_{ij^*}= g_{mj^*}
\Gamma^m_{ik} \nonumber \\
g_{ij^*, k^*} &=& \frac{\partial}{\partial \phi^{k*}} g_{ij^*} =
g_{im^*} \Gamma^{m^*}_{j^* k^*} \nonumber \\
g_{ij^*, kl^*} &=&  \frac{\partial^2}{ \partial \phi^k \partial
\phi^{*l}} g_{ij^*} = R_{ij^* kl^*} + \Gamma^p_{ik} g_{p p^*}
\Gamma^{p^*}_{j^* l^*}.  
\end{eqnarray}
and 
\begin{eqnarray}
D_{\mu} \psi^i = \partial_{\mu} \psi^i + \Gamma^{i}_{jk}
\partial_{\mu} \phi^j \psi^k \nonumber \\
D_{ij}W = W_{ij}-\Gamma^k_{ij} W_k.
\end{eqnarray}
Since the Lagrangian is invariant under the K\"ahler
transformation $K(\Phi, \Phi^*) \rightarrow K(\Phi, \Phi^*) + F(\Phi)
+ F^*(\Phi^*)$, where $F(\Phi)$ is any holomorphic function of $\Phi$,
one can choose to transform away all the holomorphic and
antiholomorphic terms in the K\"ahler potential.
\footnote{Strictly speaking, what we presented here is only the
classical symmetry transformations. At quantum level, a Jacobian will
be induced in the Lagrangian after this transformation. Such a
Jacobian is crucial to preserve local supersymmetry in the rescaled
Lagrangian\cite{Bagger:2000dh}}.
After rotating/rescaling the fields, a generic $K$ can be cast into 
canonical form at leading order:
\begin{equation}
\label{canonicalK}
K=\sum \phi_i \phi_i^* + \frac{1}{4} R_{kl^* ij^*} \phi^k \phi^i
\phi^{l*} \phi^{j^*}+ O(\phi^5), 
\end{equation}
where $R_{kl^* ij^*}$ is a function of the VEVs of certain fields and can 
be derived from the K\"ahler potential. Since $K$
has mass dimension 2, $R_{kl^* ij^*}\propto 1/M^2$, in which $M$ is a 
heavy mass scale.  If the superpotential has  the usual renormalizable 
form, then the nonrenormalizable interactions are 
\begin{eqnarray}
{\mathcal{L}}_{nonrenorm} &=& \frac{1}{4} R_{kl^* ij^*} \psi^k \psi^i
\psi^{l \dagger} \psi^{j \dagger} \nonumber \\
&+& \frac{1}{2} \left[Y_{nmk} R_{i l^* j}^k \phi^n \phi^m \phi^{l *}
\psi^i \psi^j + {\rm h.c.}  \right] \nonumber \\
&-& \frac{1}{4}R^{ij^*}_{kl^*} Y_{i ab} Y^*_{j cd}\phi^k \phi^{l *}
\phi^a \phi^b \phi^{c*} \phi^{d*} + \cdots,  
\label{nonrenl}
\end{eqnarray} 
where $R_{i l^* j}^k= g^{km^*} R_{i l^* j m^*}$, $R^{ij^*}_{kl^*}=
g^{i m^*} g^{n j^*} R_{n m^* k l^*}$ and $g^{k m^*} = \delta^{k m^*}+ \cdots$. 

It is interesting to contrast this result with the result derived 
from nonrenormalizable terms present in the superpotential. First, 
the four-fermion interactions in the effective Lagrangian defined at
a  certain
scale can never be produced by a superpotential defined at that scale, 
at least with a linear realization of supersymmetry. A nontrivial 
K\"ahler potential must be included. However, the key phrase here is ``at 
the scale where the effective Lagrangian is defined.'' If the low energy 
superpotential and K\"ahler potential are assumed to be derived as 
effective functions from a high energy theory, the same effect
can come from the superpotential of the high energy theory upon decoupling
the heavy fields. For example, consider a superpotential of the form $y 
\Phi \phi \phi$, where $\Phi$ denotes a heavy scalar which
is integrated out when deriving the low energy effective Lagrangian.
Defining $|y|^2/M^2 = R$, one can see that the above four-fermion term in
the low energy effective theory is reproduced. The procedure of
integrating out the heavy fields generates nonrenormalizable corrections
to the effective superpotential and K\"ahler potential 
of the theory \cite{Cvetic:1998ef}.  The four-fermion operator 
then originates from this effective K\"ahler potential of the theory. 
It is possible to produce the terms mentioned above with a 
nonrenormalizable
term in the superpotential but there will be noticeable differences in the
effective Lagrangian. For example, if in addition to the renormalizable
terms there is a superpotential term $a_1 \phi^i \phi^k \phi^a \phi^b +
a_2 \phi^i \phi^l \phi^c \phi^d$, several of the terms of the last line of
Eq.~(\ref{nonrenl}) can be reproduced with the proper choice of $a_1$ and
$a_2$.  However, this superpotential operator does not yield the
nonrenormalizable terms in the second line of Eq.~(\ref{nonrenl}) and 
would include a number of other terms of the form $\phi \phi \psi \psi$ 
which are not included in the set of nonrenormalizable terms generated by 
the K\"ahler potential.

A nontrivial gauge kinetic function also can lead to nonrenormalizable 
operators. The couplings involving the gauge kinetic function include the 
following terms: 
\begin{eqnarray}
{\mathcal{L}}_{{\rm gauge-kinetic}} &=& 
-\frac{{\rm Im}f}{16 \pi} 
F^a_{\mu \nu}\widetilde{F}^{a \mu \nu} 
- \frac{{\rm Re}f}{16 \pi}  \left[ F^a_{\mu \nu}
F^{a \mu \nu} -i4 \lambda^{a \dagger} \overline{\sigma}^{\mu} D_{\mu} \lambda^a 
+ 2 D^a D^a \right] \nonumber \\
&-& \frac{1}{16 \pi} \frac{\partial f }{\partial \phi_i} F_i \lambda^a
\lambda^a + {\rm h.c.} + \cdots
\label{gkin}
\end{eqnarray} 
In the above expression, $F_i$ denotes the auxiliary component of 
$\Phi_i$. If $f$ is 
simply a complex number, {\it e.g.} if ${\rm Im(f)}= \frac{\theta}{2 \pi}$, 
${\rm Re}(f)= \frac{4 \pi}{g^2}$, 
the first line of Eq.~(\ref{gkin}) is just the usual kinetic terms for the 
gauge bosons and gauginos  
terms also presented in Eq.~(\ref{susygaugeL}) and the last term of 
Eq.~(\ref{gkin}) is zero.  

However, if $f$ is a function of the chiral superfields $\Phi_i$, these
couplings are nonrenormalizable interaction terms.  In particular, the
last term of Eq.~(\ref{gkin}) is nonzero and represents a potential mass
term for the gauginos. Gauge invariance dictates that $f$ must be
contained within the symmetric product of two adjoints.  It is usually
assumed to be a singlet (see \cite{Anderson:1999ui} for alternative
possibilities within the context of GUT models and the resulting
phenomenological implications).

The issue of generating gaugino masses through nontrivial gauge kinetic 
functions is most commonly discussed in the context of supergravity, which 
we will discuss in more detail in the next section.  Here we just wish to 
note a few salient points which do not require the full 
machinery of supergravity to obtain intuition about this topic.
   
We begin with a  classic example of using models with
singlets to obtain nonvanishing gaugino masses, which is string-motivated
supergravity.  In {\it e.g.} perturbative heterotic string theory, the
superstring tree-level gauge kinetic function is of the form $f_a= S/M_S$,
where $S$ (the ``dilaton") is a singlet chiral superfield and $M_S$ is the
string scale (in the literature $S$ is typically rescaled so as
to be dimensionless).  To reproduce the standard gauge couplings, the
scalar component of the dilaton must obtain a VEV $<S> = [4 \pi / g^2 - i
\theta/ 2 \pi] M_S $.  If the $S$ field also has a nonvanishing auxiliary
component $F_S \neq 0$ and hence participates in supersymmetry breaking, a
gaugino mass term of order $F_S/M_S$ is produced.

Let us now consider models without singlets. Gauge invariance then
dictates that the most general gauge kinetic function can be written as $f
\sim \Phi \overline{\Phi} / M^2 + O(M^{-3})$. Here $\overline{\Phi}$ is not 
the complex conjugate of $\Phi$, but rather another field which transforms
under the conjugate representation. If $F_{\Phi} \sim
\Lambda_S^2$ ($\Lambda_S$ denotes the supersymmetry breaking scale) and $M\sim 
M_{Pl}$,
the gaugino mass is of order $<F_{\Phi} \overline{\phi}>/M_{Pl}^2 \sim
\Lambda_S^3/M_{Pl}^2$, which usually is too small for practical purposes.  
For this reason, in practice it is desirable to have singlets which
participate in supersymmetry breaking.  An exception to this,
however, is anomaly-mediated supersymmetry breaking (see
Section~\ref{lsoftmodelsect}).

\subsection{Nonrenormalization theorem}
\label{nonrenormapp}
In this Appendix, we discuss the validity of the supersymmetric 
nonrenormalization theorem. For concreteness, consider once again the 
Wess-Zumino model as the theory defined at a  high energy scale 
$\Lambda_X$. 
The task at hand is to determine the
form of the effective Lagrangian defined at a low scale $\mu$,
${\mathcal{L}}_{eff}(m,y, \phi,  \Lambda_X, \mu)$, after integrating out
the high energy degrees of freedom.

One can easily verify that the high energy Lagrangian
Eq.~(\ref{susyL1}) possesses two global $U(1)$ symmetries as
shown in Table~\ref{U1charge}.\footnote{The $U(1)$ transformation on an
object $\Phi$ is defined by
$\Phi \rightarrow e^{i Q_{\Phi} \alpha} \Phi$ where $Q_{\Phi}$ is the
charge of $\Phi$ under the $U(1)$ transformation. The charges are
presented in the table. The symmetries are exact in the absence of gauge 
symmetries, but in general can be anomalous if gauge fields are
present.  We will discuss the effects of anomalies later.} 
\begin{table}
\begin{tabular}{cc|c|c|c|}\cline{3-5}
 \hspace{1cm} &\hspace{4cm} &\hspace{1cm}
&$\hspace{0.5cm}U(1) \hspace{0.5cm}$ & \hspace{0.5cm} $U(1)_R$
\hspace{0.5cm} \\
\cline{3-5}
\cline{3-5}
 \hspace{1cm} &\hspace{4cm} &$\Phi_0$ &1&1\hspace{3mm} 0 \\
\cline{3-5}
\hspace{1cm} &\hspace{4cm} &$m_0$&-2&0 \\
\cline{3-5}
\hspace{1cm} &\hspace{4cm} &$y_0$&-3&-1 \\
\cline{3-5}
\end{tabular}
\caption{The charge assignments with respect to the $U(1) \times U(1)_R$ 
global symmetries discussed in the text.}
\label{U1charge}
\end{table}
The notation is that $\Phi$ denotes the complete supermultiplet, and hence  
$\phi$ and $\psi$ transform similarly under the first $U(1)$ (each with a 
charge of 1).
However, they have different charges with respect to $U(1)_R$
($Q_{\phi}=1$, $Q_{\psi}=0$). 

In this discussion the parameters $m_0$, $y_0$ and the
fields are treated on equal footing as complex variables 
which transform under the global symmetry. An arguably more physical 
approach is to regard the parameters as the VEVs of heavy
background fields (the spurion fields) which are no longer 
propagating degrees of freedom. 
 From this point of view, the parameters of the theory are the scalar
 component VEVs of certain supermultiplets (the parameters can be 
considered as chiral multiplets $M=(m, ...)$ 
and $Y=(y,...)$).  In other words, this model can be treated as a theory 
of three interacting supermultiplets in which the parameter multiplets do
not contain propagating degrees of freedom, such that their only physical 
effects are due to their nonvanishing VEVs.

In this model, the global symmetries are two $U(1)$ symmetries, presented 
in the table. The 4 $U(1)$ charges associated with $\phi_0$,
$\psi_0$, $m_0$ and $y_0$ should allow the gauge invariance of the two 
terms $m y^* \phi \phi^{* 2}$ and $m\psi\psi$ (other terms are either 
trivially symmetric or not independent). Therefore, up to an overall 
normalization factor,
there are two independent solutions. Note that the global $U(1)$
symmetries remain exact as the heavy degrees of freedom are integrated
out to obtain the low energy effective Lagrangian.  The key observation is
that it is possible to integrate out the high energy degrees of freedom in
such a way that is consistent with the symmetry, {\it i.e.}, only complete 
sets of degrees of freedom transforming into each other under the 
symmetry operation are integrated out at each step.

Consider the weak coupling limit of this
theory. Taking the limit $y \rightarrow 0$ should not yield any
singularities, as this limit corresponds to a free theory
with trivial dynamics. Taking the combined limit $y \rightarrow
0$ and $m/y
\rightarrow 0$, {\it i.e.}, taking the mass to zero and the 
coupling small, should also be a smooth limit, corresponding 
to a massless weakly interacting theory. Both of these properties play 
crucial roles in determining the
renormalization properties of the model. In summary, the requirements on 
the low energy effective
Lagrangian are as follows:
\begin{itemize}
\item It must be supersymmetric.
\item It must preserve the global symmetries.
\item It has smooth weak coupling limits.
\end{itemize}
The form of the low energy Lagrangian that satisfies these
requirements is
\begin{eqnarray}
{\mathcal{L}}_{eff}&=& {\mathcal{Z}}\times({\rm kinetic} \hspace{2mm} {\rm 
terms})+ |r_m|^2 |m_0|^2
|\phi_0|^2 + r_m m \psi_0 \psi_0 + r_m^*
m_0^*\psi_0^{\dagger} \psi_0^{\dagger} \nonumber \\
&+& r_y y_0 \phi_0 \psi_0 \psi_0 + r_y^* y_0^* \phi_0^*
\psi_0^* \psi_0^* + \frac{|r_y|^2 |y_0|^2}{4} |\phi_0|^4
\nonumber \\
&+& \frac{1}{2}r_m r_y^* m_0 y_0^* \phi_0 \phi_0^{* 2} +
\frac{1}{2}r_m^* r_y m_0^* y_0 \phi_0^* \phi_0^{2},
\label{renorm-L}
\end{eqnarray}
where $r_m$ and $r_y$ are constants (they could be functions of
$\Lambda_X$ and $\mu$). ${\mathcal{Z}}$ denotes the wavefunction 
renormalization
for the kinetic terms of the Lagrangian. This is the key step and
therefore deserves more explanation.

\begin{itemize}
\item There should be no terms generated which have inverse powers
of $y$ and $m$.
Otherwise, the theory has no smooth weak coupling limit.

\item The same $r_m$ and $r_y$ occur in different terms in order to
preserve supersymmetry, 
as can be verified by using the supersymmetry transformation rules
presented in Section~\ref{backgroundsect}.

\item No terms proportional to $\Lambda^2 |\phi|^2$ should be generated
in the low energy
effective theory, because if such a term is present,
supersymmetry requires that there must be terms proportional to $\Lambda
y^* \phi \phi^{* 2}
+ {\rm h.c.}$. However, such cutoff-dependent terms are disallowed because 
they
break the $U(1)$ global symmetry.  

\item If nonrenormalizable terms such as $|y|^2
|\phi|^6/\Lambda^2$ 
(from a superpotential term $y \Phi^4/ \Lambda$)  are present in the
theory, supersymmetry requires the presence of additional terms such
as $m^* y \phi^* \phi^3$. However, this term would break the global
symmetries and thus is forbidden.  Following similar logic, it can be
shown that no nonrenormalizable terms are generated and
Eq.~(\ref{renorm-L}) contains {\it all} the terms of the effective Lagrangian.

\item $r_m$ and $r_y$ can only be functions of $\Lambda_X$ and $\mu$.
Otherwise either the global symmetry (for $yy$ type couplings) or
supersymmetry with respect to $M$ or $Y$ (for $yy^*$ type couplings) is
broken. (This is not obvious and can be shown best using superfield
techniques. We refer the interested reader to the work of Seiberg
\cite{Seiberg:1993vc,Seiberg:1994bp} for details.)

\end{itemize}
The rescaling $\phi= \sqrt{{\mathcal{Z}}}\phi_0$ can now be done to cast 
the kinetic terms into canonical form. Therefore, in terms of the canonically
normalized variables, $m_0 \rightarrow m=m_0/Z$ and  $y_0 \rightarrow
y=y_0/({\mathcal{Z}})^{\frac{3}{2}}$. The constants $r_m$ and $r_y$ can 
be determined 
by taking weak coupling
limits  of the theory. Taking the limit $y \rightarrow 0$, one obtains a
free theory where the low energy effective Lagrangian should be the same
as the high energy one, since no renormalization and counterterms are 
needed for a free propagating theory.
By requiring the mass term of the rescaled low
energy effective theory and the original theory to be equal, the constant
$r_m$ is determined to be $r_m= {\mathcal{Z}}_{free}$, in which 
${\mathcal{Z}}_{free}$ 
denotes the
wave function renormalization in the free field limit. Next, one takes the
massless limit where the interaction $y$
is small. Since the coupling can be made arbitrarily small, the
perturbative calculations using ${\mathcal{L}}_0$ and
${\mathcal{L}}_{eff}$ must match order by order to produce the
same result. This procedure yields
$r_y= ({\mathcal{Z}}_{0})^{\frac{3}{2}}$, where ${\mathcal{Z}}_{0}$ is 
the wave 
function
renormalization for free field in the zero coupling limit. Notice both
${\mathcal{Z}}_{free}$  and ${\mathcal{Z}}_{0}$ are finite constants.
Hence, the low energy effective Lagrangian has
the same form as the original one, with the effective parameters
\begin{equation}
y=\left(\frac{{\mathcal{Z}}_{0}}{{\mathcal{Z}}(\mu)}\right)^{\frac{3}{2}}y_0, 
\hspace{1cm}
m=\left(\frac{{\mathcal{Z}}_{free}}{{\mathcal{Z}}(\mu)}\right) m_0.   
\end{equation}
Hence, the parameters of Eq.~(\ref{susyL1}) are only  
renormalized due to the wavefunction rescaling.
This provides the logarithmic corrections
that are induced by using running couplings and masses.  Thus, the
hierarchy problem previously described is absent in this supersymmetric
theory. This argument can be generalized to an interacting theory with
many chiral multiplets.

Let us now comment on what happens if the above
matter theory is coupled to gauge fields. In a supersymmetric gauge 
theory, the gauge
coupling does get renormalized, but only gets perturbative
corrections at one-loop order.\footnote{Nonperturbative corrections due
to instanton effects are present, but are  generally suppressed by
$e^{-\frac{1}{g^2}}$, where $g$ is the gauge coupling.}  
The global $U(1)$'s used to prove the nonrenormalization
theorem are now anomalous.  However, the supersymmetric
Lagrangian described above still receives no
further renormalization within perturbation theory. Once
again there are suppressed nonperturbative corrections due to
instanton effects.

\subsection{Classification of soft parameters}
\label{softapp}
In this section, a discussion of the classification of
supersymmetry-breaking terms into ``soft'' or ``hard'' breaking using
power  
counting arguments is presented. To proceed, recall the usual mass 
dimension $d(\phi)=1$ and $d(\psi)=\frac{3}{2}$ of  the bosonic and
fermionic fields. The mass dimension $d_{\mathcal{O}}$ of any operator
$\mathcal{O}$ is $d_{\mathcal{O}}={n_b}+ \frac{3}{2} n_f + $
(momentum dependence), where $n_b$ and $n_f$ are the number of bosonic and
fermionic fields appearing in the  operator. In general, momentum 
dependence can arise due to derivatives in the
operator. If an
operator $\mathcal{O}$ appears in the Lagrangian, it at most can have a
cutoff dependence to the power of $p_{\Lambda}=4-d_{\mathcal{O}}$. If the
theory is fully supersymmetric, no operator in the theory will have
any power law dependence on the cutoff (the dependence is at most
logarithmic). The problem now is: including all possible supersymmetry-breaking
operators ${\mathcal{O}}_1$, ${\mathcal{O}}_2$, {\it etc.}, are new
dangerous cutoff dependence regenerated in the Lagrangian? Suppose the   
operators ${\mathcal{O}}_1$, ${\mathcal{O}}_2$, {\it etc.}, can form loops 
with
other operators (or within themselves) to give rise to new operators
$\mathcal{O}$. These are the new contributions one can have to 
the effective Lagrangian by the insertion of those new operators. By power
counting, the newly formed operator will have at most a cutoff dependence
of power \cite{Weinberg:cr}
\begin{equation}
p_{\Lambda}=  4-d_{\mathcal{O}}-({4-d_{{\mathcal{O}}_1}})
-({4-d_{{\mathcal{O}}_2}})-\cdots.
\label{power}
\end{equation}
If $d_{\mathcal{O}}=0$, the newly generated operator ${\mathcal{O}}$ has
no field dependence. It is a cosmological constant, which is not
discussed further here. The $d_{\mathcal{O}}=1$ term is a tadpole
contribution. If $d_{\mathcal{O}}=2$, it represents a cutoff-dependent   
contribution to the scalar mass.  If $d_{\mathcal{O}}=3$ and the dimension
of the supersymmetry-breaking terms $3 \geq d_{{\mathcal{O}}_i} \geq 1$ (which is always 
true for the soft terms),
there should be no power
law dependence on the cutoff by applying Eq.~(\ref{power}). Therefore, in
this discussion, attention will be focused on $d_{\mathcal{O}}=1$ and 
$d_{\mathcal{O}}=2$. If the extra 
insertion ${\mathcal{O}}_i$ is of dimension 3,
it is necessary to discuss its contribution to both the  
$d_{\mathcal{O}}=1$ and $d_{\mathcal{O}}=2$ operators. On the other hand, 
if the extra insertion ${\mathcal{O}}_i$ is of dimension 2, it is only necessary to consider
$d_{\mathcal{O}}=1$, because any insertion of dimension 2 will eliminate
the power dependence of cutoff in the case of $d_{\mathcal{O}}=2$.

For clarity, let us use the Wess-Zumino model (allowing for the 
possibility of gauge symmetry) as an example.  The list of soft supersymmetry-breaking 
parameters are as follows: 
\begin{enumerate}
\item ${\mathcal{O}}_{A}=A \phi \phi \phi$\\
This trilinear term has mass dimension $d({\mathcal{O}})_{A}=3$. The
lowest order contribution to the tadpole diagram can be made through the
contraction between ${\mathcal{O}}_{A}$ and two operators of the form
${\mathcal{O}}_{my}=m^* y \phi^2 \phi^*$. Using
Eq.~(\ref{power}), one can compute $p_{\Lambda}=3-(4- d({\mathcal{O}}_{A}))
- 2(4-d({\mathcal{O}}_{my}))=0$. Thus, there is no dangerous tadpole
contribution. Now consider its contribution to the dimension 2 operator.
The lowest order contribution will be the contraction between
${\mathcal{O}}_{A}$ and ${\mathcal{O}}_{my}$. By power counting arguments,
this will not lead to dangerous divergences. Therefore, the trilinear
coupling is indeed soft.

\item ${\mathcal{O}}_{\lambda}=M \lambda^{a} \lambda^{a}$ \\
Terms of this type give gauginos nonzero masses and have
$d_{{\mathcal{O}}_{\lambda}}=3$. One can verify this type of term do not
generate extra dangerous tadpole contributions. The lowest order
contribution to the $d_{\mathcal{O}}=2$ operators is proportional to
${\mathcal{O}}_{\lambda}{\mathcal{O}}_{\lambda}^{\dagger}$. There must be
two insertions of ${\mathcal{O}}_{\lambda}$.  Using Eq.~(\ref{power}), one
can show $p_{\Lambda}=0$. Hence, there is no power dependence of cutoff  
generated by the inclusion of ${\mathcal{O}}_{\lambda}$, such that gaugino
mass terms are soft.

\item ${\mathcal{O}}_{m^{\prime}}=|m^{\prime}|^2 |\phi|^2$\\
This term gives masses to the scalar fields of chiral multiplets and has
mass dimension $d({\mathcal{O}}_{m^{\prime}})=2$. Therefore, it is only
necessary to
discuss its contribution to tadpole diagram, $d_{\mathcal{O}}=1$. The
lowest order contribution is the contraction
between ${\mathcal{O}}_{m^{\prime}}$ and another dimension 3 operator   
${\mathcal{O}}_{my}=m^* y \phi^2 \phi^*$. Eq.~(\ref{power})
leads to
$p_{\Lambda}=3-(4-d({\mathcal{O}}_{m^{\prime}}))-
(4-d({\mathcal{O}}_{my}))=0$. Therefore, this operator does not contribute  
to tadpole divergences.

\item ${\mathcal{O}}_{b}=b \phi \phi + {\rm h.c.}$\\
This term is dimension 2 and only has a potential
contribution to tadpole divergences. One can verify that the lowest
contribution comes from the contraction between ${\mathcal{O}}_{b}$ and a
${\mathcal{O}}_{my}$ type term, which is harmless by power counting.

\end{enumerate}
There is also a set of parameters that can give rise to potential tadpole
divergences. Such terms can be soft if there is no singlet in the theory.
In the absence of singlets, the tadpole vanishes because the one point
amplitude is not gauge invariant. These terms include the following:

\begin{enumerate}
\item $C\phi^* \phi \phi + {\rm h.c.} $\\
Two fields, $\phi$ and $\phi^*$ can contract to make this operator
into a tadpole diagram. Therefore, this operator will potentially
contribute to power law dependence of the cutoff, reintroducing the
hierarchy problem.

\item $m_F \psi \psi + {\rm h.c.}$\\
This operator can contract with $y \phi \psi \psi + {\rm h.c.}$, forming a
tadpole diagram and introducing tadpole divergences. However, this is
related to the previous one by a supersymmetric transformation.  
Therefore, one of these operators can always be eliminated by an 
appropriate redefinition of the fields.

\item $m_A \psi \lambda^{a} + {\rm h.c.}$\\
This term can also lead to tadpole divergences by contracting with $\phi^*
\lambda \psi$ type terms. However, gauge invariance requires the existence
of matter in the adjoint representation of the gauge groups for such terms
to be present. Such matter content is not present in the phenomenological 
models of interest within this review, and hence such supersymmetry-breaking 
terms will not be considered further.

\end{enumerate}

There is no gauge singlet in the MSSM, which is the main subject of this
review. Therefore, in principle one should include terms of the form
$C\phi^* \phi \phi + {\rm h.c.} $ in \lsoft. However,
they are usually omitted because there is a practical difficulty in
constructing realistic supersymmetry-breaking models that give rise to
terms of this type which are also reasonable in size.

For completeness, here are the supersymmetry-breaking   
terms which are not soft:
\begin{enumerate}
\item Terms of dimension 4.\\
supersymmetry-breaking terms with dimensionless couplings 
generically lead to dangerous divergences. Such dimension 4 
terms are of the form  $\phi \psi  \psi$,
$|\phi|^4$, {\it etc.} Power counting demonstrates that all such 
operators 
lead to quadratic divergences.

\item Terms of dimension larger than $4$.\\
This type of terms are usually suppressed by powers of given high energy 
scale. Their contribution to quadratic divergences should be no worse than
that of the dimension 4 operators.
\end{enumerate}

\setcounter{footnote}{0}
\section{Supergravity basics and the gravitino}
 \label{sec:sugra}

Although a fully consistent theory of quantum gravity coupling to matter
is yet to be determined, its effective theory at energies much lower than
the Planck scale can be derived 
(albeit nonrenormalizable) based on symmetries.  A
supersymmetric effective theory which describes the coupling between
gravity and matter is {\it supergravity}, which is a theory with local
gauged supersymmetry.\footnote{Since the supersymmetry algebra includes
the spacetime translation operator $P^{\mu}$, it includes the general
coordinate transformations when supersymmetry is gauged. Therefore, it is
natural that a locally supersymmetric theory will have gravity.}

The supergravity theory of immediate phenomenological
interest is $D=4$,
$N=1$ supergravity. In this theory, there is a new fermionic field in 
which is the superpartner of the spin 2 graviton. This field is the spin 
$\frac{3}{2}$ gravitin $\widetilde{G}^{\alpha}_m$, which has a  
spinor index denoted by $\alpha$ and a spacetime index denoted by $m$. 
The K\"ahler transformation of global supersymmetry is generalized to  a 
K\"ahler-Weyl
transformation which includes a rescaling of the superpotential (see
Appendix~\ref{sugraapp}). Therefore, any
holomorphic term $F$ can be transformed into a rescaling of the
superpotential $W \rightarrow e^{\kappa^2 F} W = W + \kappa^2 F W 
+\cdots$. Notice that all holomorphic terms in the K\"ahler   
potential will by multiplied by positive powers of $\kappa$ when
transformed into the superpotential.

The supergravity Lagrangian is general at any scale below four
dimensional Planck scale and at which a four dimensional field theory
description  of our world is still valid. 
For phenomenological analyses one typically takes the flat limit, which 
is the limit of infinite Planck scale ({\it i.e.} $\kappa \rightarrow
0$), while keeping $m_{3/2}$ fixed. 
Supersymmetry is broken at low energy scales; it is assumed to be 
spontaneously broken by the VEVs of 
certain fields at higher scales. As a result, the
gravitino, which is the gauge fermion of local supersymmetry,
will acquire a mass, just like in the Higgs mechanism which gives 
gauge bosons of the corresponding broken symmetry generators  
a nonvanishing mass. On dimensional grounds the gravitino mass is 
$c \kappa <F>$, where $c$ is some dimensionless number and $<F>$
is some VEV of mass dimension 2  which breaks supersymmetry. As
seen in the discussion of gravity-mediated supersymmetry
breaking, the gravitino mass sets the scale of the soft supersymmetry-breaking 
terms which appear in the low energy effective theory. 
The resulting Lagrangian includes a globally supersymmetric sector 
(summarized by a superpotential, a K\"ahler potential, and a gauge
kinetic function) and a set of terms which break supersymmetry
explicitly.     

\subsection{$D=4$, $N=1$ supergravity Lagrangian}
\label{sugraapp}
In this section, the $D=4$, $N=1$ supergravity
Lagrangian describing chiral matter coupling
to gravity is presented (see \cite{Bagger:1990qh} for details and the 
derivation). 
The Lagrangian is presented again with the aid of
a superpotential $W$ and a K\"ahler potential $K$:
\begin{eqnarray}
e^{-1}{\mathcal{L}}_{SUGRA}= &-&\frac{1}{2} {\mathcal{R}} - g_{ij^*}
\partial_m \phi^i \partial^m \phi^{j*} \nonumber \\
 &-&i g_{ij^*} \overline{\psi}^j \overline{\sigma}^m {\mathcal{D}}_m \psi^i +
\epsilon^{klmn} \overline{\widetilde{G}}_k \overline{\sigma_l} \widetilde{\mathcal{D}}_m
\widetilde{G}_n \nonumber \\
&-&\frac{1}{2} \sqrt{2} g_{i j^*} \partial_n \phi^{*j} \psi^i \sigma^m
\overline{\sigma}^n \widetilde{G}_m  - \frac{1}{2} \sqrt{2} g_{i j^*}
\partial_n \phi^{i} \overline{\psi}^{j*} \overline{\sigma}^m \sigma^n
\overline{\widetilde{G}}_m \nonumber \\
&+&\frac{1}{4} g_{ij^*}[i \epsilon^{klmn} \widetilde{G}_k \sigma_l
\overline{\widetilde{G}}_m + \widetilde{G}_m \sigma^n \overline{\psi}^m] \psi^i \sigma_n
\overline{\psi}^j \nonumber \\
&-&\frac{1}{8}[g_{ij^*}g_{kl^*}-2R_{ij^* kl^*}] \psi^i \psi^k
\overline{\psi}^j \overline{\psi}^l \nonumber \\
&-& \exp^{K/2} \{ W^* \widetilde{G}_a \sigma^{ab} \widetilde{G}_{b} + W
\overline{\widetilde{G}}_a \overline{\sigma}^{ab} \overline{\widetilde{G}}_{b}  \nonumber \\
&+&\frac{i}{2} \sqrt{2} D_i W \psi^i \sigma^m \overline{\widetilde{G}}_m +
\frac{i}{2} \sqrt{2} D_{i^*} W^* \overline{\psi}^i \overline{\sigma}^m
\widetilde{G}_m \nonumber \\
&+& \frac{1}{2} {\mathcal{D}}_i D_j W \psi^i \psi^j +\frac{1}{2}   
{\mathcal{D}}_{i^*} D_{j^*} W^* \overline{\psi}^i \overline{\psi}^j \}
\nonumber  \\
&-& \exp(K) [g^{ij^*} (D_i W)(D_j W)^* - 3 W W^*],
\end{eqnarray}
where $(\phi^i$ and $\psi^i)$ are the usual components of chiral
multiplets. The curved spacetime is described by the metric tensor
$g_{\mu \nu}$, and $e=\sqrt{-{\rm Det}(g_{\mu \nu})}$. There is also a
superpartner of the graviton called the gravitino, which is denoted by
$\widetilde{G}_m$. The various derivatives are defined by
\begin{eqnarray}
{\mathcal{D}}_m \psi^i&=&\partial_m \psi^i +\psi^i \omega_m +
\Gamma^i_{jk} \partial_m \phi^j \psi^k -\frac{1}{4}(K_j \partial_m
\phi^j - K_{j^*} \partial_m \phi^{j^*}) \psi^i \nonumber \\
\widetilde{\mathcal{D}}_m \widetilde{G}_n &=& \partial_m \widetilde{G}_n  +  
\widetilde{G}_n \omega_m +\frac{1}{4}(K_j \partial_m
\phi^j - K_{j^*} \partial_m \phi^{j^*}) \widetilde{G}_n \nonumber \\
D_i W &=& W_i + K_i W \nonumber \\
{\mathcal{D}}_i D_j W &=& W_{ij} + K_{ij}W + K_i D_j W + K_j D_i W - K_i
K_j W - \Gamma^k_{ij} D_k W.  \nonumber \\
\end{eqnarray}
where $\omega_m$ are spin connections.\footnote{Spin connections
arise when coupling spinors to a curved backgroud in a
covariant way. }
For simplicity, the results above are expressed in units such that 
$\kappa^2 = 8 \pi G_N = 1$. The full $\kappa^2$ dependence can be restored 
on dimensional grounds, using $\kappa^2 \propto M_{Pl}^{-2}$. For example, 
the term $\frac{1}{8}[g_{ij^*}g_{kl^*}-2R_{ij^*kl^*}]\psi^i \psi^j 
\overline{\psi}^j \overline{\psi}^l $ will be suppressed by 
$\kappa^2$.\footnote{Although these 
units are often used, one should keep the $\kappa^2$ dependence in mind 
especially when studying low energy phenomenology, in which $\kappa^2 
\rightarrow 0$.}

The K\"ahler transformation of global supersymmetry is not a symmetry of 
supergravity. The appropriate transformation is the {\it K\"ahler-Weyl} 
transformation:
\begin{equation}
K(\phi, \phi^*) \rightarrow K(\phi, \phi^*)+ F(\phi) + F^*(\phi^*),
\end{equation}
and all spinor fields are rescaled
\begin{eqnarray}
\phi_i &\rightarrow& \exp \left(\frac{i}{2}{\mathbf{Im}}F \right) \phi^i
\nonumber \\
\widetilde{G}_m &\rightarrow& \exp \left(-\frac{i}{2}{\mathbf{Im}}F \right)
\widetilde{G}_m.
\end{eqnarray}
In addition, the superpotential is rescaled as
\begin{equation}
W \rightarrow e^{-F} W,
\end{equation}
such that
\begin{equation}
D_i W \rightarrow e^{-F}D_i W.
\end{equation}  
When $<W>\neq 0$ ({\it i.e.} if supersymmetry is
broken), the superpotential can be rescaled to $1$ by choosing $F=\ln W$. 
Defining $G=K+ \ln W + \ln W^*$, the Lagrangian can be recast as a 
function only of $G$ as follows:
\begin{eqnarray}
e^{-1}{\mathcal{L}}_{SUGRA}= &-&\frac{1}{2} {\mathcal{R}} - g_{ij^*}
\partial_m \phi^i \partial^m \phi^{j*} \nonumber \\
 &-&i g_{ij^*} \overline{\psi}^j \overline{\sigma}^m {\mathcal{D}}_m \psi^i +
\epsilon^{klmn} \overline{\widetilde{G}}_k \overline{\sigma_l} \widetilde{\mathcal{D}}_m
\widetilde{G}_n \nonumber \\
&-&\frac{1}{2} \sqrt{2} g_{i j^*} \partial_n \phi^{*j} \psi^i \sigma^m
\overline{\sigma}^n \widetilde{G}_m  - \frac{1}{2} \sqrt{2} g_{i j^*}
\partial_n \phi^{i} \overline{\psi}^{j*} \overline{\sigma}^m \sigma^n
\overline{\widetilde{G}}_m \nonumber \\
&+&\frac{1}{4} g_{ij^*}[i \epsilon^{klmn} \widetilde{G}_k \sigma_l
\overline{\widetilde{G}}_m + \widetilde{G}_m \sigma^n \overline{\psi}^m] \psi^i \sigma_n
\overline{\psi}^j \nonumber \\
&-&\frac{1}{8}[g_{ij^*}g_{kl^*}-2R_{ij^* kl^*}] \psi^i \psi^k
\overline{\psi}^j \overline{\psi}^l \nonumber \\
&-& \exp^{G/2} \{ \widetilde{G}_a \sigma^{ab} \widetilde{G}_{b} +
\overline{\widetilde{G}}_a \overline{\sigma}^{ab} \overline{\widetilde{G}}_{b}  \nonumber \\
&+&\frac{i}{2} \sqrt{2} G_i \psi^i \sigma^m \overline{\widetilde{G}}_m +
\frac{i}{2} \sqrt{2} G_{i^*}  \overline{\psi}^i \overline{\sigma}^m
\widetilde{G}_m \nonumber \\
&+& \frac{1}{2} [G_{ij}+G_i G_j - \Gamma^k_{ij} G_k] \psi^i \psi^j 
\nonumber \\
&+&\frac{1}{2} [G_{i^*j^*}+G_{i^*} G_{j^*} - \Gamma^k_{i^*j^*}
G_{k^*}]\overline{\psi}^i  \overline{\psi}^j \}
\nonumber  \\
&-& \exp(G) [g^{ij^*} G_i G_{j^*} - 3 ].
\end{eqnarray}

A full account on the most general gauge interactions in the
supergravity Lagrangian is again beyond the scope of this
review. Almost all of the detail introduction to supergravity contain
treatments of this subject. We
refer interested readers to those references.  We
will just briefly comment on their properties. The most relevant gauge
interactions can be added to the supergravity in a straight forward
way. The first step is again extend all the covariant derivatives in
the supergravity Lagrangian to include gauge interaction ({\it i.e.}, 
adding
term like $T^a A^a \phi$ ) for all the matter field transform under
the gauge symmetry. All the other terms  involving gauge
fields in the globally symmetric models are also present in the
supergravity Lagrangian. The only change is that they have to be
integrated over an invariant volume form ({\it i.e.}, change all the
integral $\int d^4 x \rightarrow \int d^4 x \sqrt{-g}$ ). There are
some other changes involving the nonrenormalizable couplings with
gravitinos. However, those terms are generally of less
phenomenological importance especially in the flat limit, in which 
$M_{Pl}$ is taken to infinity while $m_{3/2}$ is held fixed.

\subsection{Supergravity potential}
\label{sugrapotapp}
Let us focus on the supergravity scalar potential, assuming that the 
chiral superfields in the theory $\Phi$ can be divided into hidden sector 
fields  $h$ and observable sector states $C_a$. As demonstrated in the 
previous subsection, the theory can be described in terms of the 
K\"{a}hler function:\footnote{
Powers of the reduced Planck mass $(\widetilde{M}_{P})$ that appear 
in the K\"{a}hler function to obtain the correct dimensions are retained 
although it is
conventional to adopt natural units and set $\widetilde{M}_{P}=1$.}
\begin{eqnarray}
 G(\phi, \overline{\phi}) = \frac{K(\phi,\overline{\phi})}{\widetilde{M}_{P}^{2}}  
  + \ln \left(\frac{W(\phi)}{\widetilde{M}_{P}^{3}} \right) +
   \ln \left( \frac{W^{\ast}(\overline{\phi})}{\widetilde{M}_{P}^{3}} \right).
 \label{eq:gkahler}
\end{eqnarray}
The K\"{a}hler potential $K(\phi,\overline{\phi})$ may be expanded in powers of 
matter states $C_{a}$ (including nonperturbative contributions):
\begin{eqnarray}
 K=\overline{K}(h,\overline{h}) +\widetilde{K}_{\overline{a}b}(h,\overline{h}) \overline{C}_{\overline{a}} 
C_{b}
  + \left[ \frac{1}{2} Z_{ab}(h,\overline{h}) C_{a} C_{b} + {\rm h.c.} \right] 
+ 
\ldots
   \label{eq:kahexp}
\end{eqnarray}
where $\widetilde{K}_{a\overline{b}}$ is the (generally nondiagonal) matter metric 
and a nonzero bilinear term $Z_{ab}$ can generate the $\mu$-term
through the Guidice-Masiero mechanism~\cite{Giudice:1988yz} subject to 
gauge-invariance.
The superpotential $W(\Phi)$ can also be expanded:
\begin{eqnarray}
 W=\hat{W}(h) + \frac{1}{2} \mu_{ab}(h) C_{a}C_{b}
  + \frac{1}{6} Y_{abc} C_{a}C_{b}C_{c} + \ldots  \label{eq:supexp}
\end{eqnarray}
Notice that it includes a trilinear Yukawa term (that will generate 
fermion masses) and a bilinear $\mu$ term. 

Several mechanisms have been proposed for supersymmetry breaking.
It is convenient to analyze this breaking by considering the F term 
contribution to the SUGRA scalar potential (here the D term contribution 
to the potential that arises from the gauge sector will be ignored). It 
can be expressed in terms of derivatives of the K\"{a}hler function 
$G(\Phi, \overline{\Phi})$, or equivalently in terms of the F term auxiliary 
fields that can acquire nonzero VEV's and trigger supersymmetry breaking.
Using Eq.~(\ref{eq:gkahler}),
\begin{eqnarray}
 V(\phi, \overline{\phi})
  = e^{G} \left[ G_{I} (K^{-1})_{I \overline{J}} G_{\overline{J}}
   - 3 \right] = F_{\overline{J}} \, K_{\overline{J} I} \, F_{I}
    - 3 e^{K} \, |W|^{2}  \label{eq:potentialf}
\end{eqnarray}
where $I,J \equiv \phi_{I}, \phi_{J} \in S,T_{i},Y_{k},C_{a}$ and
\begin{eqnarray}
 G_{I} &\equiv& \frac{\partial G}{\partial \phi_{I}}
  = \frac{W_{I}}{W} + K_{I} \\
 F_{I} &=& e^{G/2} (K^{-1})_{I \overline{J}} \, G_{\overline{J}}
\end{eqnarray}
where $(K^{-1})_{I\overline{J}}$ is the inverse of $K_{\overline{J}I}$, and
satisfies the relation $(K^{-1})_{I \overline{J}} K_{\overline{J} L} = \delta_{IL}$.  
A subscript on G denotes partial differentiation, while the same subscript
on F is just a label.  A barred subscript on an F term denotes its
conjugate field $F_{\overline{I}} \equiv \left( F_{I} \right)^{\dagger}$. There
is no distinction made here between upper and lower indices.
   
After supersymmetry breaking, the supersymmetric partner of the
Goldstone boson (Goldstino) is {\it eaten} by the massless gravitino  through
the super-Higgs mechanism.  The gravitino now has a mass given by
\begin{eqnarray}
 m_{3/2}^{2} = e^{\langle G \rangle} =
  e^{\langle K \rangle} \, |\langle W \rangle|^{2}
   = \frac{1}{3} \langle F_{\overline{J}} \, K_{\overline{J} I} \, F_{I} \rangle
\end{eqnarray}
and sets the overall scale of the soft parameters.

In the absence of F term vacuum expectation values ($\langle F_{I} \rangle
= 0 \,\, \forall \, \phi_{I}$), the locally supersymmetric vacuum is
negative $V_{SUSY}= - 3 e^{G}$.  However if one (or more) of the auxiliary
F terms acquires a nonzero VEV, the negative vacuum energy can be
(partially) canceled.  This raises the exciting possibility that the
vacuum energy, or rather the cosmological constant $V_{0}$, can be made
vanishingly small in agreement with experimental limits.  Notice that such
a possibility cannot arise in {\it global} supersymmetry, for which the potential
is positive definite and the global minimum is supersymmetry preserving.

 
The presence of nonzero F term VEVs signal that supersymmetry is broken.  As the  
F term VEVs serve 
as the order parameters of supersymmetry breaking, it is useful to express the 
soft supersymmetry-breaking terms as functions of these VEVs.  
One can define a column 
vector
of F term VEVs $F$ in terms of a matrix P and column vector $\Theta$ 
(which also includes a CP-violating phase), where $\Theta$ has unit length 
and satisfies
$\Theta^{\dagger} \Theta =1$, and P canonically normalizes the
K\"{a}hler metric $P^{\dagger} K_{\overline{J} I} P = 1$:
\begin{eqnarray}
 F &=& \sqrt{3} \, C \, m_{3/2} \, \left( P \Theta \right)  
\label{eq:genf} \\
 F^{\dagger} &=& \sqrt{3} \, C \, m_{3/2} \,
  \left( \Theta^{\dagger} P^{\dagger} \right)  \nonumber
\end{eqnarray}

\noindent Replacing the fields by their VEVs, 
Eq.~(\ref{eq:potentialf}) can be rewritten as a matrix equation:
\begin{eqnarray}
 \langle V \rangle \equiv V_{0} &=& F^{\dagger} \, K_{\overline{J} I} \, F
    - 3 m_{3/2}^{2}  \nonumber  \\
 &=& 3 C^{2} m_{3/2}^{2} \, \Theta^{\dagger} \Theta
  \left( P^{\dagger} K_{\overline{J} I} P \right) - 3 m_{3/2}^{2}
   \label{eq:potentialf2} \\
 &=& 3 m_{3/2}^{2} \left( C^{2} - 1 \right) \nonumber
\end{eqnarray}  
where $V_{0}$ is the cosmological constant and hence
$C^{2}=1+\frac{V_{0}}{3m_{3/2}^{2}}$.  Therefore, choosing a vanishingly
small cosmological constant sets $C=1$.

As an example consider a model with the dilaton $S$ and an overall moduli
field $T$ with diagonal K\"{a}hler metric.  The SUGRA potential would be
a ``sum of squares''
$V_{F} \sim |F_{S}|^{2} + |F_{T}|^{2} + \ldots -3e^{G}$
and hence $P$ is a diagonal normalizing matrix:
\begin{eqnarray}
 P_{I \overline{J}} = (K_{I \overline{I}})^{-1/2} \delta_{I \overline{J}}
\end{eqnarray}
In this special case one would recover the expressions of
\cite{Brignole:1993dj}:
\begin{eqnarray}
 F\equiv \left( 
  \begin{array}{c}
   F_{S} \\
   F_{T}
  \end{array} \right) = \sqrt{3}C \, m_{3/2} \left(
   \begin{array}{c}
    (K_{S \overline{S}})^{-1/2} \sin\theta \, e^{i\alpha_{S}} \\
    (K_{T \overline{T}})^{-1/2} \cos\theta \, e^{i\alpha_{T}}
   \end{array} \right),
\end{eqnarray}
such that dilaton-dominated (moduli-dominated) supersymmetry breaking corresponds to 
$\sin\theta=1$ $(\cos\theta =1)$.
However in the more general case, the potential includes terms that mix
different F terms.  The action of $P$ is to canonically normalize
the K\"{a}hler metric and maintain the validity of the  
parameterization.

    
Using Eqs.~(\ref{eq:kahexp}),(\ref{eq:supexp}), one can write down the
{\it unnormalized} supersymmetry-breaking masses and trilinears that arise in the
soft SUGRA potential:
\begin{eqnarray}
 V_{soft}=m^{2}_{\overline{a}b} \overline{C}_{\overline{a}} C_{b}
  + \left( \frac{1}{6} A_{abc} Y_{abc} C_{a}C_{b}C_{c} + {\rm h.c.} 
\right) + 
\ldots
\end{eqnarray}
where the K\"{a}hler metrics are in general not diagonal leading to
the noncanonically normalized soft masses
\begin{eqnarray}
 m^{2}_{\overline{a}b} &=& \left( m_{3/2}^{2}+V_{0} \right) 
\widetilde{K}_{\overline{a}b}
  - F_{\overline{m}} \left( \partial_{\overline{m}} \partial_{n} 
\widetilde{K}_{\overline{a}b}
   - \partial_{\overline{m}} \widetilde{K}_{\overline{a}c} (\widetilde{K}^{-1})_{c\overline{d}}
    \partial_{n} \widetilde{K}_{\overline{d}b} \right) F_{n}  \hspace*{1cm} \\
 A_{abc} Y_{abc} &=& \frac{\hat{W}^{\ast}}{|\hat{W}|} \, e^{\overline{K}/2} 
F_{m}
  \left[ \overline{K}_{m} Y_{abc} + \partial_{m} Y_{abc}
   - \left( (\widetilde{K}^{-1})_{d\overline{e}} \partial_{m} \widetilde{K}_{\overline{e}a}
    Y_{dbc} \right. \right. \\
  & & \hspace*{65mm}\left.\left. + (a \leftrightarrow b)
   + (a \leftrightarrow c) \right) \right]  \nonumber
\end{eqnarray}
where the subscript $m=h,C_{a}$.
Notice that a nondiagonal K\"{a}hler metric
for the matter states will generate a mass matrix between different 
fields.
The physical masses and states are obtained by transforming to the
canonically normalized K\"{a}hler metric,
\begin{eqnarray}
 \widetilde{K}_{\overline{a}b} \overline{C}_{\overline{a}} C_{b} \longrightarrow
  \overline{C}'_{\overline{a}}C'_{a}.
\end{eqnarray}
The K\"{a}hler metric is canonically normalized by a transformation
$\widetilde{P}^{\dagger} \widetilde{K} \widetilde{P} = 1$, so that
the physical canonically normalized
masses $m^{2}_{a}$ are related to the previous
noncanonical mass matrix $m^{2}_{\overline{a}b}$ by the relation
\begin{eqnarray}
 m^{2}_{a}
  =\widetilde{P}^{\dagger} m^{2}_{\overline{a}b} \widetilde{P}.
\end{eqnarray}
If the K\"{a}hler matter metric is diagonal (but not canonical)
$\widetilde{K}_{a}=\widetilde{K}_{\overline{a}b}\delta_{\overline{a}b}$
then the canonically normalized scalar masses $m_{a}^{2}$ are
simply given by 
\begin{eqnarray}
 m_{a}^{2}= m_{3/2}^{2}
  - F_{\overline{J}} F_{I} \partial_{\overline{J}} \partial_{I}
   \left( \ln \widetilde{K}_{a} \right)   \hspace*{1cm} (I,J=h,C_{a}). 
    \label{eq:scmass}
\end{eqnarray}
The soft gaugino mass associated with the gauge group $G_{\alpha}$ is:
\begin{eqnarray}
 M_{\alpha} = \frac{1}{2 \, Re f_{\alpha}} F_{I} \partial_{I} f_{\alpha}
  \hspace*{1cm} (I=S,T_{i}, Y_{k})  \label{eq:gauginomass2}
\end{eqnarray}
and the canonically normalized
supersymmetry-breaking trilinear term for the scalar fields\\
\noindent $A_{abc} Y_{abc} C_{a} C_{b} C_{c}$ is
\begin{eqnarray}
 A_{abc} = F_{I} \left[ \overline{K}_{I} + \partial_{I} \ln Y_{abc}
  - \partial_{I} \ln \left( \widetilde{K}_{a} \widetilde{K}_{b} \widetilde{K}_{c} 
\right)
   \right].
    \label{eq:trilinear2}
\end{eqnarray}

\section{MSSM basics}
\subsection{MSSM conventions: flavor mixings} 
\label{conventionsapp}
The MSSM superpotential is given by
\begin{equation}
\label{superpot}
W=\epsilon_{\alpha 
\beta}[-\hat{H}_u^{\alpha}\hat{Q}_i^{\beta}Y_u^{ij}\hat{U}^c_j
+\hat{H}_d^{\alpha}\hat{Q}_i^{\beta}Y_d^{ij}\hat{D}^c_j
+\hat{H}_d^{\alpha}\hat{L}_i^{\beta}Y_e^{ij}\hat{E}^c_j- 
\mu\hat{H}_d^{\alpha}\hat{H}_u^{\beta}],
\end{equation}
in which $\epsilon_{\alpha \beta}=-\epsilon_{\beta \alpha}$ and 
$\epsilon_{12}=1$, and the superfields are defined in the standard way 
(suppressing gauge indices):
\begin{eqnarray}
\label{superfielddef1}
\hat{Q}_i&=&(\widetilde{Q}_{L_i}, Q_{L_i})\nonumber\\
\hat{U}^c_i&=&(\widetilde{U}^c_{L_i}, U^c_{L_i})\nonumber\\
\hat{D}^c_i&=&(\widetilde{D}^c_{L_i}, D^c_{L_i})\nonumber\\   
\hat{L}_i&=&(\widetilde{E}_{L_i}, E_{L_i})\nonumber\\
\hat{E}^c_i&=&(\widetilde{E}^c_{L_i}, E^c_{L_i})\nonumber\\
\hat{H_u}&=&( H_u, \widetilde{H}_u)\nonumber\\
\hat{H_d}&=&( H_d, \widetilde{H}_d),
\end{eqnarray}  
with $i,j=1\ldots 3$ labeling family indices.
The soft supersymmetry-breaking Lagrangian ${\mathcal{L}}_{soft}$ takes the form
(dropping ``helicity" indices):
\begin{eqnarray}
\label{lsoftexpr}
-{\mathcal{L}}_{soft} &=&\frac{1}{2}\left[M_3
\lambda_{\widetilde{g}}
\lambda_{\widetilde{g}}+M_2\widetilde{W}^a\widetilde{W}^a+M_1\widetilde{B}\widetilde{B}
+{\rm h.c.} \right] \nonumber\\
&+&\epsilon_{\alpha \beta}[-b H^{\alpha}_dH^{\beta}_u-
H^{\alpha}_u\widetilde{Q}^{\beta}_i\widetilde{A}_{u_{ij}}\widetilde{U}^c_j
+H^{\alpha}_d\widetilde{Q}^{\beta}_i\widetilde{A}_{d_{ij}}\widetilde{D}^c_j
+H^{\alpha}_d\widetilde{L}^{\beta}_i\widetilde{A}_{e_{ij}}\widetilde{E}^c_j 
+{\rm h.c.}]\nonumber\\
&+&m_{H_{d}}^{2}|H_{d}|^2+m_{H_{u}}^{2}|H_{u}|^2
+\widetilde{Q}^{\alpha}_i{m^2_Q}_{ij}\widetilde{Q}_j^{\alpha *} 
\nonumber\\
&+& \widetilde{L}^{\alpha}_i{m^2_L}_{ij}\widetilde{L}_j^{\alpha *}
+\widetilde{U}^{c*}_i{m^2_U}_{ij}\widetilde{U}^c_j
+\widetilde{D}^{c*}_i{m^2_D}_{ij}\widetilde{D}^c_j
+\widetilde{E}^{c*}_i{m^2_E}_{ij}\widetilde{E}^c_j.
\end{eqnarray}
The $SU(2)$ representations of the squark, slepton, and Higgs doublets can
be expressed as follows (suppressing family indices for simplicity):
\begin{eqnarray}
\widetilde{Q}=\left(\begin{array}{c c} \widetilde{U}_L
\\ \widetilde{D}_L\end{array}\right),\;\;
\widetilde{L}=\left(\begin{array}{c c}
\widetilde{N}_L\\ \widetilde{E}_L\end{array}\right);
\end{eqnarray}
\begin{eqnarray}
H_{d}=\left(\begin{array}{c c} H_{d}^0\\H_d^-\end{array}\right),\;\;
H_{u}=\left(\begin{array}{c c} H_u^+\\H_u^0\end{array}\right).
\end{eqnarray}
The Higgs fields acquire VEVs and trigger electroweak symmetry breaking:
\begin{eqnarray}
\langle H_{d} \rangle=\left(\begin{array}{c c}
v_d\\0\end{array}\right),\;\;
\langle H_{u}\rangle=\left(\begin{array}{c c}
0\\v_u\end{array}\right),
\end{eqnarray}
in which $v_d^2+v_u^2\equiv v^2$, $\tan \beta = v_u/v_d$, and $v^2=(174\, 
{\rm GeV})^2=2 m_Z^2/(g_2^2+g^{'2})$.  
$g_2$ and $g'$ are the $SU(2)$ 
and $U(1)_Y$ gauge couplings, which satisfy $e=g_2 \sin \theta_{W} = g' 
\cos \theta_{W}$, where $e$ is the electron charge and $\theta_W$ is the 
electroweak mixing angle.  The hypercharge coupling 
$g'$ differs from the GUT normalized hypercharge coupling $g_1$ by 
$g_1=\sqrt{5/3} g'$. 

After electroweak symmetry breaking, one can show explicitly that the mass
terms of the up-type squarks (neglecting diagonal and $\mu$-dependent
electroweak corrections for now) can be expressed from the Lagrangian
given above:
\begin{equation}
V_{squark}=\widetilde{U}^Tm^2_Q\widetilde{U}^*
+\widetilde{U}^{c\,\dagger}m^2_U\widetilde{U}^c
+\widetilde{U}^{c\,\dagger}\widetilde{A}_u^{\dagger}\widetilde{U}^*v_u
+\widetilde{U}^T\widetilde{A}_u\widetilde{U}^cv_u + \ldots;
\end{equation}
In matrix notation one finds:
\begin{eqnarray}
\label{squark1}
V_{squark}=\begin{array}{cc}
(\widetilde{U}^T& \widetilde{U}^{c\,\dagger}) \end{array}
\left(\begin{array}{cc}
m^2_Q& v_u\widetilde{A}_u\\ v_u\widetilde{A}_u^{\dagger}&m^2_U \end{array}
\right )
\left (\begin{array}{c} \widetilde{U}^*\\ \widetilde{U}^c \end{array} \right) +
\ldots,
\end{eqnarray}
written in a general basis in which the Yukawa matrix of the up-type
quarks is not diagonal, such that
\begin{equation}
L_{Yuk}=v_uU^TY_uU^c+ {\rm h.c.} +\ldots .
\end{equation}
In the above $\widetilde{U}$ is a 3-component
column vector, and each element of the matrix in Eq.~(\ref{squark1}) is 
itself a $3\times 3$ matrix. The superfields are defined as follows 
(following 
Eq.~(\ref{superfielddef1}), but suppressing the $L$ index):
\begin{eqnarray}
\hat{Q}_i&=&
\left (\begin{array}{c} (\widetilde{U}_i,U_i)\\ (\widetilde{D}_i,D_i)
\end{array} \right)\nonumber\\
\hat{U}^c_i &=& (\widetilde{U}^c_i,U^c_i)\nonumber\\ 
\hat{D}^c_i &=&
(\widetilde{D}^c_i,D^c_i).
\end{eqnarray}
While $\hat{Q}_i$ contains the left-handed quarks,
$\hat{U}^c_i$ and $\hat{D}^c_i$ contain the left-handed antiquarks.
The left-handed antiquarks can be replaced by right-handed quarks
by performing a CP operation on the superfields. Since $V_{squark}$ is
real, it is possible to write $V_{squark}=V_{squark}^*$ and hence obtain 
Eq.~(\ref{squark1}) as follows:
\begin{eqnarray}
\label{squark2}
V_{squark}=\begin{array}{cc}
(\widetilde{U}^{\dagger}_L& \widetilde{U}^{\dagger}_R) \end{array}
\left(\begin{array}{cc}
m^{2*}_Q& v_u\widetilde{A}^{*}_u\\ v_u\widetilde{A}_u^{T}&m^{2*}_U
\end{array}
\right )
\left (\begin{array}{c} \widetilde{U}_L\\ \widetilde{U}_R \end{array} \right).
\end{eqnarray}
Using the standard relations for charge-conjugated fermions, one obtains
\begin{equation}
L_{Yuk}=\overline{U}_Lv_uY^*_uU_R+{\rm h.c.} + \ldots ,
\end{equation}
with the left(L) and right(R)-handed superfields defined as
\begin{eqnarray}
\hat{Q}_{L_i}&=&
\left (\begin{array}{c} (\widetilde{U}_{L_i},U_{L_i}) \\
(\widetilde{D}_{L_i},D_{L_i})
\end{array} \right)\nonumber\\
\hat{U}_{R_i} &=& (\widetilde{U}_{R_i},U_{R_i})\nonumber\\
\hat{D}_{R_i} &=&
(\widetilde{D}_{R_i},D_{R_i}).
\end{eqnarray}
The complex conjugates of the Yukawa couplings and
soft parameters appear in these expressions, which is a consequence
of replacing the left-handed antiquark by the right-handed quark 
superfields.

It is necessary to express both the quarks and squarks in terms of their
mass eigenstates. For the quarks, the diagonalization of each Yukawa
matrix requires a pair of unitary $3 \times 3$ matrices, as in the
SM:
\begin{eqnarray}
{\rm diag}(m_u,m_c,m_t)&=&V_{U_L}v_uY_u^*V_{U_R}^{\dagger}\nonumber\\
{\rm diag}(m_d,m_s,m_b)&=&V_{D_L}v_dY_d^*V_{D_R}^{\dagger};
\end{eqnarray}   
in which
\begin{eqnarray}
\label{quarkU}
\left ( \begin{array}{c} u_R \\c_R\\t_R \end{array} \right
)=V_{U_R}\left (\begin{array}{c}  U_{R_1}\\ U_{R_2}\\ U_{R_3} \end{array}
\right),
\; \; \left (\begin{array}{c} u_L \\c_L\\t_L \end{array} \right ) =V_{U_L}
\left (\begin{array}{c}  U_{L_1}\\ U_{L_2}\\ U_{L_3} \end{array}
\right),
\end{eqnarray}
\begin{eqnarray}
\label{quarkD}
\left (\begin{array}{c} d_R \\s_R\\b_R \end{array} \right )=V_{D_R}
\left (\begin{array}{c}  D_{R_1}\\ D_{R_2}\\ D_{R_3} \end{array}
\right),
\;
\; \left (\begin{array}{c} d_L \\s_L\\b_L \end{array} \right
)=V_{D_L}
\left (\begin{array}{c}  D_{L_1}\\ D_{L_2}\\ D_{L_3} \end{array}
\right ).
\end{eqnarray}
In the above equations, the fields on the L.H.S. such as $(u_L
\,c_L\,t_L)$, {\it etc.} denote the mass eigenstates, while $U_L$, {\it etc.}
denote the gauge eigenstates. The Cabibbo-Kobayashi-Maskawa matrix is 
$V_{CKM} =
V_{U_L} V^{\dagger}_{D_L}$. The squarks are diagonalized by pairs of $3
\times 6$ matrices as follows:
\begin{eqnarray}
{\rm diag}(m^2_{\widetilde{u}_1} \ldots
m^2_{\widetilde{u}_6})= \left (\begin{array}{cc}
\Gamma^{\dagger}_{U_L}&\Gamma^{\dagger}_{U_R} \end{array} \right )
m^2_{\widetilde{U}} \left (\begin{array}{c} \Gamma_{U_L}
\\ \Gamma_{U_R} \end{array} \right )
\end{eqnarray}
\begin{eqnarray}
{\rm diag}(m^2_{\widetilde{d}_1} \ldots
m^2_{\widetilde{d}_6})= \left (
\begin{array}{cc}
\Gamma^{\dagger}_{D_L}&\Gamma^{\dagger}_{D_R} \end{array} \right )
m^2_{\widetilde{D}} \left (\begin{array}{c} \Gamma_{D_L}
\\ \Gamma_{D_R} \end{array} \right )
\end{eqnarray}
in which $m_{\widetilde{U}}^2$ is defined by Eq.~(\ref{squark2}) and 
$m^2_{\widetilde{D}}$ can be obtained from Eq.~(\ref{squark2}) with the 
replacements $U\rightarrow D$ and $v_u \rightarrow v_d$.  The rotation 
matrices $\Gamma_{U_{L,R}}$, $\Gamma_{D_{L,R}}$ are defined as
\begin{eqnarray}
\label{diagUsq} 
\left ( \begin{array}{c} \widetilde{U}_{L_1}\\
\widetilde{U}_{L_2}\\ \widetilde{U}_{L_3} \\  \widetilde{U}_{R_1}\\
\widetilde{U}_{R_2}\\ \widetilde{U}_{R_3} \end{array}
\right )=
\left (\begin{array}{c}  \Gamma_{U_L}\\ \Gamma_{U_R}
\end{array}   
\right)
\left (\begin{array}{c}  \widetilde{u}_1\\
\widetilde{u}_2\\ \widetilde{u}_3\\ \widetilde{u}_4\\ \widetilde{u}_5\\ \widetilde{u}_6
\end{array}
\right),
\end{eqnarray}
\begin{eqnarray}
\label{diagDsq}
\left ( \begin{array}{c} \widetilde{D}_{L_1}\\
\widetilde{D}_{L_2}\\ \widetilde{D}_{L_3} \\  \widetilde{D}_{R_1}\\
\widetilde{D}_{R_2}\\ \widetilde{D}_{R_3} \end{array}
\right )=
\left (\begin{array}{c}  \Gamma_{D_L}\\ \Gamma_{D_R}
\end{array}
\right)
\left (\begin{array}{c}  \widetilde{d}_1\\
\widetilde{d}_2\\ \widetilde{d}_3\\ \widetilde{d}_4\\ \widetilde{d}_5\\ \widetilde{d}_6
\end{array}
\right).
\end{eqnarray}

However, it is common to rotate the quarks to their mass
eigenstate basis and rotate the squarks in exactly the same way as the
quarks.  
This is the so-called Super-CKM (SCKM) basis. It is a convenient
basis to study flavor violation process since all 
the unphysical parameters in the Yukawa matrices have already been
rotated away.
In this case, the diagonalization of the scalar mass matrices thus proceeds in
two steps. First, the squarks and sleptons are rotated in the 
same way as their fermionic superpartners (see Eq.~(\ref{quarkU}) and 
Eq.~(\ref{quarkD}) above); {\it i.e.}, we do unto squarks as we do unto 
quarks:
\begin{eqnarray}
\label{squarkUCKM}
\widetilde{U}^{SCKM}_{R}=\left ( \begin{array}{c} \widetilde{u}_{R} \\
\widetilde{c}_{R} \\ \widetilde{t}_{R} \end{array}
\right  
)=V_{U_R}\left (\begin{array}{c} \widetilde{U}_{R_1}\\ \widetilde{U}_{R_2}\\
\widetilde{U}_{R_3} \end{array}
\right),
\; \; \widetilde{U}^{SCKM}_{L}=\left (\begin{array}{c} \widetilde{u}_{L} \\
\widetilde{c}_{L} \\ \widetilde{t}_{L} \end{array}
\right
) =V_{U_L}
\left (\begin{array}{c}  \widetilde{U}_{L_1}\\ \widetilde{U}_{L_2}\\
\widetilde{U}_{L_3} \end{array} \right),
\end{eqnarray}
\begin{eqnarray}
\label{squarkDCKM}
\widetilde{D}^{SCKM}_{R}=\left ( \begin{array}{c} \widetilde{d}_{R} \\
\widetilde{s}_{R} \\ \widetilde{b}_{R} \end{array}
\right
)=V_{D_R}\left (\begin{array}{c} \widetilde{D}_{R_1}\\ \widetilde{D}_{R_2}\\
\widetilde{D}_{R_3} \end{array}
\right),
\; \; \widetilde{D}^{SCKM}_{L}
=\left (\begin{array}{c} \widetilde{d}_{L} \\
\widetilde{s}_{L} \\ \widetilde{b}_{L} \end{array}
\right
) =V_{D_L}
\left (\begin{array}{c}  \widetilde{D}_{L_1}\\ \widetilde{D}_{L_2}\\
\widetilde{D}_{L_3} \end{array} \right),
\end{eqnarray}
where in the SCKM basis the squark fields $(\widetilde{u}_{L}, \widetilde{c}_{L},
\widetilde{t}_{L})$ are the superpartners of the physical mass eigenstate
quarks $(u_L, c_L, t_L)$, respectively,
({\it i.e.} $(\widetilde{u}_{L}, {u}_{L})$ form a superfield
because both components are subject to the
same rotation, thereby preserving the superfield structure).
\begin{eqnarray}
V&=& \left(\begin{array}{cc} \widetilde{U}^{\dagger \,SCKM}_{L}
& \widetilde{U}^{\dagger \,SCKM}_{R} \end{array}
\right)
\left (
\begin{array}{cc}
(m^{2\,{\rm 
SCKM}}_{\widetilde{U}})_{LL}&(m^{2\,{\rm SCKM}}_{\widetilde{U}})_{LR}\\(m^{2\,{\rm 
SCKM}}_{\widetilde{U}})_{LR})^{\dagger}&(m^{2\,{\rm 
SCKM}}_{\widetilde{U}})_{RR}
\end{array}\right ) 
\left(\begin{array}{c}\widetilde{U}^{SCKM}_{L}\\ 
\widetilde{U}^{SCKM}_{R}\end{array}
\right) 
\end{eqnarray}
The squark fields expressed in the SCKM basis are often more convenient to 
work with, even though they are not mass eigenstates. Their $6 \times 6$ 
mass matrices are obtained from Eq.~(\ref{squark2}) by adding the 
electroweak
symmetry breaking contributions and then rotating to the SCKM
basis defined in Eqs.~(\ref{squarkUCKM}),(\ref{squarkDCKM}).
They have the following form:
\begin{eqnarray}
\label{pok1}
m^{2\,{\rm SCKM}}_{\widetilde{U}} &=&
\left( \begin{array}{cc}
(m^{2}_{\widetilde{U}})_{LL} + m_u^2
- \frac{\cos 2\beta}{6}(m_Z^2 - 4m_W^2)
&  
(m^2_{\widetilde{U}})_{LR}
- \cot\beta \mu m_u\\
(m^2_{\widetilde{U}})_{LR}^{\dagger}
- \cot\beta \mu^*  m_u&
(m^2_{\widetilde{U}})_{RR} + m_u^2
+\frac{2\cos 2\beta}{3} m_Z^2s^2_W
\\
\end{array}\right),\nonumber\\ 
m^{2\,{\rm SCKM}}_{\widetilde{D}} &=&
\left( \begin{array}{cc}
(m^2_{\widetilde{D}})_{LL} + m_d^2
- \frac{\cos 2\beta}{6}(m_Z^2 + 2m_W^2)
&
(m^2_{\widetilde{D}})_{LR}
- \tan\beta \mu m_d\\
(m^2_{\widetilde{D}})_{LR}^{\dagger}  
- \tan\beta \mu^*  m_d&
(m^2_{\widetilde{D}})_{RR} + m_d^2  
-\frac{\cos 2\beta}{3} m_Z^2s^2_W
\\
\end{array}\right),   
\label{sfmass}
\end{eqnarray}
in which  $s_W\equiv \sin\theta_W$, $\hat{\mbox{\large
1}}$ stands for the $3 \times 3$ unit matrix, and  $m_u={\rm
diag}(m_u,m_c,m_t)$, $m_d={\rm diag}(m_d,m_s,m_b)$.
The flavor-changing
entries are contained in
\begin{eqnarray}
\label{pok1a}
\begin{array}{ccc}
(m^2_{\widetilde{U}})_{LL} = V_{U_L} m^{2^*}_Q V^{\dagger}_{U_L}
\hspace{0.5cm}&
(m^2_{\widetilde{U}})_{RR} = V_{U_R} m^{2^*}_U V^{\dagger}_{U_R}
\hspace{0.5cm}&
(m^2_{\widetilde{U}})_{LR} =
v_u^*V_{U_L} \widetilde{A}_u^{*} V_{U_R}^{\dagger}
\vspace{0.2cm}\\      
(m^2_{\widetilde{D}})_{LL} = V_{D_L} m^{2^*}_Q V_{D_L}^{\dagger}
\hspace{0.5cm}&
(m^2_{\widetilde{D}})_{RR} = V_{D_R} m^{2^*}_D V_{D_R}^{\dagger}
\hspace{0.5cm}&
(m^2_{\widetilde{D}})_{LR} =
v_d^*V_{D_L} \widetilde{A}_{d}^{*} V_{D_R}^{\dagger}.
\end{array}
\end{eqnarray}  
Eq.~(\ref{pok1a}) demonstrates that all four of the matrices 
$V_{{U,D}_{L,R}}$ are needed even though the observed CKM matrix only 
constrains one combination of them.
The squarks are not yet diagonal and hence it is necessary to express 
them in terms of their mass eigenstates:
\begin{eqnarray}
{\rm diag}(m^2_{\widetilde{u}_1} \ldots
m^2_{\widetilde{u}_6})= \left (\begin{array}{cc}
\Gamma^{\dagger\,{\rm SCKM}}_{U_L}&\Gamma^{\dagger \,{\rm SCKM}}_{U_R}
\end{array} \right )  
m^{2\,{\rm SCKM}}_{\widetilde{U}} \left (\begin{array}{c} \Gamma^{{\rm
SCKM}}_{U_L}   
\\ \Gamma^{{\rm SCKM}}_{U_R} \end{array} \right )
\end{eqnarray} 
\begin{eqnarray}
{\rm diag}(m^2_{\widetilde{d}_1} \ldots
m^2_{\widetilde{d}_6})= \left (
\begin{array}{cc}
\Gamma^{\dagger\,{\rm SCKM}}_{D_L}&\Gamma^{\dagger\,{\rm SCKM}}_{D_R}
\end{array} \right )
m^{2 \,{\rm SCKM}}_{\widetilde{D}} \left (\begin{array}{c} \Gamma^{{\rm  
SCKM}}_{D_L} \\ \Gamma^{{\rm SCKM}}_{D_R} \end{array} \right ),
\end{eqnarray}  
in which
\begin{eqnarray}
\label{diagUSCKM}
\left ( \begin{array}{c} \widetilde{u}_{L}\\
\widetilde{c}_{L}\\ \widetilde{t}_{L} \\
\widetilde{u}_{R}\\
\widetilde{c}_{R}\\ \widetilde{t}_{R} \end{array}
\right )=
\left (\begin{array}{c}  \Gamma^{{\rm SCKM}}_{U_L}\\ \Gamma^{{\rm
SCKM}}_{U_R}
\end{array}
\right)
\left (\begin{array}{c}  \widetilde{u}_1\\
\widetilde{u}_2\\ \widetilde{u}_3\\ \widetilde{u}_4\\ \widetilde{u}_5\\ \widetilde{u}_6
\end{array}
\right),
\end{eqnarray}  
\begin{eqnarray}
\label{diagDSCKM}
\left ( \begin{array}{c} \widetilde{d}_{L}\\
\widetilde{s}_{L}\\ \widetilde{b}_{L} \\
\widetilde{d}_{R}\\
\widetilde{s}_{R}\\ \widetilde{b}_{R} \end{array}
\right )=
\left (\begin{array}{c}  \Gamma^{{\rm SCKM}}_{D_L}\\ \Gamma^{{\rm
SCKM}}_{D_R}
\end{array} 
\right)
\left (\begin{array}{c}  \widetilde{d}_1\\
\widetilde{d}_2\\ \widetilde{d}_3\\ \widetilde{d}_4\\ \widetilde{d}_5\\ \widetilde{d}_6
\end{array}
\right).   
\end{eqnarray}
The squark diagonalization matrices in the SCKM basis defined in
Eq.~(\ref{diagUSCKM}) and Eq.~(\ref{diagDSCKM}) are related to the squark
diagonalization matrices defined in Eq.~(\ref{diagUsq}) and 
Eq.~(\ref{diagDsq}) as follows:
\begin{eqnarray}
\Gamma^{{\rm SCKM}}_{U_L}&=&V_{U_L}\Gamma_{U_L}, \; \; \Gamma^{{\rm
SCKM}}_{U_R}=V_{U_R}\Gamma_{U_R}\nonumber\\
\Gamma^{{\rm SCKM}}_{D_L}&=&V_{D_L}\Gamma_{D_L}, \; \; \Gamma^{{\rm
SCKM}}_{D_R}=V_{D_R}\Gamma_{D_R}.
\end{eqnarray}
All of these results may be readily extended to leptons. 
In Section~\ref{fruleapp}, we present an example of two flavor mixing
which could be considered as a special case of the general mixings
presented in this section in which only the third
generation has large mixings. 
\subsection{Gaugino masses and mixings}
\label{gauginoapp}

\noindent $\bullet$ {\bf Gluinos:}
The gluino mass terms in the MSSM Lagrangian are 
\begin{equation}
-{\mathcal{L}}_{\widetilde{g}} = \frac{1}{2} (M_3 e^{i \phi_3} \lambda_{g}
\lambda_{g} + {\rm h.c.}),
\end{equation}
in which the $SU(3)_c$ index has been suppressed. The mass 
eigenstate as 
$\lambda_{g}^{\prime}$ is related to $\lambda_{g}$ by a phase rotation:
\begin{equation}
\lambda_{g}=G \lambda_{g}^{\prime}, \hspace{1cm} G=e^{-i \phi_3/2}.
\end{equation}
The gluino states can be combined into four component Majorana
spinors as follows:
\begin{equation}  
\widetilde{g} = \left(  \begin{array}{c}  \lambda_{g} \\
\overline{\lambda_{g}}\end{array}\right), \hspace{1cm}\
\widetilde{g}'= \left(  \begin{array}{c}  \lambda_{g}^{\prime} \\
\overline{\lambda_{g}}^{\prime} \end{array}\right).
\end{equation}
The following relations are useful for deriving the Feynman rules:
\begin{eqnarray}
P_R \widetilde{g} = \overline{\lambda_{g}}= G^{-1} P_R \widetilde{g}' \nonumber \\
\overline{\widetilde{g}} P_L = \lambda_{g} = G \overline{\widetilde{g}}' P_L \nonumber \\
\overline{\widetilde{g}} P_R = \overline{\lambda_{g}} = G^{-1} \overline{\widetilde{g}}' P_R 
\nonumber \\
P_L \widetilde{g} = \lambda_{g} = G P_L \widetilde{g}'.
\end{eqnarray}

\noindent $\bullet$ {\bf Charginos:}
The charginos of the MSSM are the mass eigenstates which result 
from the mixing of the charged gauginos and the charged
components of the higgsinos. The gaugino mass terms are given by 
\begin{equation}
\frac{1}{2} M_2( \widetilde{W}_1\widetilde{W}_1 + \widetilde{W}_2 \widetilde{W}_2)= M_2
\widetilde{W}^{+} \widetilde{W}^{-},
\end{equation}
in which $\widetilde{W}^{+} = \frac{1}{\sqrt{2}}(\widetilde{W}_1 - i \widetilde{W}_2)$
and $\widetilde{W}^{-} = \frac{1}{\sqrt{2}}(\widetilde{W}_1 + i
\widetilde{W}_2)$. The higgsinos form $SU(2)_L$ doublets
\begin{equation}
\widetilde{H}_d= \left(\begin{array}{c} \widetilde{H}_d^0 \\  \widetilde{H}_d^{-}
\end{array}\right)
\hspace{1cm} \widetilde{H}_u= \left(\begin{array}{c}  \widetilde{H}_u^+ \\
\widetilde{H}_u^{0}\end{array}\right).
\end{equation}
Combining the gauginos and higgsinos into charged pairs
\begin{equation}
\chi^+ = (\widetilde{W}^{+}, \widetilde{H}_u^{+}), \hspace{1cm} \chi^- =
(\widetilde{W}^{-}, \widetilde{H}_d^{-}),
\end{equation}  
their mass terms can be rewritten as
\begin{equation}
{\mathcal{L}} = -\frac{1}{2} (\chi^+, \chi^{-})
\left( \begin{array}{cc}  & X^T \\ X &   \end{array}\right)
\left( \begin{array}{c} \chi^+ \\ \chi^- \end{array}\right),
\end{equation}
where
\begin{equation}
\label{charginomat}
X= \left( \begin{array}{cc} M_2 & \sqrt{2} m_W \sin \beta \\ \sqrt{2}
m_W \cos \beta & \mu  \end{array}\right).
\end{equation}
Notice that in general $M_2$ and $\mu$ can be complex. $X$, as a
general $2\times 2 $ matrix, can be diagonalized by a biunitary
transformation:  
\begin{equation}
M_{\widetilde{C}}^{diag} = \left( \begin{array}{cc} M_{\widetilde{C}_1} &  \\ &
M_{\widetilde{C}_2} \end{array}\right) = U^* X V^{-1}.
\end{equation}
In practice, one can use $V X^{\dagger} X V^{-1} =
(M_{\widetilde{C}}^{diag})^2 $ and $U^* X X^{\dagger} U^T =
(M_{\widetilde{C}}^{diag})^2 $ to find $U$ and $V$. However, these relations 
do not fix  $U$ and $V$ uniquely, but only up to diagonal phase matrices 
$P_U$ and $P_V$.  In general, the resulting mass term is proportional to 
$P_U^* U^* X V^{-1} P_V^*$. Since $U^* X V^{-1}$ is diagonal, without loss 
of generality one can effectively set $P_U$ to the unit matrix.
The phases in $P_V$ will be fixed by the requirement that
$U^* X V^{-1} P_V^* $ give a real and positive diagonal matrix, as
required by the definition of mass eigenstates. It can be absorbed
into the definition of $V$. 
Once the mixing matrices $U$ and $V$ have been obtained, the mass 
eigenstates are given by 
\begin{equation}
\widetilde{C}_i^+ = V_{ij} \chi_j^+, \hspace{1cm} \widetilde{C}_i^- = U_{ij}
\chi_j^-.
\end{equation}
The mass eigenstates can also be combined into Dirac spinors:
\begin{equation}
\widetilde{C}_1 = \left( \begin{array}{c} \widetilde{C}_1^+ \\
\overline{\widetilde{C}}_1^- \end{array}\right), \hspace{1cm} \widetilde{C}_2 =
\left( \begin{array}{c} \widetilde{C}_2^+ \\
\overline{\widetilde{C}}_2^- \end{array}\right).
\end{equation}
In this basis, the mass terms are
\begin{equation}
{\mathcal{L}} = - (M_{\widetilde{C}_1} \overline{\widetilde{C}}_1 \widetilde{C}_1 +
M_{\widetilde{C}_2} \overline{\widetilde{C}}_2 \widetilde{C}_2 ).
\end{equation}

\noindent $\bullet$ {\bf Neutralinos:}
The neutralinos of the MSSM are the mass eigenstates which result from the 
mixing of the neutral gauginos and the neutral components of the 
higgsinos. In the basis
\begin{equation}
\chi^0 = (\widetilde{B}, \widetilde{W}_3, \widetilde{H}_d^0, \widetilde{H}_u^0),
\end{equation}
in which $\widetilde{B}$ is the superpartner of the $U(1)_Y$ gauge boson and
$\widetilde{W}^3$ is the superpartner of the neutral $SU(2)_L$ gauge boson,
the mass terms are
\begin{equation}
{\mathcal{L}} = -\frac{1}{2} (\chi^{0})^T Y \chi^0 + {\rm h.c.}, 
\end{equation}
in which
\begin{eqnarray}
\label{neutralinomat}
Y = \left(\begin{array}{cccc}
M_1 & & -m_Z c_{\beta} s_w  & m_Z s_{\beta} s_w \\
 & M_2 & m_Z c_{\beta} c_w  & -m_Z s_{\beta} c_w \\
-m_Z c_{\beta} s_w & m_Z c_{\beta} c_w & & -\mu \\
 m_Z s_{\beta} s_w & -m_Z s_{\beta} c_w & -\mu &  \\
\end{array}\right).
\end{eqnarray}
This is a $4 \times 4$ symmetric complex matrix  and can be
diagonalized by 
\begin{equation}
N^* Y N^{\dagger} = M_{\widetilde{N}}^{diag},
\end{equation}
where $N$ is a $4 \times 4$ unitary matrix. $N$
and the mass eigenvalues are determined by $N Y^{\dagger} Y N^{\dagger} =
(M_{\widetilde{N}}^{diag})^2$. However, there are phase ambiguities 
similar to those encountered in the chargino sector. The phases are again 
fixed by requiring
$P_N^* N^* Y N^{\dagger} P_N^* $ to be a real and positive diagonal
matrix.
The mass eigenstates are  $\widetilde{n}_i = N_{ij} \chi^0_j$, which can be 
combined into four component Majorana spinors:
\begin{equation}
\widetilde{N}_i = \left(\begin{array}{c}  \widetilde{n}_i \\ \overline{\widetilde{n}}_i
\end{array}\right).
\end{equation}


\subsection{MSSM Feynman rules}
\label{fruleapp}
In this section, the phenomenologically most relevant Feynman rules of the 
MSSM are presented in our notation/conventions. The Feynman rules 
displayed here include several generalizations not 
included in the classic references 
\cite{Haber:1984rc,Gunion:1984yn,Gunion:1986nh,Gunion:1988yc}. 
\begin{itemize}

\item All possible phases of the MSSM parameters are included.

\item The full flavor structure of the quark/squark sector is retained
such that the CKM matrix $V_{CKM}$ and scalar quark mixing matrices
$\Gamma^{SCKM}_{U,D}$ are included explicitly in the Feynman rules. 
Slepton mixing is also included.

\item 
The gaugino-sfermion-fermion interactions include the higgsino 
contributions, which are suppressed by small 
fermion masses.

\end{itemize}

The Feynman rules are expressed in the 
SCKM basis, in which the SM 
fermions have been rotated into their mass eigenstates and thus are 
described by their masses and mixing 
matrices ($V_{CKM}$ for the quark sector). The squarks 
and sleptons are not diagonal in the SCKM basis (see {\it e.g.} Eq.~(\ref{pok1}) 
for the quarks), and their rotation matrices  
$(\Gamma^{SCKM}_{qA})_{I\alpha}$ (the chirality $A=L,R$) enter the 
Feynman rules explicitly. $I,J = 1,2,3$ denote the 
family indices of the SM fermions (and the sfermions in the SCKM basis).  
The indices $\alpha, \beta = 1, ... ,6$ label the mass eigenstates of the 
sfermions (these indices range from $1...3$ for the sneutrinos).  
Color indices ({\it e.g.} for gluons and gluinos) are denoted by $i, j, k = 1,
2, 3$. The gluinos are labeled $\widetilde{g}^a$, where $a=1 ,..., 8$ labels
the $SU(3)$ generators. The charginos are denoted by $\widetilde{C}_i$, where
$i=1,2$ labels their mass eigenstates, and the neutralinos by
$\widetilde{N}_i$, $i=1,...,4$. $e_{f}$ denotes the charge of $f$ in
units of $e$, where $e$ is the absolute value of the electron charge.


Before considering general flavor mixing, let us warm up with the  
simple example of sfermion mixing with only one generation of 
quarks and squarks of both up and down flavors.  Using the general results 
of Appendix~\ref{conventionsapp}, the squark mass terms in this limit are 
given by
\begin{equation}
-{\mathcal{L}} = (\widetilde{q}_L^{\dagger}, \widetilde{q}_R^{\dagger})
\left( \begin{array}{cc} m^2_{LL} & m^2_{LR} \\ m^{2*}_{LR} & m^2_{RR}
\end{array}\right)
\left(\begin{array}{c} \widetilde{q}_L \\ \widetilde{q}_R \end{array}\right) =  
(\widetilde{q}_L^{\dagger}, \widetilde{q}_R^{\dagger})
m^2_{\widetilde{q}}\left(\begin{array}{c} \widetilde{q}_L \\ \widetilde{q}_R
\end{array}\right),
\end{equation}
in which 
\begin{eqnarray}
m^2_{LL} &=& {m}^2_{\widetilde{Q}} + m_q^2 + \Delta_q \nonumber \\
m^2_{RR} &=& {m}^2_{\widetilde{\overline{q}}} + m_q^2 + \Delta_{\overline{q}}
\nonumber \\
(m^2_{LR})_{u} &=& v_u A_u^* - \mu v_d y_u  \nonumber \\
(m^2_{LR})_{d} &=& v_d A_d^* - \mu v_u y_d .
\end{eqnarray}
In the above, ${m}^2_{\widetilde{Q}}$ and ${m}^2_{\widetilde{\overline{q}}}$ are the 
soft supersymmetry-breaking mass-squared parameters for the left-handed 
doublet and singlets, respectively, and the  $\widetilde{A}_q$s are the
soft trilinear scalar couplings.  $m_q^2$ is the F 
term contribution derived from the superpotential Yukawa couplings which 
give masses to the up and down quarks. The term proportional to $\mu$ is 
also an F term contribution which arises from the cross terms of the 
product $|F_H|$ (where $H$ denotes both $H_u$ and $H_d$). 
The $\Delta_q$s are D term contributions to the mass
matrix: $\Delta_u = (\frac{1}{2} - \frac{2}{3} \sin^2{\theta_W})
\cos 2 \beta m^2_Z$, $\Delta_{\overline{u}} = (\frac{2}{3} \sin^2{\theta_W})
\cos 2 \beta m^2_Z$,  $\Delta_d = (-\frac{1}{2} - \frac{1}{3} 
\sin^2{\theta_W})
\cos 2 \beta m^2_Z$, and $\Delta_{\overline{d}} = (\frac{1}{3} 
\sin^2{\theta_W})
\cos 2 \beta m^2_Z$. 

The $2\times 2$ Hermitian mass matrix is diagonalized by the unitary 
transformation $m_{\widetilde{q}}^{2,diag}= U {\mathbf{m}}^2_{\widetilde{q}}
U^{\dagger}$. Denoting the mass eigenstates as $(\widetilde{q}_1,
\widetilde{q}_2)$,
\begin{equation}
U^{\dagger} \left(\begin{array}{c} \widetilde{q}_1 \\ \widetilde{q}_2
\end{array}\right) = \left(\begin{array}{c} \Gamma_L \\ \Gamma_R 
\end{array} \right) \left(\begin{array}{c} \widetilde{q}_1 \\ \widetilde{q}_2 
\end{array}\right) =\left(\begin{array}{c} \widetilde{q}_L  \\ \widetilde{q}_R
\end{array}\right),
\label{sqrotation}
\end{equation}
where the $\Gamma$s are $1 \times 2$ row vectors (recall that in the MSSM, 
the $\Gamma$ matrices are $3 \times 6$ matrices). $U$ can be parameterized 
in terms of the angles $\theta_q$ and $\phi_q$ as follows: 
\begin{equation}
U=\left(\begin{array}{cc} \cos \theta_q & \sin \theta_q e^{i \phi_q} \\
-\sin \theta_q e^{-i \phi_q} & \cos \theta_q \\
\end{array}\right).
\end{equation}
Therefore,
\begin{eqnarray}
(\Gamma_L)_{1 \alpha} = (\cos \theta_q, -\sin \theta_q e^{i \phi_q}),
\hspace{0.5cm}
(\Gamma_R)_{1 \alpha} = (\sin \theta_q e^{-i \phi_q}, \cos \theta_q).
\end{eqnarray}

To see how these couplings enter the Feynman rules, one first uses 
Eq.~(\ref{sqrotation}) to recast the Lagrangian from the original 
$(\widetilde{q}_L, \widetilde{q}_R)$ basis to the new basis 
$(\widetilde{q}_1,\widetilde{q}_2)$.  The mixing angles 
$\theta_q$ and $\phi_q$ (which are functions of the original 
Lagrangian parameters) appear as coupling constants in the 
Lagrangian. For example, consider the  coupling $ g [\widetilde{W}^+
d_L \widetilde{u}_L^* - \overline{\widetilde{W}}^+ \overline{d}_L \widetilde{u}_L +
{\rm h.c.}] $. This is just the supersymmetric completion of the 
left-handed
charged current coupling of the SM. In the new basis, this 
term is 
\begin{eqnarray}
& &g[\widetilde{W}^+ d_L (\Gamma_L)^*_{1 \alpha} \widetilde{u}^*_{\alpha} -  
\overline{\widetilde{W}}^+ \overline{d}_L (\Gamma_L)_{1 \alpha} \widetilde{u}_{\alpha}+ 
{\rm h.c.}]
\nonumber\\
&=& g[\widetilde{W}^+ d_L (\cos \theta_u \widetilde{u}^*_1 - \sin \theta_u
e^{-i \phi_u}\widetilde{u}^*_2)  \nonumber \\
&& +\overline{\widetilde{W}}^+ \overline{d}_L (\sin \theta_u e^{i \phi_u} \widetilde{u}_1
+ \cos \theta_u \widetilde{u}_2 ) + {\rm h.c.} ],
\end{eqnarray}
where $\widetilde{u}_1$ and $\widetilde{u}_2$ are two mass eigenstates of the
scalar up quarks.

This exercise can of course be carried out in the presence of full flavor 
mixing.  We now present the most phenomenologically relevant Feynman rules 
within the general MSSM-124.\\

\noindent $\bullet$ {\bf Gaugino --- Sfermion --- fermion:}
\begin{enumerate}
\item  {\it chargino-quark-squark}

\begin{eqnarray}
{\mathcal{L}}_{q\widetilde{q}\widetilde{C}^+} &=& -g_2 [{(V^{CKM})}_{IJ} \overline{u}_I 
P_R
(U_{11} {\widetilde{C}_1} + U_{21}
{\widetilde{C}_2}){\widetilde{d}}_\alpha
{(\Gamma_{DL}^{SCKM})}_{J\alpha} \nonumber \\
&+&{(V^{CKM})^{\dagger}}_{JI}\overline{d}_J P_R (V_{11}{\widetilde{C}_1}^c
+V_{21}{\widetilde{C}_2}^c)\widetilde{u}_\alpha(\Gamma_{UL}^{SCKM})_{I\alpha}] 
\nonumber \\
&+& \frac{g_2}{\sqrt{2}m_W \cos\beta}[(V^{CKM})_{IJ}\overline{u}_I P_R
(U_{12}{\widetilde{C}_1} +
U_{22}{\widetilde{C}_2})\widetilde{d}_\alpha(\Gamma_{DR}^{SCKM})_{J\alpha}
\cdot m_J^d] \nonumber \\
&+& (V^{CKM})_{IJ}^{\dagger}\overline{d}_I
P_L(U_{12}^*{\widetilde{C}_1}^c+U_{22}^*{\widetilde{C}_2}^c)\widetilde{u}_\alpha
{(\Gamma_{UL}^{SCKM})}_{J\alpha}m_I^d \nonumber \\
&+& \frac{g_2}{\sqrt{2}m_W
\sin\beta}[(V^{CKM})_{IJ}\overline{u}_I P_L(V_{12}^*{\widetilde{C}_1}
+V_{22}^*{\widetilde{C}_2})\widetilde{d}_\alpha{(\Gamma_{DL}^{SCKM})}_{J\alpha}
m_I^u \nonumber \\
&+& {(V^{CKM})}^{\dagger}_{IJ}\overline{d}_I P_R(V_{12}{\widetilde{C}_1}^c
+V_{22}{\widetilde{C}_2}^c)\widetilde{u}_\alpha(\Gamma_{UR}^{SCKM})_{\alpha
J} m_J^u] \nonumber \\
&+& {\rm h.c.}
\end{eqnarray}

\begin{figure}
\centerline{
   \epsfxsize 3.3 truein \epsfbox {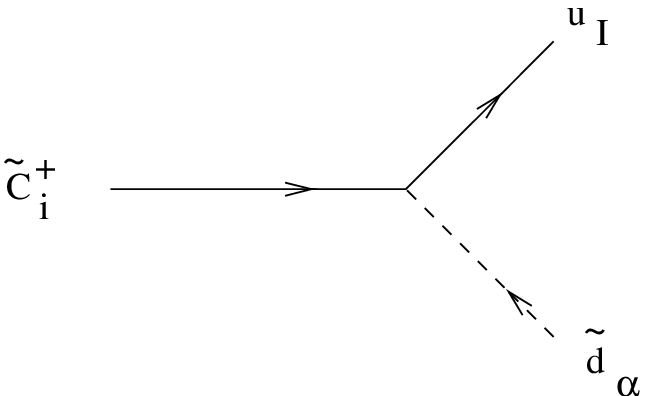}
    }
\caption{\normalsize
$-g_2(V^{CKM})_{IJ}(\Gamma_{DL}^{SCKM})_{J\alpha}U_{i1} \cdot P_R$
$+\frac{g_2}{\sqrt{2} m_W
cos\beta}(V^{CKM})_{IJ}(\Gamma_{DR}^{SCKM})_{J\alpha}m_J^d U_{i2}
\cdot P_R$ 
+$\frac{g_2}{\sqrt{2}m_W
\sin\beta}(V^{CKM})_{IJ}(\Gamma_{DL}^{SCKM})_{J\alpha}m_I^u V_{i2}^*
\cdot P_L$ 
}
\label{frule1}
\vskip 0.25 truein
\centerline{
   \epsfxsize 3.3 truein \epsfbox {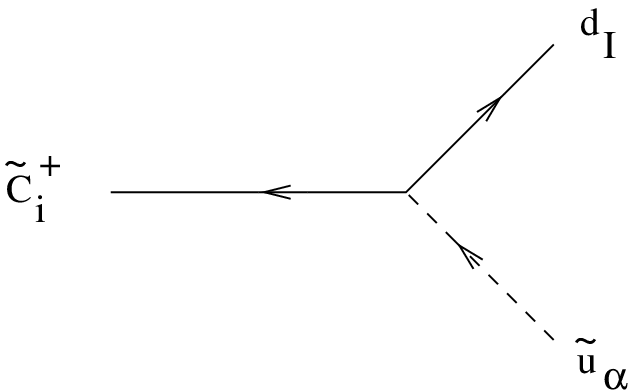}
    }
\caption{\normalsize $-g_2{{(V^{CKM})}^{\dagger}}_{IJ}
(\Gamma_{UL}^{SCKM})_{J\alpha}  V_{i1} \cdot P_ R C+ 
\frac{g_2}{\sqrt{2}m_W\cos\beta}{(V^{CKM})}^{\dagger}_{IJ}
(\Gamma_{UL}^{SCKM})_{J\alpha} m_I^d U_{i2}^* \cdot P_L C$
+$\frac{g_2}{\sqrt{2} m_W\sin\beta}{(V^{CKM})}^{\dagger}_{IJ}
(\Gamma_{UR}^{SCKM})_{J\alpha} m_J^u V_{i2} \cdot P_R C $
}
\label{frule2}
\end{figure}
%
%


\begin{figure}
\centerline{
   \epsfxsize 3.3 truein \epsfbox {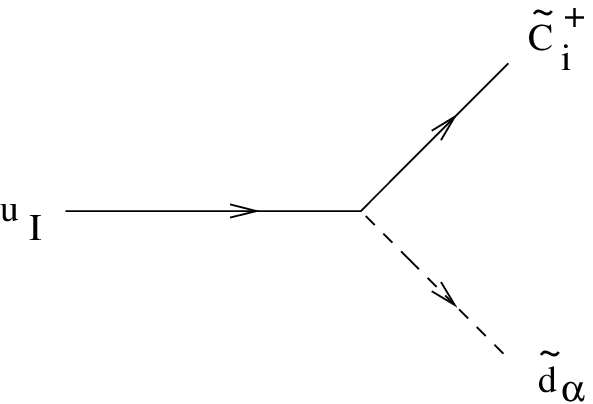}
    }
\caption{
$-g_2 (V_{IJ}^{CKM})^*(\Gamma_{DL}^{SCKM})_{J\alpha}^* U_{i1}^* \cdot 
P_L 
+\frac{g_2}{\sqrt{2}m_W \cos\beta} (V_{IJ}^{CKM})^*
(\Gamma_{DR}^{SCKM})^*_{J\alpha}m_J^d U_{22}^* \cdot P_L$ 
$+\frac{g_2}{\sqrt{2} m_W \sin\beta} (V_{IJ}^{CKM})^*
(\Gamma_{DL}^{SCKM})_{J\alpha}^* m_I^u V_{i2} \cdot P_R$ 
}
\label{frule3}
\vskip 0.25 truein
\centerline{
   \epsfxsize 3.3 truein \epsfbox {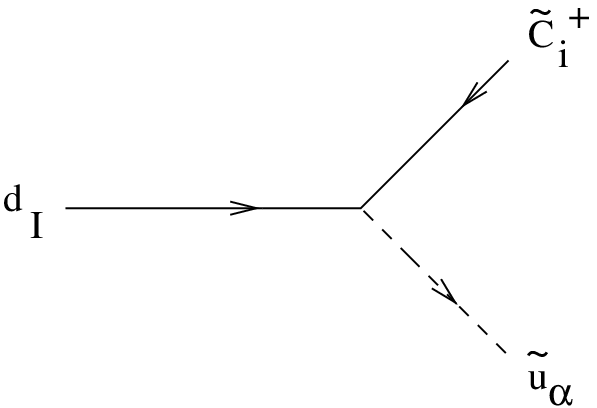}
    }
\caption{
$+g_2 V_{JI}^{CKM} (\Gamma_{UL}^{SCKM})_{J\alpha}^*
V_{i1}^* \cdot C^{-1} P_L
-\frac{g_2}{\sqrt{2}m_W \cos\beta}
V_{JI}^{CKM}(\Gamma_{UL}^{SCKM})_{J\alpha}^* m_I^d U_{i2} \cdot
C^{-1} \cdot P_R$ 
$-\frac{g_2}{\sqrt{2}m_W \sin\beta} V_{JI}^{CKM}
(\Gamma_{UR}^{SCKM})^* m_J^u V_{i2}^*\cdot C^{-1} P_L$
}
\label{frule4}
\end{figure}

\item{ \it Neutralino-quark-squark}
\begin{eqnarray}
{\mathcal{L}}_{q\widetilde{q}\widetilde{N}} &=& \sum_{q = u,d} - \sqrt{2}g_2
\big[ \overline{q}_I
P_R{\widetilde{N}}_j{(\Gamma_{qL}^{SCKM})}_{I\alpha}\widetilde{q}_\alpha[T_{3I}
N_{j2} - \tan\theta_W (T_{3I} - e_I)N_{j1} \big] \nonumber\\
&+&\sqrt{2}g_2 \tan\theta_W \overline{q}_I P_L \widetilde{N}_j
{(\Gamma_{qR}^{SCKM})}_{I\alpha} \widetilde{q}_{\alpha} [e_I N_{j1}^*] \nonumber \\
&-&\frac{g_2\;m_I^d}{\sqrt{2}m_W \cos\beta} \big[ \overline{d}_I N_{i3}^* P_L
\widetilde{N}_i{(\Gamma_{DL}^{SCKM})}_{I\alpha}\widetilde{d}_\alpha \nonumber \\
&+&\overline{\widetilde{N}}_i N_{i3} P_L d_I
{({\Gamma_{DR}}^{SCKM})}_{I\alpha}\widetilde{d}_\alpha \big] \nonumber \\
&-& \frac{g_2\;m_I^u}{\sqrt{2}m_W\sin\beta} \big[ \overline{u}_I N_{i4}^* P_L
N_i{(\Gamma_{UL}^{SCKM})}_{I\alpha}\widetilde{u}_\alpha \nonumber \\
&+& \overline{\widetilde{N}}_i N_{i4} P_L
u_I{(\Gamma_{UR}^{SCKM})}_{I\alpha}\widetilde{u}_\alpha) \big] \nonumber \\
&+& {\rm h.c.}
\end{eqnarray}

\begin{figure}
\centerline{
   \epsfxsize 3.3 truein \epsfbox {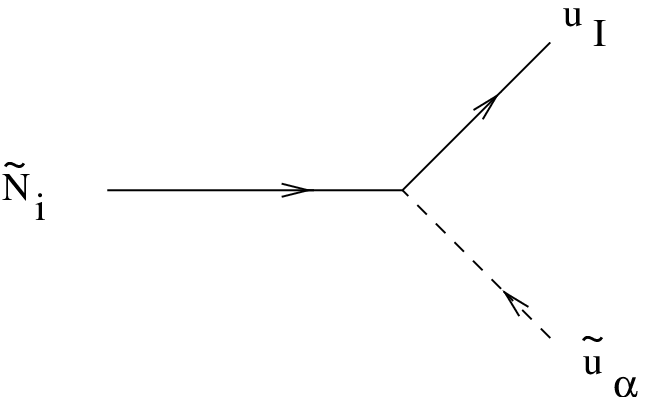}
    }
\caption{
$-\sqrt{2} g_2
{(\Gamma_{UL}^{SCKM})}_{I\alpha}[T_{3I}N_{i2}  
- \tan \theta_W(T_{3I} - e_I)N_{i1}] \cdot P_R
+ \sqrt{2} g_2 \tan \theta_W{(\Gamma_{UR}^{SCKM})}_{I\alpha}  
N^*_{i1} \cdot P_L$ -$ \frac{g_2\;m_I^u}{\sqrt{2} m_W
\sin\beta}(N^*_{i4}{(\Gamma_{UL}^{SCKM})}_{I\alpha} \cdot 
P_L
+N^*_{i4}{(\Gamma_{UR}^{SCKM})^*}_{I\alpha} P_R.)$
}
\label{frule5}
\vskip 0.25 truein
\centerline{
   \epsfxsize 3.3 truein \epsfbox {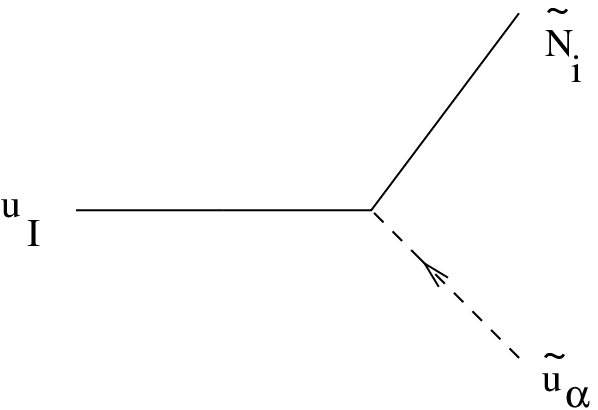}
    }
\caption{
$-\sqrt{2} g_2
{(\Gamma_{UL}^{SCKM})}_{I\alpha}^*[T_{3I}N_{i2}^*
- \tan\theta_W(T_{3I} - e_I)N_{i1}^*] \cdot P_L 
+ \sqrt{2} g_2 \tan\theta_W{(\Gamma_{UR}^{SCKM})}_{I\alpha}^*
N_{i1} \cdot P_R$ -$ \frac{g_2\;m_I^u}{\sqrt{2} m_W
\sin\beta}(N_{i4}{(\Gamma_{UL}^{SCKM})}_{I\alpha}^* \cdot P_R
+N_{i4}{(\Gamma_{UR}^{SCKM})}_{I\alpha} P_L.)$
}
\label{frule6}
\end{figure}

The processes associated with neutralino and up (s)quarks are shown in
Fig.~\ref{frule5} and Fig.~\ref{frule6}, where $T_{3I}=1/2$, $e_I = 2/3$.

\item{\it  Gaugino-lepton-slepton}

Make the substitutions

\begin{eqnarray}
&e_L& \leftrightarrow d_L\;\;\;\;\;\;e_R
\leftrightarrow d_R\;\;\;\;\;\;e \leftrightarrow d \nonumber \\
&\nu_L& \leftrightarrow U_L\;\;\;\;\;\;\;\;\;\;\;\;\;\;\;\;\;\;\;\;\;\;
 \nonumber \\
&\widetilde{e}& \leftrightarrow \widetilde{d}\;\;\;\;\;\;\;\;\; \widetilde{\nu}
\leftrightarrow \widetilde{u} \nonumber \\
&\Gamma_{EL}& \leftrightarrow \Gamma_{DL}^{SCKM},\;\;\;\;\; \Gamma_{\nu
L} \leftrightarrow \Gamma_{UL}^{SCKM} \nonumber \\
&\Gamma_{ER}& \leftrightarrow \Gamma_{DR}^{SCKM} \nonumber \\
&{\mathbf{1}}& \leftrightarrow \cdot V^{CKM} \nonumber \\
&m_I^e& \leftrightarrow m_I^d \nonumber \\
&{\mathbf{0}}& \leftrightarrow m_I^u \nonumber
\end{eqnarray}

\vspace{1cm}
\item {\it Gluino-quark-squark}
\begin{eqnarray}
{\mathcal{L}}_{q\widetilde{q}\widetilde{g}^{'a}} &=& -\sqrt{2} g_3 {T^a}_{jk}
\sum_{q = u,d}(G\overline{\widetilde{g}}^{'a} P_L
{q^k}_I{\widetilde{q}}^{j*}_\alpha{(\Gamma^{SCKM}_{qL})}_{I\alpha}^* 
\nonumber\\
&+& G^{-1}{\overline{q}^j}_I
P_R\widetilde{g}^{'a}{\widetilde{q}^k}_\alpha{(\Gamma^{SCKM}_{qL})}_{I\alpha}-
G^{-1}\overline{\widetilde{g}}^{\prime a}
P_R{q^k}_I{{\widetilde{q}}^{j*}}_\alpha
(\Gamma^{SCKM}_{qR})^*_{I\alpha}\nonumber\\
&-& G {\overline{q}^j}_I P_L
\widetilde{g}^{\prime a}{\widetilde{q}^k}_\alpha{(\Gamma^{SCKM}_{qR})}_{I\alpha}).
\end{eqnarray}
$i$, $j$... are color indices.

\begin{figure}
\centerline{
   \epsfxsize 3.3 truein \epsfbox {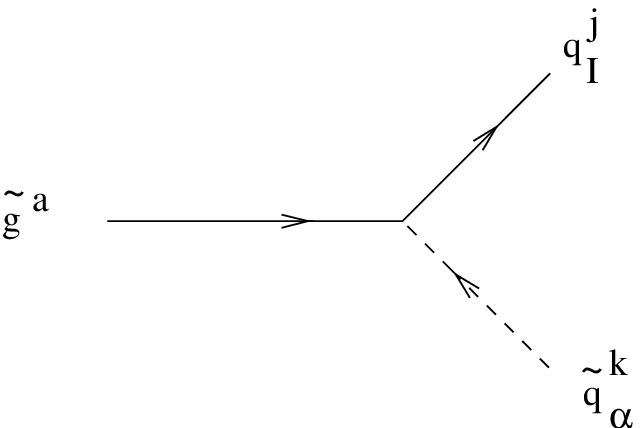}
    }
\caption{
$-\sqrt{2}g_3 {T^a}_{jk} \big[ G^{-1}(
{\Gamma^{SCKM}_{qL})}_{I\alpha} \cdot P_R
- G{({\Gamma^{SCKM}}_{qR})}_{I\alpha} P_L\big]$
}
\label{frule7}
\vskip 0.25 truein
\centerline{
   \epsfxsize 3.3 truein \epsfbox {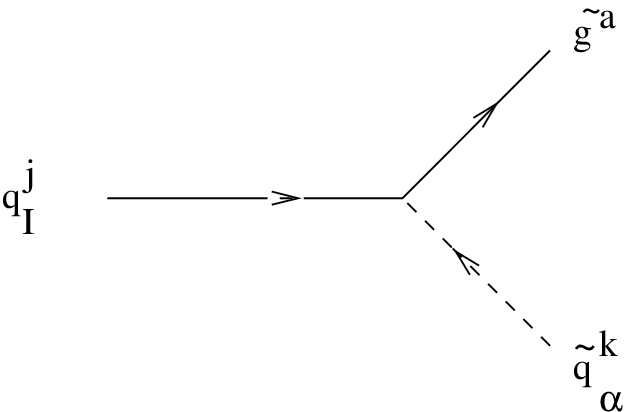}
    }
\caption{
$- \sqrt{2} g_3
{T^a}_{kj}(G{(\Gamma^{SCKM}_{qL})}_{I\alpha}^* P_L
- G^{-1}{(\Gamma^{SCKM}_{qR})}^*_{I\alpha} P_R)$
}
\label{frule8}
\end{figure}

\end{enumerate}

\noindent $\bullet$ {\bf Gaugino --- Gaugino --- Gauge boson:}

\begin{enumerate}

\item{\it Chargino-Neutralino-$W^\pm$}

\begin{eqnarray}
{\mathcal{L}}_{W^{-}\widetilde{C}_{i} \widetilde{N} } &=& g_2
W_{\mu}^-\overline{\widetilde{N}}_i \gamma^\mu(O_{ij}^L P_L + O_{ij}^R
P_R) {\widetilde{C}_j} \nonumber
\end{eqnarray}  
where
\begin{eqnarray}
O_{ij}^L &=& -\frac{1}{\sqrt{2}}N_{i4}V_{j2}^* +
N_{i2}V_{j1}^*  \nonumber \\
O_{ij}^R &=& \frac{1}{\sqrt{2}}N_{i3}^*U_{j2}+N_{i2}^* U_{j1}.
\end{eqnarray}

\begin{figure}
\centerline{
   \epsfxsize 3.3 truein \epsfbox {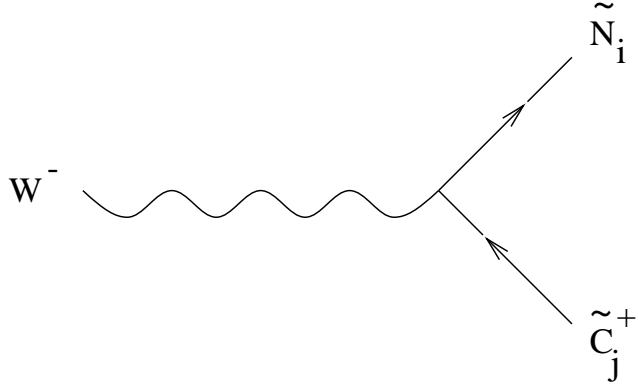}
    }
\caption{
$g_2 \gamma^{\mu}(O_{ij}^L P_L + O_{ij}^R P_R)$}
\label{frule9}
\end{figure}



\item {\it Chargino-chargino-gauge boson $(Z^0,\;\gamma)$}

a) photon $\gamma$
\begin{eqnarray}
{\mathcal{L}}_{\gamma\widetilde{C}_i \widetilde{C}_i} = -e A_\mu\overline{\widetilde{C}}_i
\gamma^\mu {\widetilde{C}_i}
\end{eqnarray}

\begin{figure}
\centerline{
   \epsfxsize 3.3 truein \epsfbox {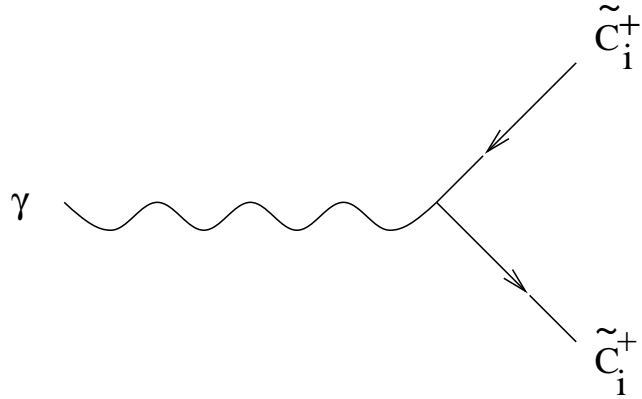}
    }
\caption{$-e\gamma^{\mu}$}
\label{frule10}
\end{figure}


b) $Z^0$

\begin{equation}
{\mathcal{L}}_{Z^0 \widetilde{C}_i \widetilde{C}_i } = \frac{g_2}{\cos\theta_W}
Z_\mu[\overline{\widetilde{C}}_i \gamma^\mu (O_{ij}^{\prime L} P_L +
O_{ij}^{\prime R} P_R){\widetilde{C}_j} ]
\end{equation}  

\begin{eqnarray}
O_{ij}^{\prime L} &=& -V_{i1}V_{j1}^* - \frac{1}{2} V_{i2}V_{j2}^* +
\delta_{ij}\sin^2\theta_W \nonumber \\
O_{ij}^{\prime R} &=& -U_{i1}^*U_{j1} - \frac{1}{2}U_{i2}^* U_{j2} +
\delta_{ij}\sin^2 \theta_W \nonumber
\end{eqnarray}

\begin{figure}
\centerline{
   \epsfxsize 3.3 truein \epsfbox {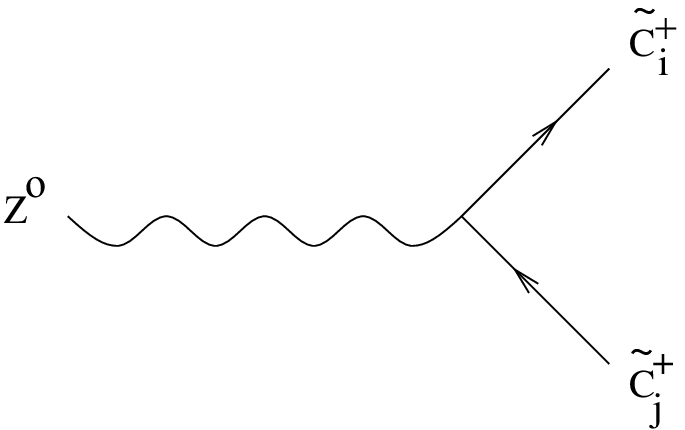}
    }
\caption{$\frac{g_2}{\cos\theta_W} \gamma^\mu[O_{ij}^{\prime
L} P_L +
O_{ij}^{\prime R} P_R]$
}
\label{frule11}
\end{figure}

\item {\it Neutralino-neutralino-gauge boson $(Z^0)$ }

\begin{eqnarray}
{\mathcal{L}}_{Z^0\widetilde{N}\widetilde{N}} = \frac{q}{2 \cos
\theta_W}{\overline{\widetilde{N}}}_i \gamma^\mu(O_{ij}^{\prime \prime L}P_L +
O_{ij}^{\prime \prime R}
P_R)\widetilde{N}_j
\end{eqnarray}
\begin{eqnarray}
O_{ij}^{\prime \prime L} &=& - \frac{1}{2} N_{i3}N_{j3}^* + \frac{1}{2}
N_{j4}N_{j4}^* \nonumber \\
O_{ij}^{\prime \prime R} &=& - O_{ij}^{\prime \prime L*} \nonumber
\end{eqnarray}  

\begin{figure}
\centerline{
   \epsfxsize 3.3 truein \epsfbox {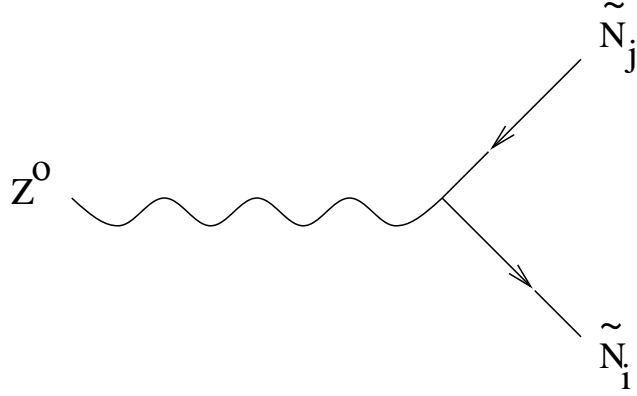}
    }
\caption{$\frac{g}{\cos\theta_W} \gamma^u(O_{ij}^{\prime \prime
L}P_L + O_{ij}^{\prime \prime R}P_R)$}
\label{frule12}
\end{figure}

\item{\it Gluino-Gluino-Gluon}
\begin{equation}
{\mathcal{L}}_{g\widetilde{G}^a\widetilde{G}^a} = \frac{i}{2}g_3
f_{abc}\overline{\widetilde{g}}^{\prime a} \gamma_\mu^\mu \widetilde{g}^{\prime b} 
G_\mu^c
\end{equation}

\begin{figure}
\centerline{
   \epsfxsize 3.3 truein \epsfbox {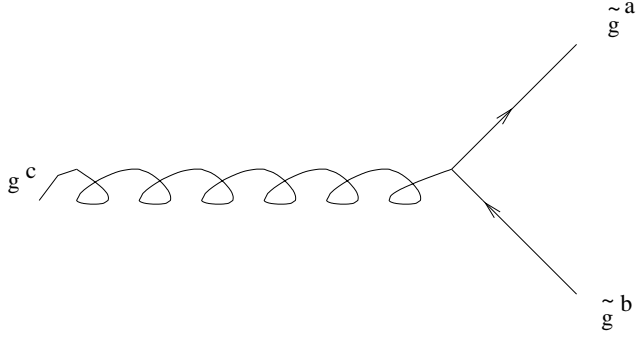}
    }
\caption{$ ig_3 f_{abc}\gamma^\mu $}
\label{frule13}
\end{figure}
\end{enumerate}

\noindent $\bullet$ {\bf Couplings between squarks and gauge bosons:}

To simplify our expressions, we define
\begin{equation}
F^1_{\alpha \beta}=(V^{CKM})_{IJ} (\Gamma^{SCKM}_{UL})^*_{I \alpha}
(\Gamma^{SCKM}_{DL})_{J \beta},
\end{equation}
and 
\begin{eqnarray}
F_{\alpha\beta}^{2I} &=&
(\Gamma_{qL}^{SCKM})_{I\alpha}^{*}(\Gamma_{gL}^{SCKM})_{I\beta} +
({\Gamma_{qR}^{SCKM}})_{I\alpha_*} {(\Gamma_{qR}^{SCKM})}_{I\beta}.
\end{eqnarray} 

\begin{enumerate}

\item{\it Scalar quark --- scalar quark --- gauge boson}

a) $W^\pm$

\begin{eqnarray}
{\mathcal{L}}_{\widetilde{qq}W^{\pm}} =
\frac{-ig_2}{\sqrt{2}}[{W_{\mu}^+}{(V^{CKM})}_{IJ}{(\Gamma_{UL}^{SCKM}})^*_{
{I\alpha}}{(\Gamma_{DL}^{SCKM})}_{J\beta}
\widetilde{u}_{\alpha}^*\stackrel{\leftrightarrow}{\partial^\mu}
\widetilde{d}_{\beta} + {\rm h.c.}] \nonumber
\end{eqnarray}
\begin{eqnarray}
(a \stackrel{\leftrightarrow}{\partial} b) = a(\partial b) - (\partial
a)b \nonumber
\end{eqnarray}

b) photon
\begin{equation}
{\mathcal{L}}_{\gamma\widetilde{q}\widetilde{q}} = - ie A_\mu e_\alpha
\widetilde{q}_{\alpha}^{*}
\stackrel{\leftrightarrow}{\partial^\mu}\widetilde{q}_\alpha
\end{equation}

\begin{figure}
\centerline{
   \epsfxsize 3.3 truein \epsfbox {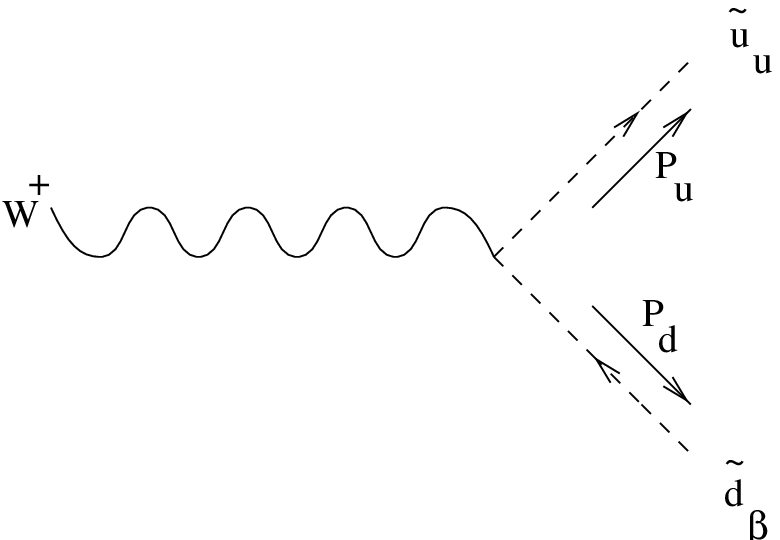}
    }
\caption{$ -i\frac{g_2}{\sqrt{2}}{(P_d - P_u)}^\mu
F_{\alpha\beta}^{1}$}
\label{frule14}
\vskip 0.25 truein
\centerline{
   \epsfxsize 3.3 truein \epsfbox {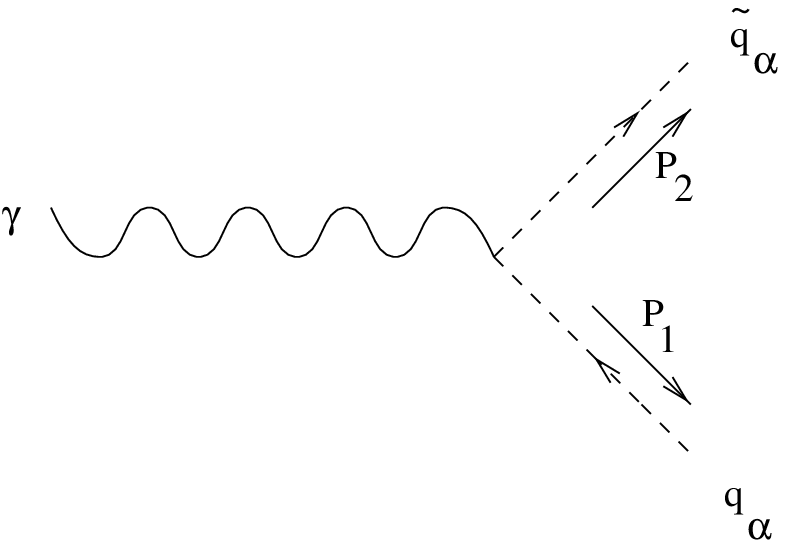}
    }
\caption{$-i e_\alpha(P_1 - P_2)^\mu$
}
\label{frule15}
\end{figure}

c) $Z^0$
\begin{eqnarray}
{\mathcal{L}}_{Z^0 \widetilde{q}\widetilde{q}} &=&
\frac{-ig_2}{\cos\theta_W}Z_\mu\widetilde{q}_{\alpha}^{*}   
\stackrel{\leftrightarrow}{\partial^\mu}
\widetilde{q}_\beta F_{\alpha\beta}^{2 I}(T_{3I} - e_{I}\sin^2\theta_W)
\end{eqnarray} 

\begin{figure}
\centerline{
   \epsfxsize 3.3 truein \epsfbox {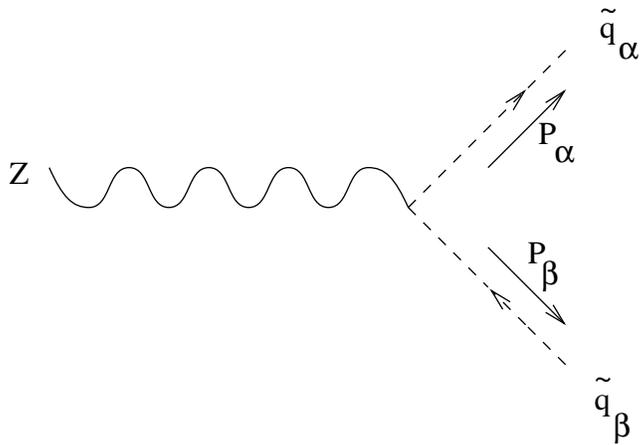}
    }
\caption{$ -i \frac{g_2}{\cos\theta_W}F_{\alpha\beta}^{2I}(P_\beta -
P_\alpha)^\mu(T_{3I} - e_I \sin^2\theta_W)$
}
\label{frule16}
\end{figure}

\item {\it Scalar quark --- scalar quark --- gauge boson --- gauge boson}

a) Electroweak
\begin{eqnarray}
{\mathcal{L}}_{AA\widetilde{q}\widetilde{q}} &=&
\frac{1}{2}g_{2}^{2}W_{\mu}^{+}W^{-\mu}(\widetilde{u}_{\alpha}^{*}
\widetilde{u}_{\beta} (\Gamma_{UL}^{SCKM})_{I\alpha}^{*}
(\Gamma_{UL}^{SCKM})_{I\beta} \nonumber \\
&+&\widetilde{d}_{\alpha}^{*} 
\widetilde{d}_\beta(\Gamma_{DL}^{SCKM})_{I\alpha}^{*}
(\Gamma_{DL}^{SCKM})_{I\beta}) \nonumber \\
&+& \frac{g_2}{\sqrt{2}} y_Q(eA^\mu - \frac{g_2 \sin^2\theta_W 
Z^\mu}{\cos\theta_W})(W_{\mu}^{+}\widetilde{u}_{\alpha}^{*}
\widetilde{d}_\beta F_{\alpha_{\beta}}^{1} + {\rm h.c.}) \nonumber \\
&+& e^2 A_\mu A^\mu e_{\widetilde{q}_{\alpha}} \widetilde{q}_{\alpha}^{*}q_{\alpha}
\nonumber \\
&+& \frac{g^2}{\cos^2\theta_W}Z_\mu Z^{\mu}
F_{\alpha\beta}^{2I}\widetilde{q}_{\alpha}^{*}\widetilde{q}_\beta(T_{3I} -
e_{I}\sin^2 \theta_W)^{2} \nonumber \\
&+& \frac{2g e}{\cos\theta_W}A_\mu Z^\mu
e_{\widetilde{q}_{\alpha}}\widetilde{q}_{\alpha}^{*}\widetilde{q}_{\beta}
F_{\alpha\beta}^{2I}(T_{3I}
- e_{I}\sin^2\theta_W), \nonumber \\
\end{eqnarray}
\begin{equation}
y_Q = -1 + 2e_{u} = 1 + 2e_{d}, \nonumber 
\end{equation}

\begin{figure}
\centerline{
   \epsfxsize 3.3 truein \epsfbox {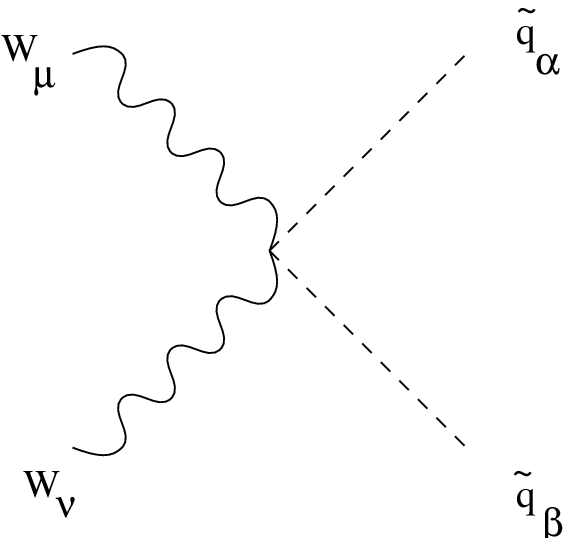}
    }
\caption{$ \frac{g_{2}^{2}}{2} \eta_{\mu\nu}
(\Gamma_{qL}^{SCKM})_{I\alpha}^{*}(\Gamma_{qL}^{SCKM})_{I\beta}$
}
\label{frule17}
\vskip 0.25 truein
\centerline{
   \epsfxsize 3.3 truein \epsfbox {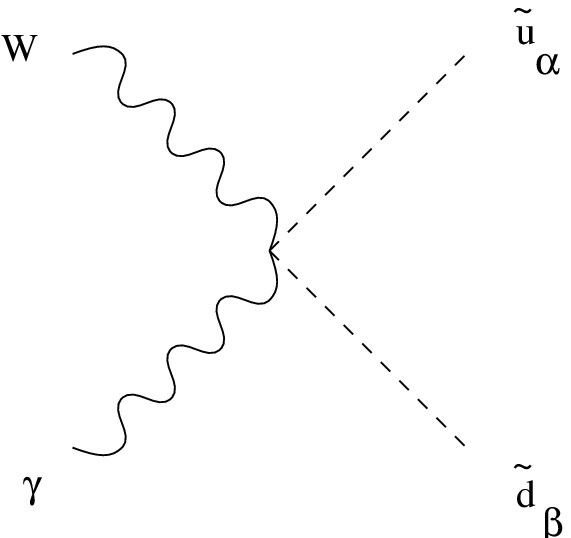}
    }
\caption{$ \frac{g_{2} e}{\sqrt{2}}y_Q
\eta_{\mu\nu}F_{\alpha\beta}^{1}$
}
\label{frule18}
\end{figure}

\begin{figure}
\centerline{
   \epsfxsize 3.3 truein \epsfbox {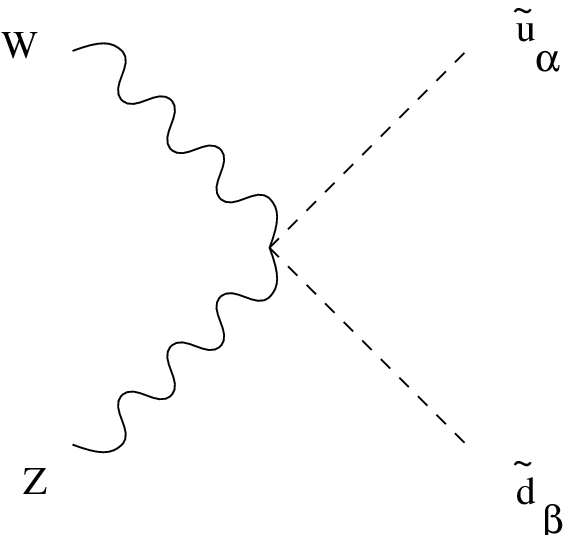}
    }
\caption{$-\frac{g_{2}^{2}}{\sqrt{2}}
\frac{y_{Q}\sin^2\theta_W}{\cos\theta_W}
\eta_{\mu\nu}F_{\alpha\beta}^{1}$ 
}
\label{frule19}
\vskip 0.25 truein
\centerline{
   \epsfxsize 3.3 truein \epsfbox {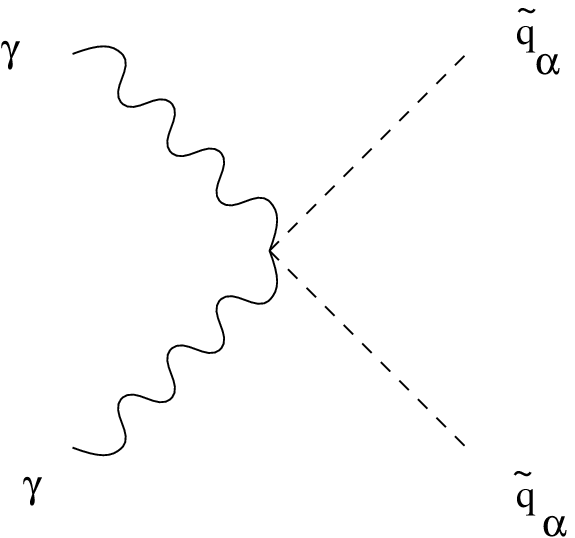}
    }
\caption{$  2e^2 e_{q_{\alpha}}^{2}g_{\mu\nu}$
}
\label{frule20}
\end{figure}

\begin{figure}
\centerline{
   \epsfxsize 3.3 truein \epsfbox {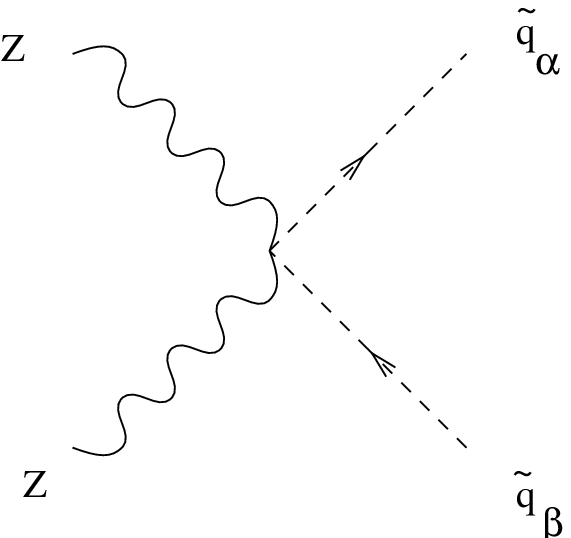}
    }
\caption{$\frac{g^2}{\cos^2\theta_W}F_{\alpha\beta}^{2I}(T_{3I} -
e_{I}  \sin^2\theta_W)^2 $ 
}
\label{frule21}
\vskip 0.25 truein
\centerline{
   \epsfxsize 3.3 truein \epsfbox {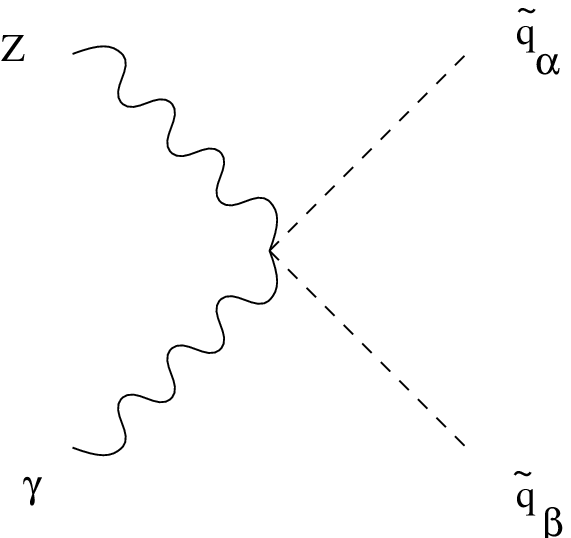}
    }
\caption{$ 
\frac{2gee_{\widetilde{q}_\alpha}}{\cos\theta_W}F_{\alpha\beta}^{2I}(T_{3I}
- e_I \sin^2\theta )$
}
\label{frule22}
\end{figure}
\begin{figure}
\centerline{
   \epsfxsize 2.3 truein \epsfbox {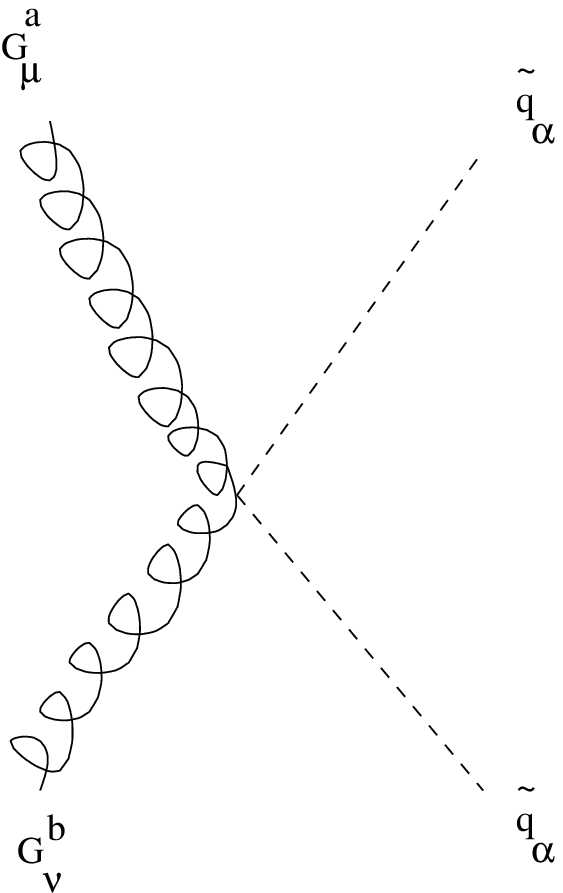}
    }
\caption{$ g_{3}^{2}(\frac{1}{3}\delta_{ab} {\mathbf{1} }+ 
d_{abc}T^c)g_{\mu\nu}
$ 
}
\label{frule23}
\vskip 0.25 truein
\centerline{
   \epsfxsize 2.3 truein \epsfbox {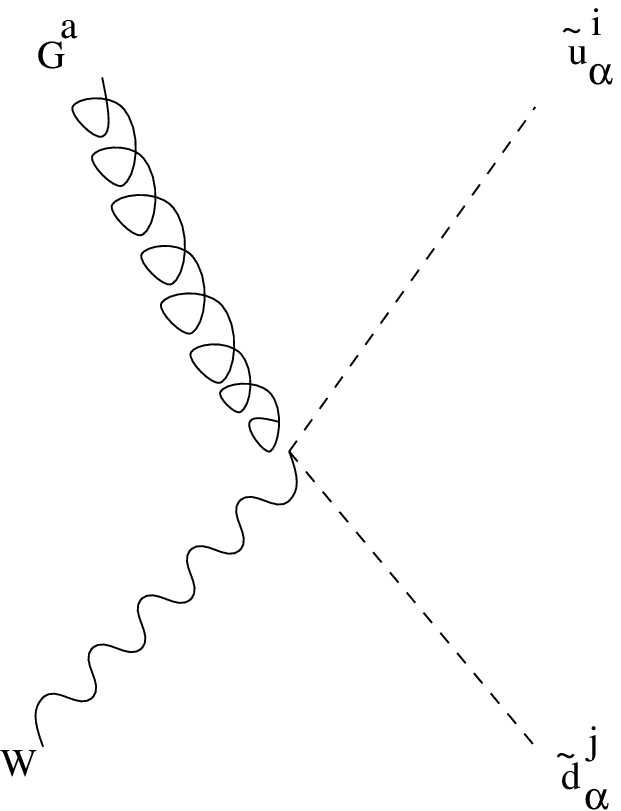}
    }
\caption{$ \sqrt{2}g_{2}g_{s}t_{ij}^{a}g_{\mu\nu}
F_{\alpha\beta}^{1}
$}
\label{frule24}
\end{figure}

\begin{figure}
\centerline{
   \epsfxsize 2.73 truein \epsfbox {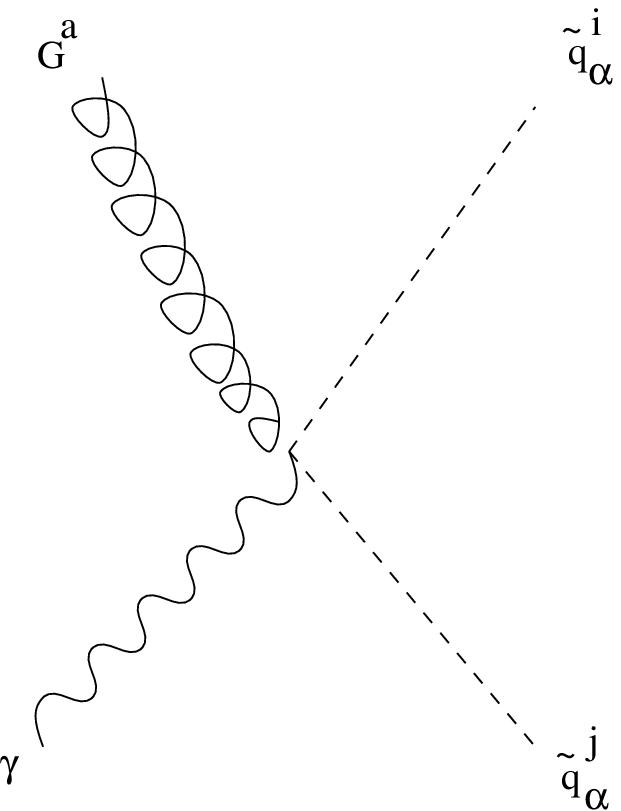}
    }
\caption{$2g_{3}e e_{\widetilde{q}_\alpha}T_{ij}^{a}  
$ 
}
\label{frule25}
\vskip 0.25 truein
\centerline{
   \epsfxsize 2.73 truein \epsfbox {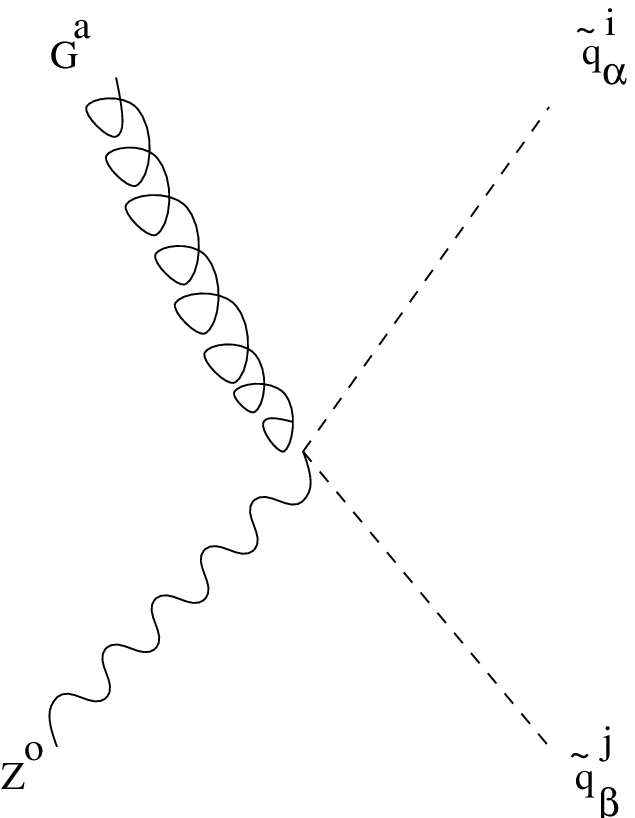}
    }
\caption{$ 
2g_3\frac{g_2}{\cos\theta_W}T_{ij}^{a}F_{\alpha\beta}^{2I}(T_{3I} -
e_{I} \sin^2 \theta_W)
$}
\label{frule26}
\end{figure}
\end{enumerate}

b) Strong Interaction
\begin{eqnarray}
{\mathcal{L}}_{\widetilde{q}\widetilde{q}G^a G^a} =
\frac{1}{6}g_{3}^{2}G_{\mu}^{a}G^{\mu
a}\widetilde{q}_{\alpha}^{*i}\widetilde{q}_{\alpha}^i
+ \frac{1}{2}g_{3}^{2}d_{abc}G_{\mu}^{a}G^{\mu
a}T_{ij}^{c}\widetilde{q}_{\alpha}^{*i}\widetilde{q}_{\alpha}^{i}.
\end{eqnarray}

\vspace{1cm}
c) Mixed Electroweak-Strong 
\begin{eqnarray}
{\mathcal{L}}_{\widetilde{qq}GA} &=& \sqrt{2}g_2 g_3
G_{a}^{\mu}(W_{\mu}^{+}T_{ij}^{a}
\widetilde{u}_{\alpha}^{i*}\widetilde{d}_{\beta}^{j}F_{\alpha\beta}^{1} +
{\rm h.c.})+ 2g_3 e A_\mu G_{a}^{\mu}e_{\widetilde{q}_\alpha}
T_{ij}^{a}\widetilde{q}_{\alpha}^{i*}\widetilde{q}_{\alpha}^{j} \nonumber \\
&+& 2g_3(\frac{g_2}{\cos\theta_W})Z_\mu
G_{a}^{\mu }T_{ij}^{a}\widetilde{q}_{\alpha}^{*}\widetilde{q}_{\beta}F_{\alpha\beta}^{2I}
(T_{3I} - e_{I}\sin^2\theta_W)
\end{eqnarray}

\subsection{Spinor handling}
\label{spinorapp}
In this section, a brief summary of the spinor conventions used here are 
presented as well as techniques needed in the calculations involving 
spinors. Similar techniques can be found in \cite{Haber:1984rc}, among 
many other places in the literature. 

Here the chiral representation is used, in which the $\gamma$ 
matrices have the form
\begin{equation}
\gamma^{\mu}= \left( \begin{array}{cc} 0&\sigma^{\mu} \\
\overline{\sigma}^{\mu} & 0
\end{array}\right),
\end{equation} 
where $\sigma^{\mu}=(1,$ \boldmath$\sigma$\unboldmath$)$ and
$\sigma^{\mu}=(1, -$\boldmath$\sigma$\unboldmath$)$. In this basis,
\begin{equation}
\gamma_{5}= i \gamma^0\gamma^1\gamma^2 \gamma^3 =\left(\begin{array}{cc}
-1&0 \\
0 & 1
\end{array}\right).
\end{equation}
The left- and right-handed projection operators are defined as follows:
\begin{equation}
P_L=\frac{1}{2}(1-\gamma_5), \hspace{1cm} P_R=\frac{1}{2}(1+\gamma_5).
\end{equation}  
A four-component Dirac spinor in this basis is written as
\begin{equation}
\Psi = \left(  \begin{array}{c} \xi_{\alpha} \\
\overline{\eta}^{\dot{\alpha}}  \end{array}\right) =
\left( \begin{array}{c}  \Psi_L \\ \Psi_R \end{array}\right)
\end{equation}
where $\xi$ and $\overline{\eta}$ are two-component Weyl spinors transforming
under the left-handed and right-handed representations of the Lorentz 
group, respectively (reflected in the use of the indices 
$\alpha$ and $\dot{\alpha}$. Upper and lower indices indicate 
that the Lorentz transformation, which is a $2 \times 2$ matrix, should be 
multiplied as a conjugate from the right or as it is from the left. The 
indices can be raised or lowered by using
\begin{equation}
\epsilon^{\alpha \beta}=-\epsilon_{\alpha\beta} = i\sigma^2 =
\left(\begin{array}{cc} 0&1 \\ -1&0 \end{array}\right).
\end{equation}  
This notation is convenient because it keeps track of the
transformation properties of the spinors. Therefore, it is easy to
construct certain products of spinors which have specific
transformation properties. A fermion bilinear which transforms as a 
Lorentz scalar is formed by the 
contraction an upper index with a lower index of the same type.  For 
example,  consider a chiral supermultiplet $(\phi, \psi, 
F)$ where $\psi$ is a
left-handed Weyl spinor. Since its mass term must be a Lorentz singlet,
it has the form $m \psi^{\alpha} \psi_{\alpha} + {\rm h.c.} = m \psi  
\psi + {\rm h.c.}$. In this notation, the $\gamma$-matrices can be written 
as
\begin{equation}
\gamma^{\mu}= \left(\begin{array}{cc} 0 & \sigma^{\mu}_{\alpha 
\dot{\beta}}
\\ \overline{\sigma}^{\mu \dot{\alpha} {\beta}} \end{array}\right)
\end{equation}
and
\begin{equation}
\sigma^{\mu \nu}=\frac{i}{2}[\gamma^{\mu}, \gamma^{\nu}] = 2i \left(
\begin{array}{cc} \hat{\sigma}^{\mu \nu \beta}_{\alpha} & 0 \\
0 & \hat{\overline{\sigma}}^{\mu \nu \dot{\alpha}}_{\dot{\beta}} \end{array} 
\right),
\end{equation}
where
\begin{eqnarray}
\hat{\sigma}^{\mu \nu \beta}_{\alpha} &=& \frac{1}{4}
(\sigma^{\mu}_{\alpha \dot{\alpha}} \overline{\sigma}^{\nu \dot{\alpha}
\beta} -\sigma^{\nu}_{\alpha \dot{\alpha}} \overline{\sigma}^{\mu \dot{\alpha}
\beta} ) \nonumber \\
\hat{\overline{\sigma}}^{\mu \nu \dot{\alpha}}_{\beta} &=& \frac{1}{4}
(\overline{\sigma}^{\mu \dot{\alpha} \alpha} \sigma^{\nu}_{ \alpha
\dot{\beta}} -\overline{\sigma}^{\nu \dot{\alpha} \alpha} \sigma^{\mu}_{\alpha  
\dot{\beta}}).
\end{eqnarray}
The original MSSM Lagrangian is usually written in terms of two-component 
spinors (because chiral supermultiplets contain Weyl spinors).
However, the four-component notation is more familiar to many people. 
Therefore, it is useful to   
establish a dictionary in order to translate back and forth between the 
two languages. This dictionary has been presented in many reviews and 
textbooks; it is presented here (along with other useful spinor 
identities) for completeness. Two-component Weyl spinors satisfy
\begin{eqnarray}
\eta \xi = \xi \eta, \hspace{1cm} \overline{\eta} \overline{\xi} = \overline{\xi}
\overline{\eta}, \hspace{1.2cm}\nonumber \\
\overline{\xi}  \overline{\sigma}^{\mu}\eta = - \eta \sigma^{\mu} \overline{\xi},
\hspace{2.2cm}
\nonumber \\
\overline{\eta} \hat{\overline{\sigma}}^{\mu \nu}\overline{\xi}=-\overline{\xi} \hat{\overline{\sigma}}^{\mu
\nu}\overline{\eta}, \hspace{1cm}
\eta \hat{\sigma}^{\mu \nu} \xi = - \xi \hat{\sigma}^{\mu \nu} \eta.
\end{eqnarray}
It is always understood that ``barred'' spinors carry dotted indices while
others carry undotted indices, and upper indices always contract with
lower ones. The four-component spinors satisfy
\begin{eqnarray}
\overline{\Psi}_1 \Psi_2 &=& \eta_1 \xi_2 + \overline{\eta_2} \overline{\xi_1}
\nonumber \\
\overline{\Psi}_1 \gamma_5 \Psi_2 &=& -\eta_1 \xi_2 + \overline{\eta_2} \overline{\xi_1}
\nonumber \\
\overline{\Psi}_1 \gamma^{\mu} \Psi_2 &=& \overline{\xi_1}
\overline{\sigma}^{\mu}\xi_2 - \overline{\eta_2} \overline{\sigma}^{\mu} {\eta_1}
\nonumber \\
\overline{\Psi}_1 \gamma^{\mu}\gamma_5 \Psi_2 &=& -\overline{\xi_1}
\overline{\sigma}^{\mu}\xi_2 - \overline{\eta_2} \overline{\sigma}^{\mu} {\eta_1}
\nonumber \\
\overline{\Psi}_1 \sigma^{\mu \nu}  \Psi_2 &=& 2i({\eta_1}
{\sigma}^{\mu \nu}\xi_2 - \overline{\eta_2} \overline{\sigma}^{\mu \nu} 
\overline{\xi_1}).
\end{eqnarray}
Projection operators can be inserted into the expressions above to obtain
\begin{eqnarray}
\overline{\Psi}_1 P_L \Psi_2 = \eta_1 \xi_2, &\hspace{0.5cm}& \overline{\Psi}_1 P_R
\Psi_2 = \overline{\eta_2} \overline{\xi_1}, \nonumber \\
\overline{\Psi}_1 \gamma^{\mu} P_L \Psi_2 = \overline{\xi}_1 \overline{\sigma}^{\mu}
\xi_2, &\hspace{0.5cm}&  \overline{\Psi}_1 \gamma^{\mu} P_R \Psi_2 = 
-\overline{\eta}_2
\overline{\sigma}^{\mu} \eta_1.
\end{eqnarray}
The following relations are also often useful, especially in the 
calculation of helicity amplitudes \cite{Hagiwara:1985yu}:
\begin{eqnarray}
\xi_1^{\dagger} \Sigma_1 \sigma^{\mu} \Sigma_2 \xi_2 \cdot
\xi_3^{\dagger} \Sigma_3 \overline{\sigma}_{\mu} \Sigma_4 \xi_4 &=&
\xi_1^{\dagger} \Sigma_1 \overline{\sigma}^{\mu} \Sigma_2 \xi_2 \cdot
\xi_3^{\dagger} \Sigma_3 {\sigma}_{\mu} \Sigma_4 \xi_4 \nonumber \\
&=&2 \xi_1^{\dagger} \Sigma_1 \Sigma_4 \xi_4  \cdot \xi_3^{\dagger}
\Sigma_3 \Sigma_2 \xi_2 \nonumber \\
\xi_1^{\dagger} \Sigma_1 \sigma^{\mu} \Sigma_2 \xi_2 \cdot
\xi_3^{\dagger} \Sigma_3 {\sigma}_{\mu} \Sigma_4 \xi_4 &=&
\xi_1^{\dagger} \Sigma_1 \overline{\sigma}^{\mu} \Sigma_2 \xi_2 \cdot
\xi_3^{\dagger} \Sigma_3 \overline{\sigma}_{\mu} \Sigma_4 \xi_4 \nonumber \\
&=&2 \xi_1^{\dagger} \Sigma_1 \Sigma_2 \xi_2  \cdot \xi_3^{\dagger}  
\Sigma_3 \Sigma_4 \xi_4 - 2 \xi_1^{\dagger} \Sigma_1 \Sigma_4 \xi_4
\cdot  \xi_3^{\dagger} \Sigma_3 \Sigma_2 \xi_2, \nonumber\\
\end{eqnarray}
where $\Sigma_i$ are arbitrary $2 \times 2$ matrices.

Charge conjugation of a four-component spinor is defined by
\begin{equation}
\Psi^c=C \overline{\Psi}^T
\end{equation}
where $C$ is the charge conjugation operator.\footnote{In the chiral
representation, $C=-i\gamma^2 \gamma^0$. However, in most calculations,
the detailed form of $C$ is not needed.} The charge conjugation operator
has the following properties:
\begin{enumerate}
\item $C^{\dagger}=C^{-1}$,
\item $C^T=-C$,
\item For the generators of the Clifford Algebra $\Gamma_i=1$, 
$i\gamma_5$,
$\gamma^{\mu}\gamma_5$, $\gamma^{\mu}$, $\sigma^{\mu \nu}$, 
$C^{\dagger}\Gamma_i C = \lambda_i \Gamma_i^{T}$, where $\lambda_i=1$ if
$1\leq i \leq 6$, and $\lambda_i=-1$ for the rest. $\Gamma_i$s
  satisfy $\gamma^0 \Gamma_i \gamma^0 = \Gamma_i^{\dagger}$.
\end{enumerate}
A Majorana spinor is defined by the condition $\Psi^c=\Psi$:
\begin{equation}
\Psi_M=\left(  \begin{array}{c} \xi_{\alpha} \\ \overline{\xi}^{\dot{\alpha}}
\end{array}\right) =\left(  \begin{array}{c} \psi_L \\ i\sigma^2 \psi_L^*
\end{array}\right).
\end{equation}
Majorana spinors satisfy
\begin{eqnarray}
\overline{\Psi}_1 \Psi_2 &=& \overline{\Psi}_2 \Psi_1 \nonumber \\
\overline{\Psi}_1 \gamma_5 \Psi_2 &=& \overline{\Psi}_2 \gamma_5 \Psi_1 \nonumber\\
\overline{\Psi}_1 \gamma^{\mu} \Psi_2 &=& -\overline{\Psi}_2 \gamma^{\mu}  \Psi_1
\nonumber\\
\overline{\Psi}_1 \gamma^{\mu} \gamma_5 \Psi_2 &=& \overline{\Psi}_2 \gamma^{\mu}
\gamma_5 \Psi_1 \nonumber\\
\overline{\Psi}_1 \sigma^{\mu \nu} \Psi_2 &=&-\overline{\Psi}_2 \sigma^{\mu \nu}  
\Psi_1.
\end{eqnarray} 

Spinors $u(p,s)$ and $v(p,s)$ which satisfy the Dirac equation, 
$(\gamma^{\mu} p_{\mu} - m ) u(p,s) =0$, and   $(\gamma^{\mu}
p_{\mu} + m ) v(p,s) =0$, also satisfy
\begin{equation}   
u(k,s)= C \overline{v}^T (k,s), \hspace{1cm} v(k,s)= C \overline{u}^T (k,s)
\label{spinor_conjg}
\end{equation}  
In calculating the scattering cross section
or decay width, one usually averages/sums over the initial/final spin
states of fermions. In doing so, one usually encounter the familiar spin
sum formula
\begin{eqnarray}
\sum_{s} u(p,s) \overline{u}(p,s) &=& \gamma^{\mu} p_{\mu} + m, \nonumber\\
\sum_{s} v(p,s) \overline{v}(p,s) &=& \gamma^{\mu} p_{\mu} - m.
\end{eqnarray} 
However, in the processes involving Majorana fermions, the following
spin sum formulae will also be useful
\begin{eqnarray}
\sum_{s} u(p,s) v^T (p,s) &=& (\gamma^{\mu} p_{\mu} + m)(-C), \nonumber\\
\sum_{s} \overline{u}^T(p,s) \overline{v}(p,s) &=& C^{\dagger}(\gamma^{\mu}
p_{\mu} - m), \nonumber \\
\sum_{s} \overline{v}^T(p,s) \overline{u}(p,s) &=& C^{\dagger}(\gamma^{\mu}
p_{\mu} + m), \nonumber \\
\sum_{s} v(p,s) u^T (p,s) &=& (\gamma^{\mu} p_{\mu} - m)(-C).
\label{spinsum_2}
\end{eqnarray}

The following simple example is useful to illustrate the spinor techniques
necessary for cross section calculations.

\begin{figure}[h!]
\vspace{1.5cm}
\centering
\epsfxsize=12.0cm
\hspace*{0in}
\epsffile{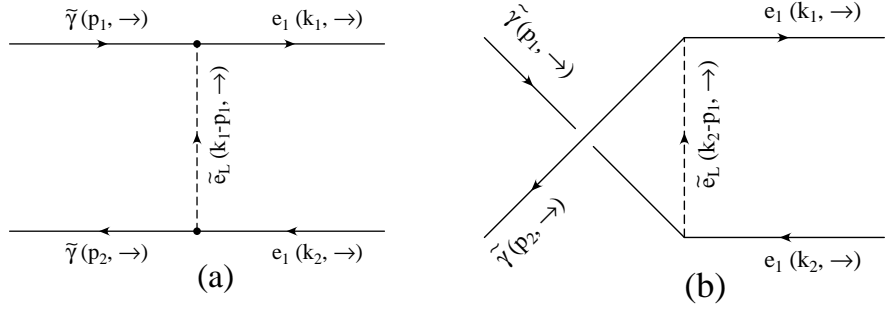}
\bigskip
\caption{The annihilation of a pair of photinos into an
electron-positron pair via a $t$-channel exchange of a left-handed
scalar electron. The arrows on the lines label the direction of fermion
number propagation.  The arrows appearing  together with the momenta
label the direction of momentum flow.}

\vspace{1.5cm}

\label{figone}
\end{figure}

Photino annihilation provides a nice example of calculating
the cross sections involving Majorana particles. It also has practical
significance, because neutralino pair annihilation through the $t$-channel
exchange of scalar fermions can be significant when calculating the 
relic density of neutralino cold dark matter. In order to derive the 
Feynman rules and
write down the amplitude, a mode expansion of the Majorana
spinors can be performed in a similar way to that of the Dirac spinors 
(just keep in mind that for Majorana spinors, there is only one type of 
creation and annihilation operator). The direction of the fermion number 
propagation is reflected in the choice of spinors $u(k,s)$ and $v(k,s)$. 
Of course, this
distinction is superficial since there is no real distinction between
fermion and antifermion for Majorana particles.  Diagram (a) is
obtained in a straightforward manner. Since the photino is a Majorana 
particle, the exchange diagram (b) is also present. The amplitudes 
are\footnote{As the focus here is on the spinor structure, the detailed 
dependence on the coupling constants is suppressed.}
\begin{eqnarray}
M_a &\propto& D_t (\overline{u}(k_1, \sigma_1) P_R u(p_1,s_1))
(\overline{v}(p_2,s_2) P_L v(k_2,\sigma_2)) \nonumber \\
M_b &\propto& -D_u (\overline{u}(k_1, \sigma_1) P_R C \overline{v}^T(p_2,s_2))  
(u^T(p_1,s_1)(-C^{\dagger}) P_L v(k_2,\sigma_2)) \nonumber \\
    &=& D_u (\overline{u}(k_1, \sigma_1) P_R u(p_2,s_2))
(\overline{v}(p_1,s_1) P_L v(k_2,\sigma_2)),
\end{eqnarray}
where $D_t = ((p_1-k_1)^2-m_{\widetilde{e}_L}^2)^{-1}$, $D_u =
((p_2-k_1)^2-m_{\widetilde{e}_L}^2)^{-1}$. To obtain the second equality
of $M_b$, Eq.~(\ref{spinor_conjg}) was used. The second expression of
$M_b$ shows manifestly that the direction of fermion number propagation  
is superficial since it is equivalent to the amplitude obtained from
reversing the arrows on the photino lines. The relative minus sign
between the two diagrams originates from the exchange of two fermion 
fields, similar to the relative minus sign of the $u$-channel 
diagram for elastic scattering of electrons in QED.  The differential 
cross section is 
\begin{eqnarray}
\frac{d\sigma}{d\Omega} 
\propto \frac{1}{4} \sum_{s_1,s_2,\sigma1, 
\sigma_2} |M_a+M_b|^2  
=\frac{1}{4} \sum_{s_1,s_2,\sigma1, \sigma_2} |M_a|^2+|M_b|^2 + M_a 
M_b^*+
M_a^* M_b.
\end{eqnarray}
$|M_a|^2$ and $|M_b|^2$ can be obtained using the standard trace 
technology
\begin{eqnarray}
\sum |M_a|^2 &\propto& D_t^2 (t-M_{\widetilde{\gamma}}^2 - m_e^2)^2
\nonumber \\
\sum |M_b|^2 &\propto& D_u^2 (u-M_{\widetilde{\gamma}}^2 - m_e^2)^2.
\end{eqnarray}
However, an amount of extra effort is needed when calculating $M_a M_b^*$.
After summing over the final spin states,
\begin{eqnarray}
\sum M_a M_b^* \propto && 
-[\overline{u}(p_2, s_2) P_L (\gamma^{\mu}
k_{1\mu}+ m_e) P_R u(p_1,s_1)] 
\nonumber \\  
&& \times [\overline{v}(p_2, s_2) (\gamma^{\mu} k_{2\mu}-m_e) P_R v(p_1,s_1)].
\end{eqnarray}
There is no obvious way to sum the spin indices except
doing it explicitly. However, one can take the transpose of the terms in
the first square bracket, which will not change the result since it is
just a number. Using the properties of charge
conjugation and the appropriate formula in Eq.~(\ref{spinsum_2}), 
\begin{eqnarray}
\sum M_a M_b^* &\propto &
 - Tr [ (\gamma^{\mu} p_{1 \mu} - M_{\widetilde{\gamma}}) P_R (\gamma^{\mu} 
k_{1\mu}-m_e) P_L
(\gamma^{\mu}p_{2\mu} - M_{\widetilde{\gamma}}) P_L (\gamma^{\mu}
k_{2\mu}-m_e) ] \nonumber \\
 &=& (s-2m_e^2) M_{\widetilde{\gamma}}^2 .
\end{eqnarray}
Since all of the couplings are real, $M_aM_b^*=M_a^*M_b$. Putting 
everything together,
\begin{eqnarray}
\frac{d \sigma}{d\Omega} \propto D_t^2 (t-M_{\widetilde{\gamma}}^2 -
m_e^2)^2 + D_u^2 (u-M_{\widetilde{\gamma}}^2 - m_e^2)^2 
-2 D_u D_t[s-2m_e^2] M_{\widetilde{\gamma}}^2.
\end{eqnarray}
In the cosmologically interesting limit where $E_{\widetilde{\gamma}} \sim
M_{\widetilde{\gamma}}$, 
\begin{equation}
\frac{d \sigma}{d \Omega} \sim
(m_e^2/(M_{\widetilde{\gamma}}-m_{\widetilde{e}_L}^2)^2. 
\end{equation}
This is an example of the general result that s-wave neutralino 
annihilation to fermion pairs is suppressed by the fermion mass.

\subsection{FCNC example}
\label{FCNCexapp}
Consider the following simple two-flavor example, 
in which the squark mass matrix is given by 
\begin{equation}
 {\mathcal{L}}= \widetilde{q}_i^{\dagger} {\mathbf{m}^2}_{ij}
\widetilde{q}_j, \hspace{1cm}
{\mathbf{m}^2}=\left( \begin{array}{cc}
 m_1^2 & \Delta \\
\Delta & m_2^2 \end{array}\right),
\end{equation}
in which $i,j=1,2$ (for simplicity here we neglect CP violation).  The 
mass matrix is diagonalized by
\begin{equation}
\Gamma_{\alpha i} \widetilde{q}_j = \widetilde{q}_{\alpha}. \hspace{0.5cm}
\Gamma {\mathbf{m}^2} 
\Gamma^{\dagger}={\mathbf{Diag}}[\widetilde{m}_{\alpha}^2],
\end{equation}
where $\alpha=1,2$ labels the mass eigenstates and 
$\widetilde{m}_{\alpha}$ denotes the mass eigenvalues.

\begin{figure}
\centerline{
   \epsfxsize 3.3 truein \epsfbox {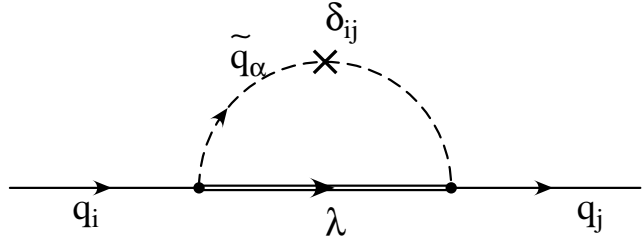}
    }
\caption{One-loop diagram which can induce FCNCs.}
\label{samplefcnc}
\end{figure}

Consider the FCNC process mediated by the gaugino-squark loop as shown in 
Figure~\ref{samplefcnc}. This diagram (which is usually called a penguin 
diagram when a gauge boson attaches to one of 
the internal lines and then
to a spectator particle) contributes to FCNC rare decays (such as 
$b\rightarrow s\gamma$) through dipole transitions; as the SM 
contributions to such processes are also loop-suppressed, the supersymmetric 
contributions are typically competitive. Recalling the form of the 
quark-squark-gaugino coupling 
\begin{equation}
{\mathcal{L}}\propto g(\overline{q}_i P_L \lambda \widetilde{q}_i +
\overline{\lambda} P_R q_i \widetilde{q}_i^*),
\end{equation}
the amplitude associated with this process is
\begin{equation}
{\mathcal{M}}_{i\rightarrow j} \propto g^2 \sum_{\alpha=1,2}
\Gamma^{\dagger}_{j\alpha} \Gamma_{\alpha i} f(x_{\alpha}),
\end{equation}
where $x_{\alpha}=\frac{\widetilde{m}_{\alpha}^2}{m_{\lambda}^2}$ and $f(x)$
is a function which arises from the loop integral. 
If $\Delta=0$ and $m_1^2=m_2^2$, $\widetilde{m}_1^2=\widetilde{m_2}^2$,
and $x_1=x_2$. In this limit, ${\mathcal{M}}_{i\rightarrow j}
\propto \sum_{\alpha=1,2} \Gamma^{\dagger}_{j\alpha} \Gamma_{\alpha i}
= 0$ if $i\neq j$. This cancellation is an example of the super-GIM
mechanism, which of course holds only in this limit.  
To approximate this process, we will assume that 
$m_1^2 \sim m_2^2 \gg \Delta $ and develop the mass insertion 
approximation. In this limit, the physical masses are 
\begin{eqnarray}
\widetilde{m}^2_1 &\sim& m_1^2 + \frac{\Delta^2}{m_1^2-m_2^2} \nonumber \\
\widetilde{m}^2_2 &\sim& m_2^2 - \frac{\Delta^2}{m_1^2-m_2^2} \nonumber \\
\widetilde{m}^2_1-\widetilde{m}^2_2 &\sim& m_1^2 - m_2^2 +
\frac{2\Delta^2}{m_1^2-m_2^ 2},
\label{mimass}
\end{eqnarray}
and the corresponding mixing matrix elements are given by   
\begin{equation}
\Gamma_{11} \sim \Gamma_{22} \sim 1 +
O(\frac{\Delta^2}{(m_1^2-m_2^2)^2}), \hspace{1cm}
\Gamma_{12}=-\Gamma_{21} \sim \frac{\Delta}{m_1^2-m_2^2}.
\label{mimixing}
\end{equation}
The loop function is then expanded as follows (using Eq.~(\ref{mimass})):
\begin{eqnarray}
f(x_1)&=& f(x_2) + f'(x_2) (x_1-x_2)+\cdots \nonumber \\
x_1-x_2 &=& \frac{\widetilde{m}_1^2-\widetilde{m}_2^2}{m_{\lambda}^2} \sim
\frac{m_1^2- m_2^2}{\sqrt{m_1^2 m_2^2}} x_2.
\end{eqnarray}
After substituting this
expansion in the amplitude for the FCNC process and using
Eq.~(\ref{mimixing}), the result is (setting $i=1$ and $j=2$)
\begin{equation}
{\mathcal{M}}_{1\rightarrow 2} \propto g^2(f(x_2) \sum_{\alpha}
\Gamma^{\dagger}_{2 \alpha} \Gamma_{\alpha 1} + x_2 f'(x_2)
\delta_{12}+ \cdots),
\end{equation}
in which the definition of the mass insertion parameter
\begin{equation}  
\delta_{12}=\frac{\Delta}{\sqrt{m_1^2 m_2^2}}
\end{equation}
has been utilized. As the first term vanishes due to the unitarity of 
the mixing matrix, the amplitude is given by
\begin{equation}
{\mathcal{M}}_{1\rightarrow 2}\propto g^2(x_2 f'(x_2)   
\delta_{12}+ \cdots).
\end{equation}
This result is straightforward to interpret.  As the mixing is small, the
mass eigenstates are approximately equal to the flavor 
eigenstates, and hence approximate flavor eigenstates are propagating in 
the loops (squarks 1 and 2 in this example). The mixing leads to an effective
interaction Lagrangian which couples different squark flavors
($\delta_{12}$ in our example) that provides nonvanishing contributions to
FCNCs.\footnote{This process can naturally be viewed as follows:  quark 1 
splits into a gaugino and squark 1; squark 1 then connects to the flavor 
changing vertex $\delta_{12}$ which switches it into squark 2. Finally, 
squark 2 combines with the gaugino into quark 2 to complete the loop. This 
intuitive picture is often useful when considering generic FCNC
processes.}

\subsection{MSSM RGEs}
\label{rgeapp}
The renormalization group equations (RGEs) for the gauge
couplings to two-loop order are
\begin{equation}
\frac{d g_a}{dt} = \frac{g_a^3}{16 \pi^2} b_a + 
\frac{g_a^3}{(16 \pi^2)^2} \left 
[\sum^3_{b=1}B^{(2)}_{ab} g_b^2 -  
\frac{1}{16 
\pi^2}\sum_{x}\frac{C^x_a}{16\pi^2}{\rm Tr}({Y_x}^{\dagger}Y_x)\right],
\label{grge2lp}
\end{equation}
where $t=\ln (\mu/M_{X})$ ($\mu$ is the $\overline{MS}$ scale
and $M_X$ is the high energy scale), $b_a=(\frac{33}{5},1,-3)$, and 
\begin{equation}
B^{(2)}_{ab} = \left( \begin{array}{ccc}  \frac{199}{25}& \frac{27}{5} 
&\frac{88}{5} \\ \frac{9}{5}&25&24\\\frac{11}{5}&9&14
\end{array}\right), 
\end{equation}
and 
\begin{equation} 
C^{u,d,e,\nu}_{a} = \left( \begin{array}{cccc}  \frac{26}{5}& 
\frac{14}{5}& \frac{18}{5}&\frac{6}{5} \\ 6&6&2&2\\4&4&0&0
\end{array}\right).      
\end{equation}  
Of course, for the MSSM, $Y_{\nu}=0$. 

The RGEs for the gaugino masses to two-loop order are (in $\overline{{\rm 
DR}}$):
\begin{eqnarray}
\frac{dM_a}{dt}&=& \frac{2g_a^2}{16 \pi^2}b_a M_{a} 
+ \frac{2 g^2_a }{(16 \pi^2)^2} 
\sum^3_{b=1}B^{(2)}_{ab} g_b^2(M_a+M_b)  \nonumber \\
&+&   \frac{2 g^2_a }{(16 \pi^2)^2} \sum_{x=u,d,e,\nu} C^x_a \left({\rm 
Tr}[Y^{\dagger}_x\widetilde{A}_x]
-M_a{\rm Tr}[Y^{\dagger}_xY_x]\right ) .
\label{gmrge2lp} 
\end{eqnarray}
The following will all be one-loop results. The RGEs for the 
superpotential Yukawa couplings are
\begin{eqnarray}
\frac{dY_u}{dt} &=&\frac{1}{16 \pi^2}
[N_q.Y_u+Y_u.N_u+(N_{H_u})Y_u]
\nonumber \\
\frac{dY_d}{dt} &=&\frac{1}{16 \pi^2}
[N_q.Y_d+Y_d.N_d+(N_{H_d})Y_d]
\nonumber \\
\frac{dY_{\nu}}{dt} &=&\frac{1}{16 \pi^2}
[N_l.Y_{\nu}+Y_{\nu}.N_{\nu}+(N_{H_u})Y_{\nu}] 
\nonumber \\
\frac{dY_e}{dt} &=&\frac{1}{16\pi^2}
[N_l.Y_e+Y_e.N_e+(N_{H_d})Y_e]
\nonumber \\
\label{YukRGEs}
\end{eqnarray}
where the wavefunction anomalous dimensions are
\begin{eqnarray}
N_q 
&=&Y_u{Y_u}^{\dagger}+Y_d{Y_d}^{\dagger}  
- (\frac{8}{3}g_3^2+\frac{3}{2}g_2^2+\frac{1}{30}g_1^2) \hat{\mbox{\large 
1}} \nonumber \\
N_u &=&2{Y_u}^{\dagger}Y_u  
- ( \frac{8}{3}g_3^2+\frac{8}{15}g_1^2 )\hat{\mbox{\large 
1}}\nonumber \\
N_d &=&2{Y_d}^{\dagger}Y_d  
- ( \frac{8}{3}g_3^2+\frac{2}{15}g_1^2)\hat{\mbox{\large 
1}}\nonumber \\
N_l &=&Y_e{Y_e}^{\dagger}+Y_{\nu}{Y_{\nu}}^{\dagger}  
- (\frac{3}{2}g_2^2+\frac{3}{10}g_1^2 )\hat{\mbox{\large 
1}}\nonumber \\
N_e &=&2{Y_e}^{\dagger}Y_e - \frac{6}{5}g_1^2 \hat{\mbox{\large 
1}}\nonumber \\
N_{\nu} &=&2{Y_{\nu}}^{\dagger}Y_{\nu} \nonumber \\
N_{H_u} &=&3{\rm 
Tr}({Y_u}^{\dagger}Y_u)+{\rm Tr}({Y_{\nu}}^{\dagger}Y_{\nu})  
- ( \frac{3}{2}g_2^2+\frac{3}{10}g_1^2 )
\nonumber \\
N_{H_d} &=&3{\rm Tr}({Y_d}^{\dagger}Y_d)+{\rm Tr}({Y_e}^{\dagger}Y_e)  
-(\frac{3}{2}g_2^2+\frac{3}{10}g_1^2) \nonumber \\
\label{anomdim}
\end{eqnarray}  
in which $\hat{\mbox{\large 1}}$ is the unit matrix. 
Similarly, the RGE for the $\mu$ parameter is 
\begin{equation}
\frac{d\mu}{dt} =\frac{1}{16\pi^2}
[N_{H_u}+N_{H_d}]\mu.
\end{equation}
The RGEs for the soft supersymmetry-breaking
trilinear parameters to one-loop order are
\begin{eqnarray}
\frac{d\widetilde{A}_u}{dt} &=&\frac{1}{16\pi^2}
[N_q.\widetilde{A}_u+\widetilde{A}_u.N_u+(N_{H_u})\widetilde{A}_u
+2P_q.Y_u+2Y_u.P_u+2(P_{H_u})Y_u]
\nonumber \\  
\frac{d\widetilde{A}_d}{dt} &=&\frac{1}{16\pi^2}
[N_q.\widetilde{A}_d+\widetilde{A}_d.N_d+(N_{H_d})\widetilde{A}_d
+2P_q.Y_d+2Y_d.P_d+2(P_{H_d})Y_d]
\nonumber \\
\frac{d\widetilde{A}_{\nu}}{dt} &=&\frac{1}{16\pi^2}
[N_l.\widetilde{A}_{\nu}+\widetilde{A}_{\nu}.N_{\nu}+(N_{H_u})\widetilde{A}_{\nu}
+2P_l.Y_{\nu}+2Y_{\nu}.P_{\nu}+2(P_{H_u})Y_{\nu}]
\nonumber \\  
\frac{d\widetilde{A}_e}{dt} &=&\frac{1}{16\pi^2}
[N_l.\widetilde{A}_e+\widetilde{A}_e.N_e+(N_{H_d})\widetilde{A}_e
+2P_l.Y_e+2Y_e.P_e+2(P_{H_d})Y_e]  
\nonumber \\
\label{TrilinearRGEs}
\end{eqnarray}
where
\begin{eqnarray}
P_q 
&=& (\frac{8}{3}g_3^2M_3+\frac{3}{2}g_2^2M_2+ 
\frac{1}{30}g_1^2M_1 )\hat{\mbox{\large 1}}  
+\widetilde{A}_u{Y_u}^{\dagger}+\widetilde{A}_d{Y_d}^{\dagger}  \nonumber \\
P_u &=& (\frac{8}{3}g_3^2M_3+ \frac{8}{15}g_1^2M_1 )\hat{\mbox{\large 1}}  
+2{Y_u}^{\dagger}\widetilde{A}_u  \nonumber \\
P_d &=&(\frac{8}{3}g_3^2M_3+ \frac{2}{15}g_1^2M_1 )\hat{\mbox{\large 1}}
+2{Y_d}^{\dagger}\widetilde{A}_d  \nonumber \\
P_l &=&(\frac{3}{2}g_2^2M_2+ \frac{3}{10}g_1^2M_1 )\hat{\mbox{\large 1}}
+\widetilde{A}_e{Y_e}^{\dagger}+\widetilde{A}_{\nu}{Y_{\nu}}^{\dagger}  \nonumber 
\\
P_e &=& \frac{6}{5}g_1^2M_1 \hat{\mbox{\large 1}}
+2{Y_e}^{\dagger}\widetilde{A}_e \nonumber \\   
P_{\nu} &=&2{Y_{\nu}}^{\dagger}\widetilde{A}_{\nu} \nonumber \\
P_{H_u} &=&(\frac{3}{2}g_2^2M_2+\frac{3}{10}g_1^2M_1)
+3{\rm Tr}({Y_u}^{\dagger}\widetilde{A}_u)+{\rm 
Tr}({Y_{\nu}}^{\dagger}\widetilde{A}_{\nu})  
\nonumber \\
P_{H_d} &=&(\frac{3}{2}g_2^2M_2+\frac{3}{10}g_1^2M_1)
+3{\rm Tr}({Y_d}^{\dagger}\widetilde{A}_d)+{\rm 
Tr}({Y_e}^{\dagger}\widetilde{A}_e)  
\nonumber 
\\
\end{eqnarray}
The $b$ term RGE is 
\begin{equation}
\frac{db}{dt}=\frac{1}{16 \pi^2} [(N_{H_u}+N_{H_d})b+2(P_{H_u}+P_{H_d})\mu].
\end{equation}  
The RGEs for the soft supersymmetry-breaking scalar mass-squared
parameters are as follows:
\begin{eqnarray}
\frac{dm_Q^2}{dt} &=&\frac{1}{8\pi^2}
[
-2(\frac{8}{3}g_3^2|M_3|^2+\frac{3}{2}g_2^2|M_2|^2+\frac{1}{30}g_1^2|M_1|^2
-\frac{1}{10}g_1^2S)\hat{\mbox{\large 1}}
\nonumber \\  
&+&(\frac{1}{2}Y_u{Y_u}^{\dagger}m_Q^2+\frac{1}{2}m_Q^2Y_u{Y_u}^{\dagger}
+Y_um_U^2{Y_u}^{\dagger}+(m_{H_u}^2)Y_u{Y_u}^{\dagger}+
\widetilde{A}_u{\widetilde{A}_u}^{\dagger})
\nonumber \\
&+&(\frac{1}{2}Y_d{Y_d}^{\dagger}m_Q^2+\frac{1}{2}m_Q^2Y_d{Y_d}^{\dagger}
+Y_dm_D^2{Y_d}^{\dagger}+(m_{H_d}^2)Y_d{Y_d}^{\dagger}+
\widetilde{A}_d{{\widetilde{A}}_d}^{\dagger})
]
\nonumber \\
\frac{dm_U^2}{dt} &=&\frac{1}{8\pi^2}
[
-2(\frac{8}{3}g_3^2|M_3|^2+\frac{8}{15}g_1^2|M_1|^2+ 
\frac{2}{5}g_1^2S)\hat{\mbox{\large 1}}
\nonumber \\
&+&2(\frac{1}{2}{Y_u}^{\dagger}Y_um_U^2+\frac{1}{2}m_U^2{Y_u}^{\dagger}Y_u
+{Y_u}^{\dagger}m_Q^2{Y_u}+(m_{H_u}^2){Y_u}^{\dagger}Y_u+
{\widetilde{A}_u}^{\dagger}\widetilde{A}_u)]
\nonumber \\  
\frac{dm_D^2}{dt} &=&\frac{1}{8\pi^2}
[
-2(\frac{8}{3}g_3^2|M_3|^2+\frac{2}{15}g_1^2|M_1|^2- 
\frac{1}{5}g_1^2S)\hat{\mbox{\large 1}}
\nonumber \\
&+&2(\frac{1}{2}{Y_d}^{\dagger}Y_dm_D^2+\frac{1}{2}m_D^2{Y_d}^{\dagger}Y_d
+{Y_d}^{\dagger}m_Q^2{Y_d}+(m_{H_d}^2){Y_d}^{\dagger}Y_d+
{\widetilde{A}_d}^{\dagger}\widetilde{A}_d)] 
\nonumber \\
\frac{dm_L^2}{dt} &=&\frac{1}{8\pi^2}
[
-2(\frac{3}{2}g_2^2|M_2|^2+\frac{3}{10}g_1^2|M_1|^2+ 
\frac{3}{10}g_1^2S)\hat{\mbox{\large 1}}
\nonumber \\
&+&(\frac{1}{2}Y_e{Y_e}^{\dagger}m_L^2+\frac{1}{2}m_L^2Y_e{Y_e}^{\dagger}
+Y_em_E^2{Y_e}^{\dagger}+(m_{H_d}^2)Y_e{Y_e}^{\dagger}+
\widetilde{A}_e{\widetilde{A}_e}^{\dagger})
\nonumber \\
&+&(\frac{1}{2}Y_{\nu}{Y_{\nu}}^{\dagger}m_L^2+
\frac{1}{2}m_L^2Y_{\nu}{Y_{\nu}}^{\dagger}
+Y_{\nu}m_N^2{Y_{\nu}}^{\dagger}+(m_{H_u}^2)Y_{\nu}{Y_{\nu}}^{\dagger}+
\widetilde{A}_{\nu}{{\widetilde{A}}_{\nu}}^{\dagger})
]
\nonumber \\
\frac{dm_E^2}{dt} &=&\frac{1}{8\pi^2}
[
-2(\frac{6}{5}g_1^2|M_1|^2- \frac{3}{5}g_1^2S)\hat{\mbox{\large 1}}
\nonumber \\
&+&2(\frac{1}{2}{Y_e}^{\dagger}Y_em_E^2+\frac{1}{2}m_E^2{Y_e}^{\dagger}Y_e
+{Y_e}^{\dagger}m_L^2{Y_e}+(m_{H_d}^2){Y_e}^{\dagger}Y_e+
{\widetilde{A}_e}^{\dagger}\widetilde{A}_e)]
\nonumber \\
\frac{dm_N^2}{dt} &=&\frac{1}{8\pi^2}
[2(\frac{1}{2}{Y_{\nu}}^{\dagger}Y_{\nu}m_N^2+
\frac{1}{2}m_N^2{Y_{\nu}}^{\dagger}Y_{\nu}
+{Y_{\nu}}^{\dagger}m_L^2{Y_{\nu}}+(m_{H_u}^2){Y_{\nu}}^{\dagger}Y_{\nu}+
{\widetilde{A}_{\nu}}^{\dagger}\widetilde{A}_{\nu})]
\nonumber \\
\frac{dm_{H_u}^2}{dt} &=&\frac{1}{8\pi^2}   
[-2(\frac{3}{2}g_2^2|M_2|^2+\frac{3}{10}g_1^2|M_1|^2- 
\frac{3}{10}g_1^2S)
\nonumber \\
&+&3({\rm Tr}(Y_um_Q^2{Y_u}^{\dagger})+{\rm Tr}(Y_um_U^2{Y_u}^{\dagger})
+(m_{H_u}^2){\rm Tr}(Y_u{Y_u}^{\dagger})
+{\rm Tr}(\widetilde{A}_u{\widetilde{A}_u}^{\dagger}))
\nonumber \\
&+&({\rm 
Tr}(Y_{\nu}m_L^2{Y_{\nu}}^{\dagger})+{\rm 
Tr}(Y_{\nu}m_N^2{Y_{\nu}}^{\dagger})
+(m_{H_u}^2){\rm Tr}(Y_{\nu}{Y_{\nu}}^{\dagger})
+{\rm Tr}(\widetilde{A}_{\nu}{\widetilde{A}_{\nu}}^{\dagger}))
]
\nonumber \\
\frac{dm_{H_d}^2}{dt} &=&\frac{1}{8 \pi^2}
[
-2(\frac{3}{2}g_2^2|M_2|^2+\frac{3}{10}g_1^2|M_1|^2+\frac{3}{10}g_1^2S)
\nonumber \\
&+&3({\rm Tr}(Y_dm_Q^2{Y_d}^{\dagger})+{\rm Tr}(Y_dm_D^2{Y_d}^{\dagger})
+(m_{H_d}^2){\rm Tr}(Y_d{Y_d}^{\dagger})
+{\rm Tr}(\widetilde{A}_d{\widetilde{A}_d}^{\dagger}))
\nonumber \\
&+&({\rm Tr}(Y_em_L^2{Y_e}^{\dagger})+{\rm Tr}(Y_em_E^2{Y_e}^{\dagger}) 
+(m_{H_d}^2){\rm Tr}(Y_e{Y_e}^{\dagger})
+{\rm Tr}(\widetilde{A}_e{\widetilde{A}_e}^{\dagger}))
]
\nonumber \\
\label{softRGEs}
\end{eqnarray}
where
\begin{equation}
S=m_{H_u}^2-m_{H_d}^2+{\rm Tr}(m_Q^2-m_L^2-2m_U^2+m_D^2+m_E^2).
\end{equation}

The above RGEs have been presented in full generality within the MSSM.  
However, given the hierarchical form of the Yukawa matrices it is often 
useful to express the RGEs in terms of the leading third family couplings. 
To leading order, the Yukawa couplings (dropping $Y_{\nu}$) are then given 
by 
\begin{equation}
\label{thirdfamilyYs}
Y_u\approx\left( \begin{array}{ccc}
 0 & & \\
  & 0&  \\
 & &Y_t \end{array}\right),\; \;
Y_d\approx\left( \begin{array}{ccc}
 0 & & \\
  & 0& \\
 & &Y_b \end{array}\right),\; \;
Y_e\approx\left( \begin{array}{ccc}
 0 & & \\
  & 0& \\
 & &Y_{\tau} \end{array}\right). 
\end{equation}
The Yukawa RGEs for the third family couplings 
$Y_{t,b,\tau}$ can then be expressed as follows:
\begin{eqnarray}
\label{thirdfamilyYrges}
\frac{dY_t}{dt}&=& \frac{1}{16 \pi^2}Y_t[6|Y_t|^2+|Y_b|^2-(\frac{16}{3}g_3^2+ 
3g_2^2+\frac{13}{15}g_1^2)]\\
\frac{dY_b}{dt}&=& \frac{1}{16 \pi^2}Y_b[6|Y_b|^2+|Y_t|^2+|Y_{\tau}|^2- 
(\frac{16}{3}g_3^2+ 3g_2^2+\frac{7}{15}g_1^2)]\\
\frac{dY_{\tau}}{dt}&=& \frac{1}{16 \pi^2}Y_{\tau} [4|Y_{\tau}|^2+3|Y_b|^2- 
(3g_2^2+\frac{9}{5}g_1^2)],
\end{eqnarray}
and the RGE for the $\mu$ parameter is
\begin{equation}
\frac{d\mu}{dt}= \frac{1}{16 \pi^2}\mu [3|Y_t|^2+3|Y_b|^2+|Y_{\tau}|^2-
(3g_2^2+\frac{3}{5}g_1^2)].
\end{equation}

Similarly, one can assume that the $\widetilde{A}$ parameters have a similar 
hierarchical structure to the Yukawas:
\begin{equation}
\label{factorizedAs}
\widetilde{A}_u\approx\left( \begin{array}{ccc}
 0 & & \\
  & 0&  \\
 & &\widetilde{A}_t\equiv A_t Y_t \end{array}\right),\,
\widetilde{A}_d\approx\left( \begin{array}{ccc}
 0 & & \\
  & 0& \\
 & &\widetilde{A}_b\equiv A_b Y_b \end{array}\right),\, 
\widetilde{A}_e\approx\left( \begin{array}{ccc}
 0 & & \\
  & 0& \\
 & &\widetilde{A}_{\tau}\equiv A_{\tau} Y_{\tau} \end{array}\right). 
\end{equation}
The RGEs for $A_{t,b,\tau}$ are then given by
\begin{eqnarray}
\label{thirdfamilyArges}
\frac{dA_t}{dt}&=&\frac{1}{8\pi^2}[6|Y_t|^2A_t+ |Y_b|^2 A_b+ 
(\frac{16}{3}g_3^2M_3+ 3g_2^2M_2+\frac{13}{15}g_1^2M_1)]\\
\frac{dA_b}{dt}&=& \frac{1}{8\pi^2}[6|Y_b|^2A_b+ |Y_t|^2 A_t+ 
 |Y_{\tau}|^2A_{\tau}\nonumber \\
&+&(\frac{16}{3}g_3^2M_3+ 3g_2^2M_2+\frac{7}{15}g_1^2)M_1]\\
\frac{dA_{\tau}}{dt}&=& \frac{1}{8\pi^2}[4|Y_{\tau}|^2A_{\tau}+3
  |Y_b|^2A_b+
(3g_2^2M_2+\frac{9}{5}g_1^2M_1)],
\end{eqnarray}
and the RGE for $B\equiv b/\mu$ is 
\begin{equation}
\frac{dB}{dt}=\frac{1}{8 \pi^2}[3 |Y_t|^2A_t+3 |Y_b|^2A_b+ |Y_{\tau}|^2 
A_{\tau}+
(3g_2^2M_2+\frac{3}{5}g_1^2M_1)].
\end{equation}
Finally, let us consider the soft mass-squared parameters in this limit.  
If the soft mass-squares $m^2_{\alpha=Q,u,d,L,e}$ are flavor diagonal at a 
given (usually high) scale, at any scale they remain approximately 
diagonal with the first and second family entries nearly degenerate: 
\begin{equation}
\label{softmassapprox}
m^2_{\alpha}\approx \left( \begin{array}{ccc}
 m^2_{\alpha \,1} & & \\
  & m^2_{\alpha \,1}&  \\
 & &m^2_{\alpha \,3} \end{array}\right),
\end{equation}
with $m^2_{\alpha \,3}\neq m^2_{\alpha \,1}$.  This can be seen 
from the form of the RGEs for the first and second family entries in this 
limit:
\begin{equation}
\frac{dm^2_\alpha}{dt}=\frac{-1}{16\pi^2}\sum_{a=1,2,3} 8 
g_a^2C^{\alpha}_a|M_a|^2,
\end{equation}
in which the $C^{\alpha}_a$ are the quadratic Casimir invariants which 
occur in the corresponding anomalous dimensions in Eq.~(\ref{anomdim}). 
The RGEs for the third family entries and $m^2_{H_{u,d}}$ include 
nontrivial dependence on the third family Yukawas:
\begin{eqnarray}
\label{thirdfamilym2rges}
\frac{dm^2_{Q_3}}{dt} &=&\frac{1}{8 \pi^2}
[
\frac{1}{15}g_1^2|M_1|^2)  
(|Y_t|^2(m_{Q_3}^2+m_{U_3}^2+m_{H_u}^2+|A_t|^2)
+|Y_b|^2(m_{Q_3}^2+m_{D_3}^2+m_{H_d}^2+|A_b|^2))
\nonumber\\&-& 
(\frac{16}{3}g_3^2|M_3|^2+3g_2^2|M_2|^2] \\
\frac{dm^2_{U_3}}{dt} &=&\frac{1}{8\pi^2}
[
(2|Y_t|^2(m_{Q_3}^2+m_{U_3}^2+m_{H_u}^2+|A_t|^2)
(\frac{16}{3}g_3^2|M_3|^2+\frac{16}{15}g_1^2|M_1|^2)]
\\
\frac{dm^2_{D_3}}{dt} &=&\frac{1}{8\pi^2}
[(2|Y_b|^2(m_{Q_3}^2+m_{D_3}^2+m_{H_d}^2+|A_b|^2))]
-(\frac{16}{3}g_3^2|M_3|^2+\frac{4}{15}g_1^2|M_1|^2)
\\
\frac{dm^2_{L_3}}{dt} &=&\frac{1}{8\pi^2}
[(|Y_{\tau}|^2(m_{L_3}^2+m_{E_3}^2+m_{H_d}^2+|A_{\tau}|^2))
-(3g_2^2|M_2|^2+\frac{3}{10}g_1^2|M_1|^2)]\\
\frac{dm^2_{E_3}}{dt} &=&\frac{1}{8\pi^2}
[(2|Y_{\tau}|^2(m_{L_3}^2+m_{E_3}^2+m_{H_d}^2+|A_{\tau}|^2))
-\frac{12}{5}g_1^2|M_1|^2]\\
\frac{dm^2_{H_u}}{dt} &=&\frac{1}{8\pi^2}
[(3|Y_t|^2(m_{Q_3}^2+m_{U_3}^2+m_{H_u}^2+|A_t|^2)]
-(3g_2^2|M_2|^2+\frac{3}{5}g_1^2|M_1|^2)\\
\frac{dm^2_{H_d}}{dt} &=&\frac{1}{8 \pi^2}
[(3|Y_b|^2(m_{Q_3}^2+m_{D_3}^2+m_{H_d}^2+|A_b|^2)
+|Y_{\tau}|^2(m_{L_3}^2+m_{E_3}^2+m_{H_d}^2+|A_{\tau}|^2))
\nonumber\\&-&(3g_2^2|M_2|^2+\frac{3}{5}g_1^2|M_1|^2)].
\end{eqnarray}


\vfill\eject

\bigskip
\medskip

\bibliographystyle{unsrt}

\vfill\eject

\end{document}